\documentclass[prfluids,preprint,superscriptaddress]{revtex4-2}
\usepackage[utf8]{inputenc}
\usepackage{amsmath}
\usepackage{amsfonts}
\usepackage{amssymb}
\usepackage[usesnames,dvipsnames]{xcolor}
\usepackage{pifont}
\usepackage{indentfirst}
\usepackage{graphicx}
\usepackage{subcaption}
\usepackage{multirow}

\begin{document}

\title{
    Progress Report on Numerical Modeling of a Prototype Fuel Cell: 1.
    Comparison between Mathematical Formulations
}

\author{Otávio Beruski}
\email[]{oberuski@alumni.usp.br}
\affiliation{Instituto de Pesquisas Energéticas e Nucleares,
IPEN/CNEN-SP, 05508-000, São Paulo, São Paulo, Brazil}

\author{Ivan Korkischko}
\affiliation{Instituto de Pesquisas Energéticas e Nucleares,
IPEN/CNEN-SP, 05508-000, São Paulo, São Paulo, Brazil}

\author{Thiago Lopes}
\affiliation{Instituto de Pesquisas Energéticas e Nucleares,
IPEN/CNEN-SP, 05508-000, São Paulo, São Paulo, Brazil}

\author{Fabio Coral Fonseca}
\affiliation{Instituto de Pesquisas Energéticas e Nucleares,
IPEN/CNEN-SP, 05508-000, São Paulo, São Paulo, Brazil}

\begin{abstract}
    Progress on the numerical modeling of a prototype fuel cell is reported.
Some known limitations of the previously published Alpha model are addressed, 
and the numerical uncertainty due to discretization of the improved model, 
Beta, was estimated. In this part, the Beta model is compared to Alpha, where
significant albeit small differences are seen. One noteworthy difference lies
in the reactant usage,  where the Beta model shows a better fit to available
experimental data. Shortcomings of the improved model are discussed, paving the
way forward, while a discrepancy with previous results is addressed, further
suggesting the use of the Darcy-Brinkman over Stokes-Darcy formulation for free
and porous media flow. Furthermore, a parametric study is carried out,
constraining plausible values of the reaction rate constants to $10<k_1\approx
10^2<10^3\ \mathrm{s^{-1}}$, for adsorption, and $k_2\gtrapprox 10^3\ \mathrm{s
^{-1}}$ for decomposition reactions. Additional opportunities for validation
are identified, viz. the stoichiometry and reactant partial pressure profiles
along the catalyst layer. Nevertheless, given the uncertainties in the
numerical data and the available experimental data, the results lack validation
power, highlighting the need for additional experimental data and improved
precision for the numerical data.
\end{abstract}


\maketitle

\section{\label{sec:int}Introduction}

    Research in fuel cell modeling has come a long way since the early work
of Giner and Hunter\cite{giner69} and Cutlip\cite{cutlip75}. Constraints on
resources and numerical techniques allowed only highly simplified models with
low dimensionality and relatively coarse spatial and temporal resolution (see
for instance the overview in \cite{weber04}). The research field has now
reached a level of maturity where relatively robust models are routinely
coupled to empirical development, providing greater detail than most
experiments are able to achieve (\cite{bavarian10,hajimolana11,weber14} and
references therein). Fuel cells are known to harbor an awe-inspiring, perhaps
frightening, set of coupled non-linear physical processes, covering a wide
range of length and time scales\cite{andersson10,barbir13,weber14}. Thus,
despite the level of detail found in modern models, in both physical processes
and resolution, accurate models still prove too demanding in most situations.

    In order to deal with such complexity, not to mention the idiosyncrasies
of each device, it is common to approximate, or even neglect, some phenomena,
a notable case being two-phase flow in polymer electrolyte fuel cells (PEFCs).
In this case, it is common to work in conditions where two-phase flow is
minimized, such as high temperatures and stoichiometries (the ratio between
molar influx and usage for a given reactant), with numerical models then
ignoring the effects of two-phase flow\cite{pourmahmoud17,liu18,darling18,
chevalier19}. On the other hand, in order to deal with it, one strategy has
been to focus on the two-phase flow in porous media, using generalized Darcy's
law for each phase\cite{weber14,zenyuk16}, while another has been to use
multiphase mixture models\cite{weber14,shimpalee17,bednarek17}. Of course, more
complex models exist, for instance the ones compared in \cite{zhao19} for
porous media, and phase field, level set, and volume-of-fluid for fully
resolved two-phase flows. However these are in general transient-only models,
computationally intensive and usually do not consider phase transition,
although this point has developed rapidly\cite{pan19,yi19}.

    Another common source of approximations is the porous media, in particular
the catalyst layer (CL). Apart from the possibility of two-phase flow, which is
particular to some devices, the CL always show the complexities of dealing with
reacting flow in porous media\cite{andersson10,weber14}: Knudsen diffusivity is
usually important; the reaction mechanisms involves several steps and
intermediate species; the heterogeneous kinetics demands knowledge of the
so-called triple-phase boundary (where reactant, electrolyte and electrode meet
simultaneously) and its distribution within the porous medium; not to mention
the effects of the overarching porous structure on mass, charge and heat
transfer due to the distribution of pore radii and tortuosity of the pore
network. Research on porous media is of course not restricted to fuel cells,
and valuable input has been provided from other fields (see for instance
\cite{whitaker86,ochoa95a,ochoa95b} and more recently \cite{bakhshian16,
dalbe18,qiao19}).

    Considering the interrelated complexities of the inner working of fuel
cells, as briefly described above, a promising approach would be to investigate
relatively simpler systems, with fewer or uncoupled phenomena, in order to
acquire a firmer grasp on the common physical processes underlying both.
Significant efforts are directed towards such approach, widely attempted and
done all over the scientific community, however it is not always
straightforward to find a suitable prototype. Half-cells are just an example,
where part of the device is neglected in favor of a set of processes of
interest, commonly employed in basic electrochemistry research. For fuel cells,
half-cells are used to some extent, however they are not only somewhat
difficult to operate, they still exhibit most of the complexity of a full
device. In this context, a prototype half-cell system has been developed that
puts aside both charge and heat transfer\cite{lopes15}, focusing on single-phase
fluid flow and species transfer of a PEFC-like device. In a similar fashion, a
setup allowing direct visualization of reactant transport with liquid flow has
been developed for redox flow batteries\cite{garcia18}. Work on such prototypes
allows greater control and a clearer view of the processes underlying mass
transfer in fuel cells and similar devices, for both experimental\cite{lopes19}
and computational\cite{beruski17} research.

    We build on the previously published model for the aforementioned
prototype fuel cell cathode\cite{beruski17}, with the main goals of improving
the mathematical framework, and therefore its accuracy, and estimating the
discretization error, thus assessing its precision. In this
part\footnotemark[1], known limitations of the original model have been
addressed, such as the use of Fick's law for species diffusion and a
homogeneous reaction model. A comparison between formulations and to existing
experimental data is then carried out, and a parametric study on the reaction
rate constants was performed in order to assess the behavior of the response
variables as function of these degrees of freedom. Shortcomings of the new
model are addressed and discussed, for instance by considering the results of
Part 2 on the discretization error and validation, and bounds for reasonable
values for the reaction rate constants are drawn. In this way, this manuscript
is organized as follows. Section \ref{sec:met} provides an overview
of the original model's mathematical formulation and the improvements made,
as well as the details for the comparison between models and parametric study.
Section \ref{sec:res} first presents the comparison between model formulations,
followed by a discussion addressing shortcomings pointed out during the
development of the work. Afterwards, the parametric study on reaction rate
constants is presented, pointing out instances of validation that might be used
in future work and establishing the range of values that reasonably reproduce
the known experimental response. Finally, Section \ref{sec:sum} concludes with
a summary of the results and brief remarks on future studies.

\footnotetext[1]{A joint version of parts 1 and 2 can be found with DOI
arXiv:2002.04519.}

\section{\label{sec:met}Methods}

    We first present a brief description of the experimental device that is
simulated by our model, first described by Lopes \emph{et al.}\cite{lopes15},
followed by the computational methods used for the development of this work.

    The device is a prototype polymer electrolyte fuel cell (PEFC), more
specifically a cathodic half-cell, which allows \emph{in situ} observation of
reactive fluid flow in a porous medium. Ozone-enriched air is used as tracer in
a carrier fluid, and a coumarin-based dye is used as the sensor to map the
local concentration of $\mathrm{O_3}$. The $\mathrm{O_3}$ interacts with the dye
anchored on silica particles, akin to electrocatalysts dispersed in a porous
layer, resulting in the emission of photons and the degradation of both
reactant and dye. Measuring the light emission from this “catalyst layer”
allows determination of the local $\mathrm{O_3}$ concentration and, thus, the
species and fluid dynamics. Global variables, such as pressure drop and
total reactant drop, provide additional information on the inner workings of
the device.

\begin{figure}
    \centering
    \begin{subfigure}{0.45\textwidth}
        \includegraphics[width=\textwidth]{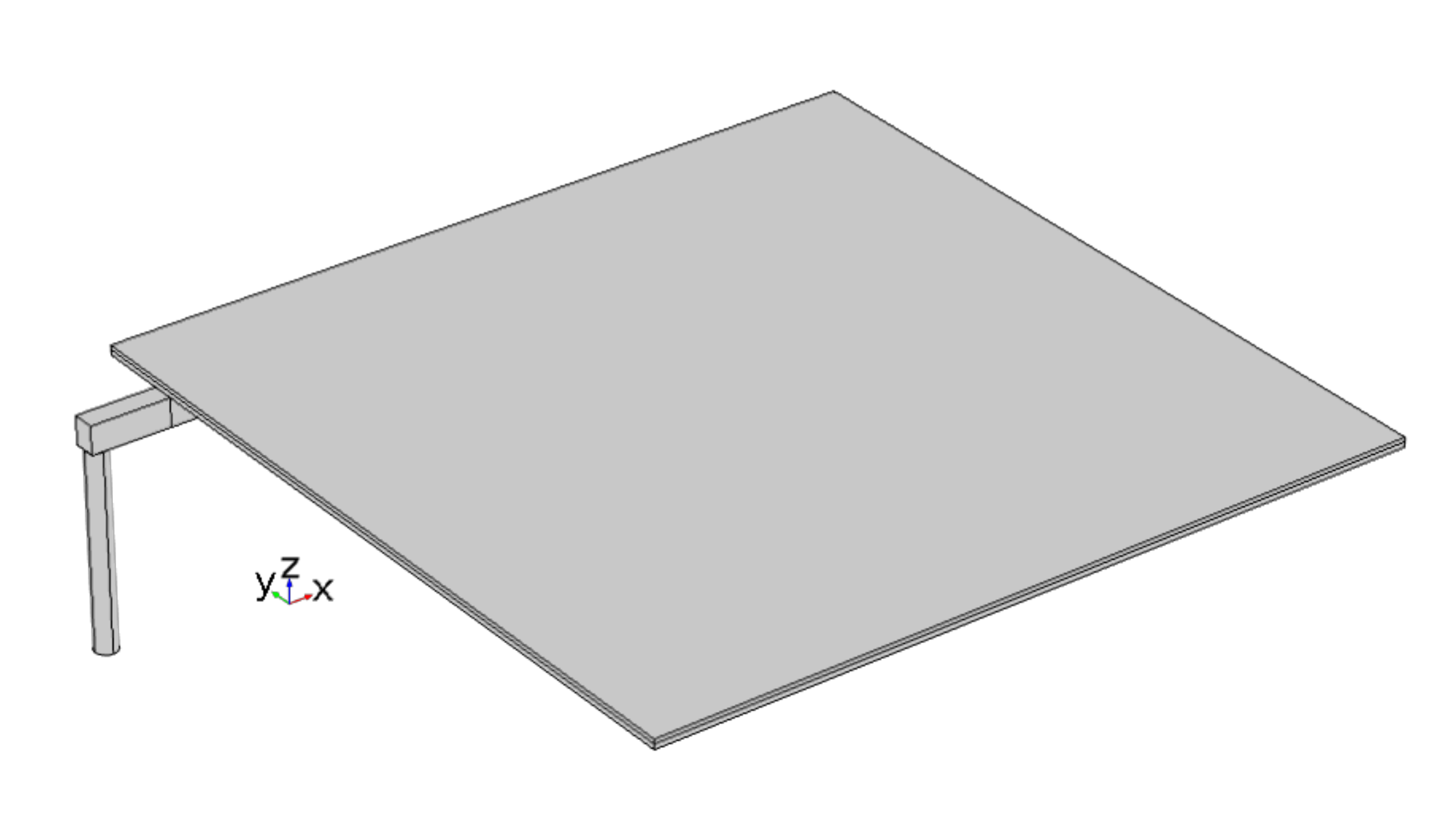}
        \caption{}
    \end{subfigure}
    ~
    \begin{subfigure}{0.45\textwidth}
        \includegraphics[width=0.8\textwidth]{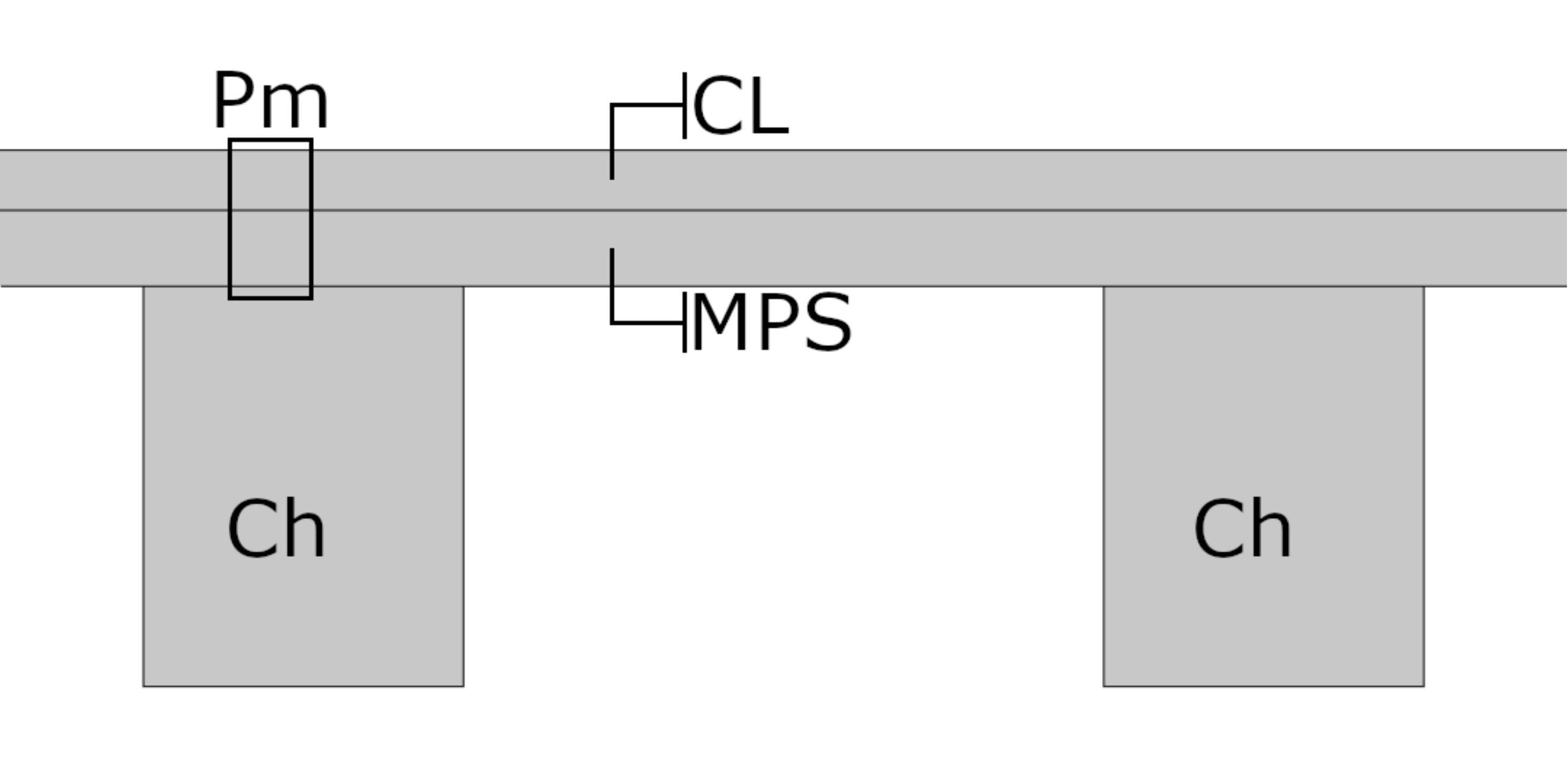}
        \caption{}
    \end{subfigure}

    \begin{subfigure}{0.45\textwidth}
        \includegraphics[width=\textwidth]{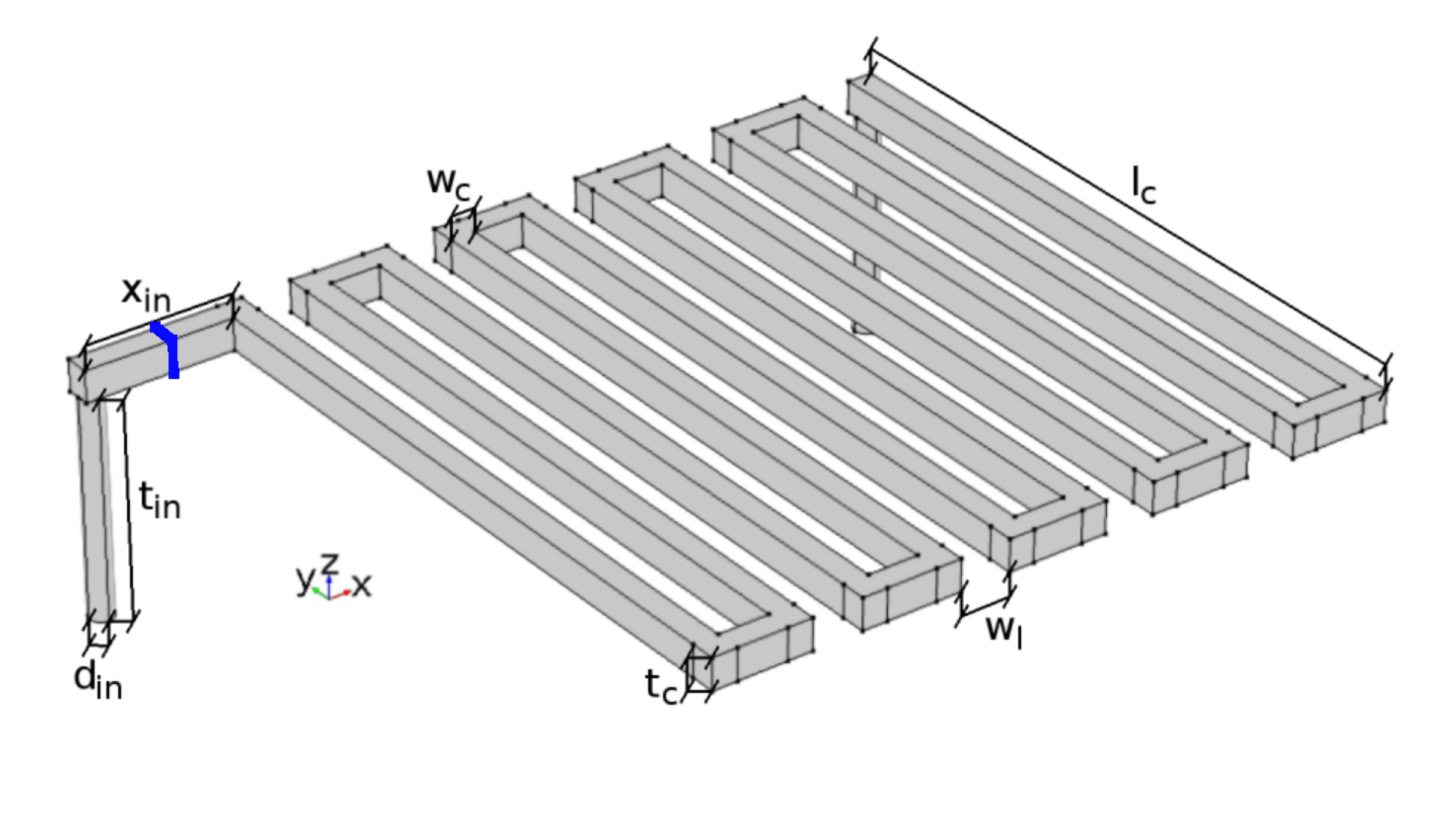}
        \caption{}
        \label{fig:geo_ch}
    \end{subfigure}
    ~
    \begin{subfigure}{0.45\textwidth}
        \includegraphics[width=0.8\textwidth]{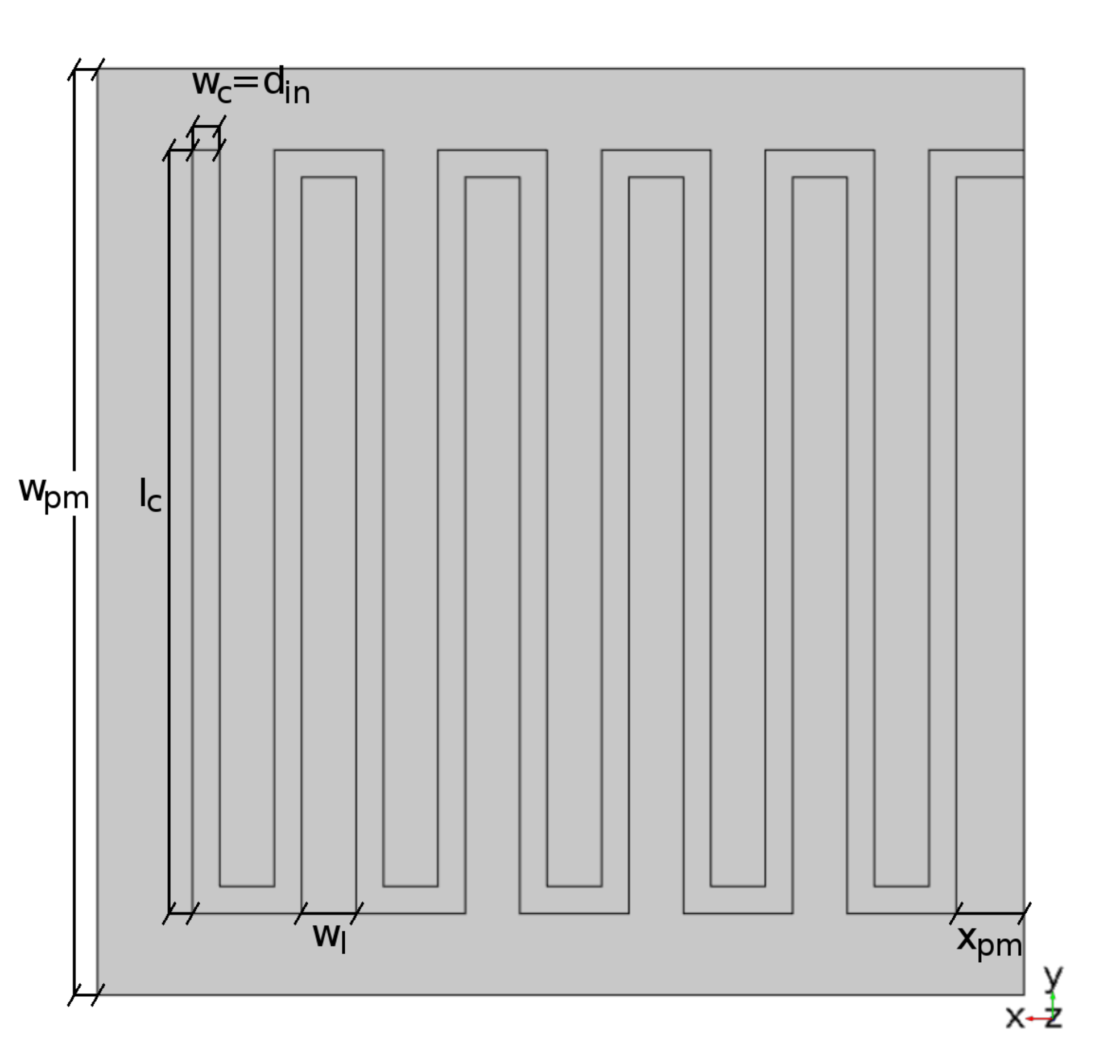}
        \caption{}
        \label{fig:geo_pm}
    \end{subfigure}
    \caption{
        Geometry used in the proto-cell simulations: \textbf{(a)} full geometry,
        showing the flow channel (Ch) and porous media (Pm) domains;
        \textbf{(b)} cross-section in the $xz$ plane, showing the structure
	along the $z$ axis; \textbf{(c)} Ch domain, with the segmentation edges
        shown in blue (see text), and \textbf{(d)} lower boundary of the MPS
        domain, showing the projection of the Ch domain.
    }
    \label{fig:geo}
\end{figure}

    Thus, considering the experimental device, the proposed computational model
covers momentum and species transport, while assuming thermal equilibrium and
steady state. The computational domains included are: i) the flow channel (Ch),
in this case a single serpentine geometry, ii) the porous transport layer, here
comprising only a macroporous substrate (MPS) made of carbon paper
(representing Toray TGH-060 with a 10\% PTFE hydrophobic treatment), and iii)
the catalyst layer (CL), where the catalyst and substrate particles are
considered homogeneously distributed (representing Sigma Aldrich Nano Silica
Gel on TLC plates, see \cite{lopes15} for details). Domains ii) and iii)
comprise the porous media domains (Pm). Figure \ref{fig:geo} shows the domains'
disposition and Table \ref{tab:geo} presents the geometrical parameters.

    The setup used for all simulations was a workstation with two
Intel$^\circledR$ Xeon$^\circledR$ E3 processors and 128GB of RAM, operated
with a 64 bits Debian9 distribution, Linux kernel v. 4.9.0-4. The software used
was the commercial package COMSOL Multiphysics$^\circledR$, v.5.1.0.234, along
with the Batteries and Fuel Cell, CFD, and Chemical Engineering modules. All
data handling and processing was done using GNU Octave v.4.2.1\cite{octave},
while image processing was done using GIMP v.2.10.14\footnotemark[2].

\footnotetext[2]{Available at https://www.gimp.org/.}

\begin{table}[h]
    \centering
    \caption{Geometrical parameters of the computational domains.}
    \label{tab:geo}
    \begin{tabular}{c c c}
        \hline
        \textbf{Parameter} & \textbf{Symbol} & \textbf{Value} \\
        \hline
        Channel width & $w_{\mathrm{c}}$ & $0.8\ \mathrm{mm}$\\
        Land width & $w_{\mathrm{l}}$ & $1.6\ \mathrm{mm}$\\
        Channel depth & $t_{\mathrm{c}}$ & $1.0\ \mathrm{mm}$\\
        Channel section length & $l_{\mathrm{c}}$ & $22.4\ \mathrm{mm}$\\
        Inlet/outlet diameter & $d_{\mathrm{in}}$ & $0.8\ \mathrm{mm}$\\
        Inlet/outlet length & $t_{\mathrm{in}}$ & $7.0\ \mathrm{mm}$\\
        Inlet offset & $x_{\mathrm{in}}$ & $4.7752\ \mathrm{mm}$\\
        Porous media edge & $w_{\mathrm{pm}}$ & $(l_{\mathrm{c}}+x_{\mathrm{in}})$\\
	    Porous media offset & $x_{\mathrm{pm}}$ & $(x_{\mathrm{in}}-w_{\mathrm{w}})/2$\\
        MPS thickness & $t_{\mathrm{MPS}}$ & $190\ \mathrm{\mu m}$\\
        CL thickness & $t_{\mathrm{CL}}$ & $150\ \mathrm{\mu m}$\\
        \hline
    \end{tabular}
\end{table}

\subsection{\label{ssec:math}Mathematical Formulation}

\subsubsection{\label{sssec:math_old}Original model: Alpha}

    The original mathematical formulation, henceforth denominated \emph{Alpha},
can be found in \cite{beruski17}, but a brief description will be given below
followed by the improvements which are the focus of this work.

    The momentum transport is described using the (compressible) Darcy-Brinkman
formulation (DB), which covers both free and porous media flow\cite{lebars06},
coupled to conservation of mass:
\small
\begin{eqnarray}
    \frac{\rho}{\epsilon}\left(\mathbf{u}\cdot\nabla\right)\left(\frac{\mathbf{u}}{\epsilon}\right) & = & \nabla\left[-P\mathbf{I} + \frac{\mu}{\epsilon}\left(\nabla\mathbf{u}+\left(\nabla\mathbf{u}\right)^{\mathrm{T}}\right) - \frac{2\mu}{3\epsilon}\left(\nabla\cdot\mathbf{u}\right)\mathbf{I}\right] - \frac{\mu}{\kappa}\mathbf{u}, \label{eq:db} \\
    \nabla\cdot\left(\rho\mathbf{u}\right) & = & 0, \label{eq:c_of_m}
\end{eqnarray}
\normalsize
\noindent
where $\rho$ and $\mu$ are the fluid's density and kinematic viscosity,
respectively, $\mathbf{u}$ is the velocity field vector, $P$ is the relative
pressure, and $\epsilon$ and $\kappa$ are the domain's porosity and
permeability, respectively ($\epsilon = 1$ and $\kappa \rightarrow \infty$ for
free flow). As shown in \cite{beruski17}, DB provides a better description of
this system when compared to a Stokes-Darcy approach, i.e. manually coupling
the Navier-Stokes equation for the Ch domain and Darcy's law for porous media.
The parameters are shown in Table \ref{tab:par}, chosen in order to describe
the materials and conditions used (see \cite{lopes15} for additional
information). A normal inflow velocity, as inlet, and a constant pressure
boundary conditions, for the outlet, were used.

    The species transport was originally modeled using a diluted-species
approach, coupling the advection-reaction equation:
\begin{eqnarray}
    \nabla\cdot\mathbf{J}_{\mathrm{O_3}} + \left(\mathbf{u}\cdot\nabla\right)C_{\mathrm{O_3}} & = & R_{\mathrm{O_3}},
\end{eqnarray}
\noindent
to Fick's law of diffusion, which defines the molar diffusive flux vector:
\begin{eqnarray}
    \mathbf{J}_{\mathrm{O_3}} & = & -D_{\mathrm{O_3}}\nabla C_{\mathrm{O_3}}, \label{eq:fl}
\end{eqnarray}
\noindent
where $D_{\mathrm{O_3}}$ and $C_{\mathrm{O_3}}$ are the diffusion coefficient
and the concentration of ozone, respectively. Given the inlet concentration of
the species of interest, $C_\mathrm{O_3,in}$, of the order $10^{3}\
\mathrm{ppm}$, such an approach was considered reasonable\cite{beruski17}. The
diffusion was corrected for porous media transport:
\begin{eqnarray}
    D^{\mathrm{Pm}}_{\mathrm{O_3}} & = & f_{\mathrm{Pm}}D_{\mathrm{O_3}}, \label{eq:Dpm}
\end{eqnarray}
\noindent
where $f_{\mathrm{Pm}} = f_{\mathrm{Pm}}\left(\epsilon,\tau\right)$, with
$\tau$ being the medium's tortuosity. A common form for $f_{\mathrm{Pm}}$ is:
\begin{eqnarray}
    f_{\mathrm{Pm}} & = & \frac{\epsilon}{\tau},
\end{eqnarray}
\noindent
which might be readily used whenever $\tau$ is available, as is the case of the
MPS domain. Otherwise, a porous medium model might be used, typically with the
form $\tau = \tau\left(\epsilon \right)$. In this case, for the CL domain, the
Millington \& Quirk model was used\cite{quirk61}, as implemented in the
software, giving $\tau = \epsilon^{1/2}$. The reaction term was defined by a
homogeneous first-order reaction:
\begin{eqnarray}
    R_{\mathrm{O_3}} & = & -k_{\mathrm{app}}C_{\mathrm{O_3}}, \label{eq:sink_alpha}
\end{eqnarray}
\noindent
where the apparent reaction rate constant, $k_{\mathrm{app}}$, is a free
parameter to be adjusted against experimental data. Parameters are shown in
Table \ref{tab:par}.

\begin{table}[h]
    \centering
    \caption{Parameterization of both models.}
    \label{tab:par}
    \begin{tabular}{c c c c}
        \hline
        \textbf{Parameter} & \textbf{Symbol} & \textbf{Value} & \textbf{Reference}\\
        \hline
        Inlet concentration of $\mathrm{O_3}$ & $C_{\mathrm{O_3,in}}$ & $1200\times 10^{-6}\ \mathrm{mol\ m^{-3}}$ & \cite{lopes15,beruski17} \\
        Reference pressure & $P_{\mathrm{ref}}$ & $1.027\ \mathrm{bar}$ & \cite{beruski17} \\
        Cell back pressure & $P_{\mathrm{out}}$ & $1.0994\ \mathrm{bar} - P_{\mathrm{ref}}$ & \cite{beruski17} \\
        Carbon paper porosity & $\epsilon_{\mathrm{MPS}}$ & $0.801$ & \cite{fishman10} \\
        Carbon paper permeability & $\kappa_{\mathrm{MPS}}$ & $9.18\times 10^{-12}\ \mathrm{m}^2$ & \cite{fishman11} \\
        Carbon paper tortuosity & $\tau_{\mathrm{MPS}}$ & $1.199$ & \cite{fishman11} \\
        Catalyst layer porosity & $\epsilon_{\mathrm{CL}}$ & $0.497$ & \cite{beruski17} \\
        Catalyst layer permeability & $\kappa_{\mathrm{CL}}$ & $8.82\times 10^{-11}\ \mathrm{m}^2$ & \cite{beruski17} \\
        $\mathrm{O_3}$ diffusion coefficient in N$_2$ & $D_{O_3}$ & $0.16\ \mathrm{cm}^2\mathrm{s}^{-1}$ & \cite{ono04,massman98} \\
        $\mathrm{O_3}$ apparent reaction rate constant\footnotemark[1] & $k_{\mathrm{app}}$ & $256.15\ \mathrm{s}^{-1}$ & \cite{beruski17} \\
        \hline
    \end{tabular}
\footnotetext[1]{Parameter used only in the original (Alpha) model.}
\end{table}

    Finally, for the sake of reference, the mesh used in \cite{beruski17}
followed a custom-made procedure, based on the software's recommendations, with
ca. $1.7\times 10^6$ domain elements. The set of equations was solved using a
two-step segregated solver, with each step using an implementation of the
PARDISO\cite{pardiso} linear solver, to a relative tolerance of $10^{-3}$.
A parametric solver was used to vary the inlet flow rate, with initial value
$Q = 200\ \mathrm{cm^3\ min^{-1}}$ and a step of $\delta Q = 50\ \mathrm{cm^3\
min^{-1}}$ up to $450\ \mathrm{cm^3\ min^{-1}}$, where each solution was used
as the next step's initial value.

\subsubsection{\label{sssec:math_new}Improved model: Beta}

    The improvements were largely implemented in the species transport
formulation, with a closely related addition of surface cover kinetics, with two
relatively minor changes in momentum transport and one in the model geometry.
These modifications make up the \emph{Beta} model. Additional parameters
relevant to the Beta model are given in Table \ref{tab:par2}.

\begin{table}[h]
    \centering
    \caption{Parameters relevant to the improved (Beta) model.}
    \label{tab:par2}
    \begin{tabular}{c c c c}
        \hline
        \textbf{Parameter} & \textbf{Symbol} &\textbf{Value} & \textbf{Reference}\\
        \hline
        Inlet molar fraction of $\mathrm{O_3}$ & $\chi_{\mathrm{O_3,in}}$ & $1200\ \mathrm{ppm}$ & \cite{lopes15} \\
        Average molar mass of dry air & $M_{\mathrm{air}}$ & $28.96546\ \mathrm{g\ mol^{-1}}$ & \cite{nist08}\\
        $\mathrm{O_3}$ adsorption reaction rate constant & $k_1$ & $100\ \mathrm{s^{-1}}$ & see text \\
        $\mathrm{O_3(ads)}$ decomposition reaction rate constant & $k_2$ & $10\ \mathrm{s^{-1}}$ & see text \\
        Average silica particle radius & $r_p$ & $6.5\ \mathrm{\mu m}$ & see text \\
        Quantity of dye deposited on the CL & $\Gamma_{\mathrm{dye}}$ & $3\ \mathrm{\mu mol\ cm^{-2}}$ & \cite{lopes15} \\
        \hline
    \end{tabular}
\end{table}

    The main change in the model was the use of a concentrated-species approach
to the fluid, which uses the following statement of species conservation, i.e.
the advection-reaction equation:
\begin{eqnarray}
    \nabla\cdot\mathbf{J}_i + \rho\left(\mathbf{u}\cdot\nabla\right)\omega_i & = & R_i, \label{eq:mf_ar}
\end{eqnarray}
\noindent
where $\omega_i$ is the mass fraction of the $i$-th species, while the
\emph{mass} diffusive flux is given by the Maxwell-Stefan (MS) model:
\begin{eqnarray}
    \mathbf{J}_i & = & -\left(\rho\omega_i\sum_k D_{ik}\mathbf{d}_k\right) \label{eq:ms} \\
    \mathbf{d}_k & = & \nabla\chi_k + \frac{\left[\left(\chi_k - \omega_k\right)\nabla P_A\right]}{P_A},
\end{eqnarray}
\noindent
where $D_{ik}$ are multi-component diffusivities, $\mathbf{d}_k$ is a so-called
diffusional driving force, $\chi_k$ is the molar fraction and $P_A$ is the
absolute pressure. According to the User Guide for COMSOL's Batteries and Fuel
Cells module, in the version used the $D_{ik}$ are multi-component Fick
diffusivities, which are obtained from the MS diffusion matrix by solving the
following relation:
\begin{eqnarray}
    \frac{\chi_i\chi_k}{D^{\mathrm{MS}}_{ik}} & = & -\omega_i\omega_k\frac{\sum_{j\neq i} \left(\mathrm{adj} B_i\right)_{jk}}{\sum_{j\neq i}D_{ij}\left(\mathrm{adj} B_i\right)_{jk}},
\end{eqnarray}
\noindent
where $D^{\mathrm{MS}}_{ik}$ are the binary MS diffusion coefficients,
$\left(B_i\right)_{kj} = D_{kj}-D_{ij}$, for $i\neq j$, and $\mathrm{adj}$
stands for the adjoint operation $\mathrm{adj}(A_{ij}) = A^*_{ji}$,
where $^*$ is the complex conjugate. Both the MS and multi-component Fick
diffusion matrices are symmetric. Furthermore, for two- and three-component
mixtures, the software has analytical expressions for $D_{ik}$ implemented,
while for four components or more a numerical procedure is used. In addition
to multi-component diffusion, the Knudsen regime was implemented for porous
media transport:
\begin{eqnarray}
    D^{\mathrm{MS}}_{ij} = \left[\frac{1}{D_{ij}^{\mathrm{Pm}}} + \frac{1}{D_{ij}^K}\right]^{-1}. \label{eq:dcs}
\end{eqnarray}
\noindent
Here two effects are added in a parallel-resistances fashion: i) the porous
media diffusion coefficient, $D_{ij}^{\mathrm{Pm}}$, which follows Eq.
\ref{eq:Dpm}; and ii) the Knudsen regime diffusivity, obtained through kinetic
theory:
\begin{eqnarray}
    D_i^K & = & \frac{d_p}{3}\sqrt{\frac{8RT}{\pi M_i}}, \label{eq:Kdiff}
\end{eqnarray}
\noindent
where $d_p$ is the average pore diameter, commonly used instead of the mean
free path in porous media, $R$ is the gas constant, $T$ is the temperature
and $M_i$ is the molar mass. In this case, the average pore diameter is
obtained using the Millington \& Quirk model for cemented particle
beds\cite{quirk61} for simplicity. It will be noticed that Eq. \ref{eq:Kdiff}
is written for a single species, and it is not clear how a binary Knudsen
diffusion coefficient should be expressed. One option would be to calculate an
average molar mass between a pair of species, weighed by their relative molar
fractions. Another would be to sum the individual coefficients in a similar
fashion as Eq. \ref{eq:dcs}, weighed by their relative molar fractions. While
this may be worthwhile to investigate, it is neglected at this point, as
discussed below.

    Considering the formulation described, a few issues should be addressed.
First, by switching to the MS model, the $D_{\mathrm{O_3}}$ used in Eq.
\ref{eq:fl} is no longer valid, being necessary appropriate MS binary
diffusivities instead. Due to the difficulty of obtaining these coefficients,
and given that the concentration of ozone is small, whereas that of $\mathrm{N
_2}$ and $\mathrm{O_2}$ are approximately constant, the value given in Table
\ref{tab:par} is still used. Second, given the constant relative concentrations
of $\mathrm{N_2}$ and $\mathrm{O_2}$, and to minimize computational costs, the
components of the mixture were $\mathrm{O_3}$ and ``air'', the latter with a
molar mass of $M_{\mathrm{air}}$ (see Table \ref{tab:par2}). A brief discussion
on this issue will be given in Section \ref{sssec:disc_AxB}. See also Section
SII of the Supplemental Material\footnotemark[3] (SM) for a comparison between
results using $\mathrm{O_3/O_2/ N_2}$ or $\mathrm{O_3/air}$ mixtures.

\footnotetext[3]{Supplemental Material is available at
https://arxiv.org/abs/2002.04519.}

    Finally, it is not clear how the different phenomena affecting diffusion
couple together, such as shown in Eq. \ref{eq:dcs}. For instance, whether it is
appropriate to correct the free diffusion coefficient before calculating the
multi-component Fick diffusivities, or, given the formulation and software
limitations, how to correctly couple the individual Knudsen diffusivities to
the Maxwell-Stefan diffusivities. The approach used here relies on the
approximation of the constant $\mathrm{N_2/O_2}$ mixture as ``air'', which
results in a single entry in the MS diffusion matrix, then assuming ozone as
being the sole contribution to the term ($M_i = M_{\mathrm{O_3}}$). It is also
understood that the porous media affect species transport despite the
contribution of the Knudsen regime, thus the application of the correction
factor in Eq. \ref{eq:dcs}. Section \ref{sssec:disc_AxB} provides some
discussion around this issue. Section SIII of the SM\footnotemark[3] provides a
comparison with a few possibilities regarding this issue, in order to
illustrate the expected effects in the system under study.

    Back to the mathematical formulation, the reaction involving ozone
degradation, and hence light emission, was modified to include the interaction
with the dye molecule in an adsorption-desorption step, then leading to
decomposition:
\begin{eqnarray}
    \mathrm{O_3(g)} + \mathrm{dye} & \overleftarrow{\rightarrow} & \mathrm{O_3(ads)} \\
    \mathrm{O_3(ads)} & \rightarrow & \mathrm{products} + h\nu,
\end{eqnarray}
\noindent
where the dye molecule is 7-diethylamino-4-methylcoumarin\cite{lopes15}. In
this way, a new variable was defined, $\theta_{\mathrm{O_3}}$, representing the
surface coverage of ozone, with kinetics modeled by the differential equation:
\begin{eqnarray}
    \frac{\mathrm{d}\theta_{\mathrm{O_3}}}{\mathrm{d}t} & = & k_1 c_{\mathrm{O_3}}\chi_{\mathrm{O_3}}\left(1-\theta_{\mathrm{O_3}}\right) - k_{-1}\theta_{\mathrm{O_3}} - k_2\theta_{\mathrm{O_3}}, \label{eq:hk}
\end{eqnarray}
\noindent
where $c_{\mathrm{O_3}}$ is the ratio of the appropriate activity coefficients
for bulk and adsorbed $\mathrm{O_3}$, $k_1$ and $k_{-1}$ are the forward and
backward reaction rate constants for the adsorption-desorption reaction,
respectively, and $k_2$ is the reaction rate constant for the decomposition
reaction. Eq. \ref{eq:hk} can be written as function of alternative variables
concerning $\mathrm{O_3}$, such as mass fraction or density, with the
appropriate factors incorporated in the reaction rate constant. To couple with
Eq. \ref{eq:mf_ar}, one may write:
\begin{eqnarray}
    A_v\Gamma^*_s\frac{\mathrm{d}\theta_{\mathrm{O_3}}}{\mathrm{d}t} & = & R_{\mathrm{O_3,ads}}, \label{eq:sink_ads}
\end{eqnarray}
\noindent
which corresponds to the total reaction rate for the adsorbed ozone over the
CL domain, in $\mathrm{mol\ s^{-1}}$, given that $A_v$ is the total surface
area by unit volume and $\Gamma^*_s$ is the total surface concentration of
active sites, i.e. dye molecules (see below). Thus, by excluding the last term
of Eq. \ref{eq:hk}, one may write the sink term for bulk ozone as:
\begin{eqnarray}
    R_{\mathrm{O_3}} & = & -M_{\mathrm{O_3}}R_{\mathrm{O_3,ads}} - M_{\mathrm{O_3}}A_v\Gamma^*_s k_2\theta_{\mathrm{O_3}}
\end{eqnarray}
\noindent
since the source/sink term in Eq. \ref{eq:mf_ar} is given in units of
$\mathrm{kg\ m^{-3}\ s^{-1}}$. For simplicity, the desorption term was
neglected, thus leaving the sink term as:
\begin{eqnarray}
    R_{\mathrm{O_3}} & = & -M_{\mathrm{O_3}}A_v\Gamma^*_s k_1 \chi_{\mathrm{O_3}}\left(1-\theta_{\mathrm{O_3}}\right), \label{eq:sink}
\end{eqnarray}
\noindent
where it was further assumed that $c_{\mathrm{O_3}} \approx 1$. The total
surface concentration of adsorption sites can be obtained with:
\begin{eqnarray}
    \Gamma_s^* & = & \frac{\Gamma_{\mathrm{dye}}}{t_{\mathrm{CL}}A_v},
\end{eqnarray}
\noindent
where $\Gamma_{\mathrm{dye}}$ is the reported\cite{lopes15} quantity of dye
deposited over a given geometric area of CL with thickness $t_{\mathrm{CL}}$.
The $A_v$ can, in principle, be determined experimentally, but for now it is
calculated using a simple model:
\begin{eqnarray}
    A_v & = & \frac{3\epsilon_{\mathrm{CL}}}{r_p},
\end{eqnarray}
\noindent
where $r_p$ is the average particle radius, in this case the silica particles
where the dye were anchored on (Nano Silica Gel, Sigma Aldrich).

    Regarding the minor changes in momentum transport, the first concerns the
inclusion of a mass source term to Eqs. \ref{eq:db} and \ref{eq:c_of_m}
accounting for the reaction in the CL domain:
\begin{eqnarray}
    S_M & = & R_{\mathrm{O_3}}. \label{eq:mt_sink}
\end{eqnarray}
\noindent
The momentum formulation written for the Beta model is thus written as:
\small
\begin{eqnarray}
    \frac{\rho}{\epsilon}\left(\mathbf{u}\cdot\nabla\right)\left(\frac{\mathbf{u}}{\epsilon}\right) & = & \nabla\left[-P\mathbf{I} + \frac{\mu}{\epsilon}\left(\nabla\mathbf{u}+\left(\nabla\mathbf{u}\right)^{\mathrm{T}}\right) - \frac{2\mu}{3\epsilon}\left(\nabla\cdot\mathbf{u}\right)\mathbf{I}\right] + \left(\frac{S_M}{\epsilon^2} - \frac{\mu}{\kappa}\right)\mathbf{u} \\
    \nabla\cdot\left(\rho\mathbf{u}\right) & = & S_M,
\end{eqnarray}
\normalsize
\noindent
where $S_M$ is defined by Eq. \ref{eq:mt_sink} in the CL domain only, being
zero otherwise. While the inclusion of a source term may be considered a major
change theoretically, it is deemed minor in this case since the magnitude of
$S_M$ is expected to be quite small when compared to the remaining terms in the
DB formulation. The second change refers to the inlet boundary condition, where
a (standard) mass flow rate condition is now being used:
\begin{eqnarray}
    -\int_{\partial\Omega}\frac{\rho}{\rho_{\mathrm{std}}}\left(\mathbf{u}\cdot\mathbf{n}\right)\mathrm{d}S & = & Q,
\end{eqnarray}
\noindent
with dry air at $T_{\mathrm{std}}=273.15\ \mathrm{K}$ and $P_{\mathrm{std}}=1\
\mathrm{atm}$ defining the standard density $\rho_{\mathrm{std}} =
P_{\mathrm{std}}M_{\mathrm{air}}/RT_\mathrm{std}$. Finally, given the use of a
concentrated-species approach to species transport, the fluid composition is
expected to play a role in its properties. However, for simplicity, the density
is still given by the ideal gas law, while the viscosity is given by a
constitutive relation for air, provided by the software:
\begin{eqnarray}
    \mu_{\mathrm{air}}\left(T\right) & = & -8.38278\times 10^{-7} + 8.35717342\times 10^{-8}T  - 7.69429583\times 10^{-11}T^2 \nonumber\\
    & & + 4.6437266\times 10^{-14}T^3 - 1.06585607\times 10^{-17}T^4,
\end{eqnarray}
\noindent
where $\mu_{\mathrm{air}}$ has units of $\mathrm{Pa\ s}$ and $T$ of
$\mathrm{K}$.

    The final minor change is related to geometry of the domains. The full
geometry was an assembly of two distinct geometrical entities: the Ch and the
Pm domains. The Ch domain follows the description in \cite{beruski17}, with the
following modifications: i) the inlet and outlet sections where extended to
fully cover the supporting plates existing in the experimental prototype ($7\
\mathrm{mm}$)\cite{lopes15}, and ii) the upper boundary, partly in contact with
the MPS, was segmented in two, corresponding to the section in contact with the
MPS and the remaining part, closer to the inlet. The Pm domains are simply
parallelepipeds, with the addition of a copy of the Ch upper boundary at the
lower boundary of the MPS, representing the part of the boundary in contact
with the Ch domain. Figures \ref{fig:geo_ch} and \ref{fig:geo_pm} illustrates
the above points. As a consequence of these changes, explicit coupling between
the geometric entities was needed to ensure the proper continuity of the
variables, i.e. the  flow field, $\mathbf{u}$, and ozone mass fraction, $\omega
_{\mathrm{O_3}}$. This was accomplished via an identity pair. Since the
relevant boundaries were sectioned to properly match one another (at the Ch
and MPS domains), no fallback features were necessary.

    The mesh and solver schemes are described in more detail in Part 2, as
well as the experimentation and mesh convergence study performed for this
model. In this part of the work, three meshes  were used with decreasing
spatial resolution, termed meshes \#1 to \#3 ($\sigma = 12,6, 4$), with the
following outline:
\begin{enumerate}
    \item Meshing of the Ch domain with tetrahedral elements, with scaling
    factor $\sigma$.
    \item Copying the upper boundary of the Ch mesh to the projection onto the
    lower MPS boundary.
    \item Meshing the remainder of the lower MPS boundary with triangular
    elements, with scaling factor $1$.
    \item Extruding the lower boundary elements throughout the Pm domains,
    with $\sigma$ elements in each domain.
\end{enumerate}
\noindent
The final solution, shown in Section \ref{sec:res}, used mesh \#1, with ca.
$8.86 \times 10^6$ domain elements. The solver scheme was a two-step solution
using meshes \#2 and \#1, with \#2 at inlet flow rate $Q = 200\ \mathrm{cm^3\
min^{ -1}}$ as the initial values for mesh \#1. The general outline for the
solver for each step was the following: a segregated 3-steps solver was
employed, with step i) solving for $\theta_{\mathrm{O_3}}$, step ii) solving for
$\mathbf{u}$ and $P$, and step iii) solving for $\omega_{\mathrm{O_3}}$. Step
i) employed a direct solver based on MUMPS\cite{MUMPS:1,MUMPS:2}, step ii)
employed an iterative solver based on GMRES\cite{saad86} with a geometric
multigrid scheme as pre-conditioner, and step iii) employed the same
GMRES-based iterative solver however using the domain decomposition
method\cite{toselli05}. For the first solution, using mesh \#2, mesh \#3 was
used as the coarse multigrid level, while for the second solution, with mesh
\#1, both meshes \#2 and \#3 were used. Finally, for the second step, a
parametric solver was employed for $Q$, with $200\leq Q\leq 450\ \mathrm{cm^3\
min^{-1}}$ and a step of $\delta Q = 50\ \mathrm{cm^3\ min^{-1}}$.

\subsection{\label{ssec:comp}Model Comparison}

    The models described in Section \ref{ssec:math} were compared to each
other. Similarly to the mesh convergence analysis described in Part 2,
the response variables were chosen for compatibility with the experimental
device, envisioning future validation rounds against empirical data. The chosen
quantities are the following:
\begin{itemize}
    \item the ratio $K'$ between reactant consumption $\Delta\chi_{\mathrm{O_3}}$
    and apparent reaction rate $R'_{\mathrm{O_3}}$;
    \item the stoichiometry $\lambda$ and apparent stoichiometry $\lambda'$;
    \item ozone partial pressure and normalized reaction rate profiles,
    $P_{\mathrm{O_3}}(x)$ and $\bar{R}_{\mathrm{O_3}}(x)$ respectively, obtained
    at the upper surface of the CL, along the $x$ axis;
    \item ozone partial pressure and normalized reaction rate surfaces,
    $\mathbf{P}_{\mathrm{O_3}}$ and $\mathbf{\bar{R}}_{\mathrm{O_3}}$
    respectively, obtained at the upper surface of the CL domain.
\end{itemize}
\noindent
Since the models differ mainly in the species transport formulations, in
particular concerning the reaction kinetics, the ratio $K'$, stoichiometries
$\lambda$ and $\lambda'$, and $\bar{R}_{\mathrm{O_3}}$ profiles and surfaces
were primarily used for comparison. The individual variables, $\Delta\chi
_{\mathrm{O_3}}$ and $R'_{\mathrm{O_3}}$, as well as $P_{\mathrm{O_3}}$ profiles
and surfaces, were also compared for the sake of completeness, however the
differences in formulation and parameterization should be taken in account when
analyzing the results. Variables solely related to the flow field, such as
pressure drop and flow speed profile, were not expected to show significant
changes and are shown in Section SIV.A of the SM\footnotemark[3].

    Another consequence of the difference in species transport formulation is
that $C_{\mathrm{O_3}}$ is not the same for both models. This is because the
Alpha model uses a diluted species approach that does not considers the solvent
explicitly, thus molar fraction is not an available variable. However, since
the excess of solvent with respect to the solute does not affect the results,
using an inlet concentration of $C^{(\alpha)}_{\mathrm{O_3,in}} = 1200\times
10^{-6}\ \mathrm{mol\ m^{-3}}$ is essentially the same as saying that there is
$1200\times 10^{-6}$ moles of $\mathrm{O_3}$ for each mole of solvent. Thus, in
general, one may use $\chi^{(\alpha)}_{\mathrm{O_3}} = C^{(\alpha)}
_{\mathrm{O_3}}/C^*$, where $C^*=1\ \mathrm{mol\ m^{-3}}$. While the work
reported in \cite{beruski17} corroborates this approach, the Alpha and Beta
models are not directly comparable even when disregarding formulation
differences, where the latter explicitly uses $\chi^{(\beta)}_{\mathrm{O_3,in}}
= 1200\ \mathrm{ppm}$. Therefore, variables which depend on molar
concentrations, notably $R'_{\mathrm{O_3}}$ and consequently $\Delta\chi
_{\mathrm{O_3}}/R'_{\mathrm{O_3}}$, have to be processed before comparing them.

\begin{figure}
    \centering
    \begin{subfigure}{0.4\textwidth}
        \includegraphics[width=\textwidth]{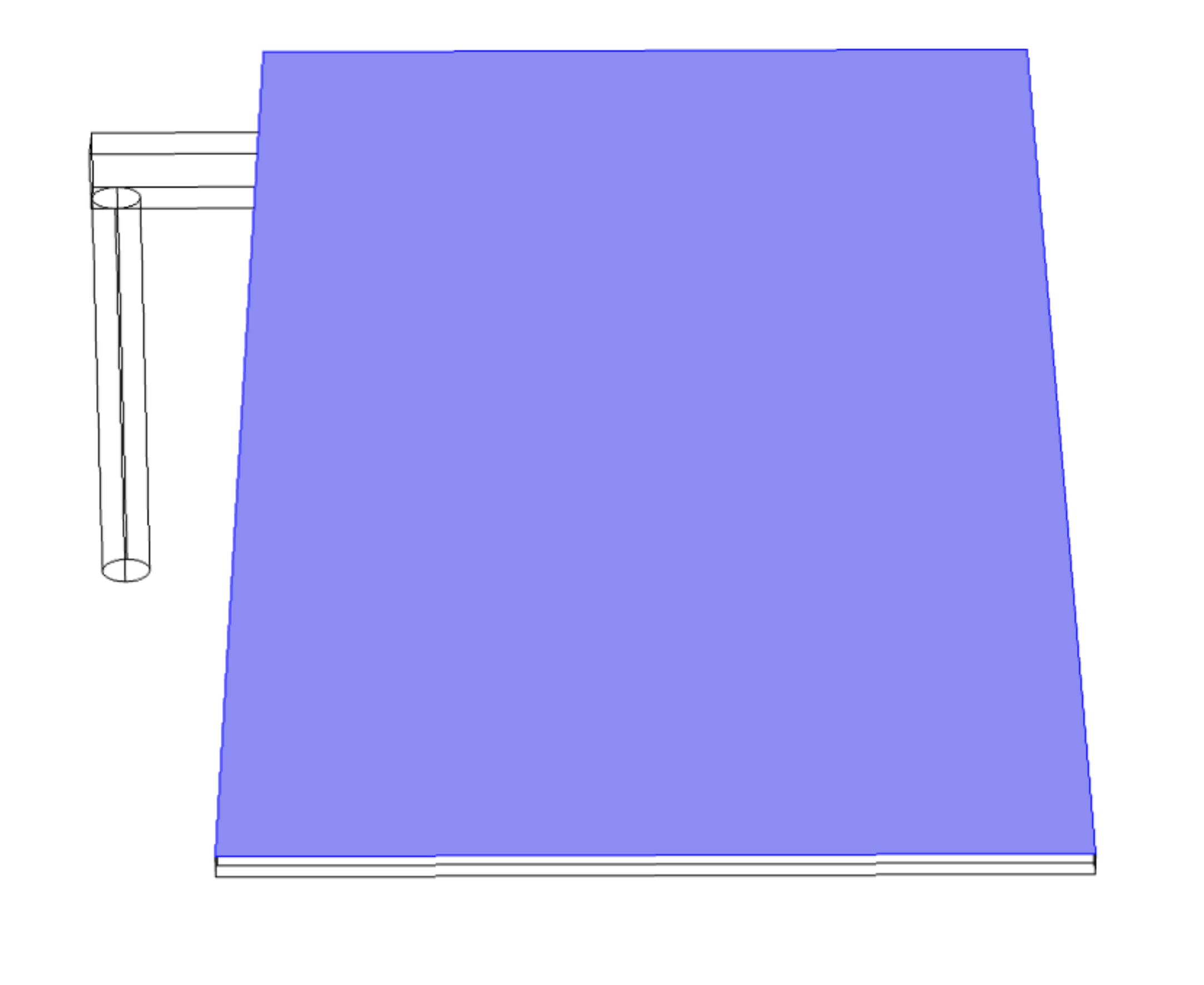}
        \caption{}
        \label{fig:surf}
    \end{subfigure}
~
    \begin{subfigure}{0.4\textwidth}
        \includegraphics[width=\textwidth]{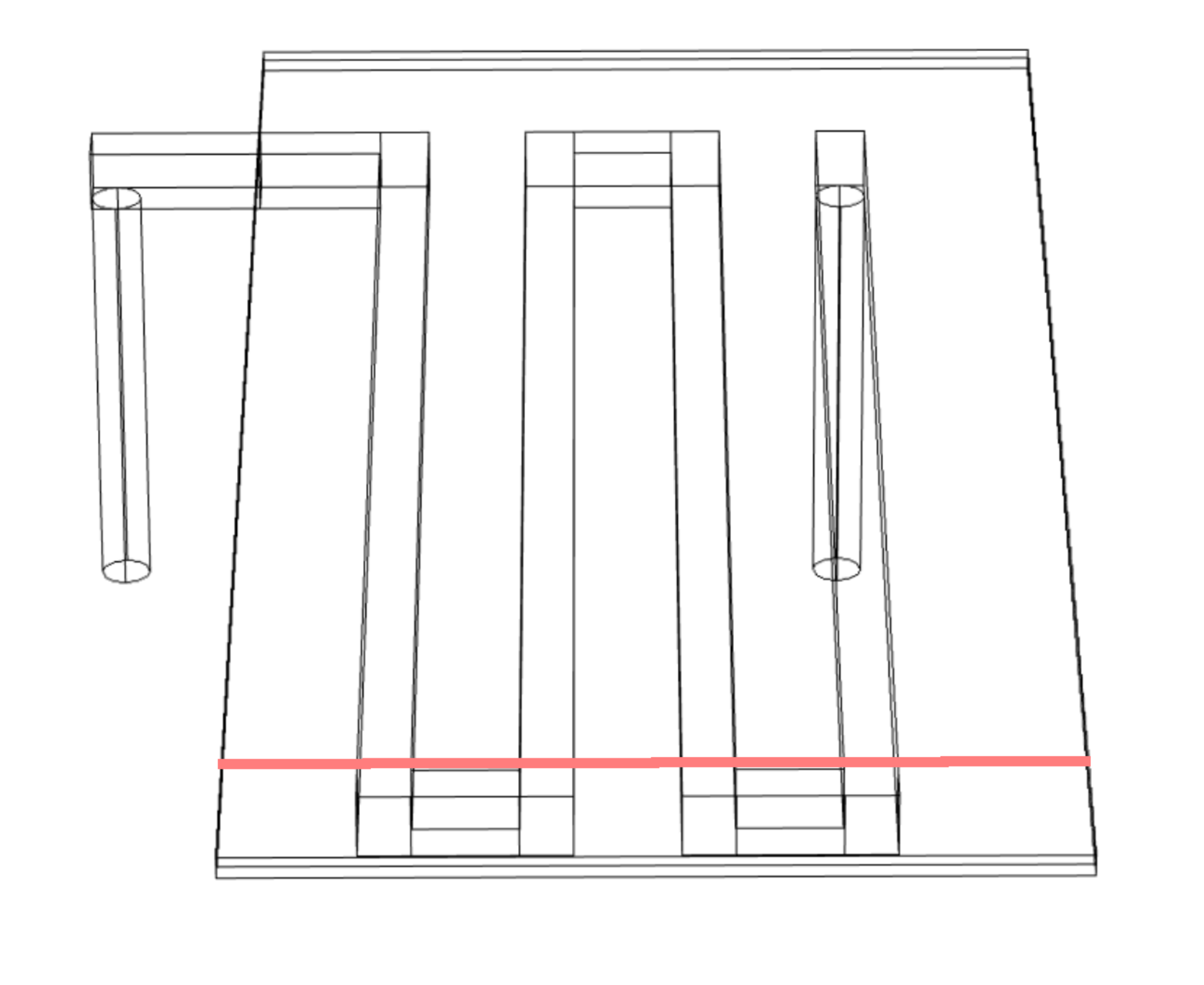}
        \caption{}
        \label{fig:cl_line}
    \end{subfigure}
    \caption{
        Geometrical entities used to obtain the variables relevant for mesh
        convergence, shown for the reduced geometry: \textbf{(a)}
        \textcolor{blue}{surface} used to obtain the apparent reaction rate,
        corresponding to the upper boundary of the CL domain; and \textbf{(b)}
        \textcolor{red}{line} used to track the ozone partial pressure profile
        atop the CL domain.
    }
    \label{fig:var_track}
\end{figure}

    A brief description of the acquisition of the response variables will now
be given. Figure \ref{fig:var_track} illustrates the relevant geometrical
entities used for the acquisition of the data. While a reduced geometry is
shown (used in Part 2 and Section \ref{ssec:ks}, below), the same were used
here, appropriately scaled. $\Delta\chi_{\mathrm{O_3}}$ is simply the difference
in value between inlet and outlet averages of $\chi_{\mathrm{O_3}}$, thus
$\Delta\chi_{\mathrm{O_3}}^{(\alpha)} = \Delta C_{\mathrm{O_3}}/C^*$, while
$\Delta\chi_{\mathrm{O_3}}^{(\beta)} = \Delta\chi_{\mathrm{O_3}}$. $R'
_{\mathrm{O_3}}$ is given by integration of the decomposition term for
$\mathrm{O_3}$ over the upper CL upper boundary. For Alpha, $R_{\mathrm{O_3,
dec}}$ is the sink term of Eq. \ref{eq:sink_alpha}; while for Beta, it is
$R^{(\beta)}_{\mathrm{O_3,dec}}=A_v\Gamma^*_s k_2\theta_{\mathrm{O_3}}$, which
is the volume-averaged decomposition term for $\theta_{\mathrm{O_3}}$. Since
$C^{(\alpha)}_{\mathrm{O_3}} \neq C^{(\beta)}_{\mathrm{O_3}}$, $R'
_{\mathrm{O_3}}$ was normalized by $C_{\mathrm{O_3,in}}$. Thus:
\begin{eqnarray}
    R'_{\mathrm{O_3}} & = & \frac{\int_{\partial\Omega} R_{\mathrm{O_3,dec}}\left(\mathbf{r}\right)\mathrm{d}A}{C_{\mathrm{O_3,in}}},
\end{eqnarray}
\noindent
where $\partial\Omega$ is the upper CL boundary (Fig. \ref{fig:surf}). The
stoichiometry $\lambda$ is defined as:
\begin{eqnarray}
    \lambda & = & \frac{\dot{n}_{\mathrm{O_3,in}}}{R_{\mathrm{O_3}}},
\end{eqnarray}
\noindent
where $\dot{n}_{\mathrm{O_3,in}}$ is the inlet molar rate of $\mathrm{O_3}$,
obtained in a simplified manner as $\dot{n}_{\mathrm{O_3,in}} = QC
_{\mathrm{O_3,in}}$; while $R_{\mathrm{O_3}}$ is the total $\mathrm{O_3}$
decomposition rate in the CL domain:
\begin{eqnarray}
    R_{\mathrm{O_3}} & = & \int_{\Omega}R_{\mathrm{O_3,dec}}\left(\mathbf{r}\right)\mathrm{d}V,
\end{eqnarray}
\noindent
with $\Omega$ being the CL domain. The apparent stoichiometry $\lambda'$ is
given by $\lambda'= \Delta\chi_{\mathrm{O_3}}/\chi_{\mathrm{O_3,in}}$, and is
an approximation of $\lambda$ more easily obtained experimentally. The $P
_{\mathrm{O_3}}(x)$ and $\bar{R}_{\mathrm{O_3}}(x)$ profiles were obtained at
the upper CL boundary (Fig. \ref{fig:cl_line}), along a line parallel to the
$x$ axis, passing over the turn sections of the flow channel. The $\mathbf{P}
_{\mathrm{O_3}}$ and $\mathbf{\bar{R}}_{\mathrm{O_3}}$ surfaces were obtained
at the upper CL boundary (Fig. \ref{fig:surf}). For both models, $P
_{\mathrm{O_3}} = \chi_{\mathrm{O_3}}P_A$, and thus $P^{(\alpha)}_{\mathrm{O_3}}
= C^{(\alpha)}_{\mathrm{O_3}}P_A/C^*$; while $\bar{R}_{\mathrm{O_3}}$ is simply
the decomposition term for a given model, $R_{\mathrm{O_3,dec}}\left(\mathbf{r}
\right)$, normalized by $R_{\mathrm{O_3}}V_{\mathrm{CL}}$, where $V
_{\mathrm{CL}}$ is the geometrical volume of the CL domain.

    Finally, in order to ascertain that differences between models are not due
to mesh influence, both Alpha and Beta models were run using the mesh chosen
after the convergence analysis (see Part 2), briefly described at the end of
Section \ref{sssec:math_new}. Thus, differences are expected between the
results presented by Alpha in Section \ref{sec:res} and those reported in
\cite{beruski17}.

\subsection{\label{ssec:ks}Parametric Study of Reaction Rate Constants}

    Similarly to what has been previously reported for the Alpha
model\cite{beruski17}, a parametric study has been carried out for the reaction
rate constants related to the adsorption-decomposition reactions of ozone, $k
_1$ and $k_2$. This was done in order to assess the dependency of the chosen
response variables to the order of magnitude of these parameters, arguably the
major source of uncertainty in the Beta model. In this way, while a proper
validation of the model would demand well-defined parameters, it may be
possible to establish a range of $k_1$ and $k_2$ values where the model
reasonably reproduce the experimental data available. This would, in turn,
provide a rough prediction of the reaction rate constants, while at the same
time constraining the model concerning its allowed behavior when compared to
independently measured experimental variables.

    The parametric study was carried using the Beta model, as described in
Section \ref{sssec:math_new}, using the mesh and solvers described at the end
of said section. The geometry was, however, reduced, in order to minimize
time and computational resources spent on the simulations. This reduced
geometry, shown in Fig. \ref{fig:var_track}, was used for the mesh convergence
study, described in Part 2. Since we are interested in relative changes in
behavior due to changes in both $k_1$ and $k_2$ values, it is expected that
the same conclusions hold for the full geometry. The study covers a range of 5
orders of magnitude, for both rate constants, centered on the value given in
Table \ref{tab:par2}. Thus, the parameters assumed the values given by the sets:
\begin{eqnarray}
    k_1 & \in & \left[1,10,10^2,10^3,10^4\right]\ \mathrm{s^{-1}}\ \mathrm{and} \nonumber \\
    k_2 & \in & \left[10^{-1},1,10,10^2,10^3\right]\ \mathrm{s^{-1}}, \nonumber
\end{eqnarray}
\noindent
where each parameter was varied independently of the other. For each doublet
$\left\{k_1,k_2\right\}$, an auxiliary parametric study was used such that the
inlet flow rate assumed the values $Q \in \left[250,350,450\right]\
\mathrm{cm^3\ min^{-1}}$. The response variables analyzed are the same used for
model comparison, described in Section \ref{ssec:comp}.

\section{\label{sec:res}Results}

    As mentioned in Section \ref{sec:int}, first a comparison between model
formulations will be carried, along with comparison to experimental data when
available. Afterwards, some shortcomings pointed out during the development of
the work will be addressed and discussed, along with further improvements that
might be important for the model. Finally, the parametric study of reaction
rate constants will be presented and its implications for model validation
discussed.

\subsection{\label{ssec:comp_res}Model Comparison}

    The improvements described on Section \ref{sssec:math_new} were made based
on well-known limitations of the original model, regarding the physical
phenomena described by it. Thus, the Beta model is expected to more accurately
replicate the physical phenomena underlying the experimental device. While a
better fit to experimental data is expected, it is not at all guaranteed and
neither is the current goal of this work. We focus here on the description of
the Beta model, and its comparison with Alpha, as a reference frame for future
comprehensive validation against experimental data. A brief discussion on this
point will be given in Section \ref{sssec:disc_AxB}.

\begin{figure}
    \centering
    \begin{subfigure}{0.45\textwidth}
        \includegraphics[width=\textwidth]{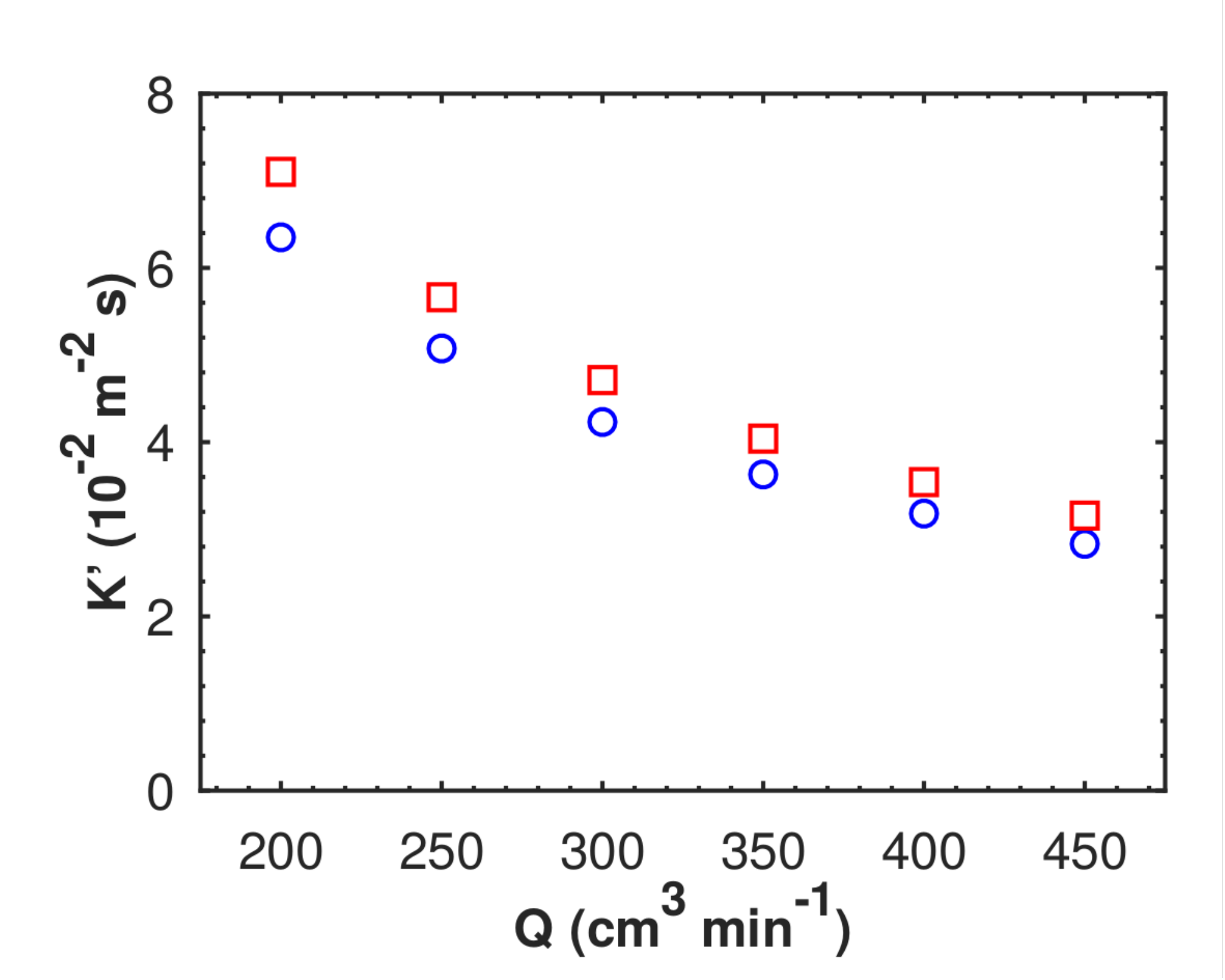}
        \caption{}
        \label{fig:AxB_K}
    \end{subfigure}
    ~
    \begin{subfigure}{0.45\textwidth}
        \includegraphics[width=\textwidth]{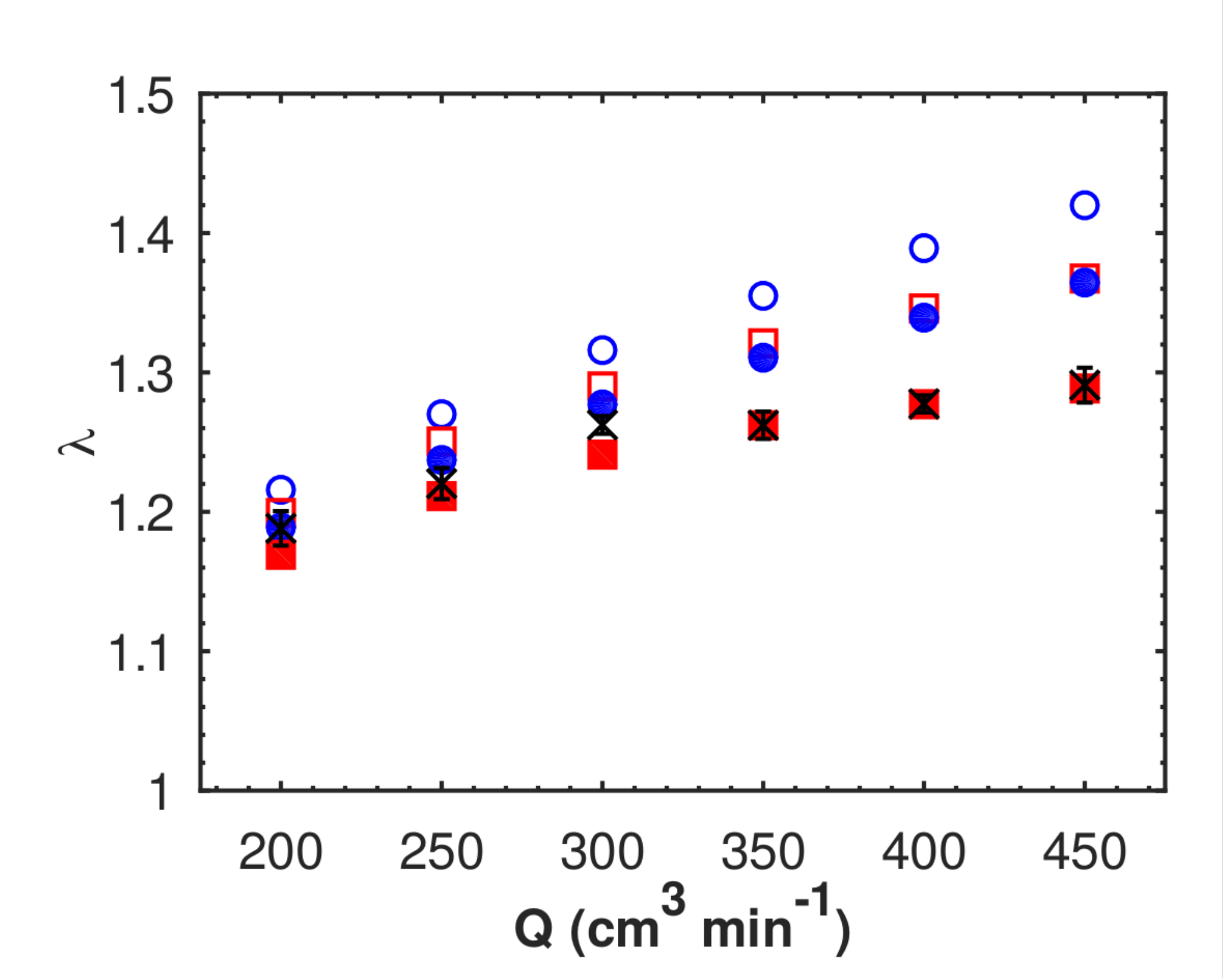}
        \caption{}
        \label{fig:AxB_lambda}
    \end{subfigure}

    \begin{subfigure}{0.45\textwidth}
        \includegraphics[width=\textwidth]{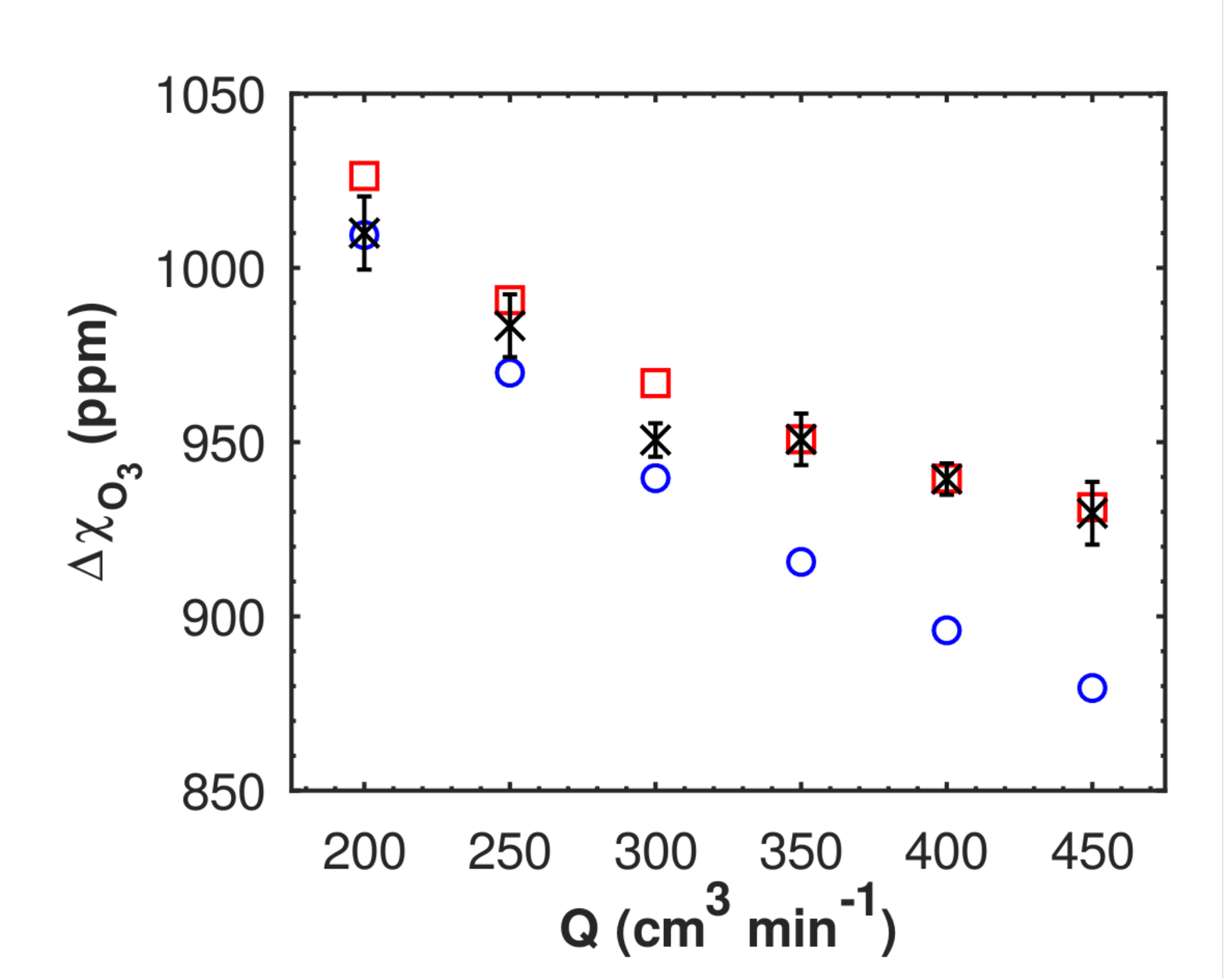}
        \caption{}
        \label{fig:AxB_dchi}
    \end{subfigure}
    ~
    \begin{subfigure}{0.45\textwidth}
        \includegraphics[width=\textwidth]{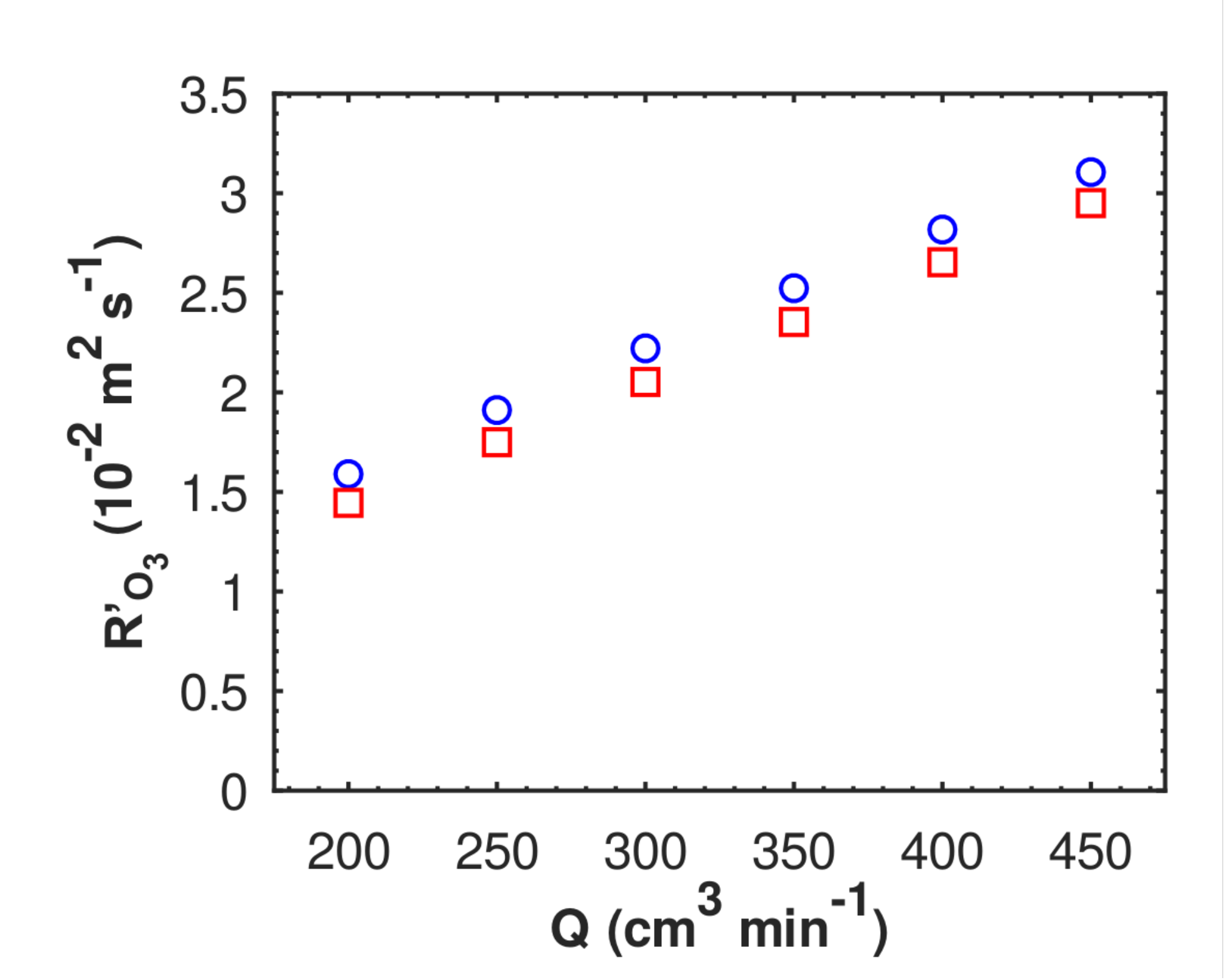}
        \caption{}
        \label{fig:AxB_Rapp}
    \end{subfigure}
    \caption{
        Scalar variables as function of the inlet flow rate for the Alpha
        (\textcolor{blue}{$\circ$}) and Beta (\textcolor{red}{$\Box$}) models:
        \textbf{(a)} $K'$, \textbf{(b)} $\lambda$ (empty) and $\lambda'$
        (full), \textbf{(c)} $\Delta\chi_{\mathrm{O_3}}$, and
        \textbf{(d)} $R'_{\mathrm{O_3}}/C_{\mathrm{O_3,in}}$. Experimental
        data\cite{beruski17} ($\times$) is also shown in \textbf{(b)} and
        \textbf{(c)} as reference, with error bars representing one standard
        deviation.
    }
    \label{fig:AxB_scalars}
\end{figure}

    Starting with the scalar response variables, Figure \ref{fig:AxB_scalars}
shows the ratio $K' = \Delta\chi_{\mathrm{O_3}}/R'_{\mathrm{O_3}}$, its
individual variables, and the stoichiometries $\lambda$ and $\lambda'$. As
mentioned in Section \ref{ssec:comp}, the variables concerning the flow field 
are left to the SM\footnotemark[3] (Section SIV.A, Figs. S10 and S11). The most
straightforward comparisons between models are given by $K'$ and the
stoichiometries, as they normalize the differences in reaction modeling and
species transport formulation. Concerning $K'$, Fig. \ref{fig:AxB_K}, it can be
observed that the Beta model shows a slightly higher rate of variation with $Q$
than the Alpha model, suggesting that the changes implemented do affect the
results qualitatively. Concerning the stoichiometries, Fig.
\ref{fig:AxB_lambda}, for both $\lambda$ and $\lambda'$ the Alpha model
predicts a higher value at all $Q$, also displaying a larger rate of increase.
It is noticeable that, for $\lambda'$, the Beta model reproduces the available
experimental data quite well, particularly at high $Q$. The reactant drop
$\Delta\chi_{\mathrm{O_3}}$, shown in Fig. \ref{fig:AxB_dchi}, corroborates
this as expected, showing a closer fit to the experimental data from the Beta
model, while the Alpha model predicts an almost linear decrease in $\Delta\chi
_{\mathrm{O_3}}$ with $Q$. The inlet-normalized apparent reaction rate $R'
_{\mathrm{O_3}}/C_{\mathrm{O_3,in}}$, shown in Fig. \ref{fig:AxB_Rapp}, shows
little difference between models, with the Alpha model predicting slightly
higher values.

    A possible interpretation of Fig. \ref{fig:AxB_scalars} is that $Q$, and
thus the convective transport, play a larger role in reactant transport in the
Alpha model. Larger values of $R'_{\mathrm{O_3}}/C_{\mathrm{O_3,in}}$ for the
Alpha model, despite lower $\Delta\chi_{\mathrm{O_3}}$ ones, might suggest that
reactants are brought to the upper CL surface more efficiently. On the other
hand, the faster drop of $\Delta\chi_{\mathrm{O_3}}$ with $Q$, as well as the
higher values of $\lambda$, suggests that the reactants are also removed more
efficiently from the CL. Comparison with available data, however, suggests this
is not the case in the experimental device, and that the Beta model more
closely captures this convective contribution.

    Two points should be kept in mind, however. The first being that the
reaction rate constants are degrees of freedom in both models, and thus
variables that depend on it must be analyzed with care, such as $\Delta\chi
_{\mathrm{O_3}}$ and $\lambda'$. Section \ref{ssec:ks_res} provides a
discussion over this issue. The second point is the matter of magnitude of
$C_{\mathrm{O_3}}$, as pointed out in Section \ref{ssec:comp}. While the
overall effect is mainly a shift in order of magnitude in $K'$ and $R'
 _{\mathrm{O_3}}$, this suggests that the Alpha model would not reproduce the
 raw data obtained from the experimental technique used for validation, i.e.
 the chemiluminescence image from where the $\mathrm{O_3}$ partial pressure
 surfaces are obtained\cite{lopes15}. Thus, in the absence of measurements of
 absolute $\mathrm{O_3}$ concentration in the experimental device, a change
 towards dealing directly with molar fractions was inevitable. The Beta model
 therefore has the advantage over Alpha, mathematical frameworks aside, of
 being able to directly employ the experimental data available for model
 validation.

    Finally, a brief note on the behaviors seen in Fig. \ref{fig:AxB_dchi} for
$\Delta\chi_{\mathrm{O_3}}\left(Q\right)$ is warranted. The resulting curves
for the Alpha and Beta models are in contrast to the conclusion drawn in
\cite{beruski17}: when comparing between Stokes-Darcy (SD) and Darcy-Brinkman
(DB) formulations, the latter showed a closer fit to experimental data,
resembling the current Beta model; where the current Alpha model more closely
resembles the SD formulation of \cite{beruski17}. That is a testimony of how
important grid convergence studies are, as it is clear now that the correlation
between experiment and the Alpha model is dependent on the grid. This point
will be briefly discussed in Section \ref{sssec:disc_AxB}.

\begin{figure}
    \centering
    \begin{subfigure}{0.45\textwidth}
        \includegraphics[width=\textwidth]{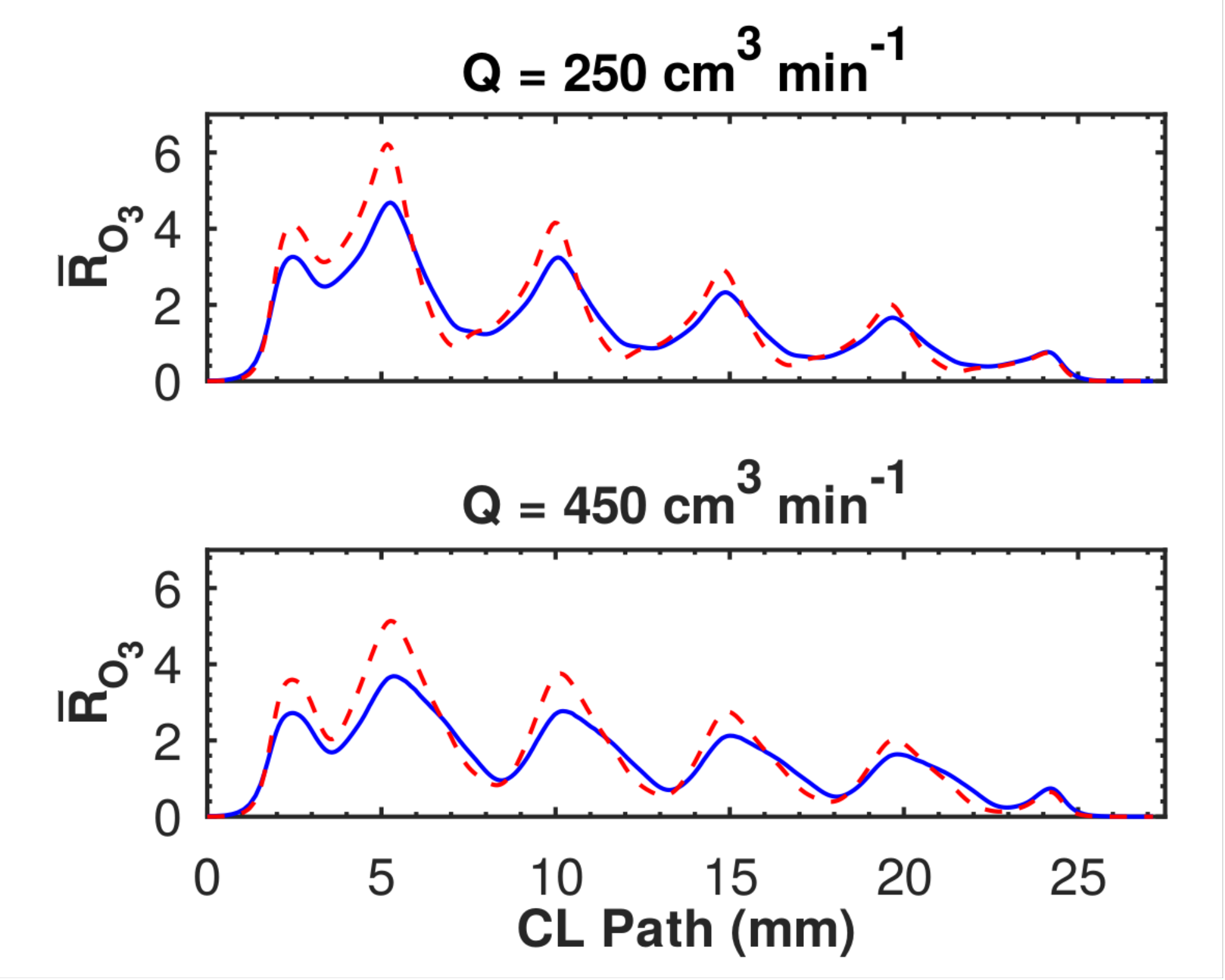}
        \caption{}
        \label{fig:AxB_Ro3x}
    \end{subfigure}
    ~
    \begin{subfigure}{0.45\textwidth}
        \includegraphics[width=\textwidth]{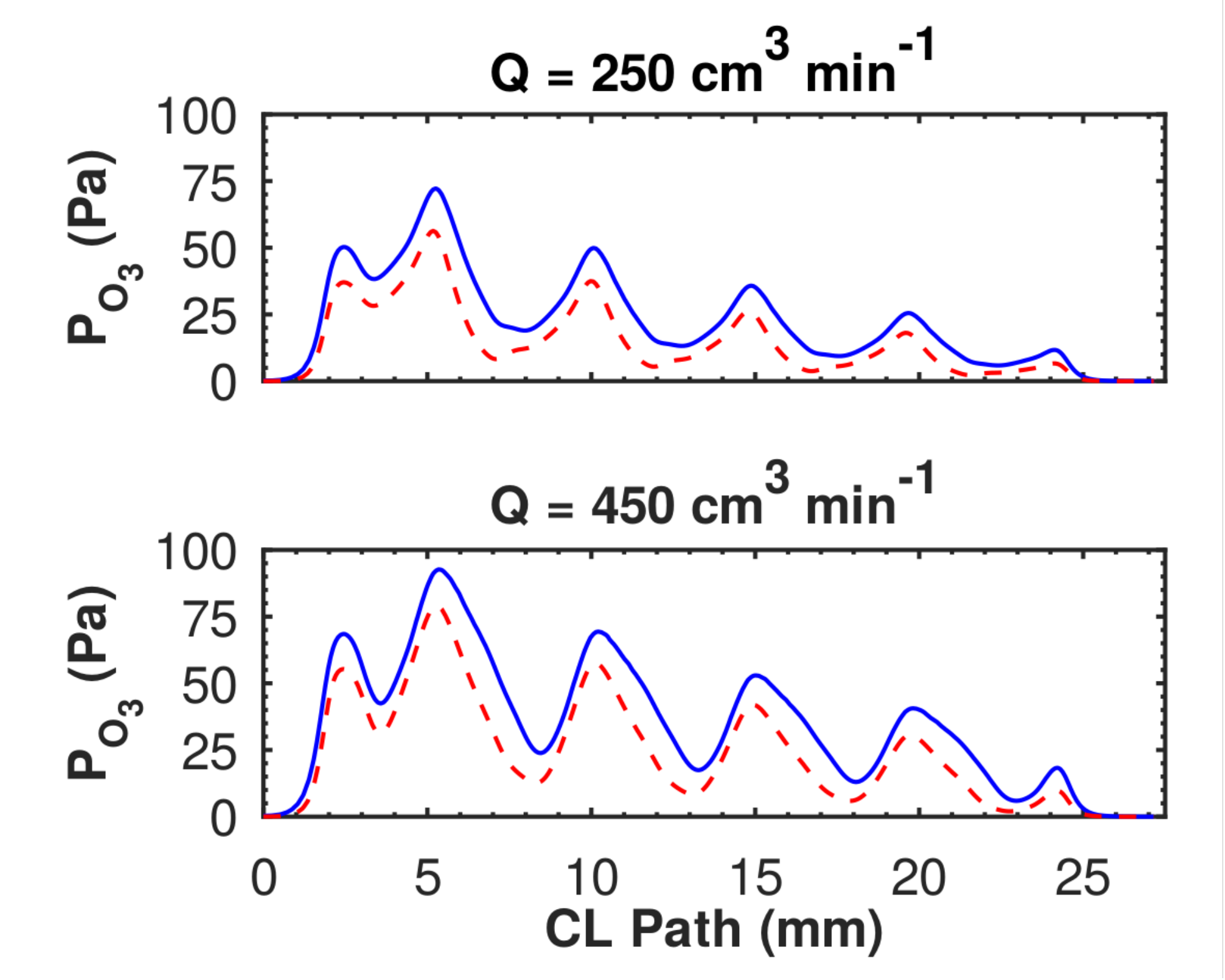}
        \caption{}
        \label{fig:AxB_Po3x}
    \end{subfigure}
    \caption{
        \textbf{(a)} Normalized reaction rate and \textbf{(b)} ozone partial
        pressure profiles, at $Q = 250$ (top) and $450\ \mathrm{cm^3\
        min^{-1}}$ (bottom), for \textcolor{blue}{Alpha} (full) and
        \textcolor{red}{Beta} (dashed) models.
    }
    \label{fig:AxB_prfls}
\end{figure}

    Moving on with the comparison between Alpha and Beta models, using the DB
formulation, we turn to the profiles atop the CL domain. Figure
\ref{fig:AxB_prfls} shows both $\bar{R}_{\mathrm{O_3}}$ and $P_{\mathrm{O_3}}$
profiles along the $x$ axis, for $Q = 250$ and $450\ \mathrm{cm^3\ min^{-1}}$.
Considering first $\bar{R}_{\mathrm{O_3}}$, Fig. \ref{fig:AxB_Ro3x}, in both
cases differences are readily seen between models, with Beta showing narrower
peaks and more pronounced structures between peaks. Analyzing one given peak,
it is noticeable how the increase in $\bar{R}_{\mathrm{O_3}}$ and peak position
are virtually the same for both models. The difference thus lies mainly with
the decay in $\bar{R}_{\mathrm{O_3}}$ after each turn in the Fc, with the Alpha
model showing slower decay and, consequently, larger ozone bypass between Fc
sections, directly over the land and through the Pm domains. The structures
between peaks are likely to be present in both models, seen more prominently
for $Q = 250\ \mathrm{cm^3\ min^{-1}}$ as unresolved local maxima between
$\bar{R}_{\mathrm{O_3}}$ peaks. These are likely to be connected to
recirculation at the first corner of each turn of the Ch domain. The faster
decay predicted by the Beta model allows a better view of these local maxima,
although only the pre-peak region, while the slower decay predicted by the
Alpha model allows only inference of the presence of these local maxima
\emph{via} the $\bar{R}_{\mathrm{O_3}}$ ``plateau'' between peaks. A similar
effect is seen for the Beta model at $Q = 450\ \mathrm{cm^3\ min^{-1}}$, where
the presence of the smaller peaks can be inferred despite being unresolved. For
the Alpha model, on the other hand, the presence of the smaller peaks seems to
have vanished.

    Considering now the $P_{\mathrm{O_3}}$ profiles, Fig. \ref{fig:AxB_Po3x},
for a qualitative comparison, albeit no less important due to the interest on
reactant partial pressure on fuel cells. These profiles are directly related to
the $\bar{R}_{\mathrm{O_3}}$ ones, and as such, it is seen that $P
_{\mathrm{O_3}}$ are lower for the Beta model. It can be seen by comparing
Figs. \ref{fig:AxB_Ro3x} and \ref{fig:AxB_Po3x} that despite the adsorption
step implemented in the Beta model, for the overall ozone decomposition
reaction, the $P_{\mathrm{O_3}}$ profiles are still proportional to $\bar{R}
_{\mathrm{O_3}}$ and, consequently, the non-normalized reaction rate $R
_{\mathrm{O_3}}$. Thus, the points discussed above translate almost directly to
this case, in particular the presence of the local maxima between peaks for the
Beta model at $Q = 250\ \mathrm{cm^3\ min^{-1}}$, albeit not as pronounced as
in Fig. \ref{fig:AxB_Ro3x}. The overall lower values of $P_{\mathrm{O_3}}$
predicted by the Beta model are also in line with the results shown in Fig.
\ref{fig:AxB_scalars}, thus corroborating that the differences between profiles
are not just a matter of rate constant values. To strengthen the case, Fig.
S12 shows the same $P_{\mathrm{O_3}}$ profiles normalized by their respective
$R'_{\mathrm{O_3}}$ values. In this case, it can be seen that the difference
are even larger, further corroborating the idea that changes in $P_{\mathrm{O
_3}}$ at the CL are due to more than just the magnitude of reaction kinetics.

    It should be kept in mind that, in the case of the normalized reaction rate
profiles, differences seen in Fig. \ref{fig:AxB_Ro3x} are expected to be mainly
a consequence of the different mathematical frameworks. Although some influence
on the magnitude of the reaction rate constants is to be expected, as is
discussed in Section \ref{ssec:ks_res}, the close results shown in Fig.
\ref{fig:AxB_scalars} suggest that such influence is relatively small in this
case. Nevertheless, these profiles join the scalar values shown in Fig.
\ref{fig:AxB_scalars} as potential validation points, either as $\bar{R}
_{\mathrm{O_3}}$ or $P_{\mathrm{O_3}}$ given the clear proportionality seen
in Fig. \ref{fig:AxB_prfls}. However, it should be noticed that experimental
data currently available does not allow distinction between the Alpha and Beta
models (see Fig. 4 of \cite{beruski17}). Such an increase in precision is
currently being sought.

\begin{figure}
    \centering
    \begin{subfigure}{0.3\textwidth}
        \includegraphics[width=\textwidth]{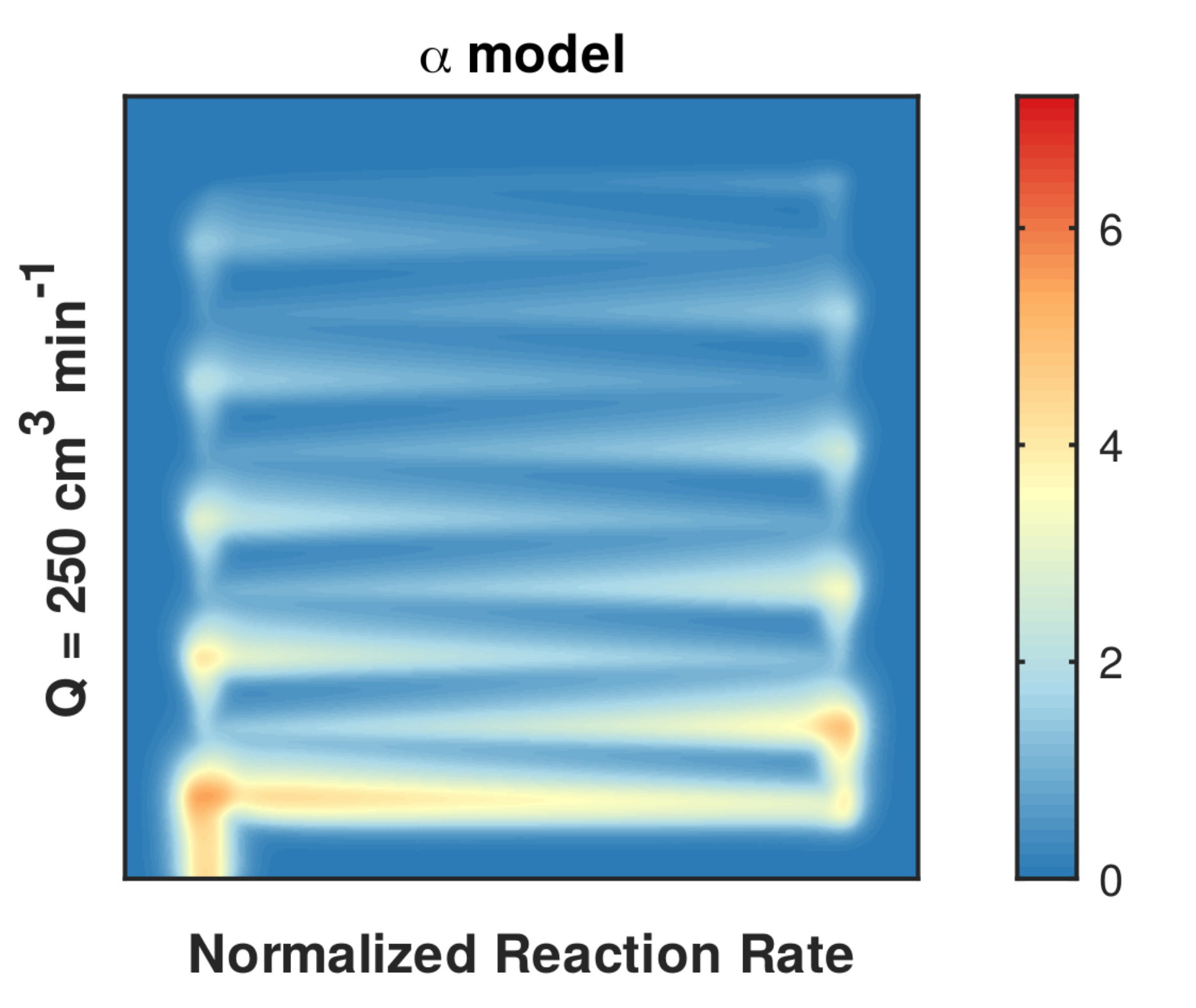}
        \caption{}
    \end{subfigure}
    ~
    \begin{subfigure}{0.3\textwidth}
        \includegraphics[width=\textwidth]{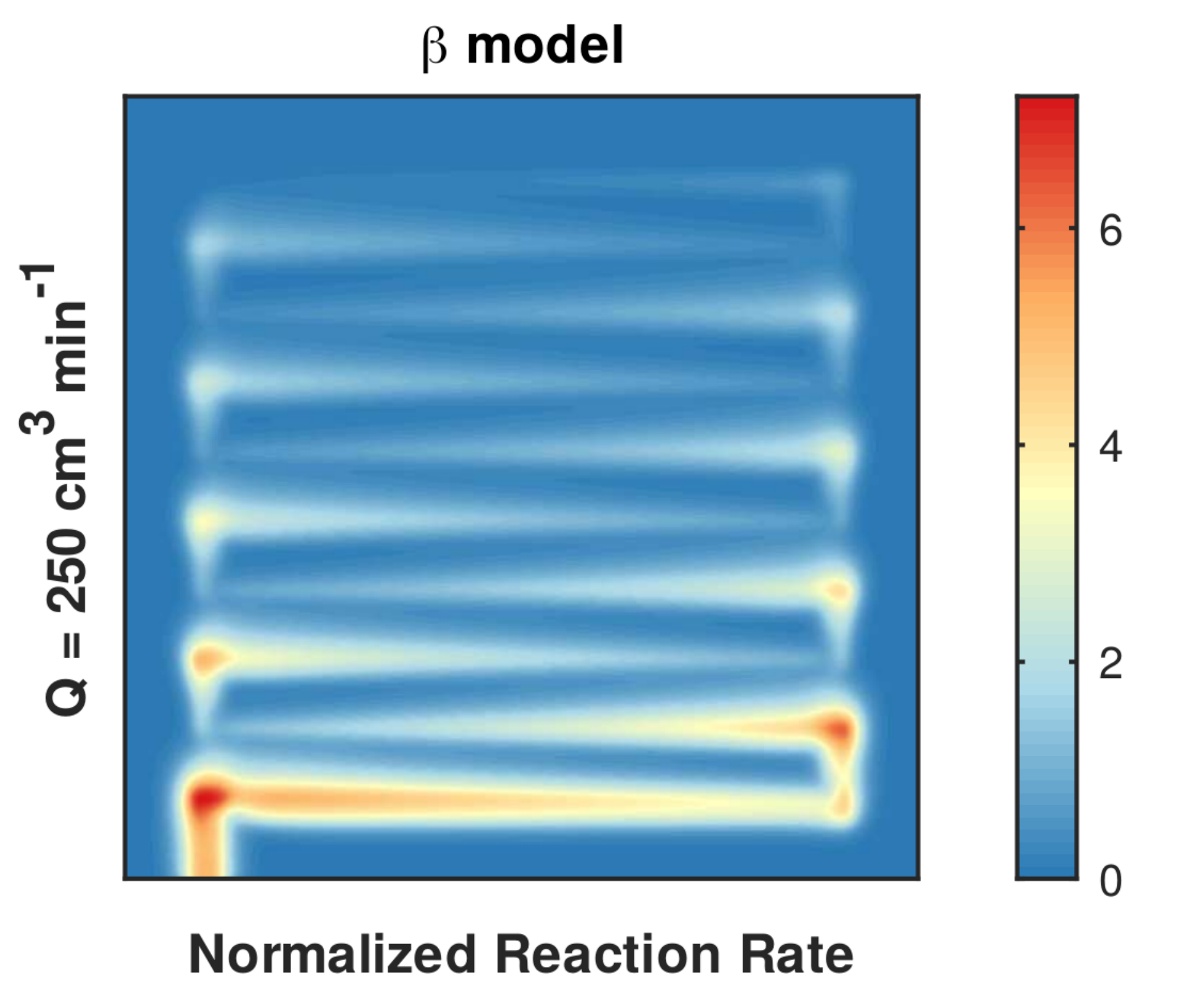}
        \caption{}
    \end{subfigure}
    ~
    \begin{subfigure}{0.3\textwidth}
        \includegraphics[width=\textwidth]{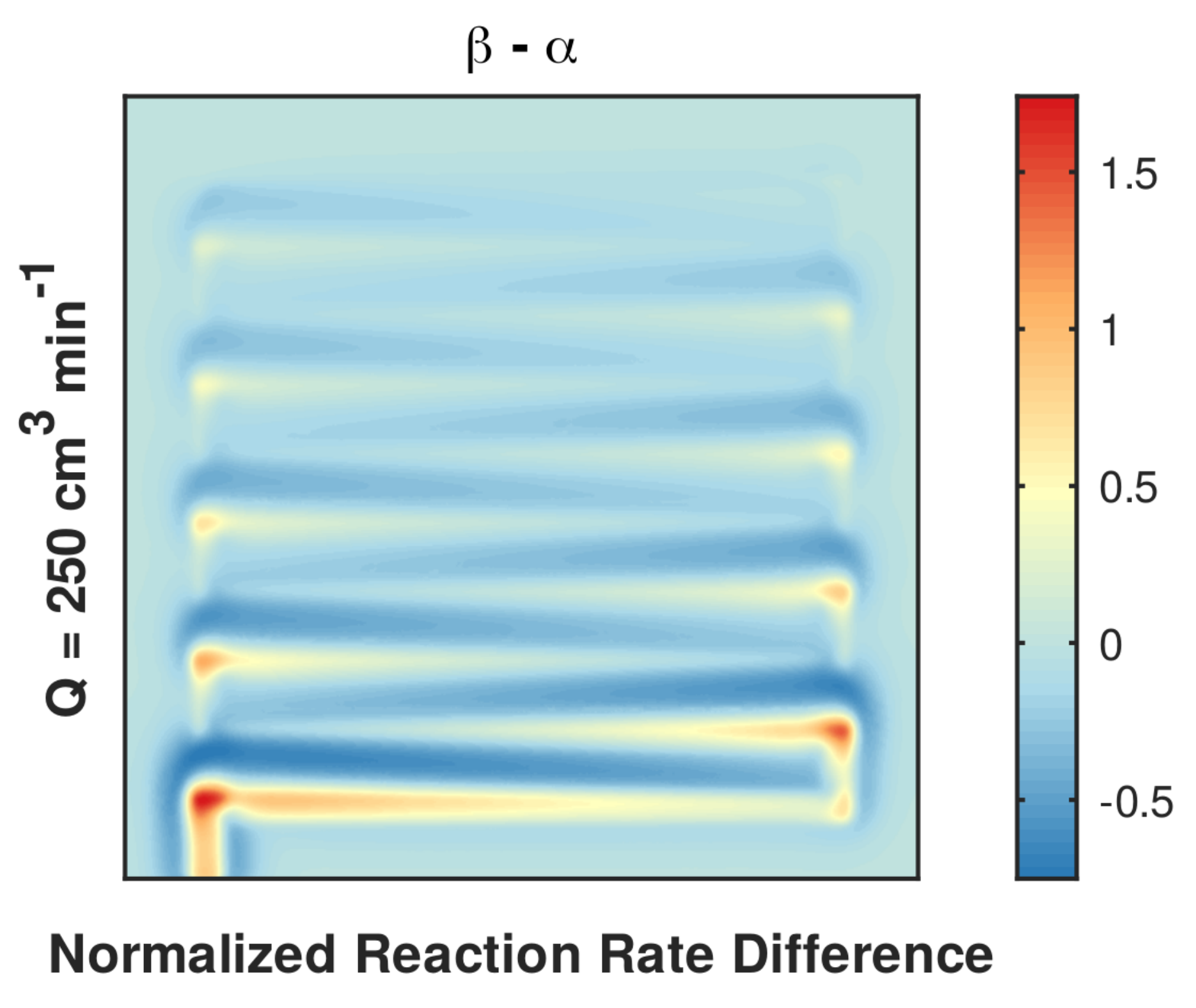}
        \caption{}
        \label{fig:AxB_Rxydiff_250}
    \end{subfigure}

    \begin{subfigure}{0.3\textwidth}
        \includegraphics[width=\textwidth]{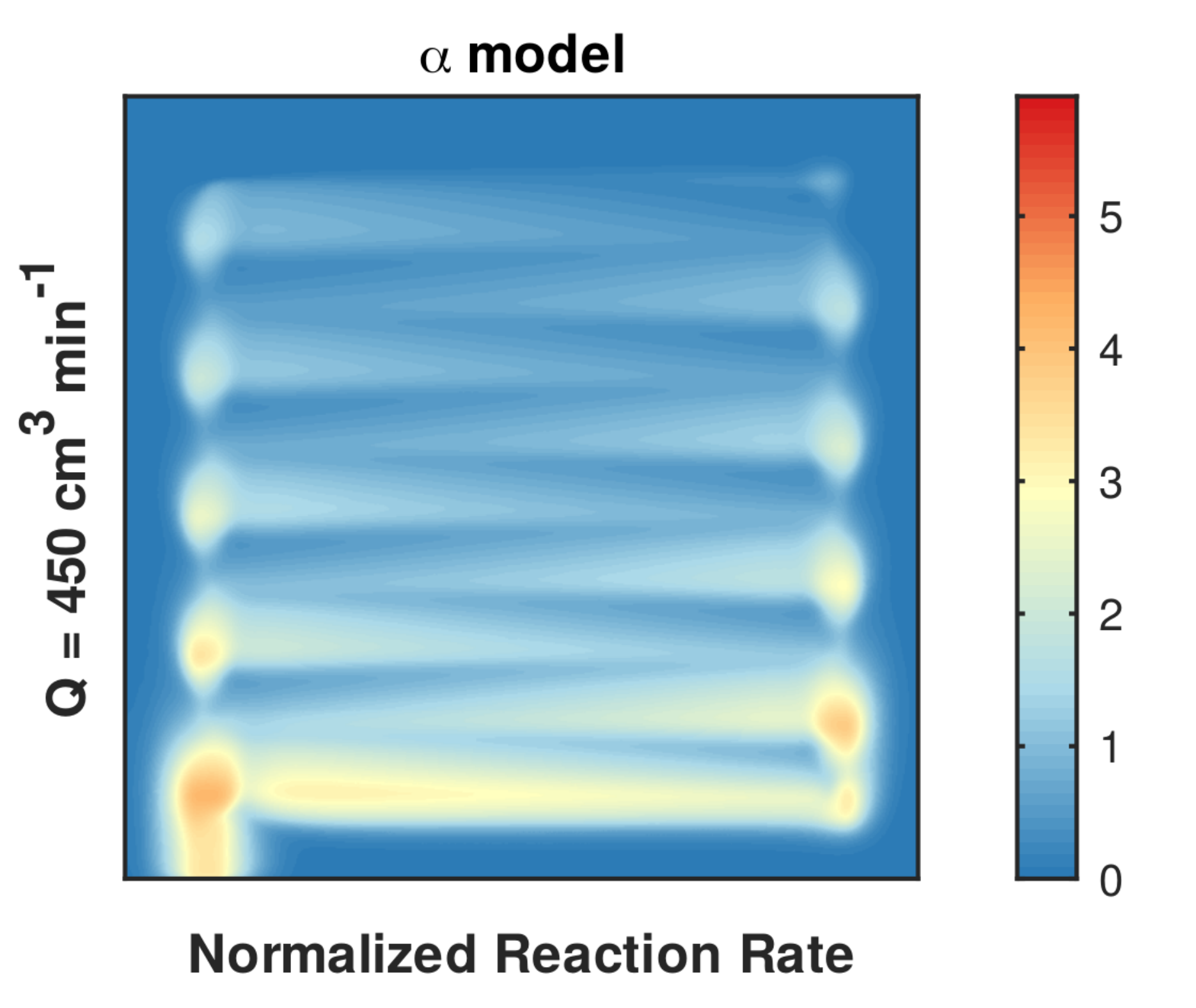}
        \caption{}
    \end{subfigure}
    ~
    \begin{subfigure}{0.3\textwidth}
        \includegraphics[width=\textwidth]{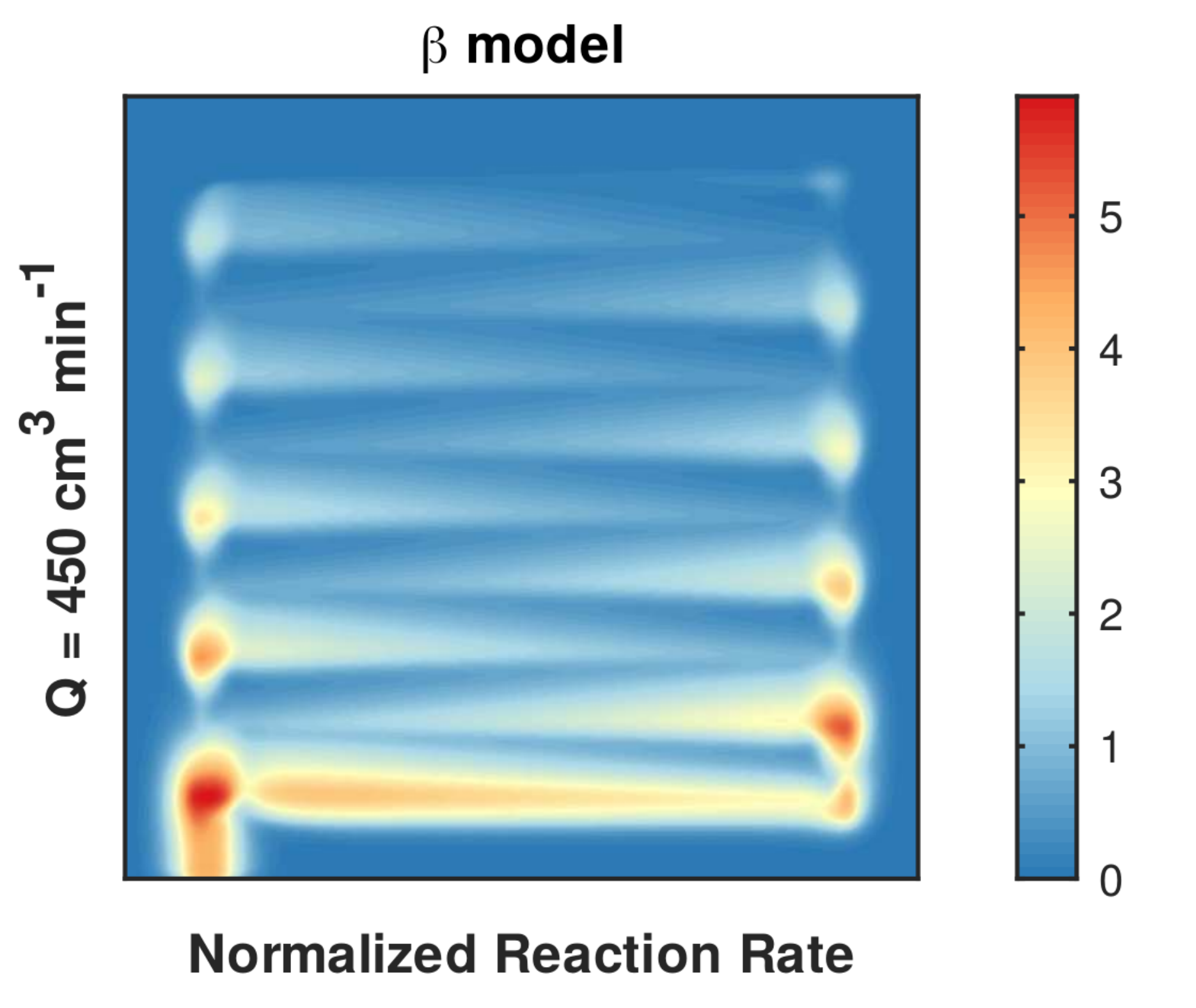}
        \caption{}
    \end{subfigure}
    ~
    \begin{subfigure}{0.3\textwidth}
        \includegraphics[width=\textwidth]{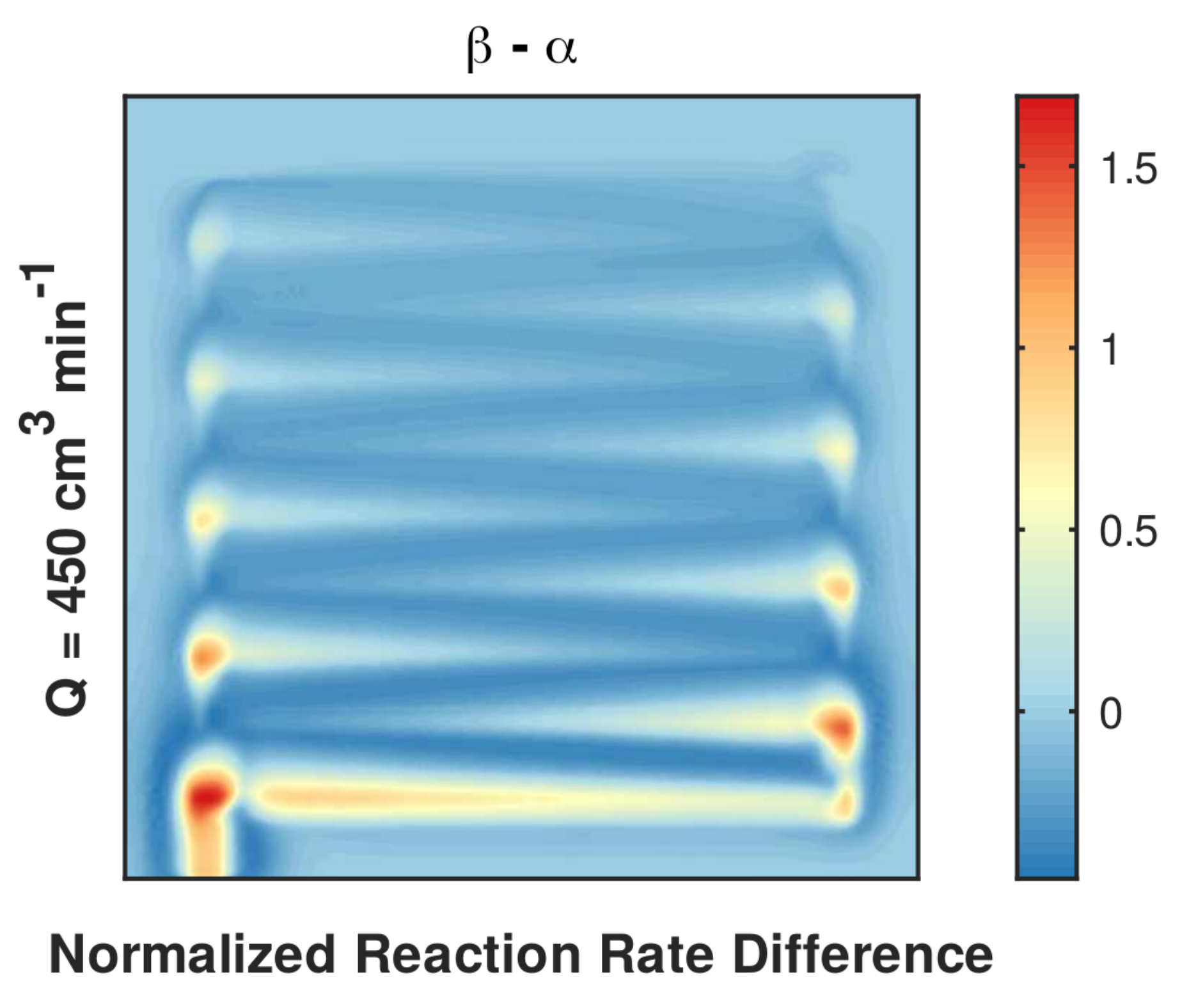}
        \caption{}
        \label{fig:AxB_Rxydiff_450}
    \end{subfigure}
    \caption{
        Normalized reaction rate surfaces for the Alpha (left column) and Beta
        (central column) models, and the difference between Beta and Alpha
        models (right column), for $Q = 250$ (top row) and $450\ \mathrm{cm^3\
        min^{-1}}$ (bottom row).
    }
    \label{fig:AxB_Ro3xy}
\end{figure}

    Carrying on, we proceed to analyze spatially-resolved data available for
both models. Figure \ref{fig:AxB_Ro3xy} shows the surfaces for the same values
of $Q$ as Fig. \ref{fig:AxB_Ro3x}, as well as the differences between Beta and
Alpha models. As expected from Fig. \ref{fig:AxB_Ro3x}, the Beta model shows
larger $\bar{R}_{\mathrm{O_3}}$ values over the corners of the Ch domain.
Similarly, the Alpha model shows larger plumes of $\bar{R}_{\mathrm{O_3}}$ at
the same regions. In addition, it is now possible to observe that this is the
case all over the CL upper surface, being particularly evident in Figs.
\ref{fig:AxB_Rxydiff_250} and \ref{fig:AxB_Rxydiff_450}. The increase in $Q$
is followed by an increased difference in plume spread and the regions where it
is evident, with slightly reduced differences in magnitude. Concerning the
differences between peaks seen in Fig. \ref{fig:AxB_Ro3x}, $\bar{\mathbf{R}}
_{\mathrm{O_3}}$ surfaces provide little additional information, aside from
the confirmation that it happens in all turn sections of the Ch domain.

\begin{figure}
    \centering
    \begin{subfigure}{0.3\textwidth}
        \includegraphics[width=\textwidth]{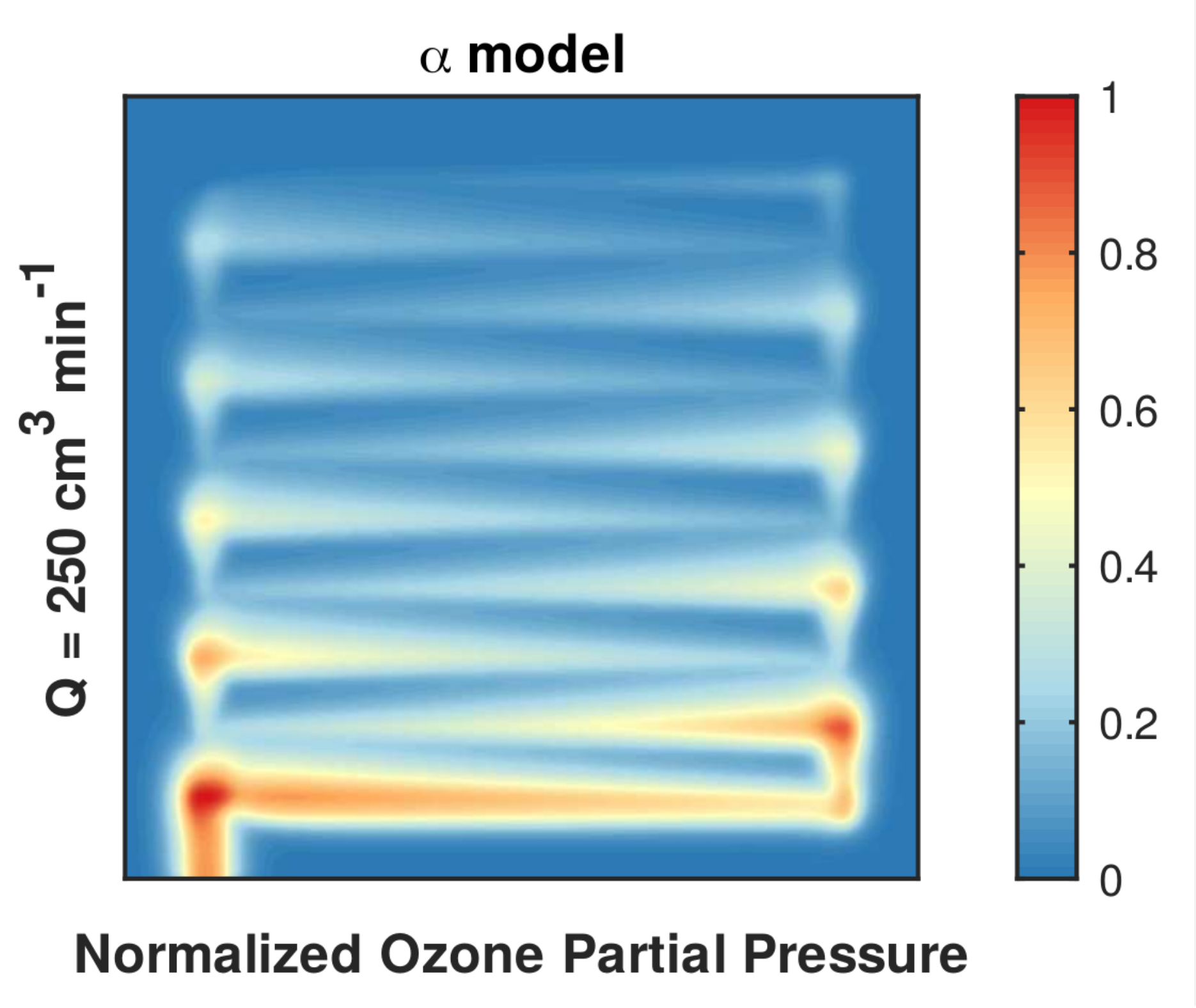}
        \caption{}
    \end{subfigure}
    ~
    \begin{subfigure}{0.3\textwidth}
        \includegraphics[width=\textwidth]{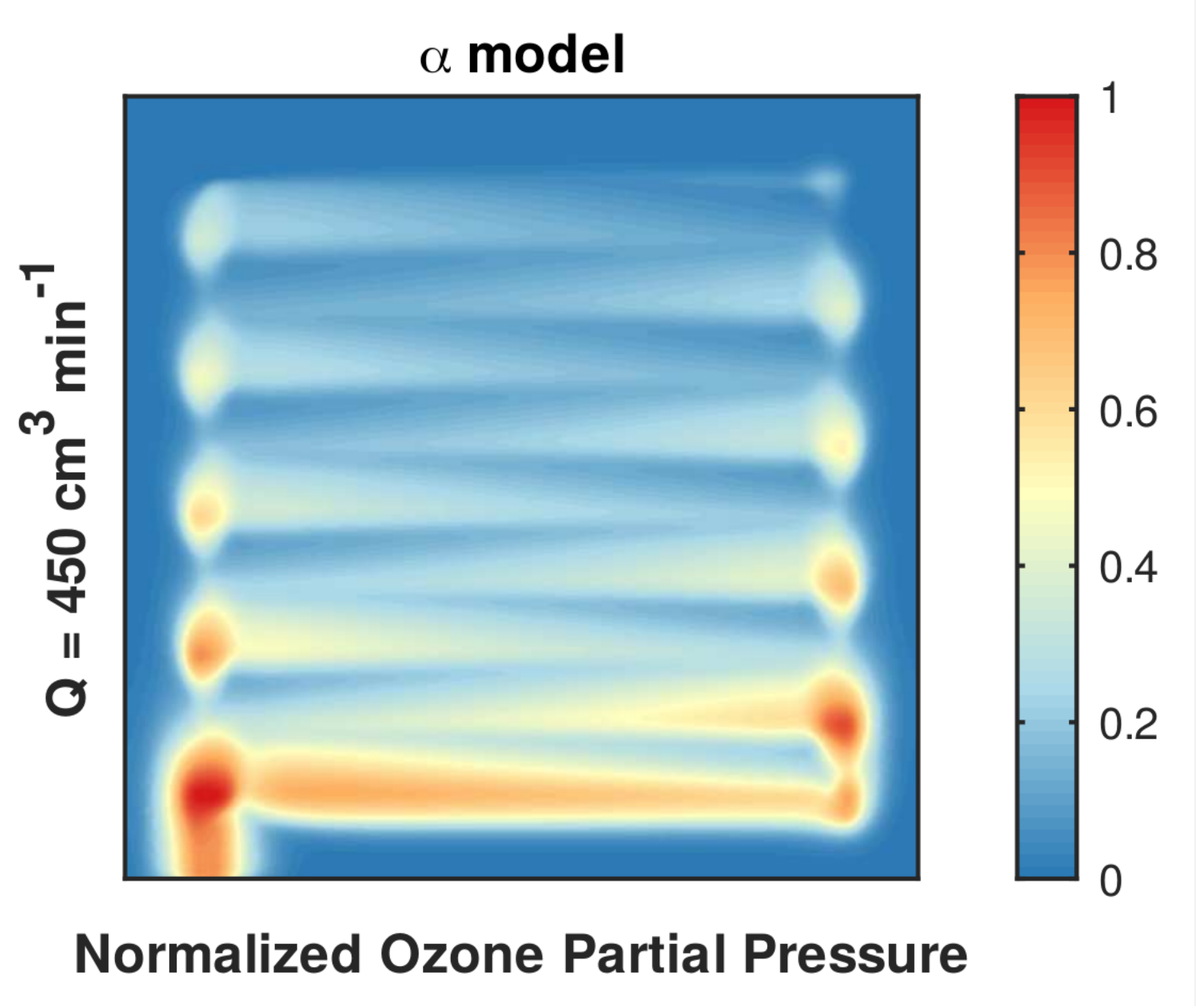}
        \caption{}
    \end{subfigure}

    \begin{subfigure}{0.3\textwidth}
        \includegraphics[width=\textwidth]{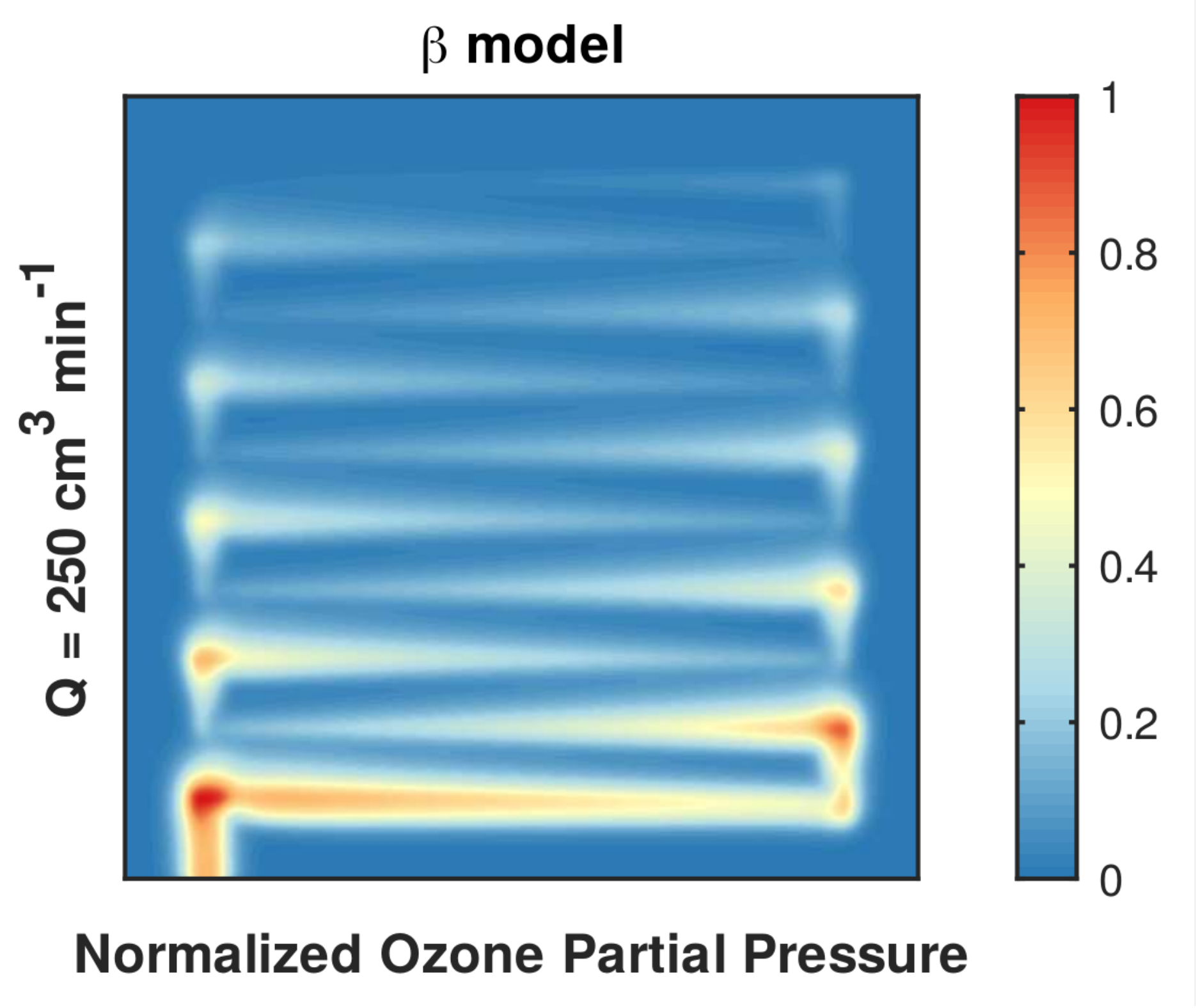}
        \caption{}
    \end{subfigure}
    ~
    \begin{subfigure}{0.3\textwidth}
        \includegraphics[width=\textwidth]{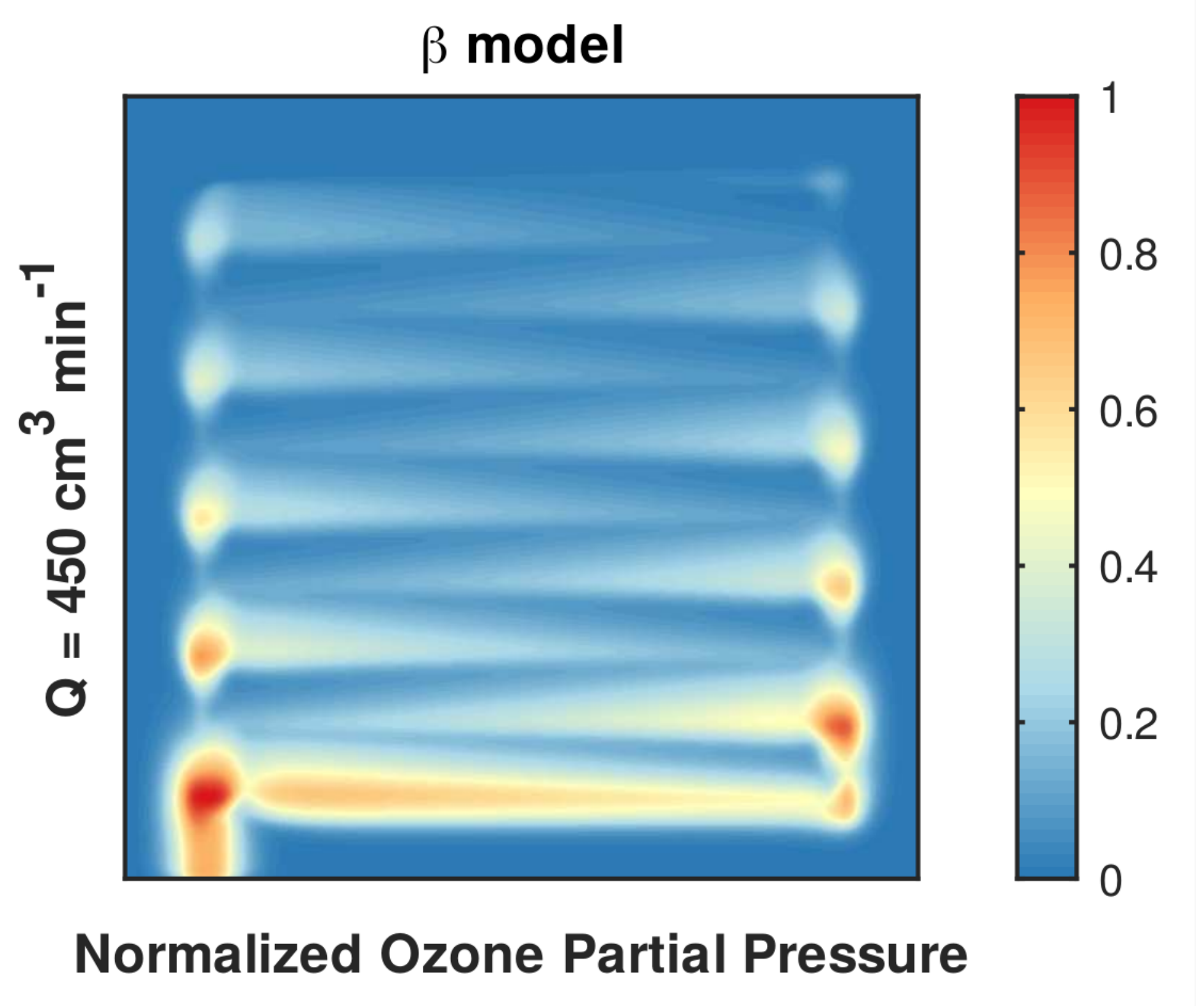}
        \caption{}
    \end{subfigure}
    \caption{
        Normalized ozone partial pressure surfaces for the Alpha (top row) and
        Beta (bottom row) models, for $Q = 250$ (left column) and $450\
        \mathrm{cm^3\ min^{-1}}$ (right column).
    }
    \label{fig:AxB_Po3xy}
\end{figure}

    For the final comparison, the $\mathbf{P}_{\mathrm{O_3}}$ surfaces are
shown, for selected values of $Q$, in Figure \ref{fig:AxB_Po3xy}. It should
be kept in mind that comparisons between Alpha and Beta models using $P
_{\mathrm{O_3}}$ are qualitative in nature, thus it was chosen to normalize
the color scales in Fig. \ref{fig:AxB_Po3xy} by their respective maximum
values, in order to avoid quantitative comparisons. Similarly to Fig.
\ref{fig:AxB_prfls}, a direct relation between $\bar{\mathbf{R}}
_{\mathrm{O_3}}$ and $\mathbf{P}_{\mathrm{O_3}}$ surfaces is seen, just as
expected. Larger ozone plumes are also seen for the Alpha model, particularly
for $Q=450\ \mathrm{cm^3\ min^{-1}}$. Aside from that, as expected
from Fig. \ref{fig:AxB_Po3x}, there is little difference between the models.
A surface of differences in $P_{\mathrm{O_3}}$ could be instructive, as in
Fig. \ref{fig:AxB_Ro3xy}, however as pointed out the difference in reaction
kinetics makes such data misleading, and thus is better left out from the
analysis.

    Figs. \ref{fig:AxB_Ro3xy} and \ref{fig:AxB_Po3xy}, in particular Figs.
\ref{fig:AxB_Rxydiff_250} and \ref{fig:AxB_Rxydiff_450}, provide additional
insight over a discrepancy between experimental and numerical data reported in
\cite{beruski17}. It was seen that, for a $k_{\mathrm{app}}$ value that
reasonably matched experimental results, the preferred model predicted a
narrower ozone plume in much of a given horizontal section of the Ch domain.
Here we see that the Beta model predicts a narrower plume, when compared to
Alpha, deepening the discrepancy between numerical and experimental systems.
However, as pointed out above, the precision of the experimental
spatially-resolved data offers relatively small constraints to the
computational models, thus an improved setup is being sought in order to allow
robust comparison with numerical data and, consequently, further improvements
to the models.

\subsubsection{\label{sssec:disc_AxB}Discussions}

    Given the comparisons made above, it is clear that, in general, the
differences between the Alpha and Beta models are relatively small, in
particular with few qualitative differences. The most relevant difference
seen, in Fig. \ref{fig:AxB_dchi}, lacks validation power due to the degree of
freedom assigned to the reaction rate constants. To complicate matters, the
estimated error for the Beta model (un)comfortably puts both experimental and
the Alpha model's data within uncertainty bounds (see Figs. S13 and S14),
specially more so if one assumes the discretization error of Alpha is of the
same magnitude as Beta. Variables that show error compensation and thus would
be useful for validation, such as $K' = \Delta\chi_{\mathrm{O_3}}/R'_{\mathrm{O
_3}}$ and $R'_{\mathrm{O_3}}$, lack experimental data. Therefore, two major
issues outside the mathematical framework need to be further considered before
actual validation may be pursued: i) discretization error, i.e. the mesh
refinement, and ii) the availability of experimental data. While both demand
time and resources, they are currently being sought.

    In the meantime, there are some shortcomings and approximations in the
Beta model that deserve some clarifications. These have been brought up before,
namely the Fickean diffusion coefficient for $\mathrm{O_3}$, the use of an
$\mathrm{O_3/air}$ mixture, and the coupling between free, porous media and
Knudsen diffusion regimes. In addition, there is the insight regarding the
behavior of $\Delta\chi_{\mathrm{O_3}}$ when compared to the experimental data
and previously published results\cite{beruski17}. Concerning the first point,
we are not aware of the existence of measurements or calculations of the binary
diffusion coefficients for ozone in any mixture. Following the rationale
provided in Section \ref{sssec:math_new}, i.e. that the concentration of
$\mathrm{O_3}$ is small and the relative concentrations of $\mathrm{N_2}$ and
$\mathrm{O_2}$ are constant, and given such limitation, it seems that there is
little else to be done at this moment other than directly using the
single-component diffusion coefficient.

    Regarding the mixture, it was mentioned in Section \ref{sssec:math_new}
that the Beta model uses a simplified formulation for the fluid's species. As
reported in \cite{lopes15}, the experimental device used
$\mathrm{O_3}$-enriched air as working fluid, and thus a first approach would
be to model its species as a $\mathrm{O_3/O_2/N_2}$ fluid. However it is known
that only the $\mathrm{O_3}$ is reactive in the experimental conditions used,
and that $\chi_{\mathrm{O_3}}\sim 10^{-3}$ at the inlet. Thus one might
consider that $\chi_{\mathrm{O_2}}$ and $\chi_{\mathrm{N_2}}$ are approximately
constant. Indeed, this is the same rationale behind the diluted species
approach used previously\cite{beruski17}. This has two implications, namely i)
the fluid's properties, i.e. viscosity and density, are largely due to
$\mathrm{O_2}$ and $\mathrm{N_2}$, and thus very much like air; and ii) the
chemical driving forces acting on $\mathrm{O_3}$ are mainly given by $\nabla
\chi_{\mathrm{O_3}}$ in a bath of air. Therefore, an alternative would be to
model the fluid using effective ``air'' molecules along with $\mathrm{O_3}$. In
this way, one saves computational resources while at the same time reducing the
hardships of measuring, or calculating, the binary diffusion coefficients for
each pair of species in the fluid at varied concentrations.

    Additional simulations were carried out using the proper $\mathrm{O_3/O_2/
N_2}$ mixture, however still using the same diffusion coefficient of Table
\ref{tab:par} for $\mathrm{O_3-O_2}$ and $\mathrm{O_3-N_2}$ pairs. Section SII
of the SM\footnotemark[3] provides additional information on the simulations,
as well as the results (Figs. S1 to S3). While a reduced geometry was used for
these simulations (shown in Fig. \ref{fig:var_track}), they are compared
against results for the $\mathrm{O_3/air}$ mixture using the same geometry. For
all cases, scalar variables, $\mathrm{O_3}$ profiles and surfaces, the
differences are negligible, particularly when considering the estimated
discretization errors. Notably, Fig. S3 shows the difference in $\mathbf{P}
_{\mathrm{O_3}}$ between mixtures, where it is shown that, while a pattern can
be seen, indicating a clear physical effect, the differences in $\mathrm{O_3}$
partial pressure are $\sim 0.1\ \mathrm{Pa}$, which amounts to $\sim 0.1\%$ of
the values observed. Thus, considering both the error estimates and the order
of magnitude of the experimentally relevant variables, it is safe to assert
that, for the moment, the $\mathrm{O_3/air}$ approximation is adequate.

    Concerning the coupling between diffusion modes, it was pointed out in
Section \ref{sssec:math_new} that it was not entirely clear at this point.
By that it is meant the mathematical relation that expresses the effective
diffusion coefficient that would be measured in a macroscopic device, $D_i$.
Here it was chosen to apply a correction to the free diffusion coefficient,
due to the porous media, which was then coupled to Knudsen diffusivity.
Physically speaking, it means that regardless of the distribution of
characteristic lengths in the pore network, the existence of such network and
its tortuosity affects molecular diffusion, while the Knudsen regime exists
alongside it. In appropriate circumstances, the Knudsen regime becomes dominant
($\mathrm{Kn} \sim 1$ and above), and the macroscopic description of the porous
medium, using porosity and tortuosity, breaks down.

    In order to assess the importance of this effect in the Beta model,
additional numerical simulations were carried with the reduced geometry
mentioned above and the $\mathrm{O_3/air}$ mixture. This was chosen, in
addition to the justification presented above, as it becomes straightforward
to apply different coupling schemes and to verify the contributions of each
diffusion mode, given that in such case there is only one entry in the
Maxwell-Stefan diffusion matrix, and thus a single generalized Fick diffusivity
is calculated. Additional information can be found in Section SIII of the
SM\footnotemark[3], and the simulations' results are shown in Figs. S4 to S9.
Five settings were tested, although due to limitations of the software, only
two couplings involving all processes were considered, namely the one described
in Section \ref{sssec:math_new} and one were the porous media correction is
applied after coupling free and Knudsen diffusivities. The results show that
the most important effect, for the simulated device, is the porous media
correction. This is in line with previous work\cite{beruski17}, where it was
argued that the low Knudsen numbers ($\mathrm{Kn} \sim 10^{-3}-10^{-2})$)
calculated for the device implied small contributions from Knudsen diffusion.
Indeed, Figs. S8 and S9 shows the differences in $\mathbf{P}_{\mathrm{O_3}}$
between the chosen coupling, Eq. \ref{eq:dcs}, and the other four settings,
where a clear spatial pattern and larger partial pressure difference are seen
when the alternative setting lacks porous media corrections. The differences
due to a lack of Knudsen diffusion are of the same order as the difference
between both settings including all effects ($\Delta P_{\mathrm{O_3}}\sim 0.01\
\mathrm{Pa}$), however showing a noisy spatial pattern, while the latter shows
a clear pattern. Considering the brief discussion given in \cite{beruski17},
concerning differences between concentrated and diluted species approaches,
these results suggest that concentration and pressure gradients, considered in
the Maxwell-Stefan formulation, are also more important than Knudsen diffusion
for the experimental device under consideration.

    Finally, a brief discussion on the difference in behavior between Alpha and
Beta models for $\Delta\chi_{\mathrm{O_3}}\left(Q\right)$. As pointed out in
Section \ref{ssec:comp_res}, the behavior of the Alpha model resembles the one
using the SD formulation in \cite{beruski17}, while the Beta model resembles
the one for the DB formulation. To be clear, here both Alpha and Beta models
use the DB formulation for free and porous media flow. This raises the question
if, given the new mesh and solver schemes, the DB is actually superior to the
SD formulation as previously claimed.

    This was investigated, and it is given in Section SIV.C of the
SM\footnotemark[3], with results shown in Figs. S15 to S20. Briefly, it is seen
that, for the current mesh, the fluid flow formulation affects the Alpha and
Beta similarly, with overall differences between SD and DB formulation being
small and following the outline previously established in \cite{beruski17}. For
the scalar response variables (Figs. S15 and S16), the SD formulation predicts
slight lower values for $K'$, $\Delta P/P_{\mathrm{in}}$, and $\Delta P$, while
predicting slightly higher values for $R'_{\mathrm{O_3}}$, where it should be
noticed that it affected the Beta model slightly more strongly, indicating
larger differences at higher $Q$. Noteworthy results are seen for $\lambda$ and
$\Delta\chi_{\mathrm{O_3}}$. For $\lambda$, the SD formulation predicts lower
values for both Alpha and Beta models as well, however the differences are
significantly larger than for other response variables, with the Beta model
showing larger differences for higher $Q$. Considering the error estimates in
Fig. S14, even relatively small gains in precision and additional data for
higher $Q$ might be able to provide significant distinction between models. For
$\Delta\chi_{\mathrm{O_3}}$, and consequently $\lambda'$, the SD formulation
affects each model in slightly different ways: for the Beta model, again a
slight increase in values is seen; however, for the Alpha model, a small
increase in the slope of the curve is seen. In this way, the difference between
formulations, and models using a given formulation, increases with $Q$. This
provides another opportunity to allow significant distinction between
formulations and/or models in the future with relatively small investment in
precision and larger parameter ranges.

    Considering now the spatially-resolved response variables, notable
differences are seen for $U(z)$ and $P_{\mathrm{O_3}}$. Concerning the $U$
$z$-profiles (Fig. S17), in the Pm domains there is a significant mismatch
between the SD and DB formulations. A comparison with Figure 5 of
\cite{beruski17} will not only show a much better-resolved $U$ profile for all
frameworks, but also closer $U$ values between formulations. The discrepancy
lies in the discontinuities in $U$ at domain interfaces and the profile inside
the Pm domains. The discontinuities between the Ch and MPS domains as well as
the difference in profile close to the upper CL boundary were already noted in
\cite{beruski17}. The boundary conditions involved in both the coupling between
free and porous media flow and with domain walls knowingly differ between the
SD and DB formulations\cite{ochoa95a,ochoa95b,lebars06}, and given deeper
scrutiny might be improved or even reconciled. The $U$ jump between different
materials, however, is unlikely to be a proper description of reality, or an
artifact due to a coarse mesh, as there is evidence of proper continuity in $U$
when using Darcy's law (see for instance \cite{ye15}). This is possibly an
artifact of how the software couples free and porous media flow, perhaps
dealing with differences in porosity and permeability in the same way no matter
the case. Nevertheless, these results are interpreted as evidence that the SD
formulation, as it has been applied here and using the COMSOL
Multiphysics$^\circledR$ software, provides a poorer description of coupled
free and porous media flow than the DB formulation.

    Considering the $P_{\mathrm{O_3}}$ and $\bar{R}_{\mathrm{O_3}}$
profiles (Fig. S18), the most prominent differences are well in line with
\cite{beruski17}, with the SD formulation showing overall values
of $P_{\mathrm{O_3}}$ and larger plumes after turns for both models. It is
noteworthy that all frameworks show distinctions that might be amenable for
validation, for instance employing the SD formulation with the Beta model
raises the overall $P_{\mathrm{O_3}}$ values to those of the Alpha model
employing the DB formulation. However, the local minima in the profiles, i.e.
the sections after a given turn and approaching the following one, differ
between fluid flow formulations: the DB approach predicts a lower, short minima
right after the local maximum, followed by a plateau preceding the next rise;
the SD formulation, on the other hand, predicts a single, longer minimum
between peaks. This, it should be reminded, for the a given model, say Beta,
and same values for the reaction rate constants. This follows after the results
of \cite{beruski17}, with clearer differences that strengthen the hypothesis
that they are caused by the fluid flow formulation, and the additional
information that they are not artifacts of a diluted-mixture approach to
species transport. With increased computational and experimental precision,
these results will prove to be important in further validating and improving
the mathematical models. The same reasoning seems to apply to the $\bar{R}
_{\mathrm{O_3}}$ profiles, although smaller differences are seen between the
Beta model using the DB formulation and the Alpha model using either fluid flow
formulation. The most notable difference in this case lie on local minima
between peaks, where, for the Beta model at higher $Q$, there is a significant
shift forward when changing from the DB to the SD formulation. The differences
are rather small, however, and it is unclear if it would provide any
advantage over the $P_{\mathrm{O_3}}$ profiles. It could be argued that the
$\bar{R}_{\mathrm{O_3}}$ profiles emphasize changes in reaction kinetics,
mixing the effects with those of changes in the fluid flow formulation. In this
way, these may be more suited to investigate the reaction kinetics after the
debate between SD and DB formulation is settled.

    Taking into account the discussions above concerning known limitations of
the Beta model, it is evident that despite the advances achieved in the
mathematical formulation, there are additional points that demand further
improvements. A Gamma model would need to consider the anisotropy and
inhomogeneity of porous media, as well as solve the conundrum involving Knudsen
diffusivity. Perhaps the dusty-gas model used in solid oxide fuel cell modeling
might solve the latter\cite{andersson10,hajimolana11}. Ideally, in order to
further reduce the gap between the prototype PEFC and an actual device, heat
transport should also be considered, as well as two-phase flow. That would
demand modifications in the experimental setup, as well as additional data on
the properties of the materials and species involved. Moving away from
phenomena under broad scrutiny, there is the discrepancy between ozone plume
sizes to solve, briefly discussed in Section \ref{ssec:comp_res} (see also Part
II and \cite{beruski17}). In this case, two hypothesis are currently under
consideration, i) inhomogeneous mechanical deformation of the MPS, and
ii) molecular slip velocities. The first stems from the way PEFC devices are
usually sealed, using screws distributed around the core of the device, while
the second is the breakdown of the no-slip boundary condition, which asserts
that $\mathbf{u}=0$ at the interface between solid and fluid. It appears that
both are known to play a role in PEFCs, and some work has been done on both for
different reasons (see \cite{nitta07,kleemann09,elkharouf12,bakhshian16} as
well as \cite{millichamp15} and references therein; and \cite{priezjev05,qian05,
wu08,ho11}, respectively). How important would they be in actual fuel cells,
however, appears little understood, as the complexity of including non-linear
solid mechanics and the molecular interplay between fluid and solid species in
already hard-to-solve differential equations is likely a powerful factor
stymieing such research. Nevertheless, until it is done and calculated, one can
only speculate.

\subsection{\label{ssec:ks_res}Parametric Study of Reaction Rate Constants}

    For our final task, as described in Section \ref{ssec:ks}, a parametric
study on the reaction rate constants, $k_1$ and $k_2$, associated with ozone
adsorption and decomposition reactions respectively, was carried out. As
pointed out in Section \ref{ssec:comp_res}, when the discussing the difference
in behavior of $\Delta\chi_{\mathrm{O_3}}$ for the Alpha and Beta models (Fig.
\ref{fig:AxB_scalars}), these are degrees of freedom in the Beta model, as is
$k_{\mathrm{app}}$ for the Alpha model, and no additional empirical or
theoretical information is currently known by the authors. Hence the need
for an assessment of the solution of the computational models as function of
these parameters, as was done for the Alpha model in \cite{beruski17}.

    We start the analysis with a consideration on the computational models
under analysis. Such a parametric study is dependent on the mesh used for the
calculations. This is clear from the changes in $\Delta\chi_{\mathrm{O_3}}$,
as pointed out in Section \ref{sssec:disc_AxB}, as well as from the grid
convergence study carried out in Part 2. For instance, when considering the
parametric study carried out for $k_{\mathrm{app}}$ for the Alpha
model\cite{beruski17}, it was shown that, for the mesh used then, values close
to $k_{\mathrm{app}} = 250\ \mathrm{s^{-1}}$ are a close fit to the
experimental data available\cite{beruski17}. This value was later refined to
the one currently in use, $k_{\mathrm{app}} = 256.15\ \mathrm{s^{-1}}$.
However, as seen in Fig, \ref{fig:AxB_scalars}, when using a finer mesh
$\Delta\chi_{\mathrm{O_3}}$ changes both quali- and quantitatively, thus a
more appropriate value for $k_{\mathrm{app}}$ could be calculated.

    With that in mind, we begin with a preliminary result: the Beta model did
not converge for $k_1 = 10^4\ \mathrm{s^{-1}}$, already showing difficulties
for $10^3\ \mathrm{s^{-1}}$. While this could potentially be solved by
adjusting the numerical solvers, results shown below suggest that it was due to
excessive consumption of ozone, which could lead to unrealistic solutions. In
this way, the parametric study was able to probe the following ranges:
$k_1 \in \left[1,10,10^2,10^3\right]\ \mathrm{s^{-1}}$ and $k_2\in\left[10^{-1},
1,10,10^2,10^3\right]\ \mathrm{s^{-1}}$.

\begin{figure}
    \centering
    \begin{subfigure}{0.45\textwidth}
        \includegraphics[width=\textwidth]{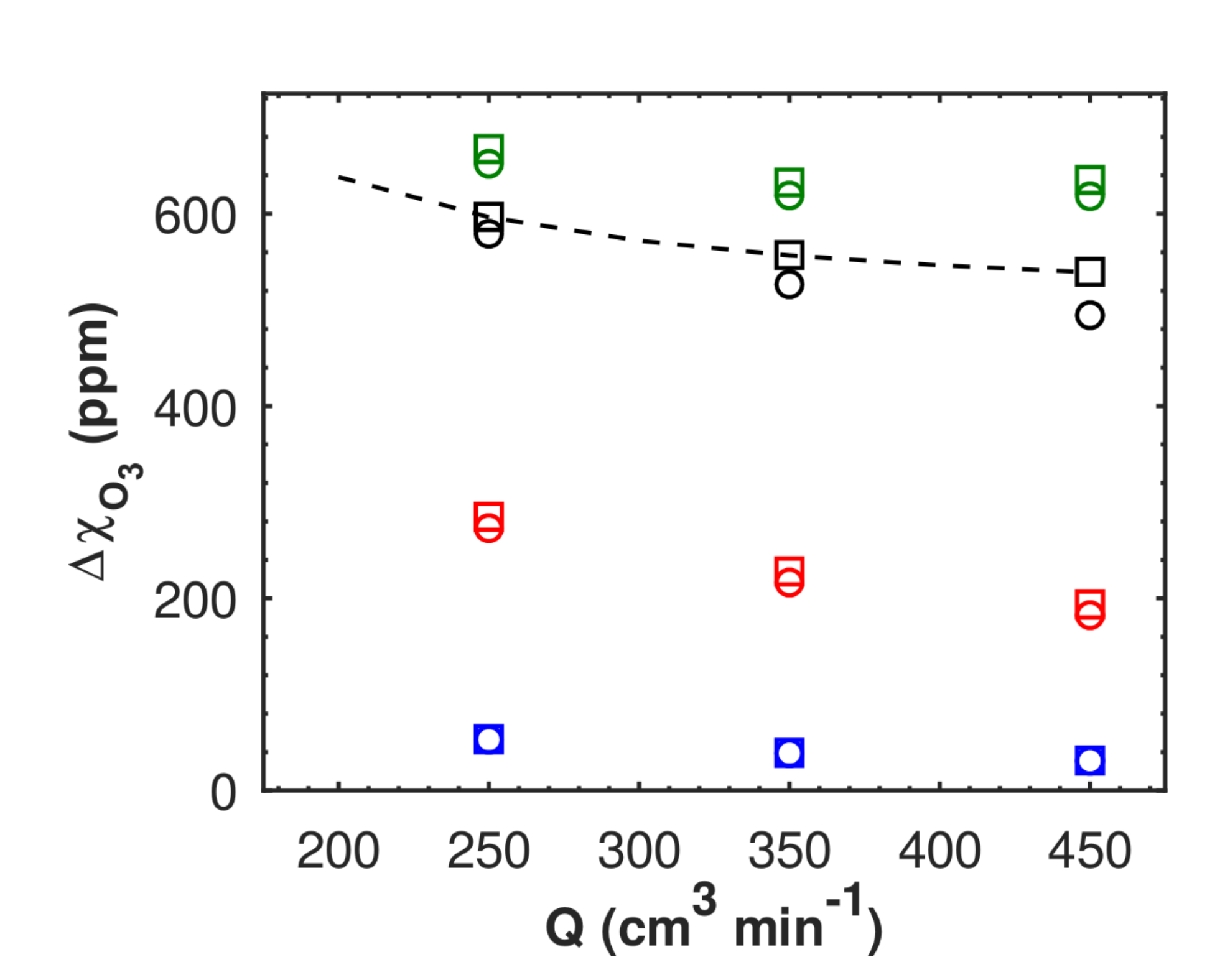}
        \caption{}
        \label{fig:ks_dchi}
    \end{subfigure}
    ~
    \begin{subfigure}{0.45\textwidth}
        \includegraphics[width=\textwidth]{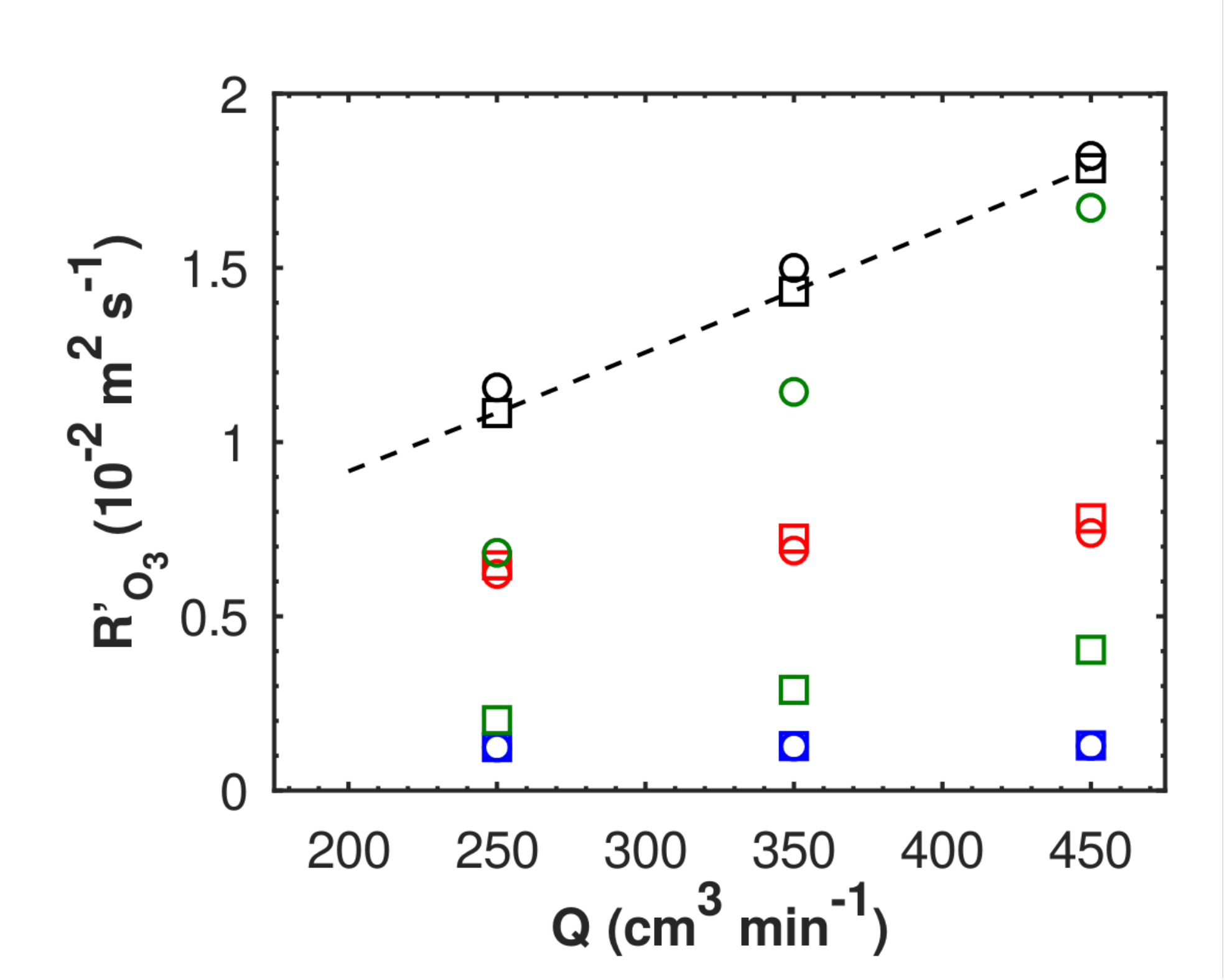}
        \caption{}
        \label{fig:ks_Rapp}
    \end{subfigure}

    \begin{subfigure}{0.45\textwidth}
        \includegraphics[width=\textwidth]{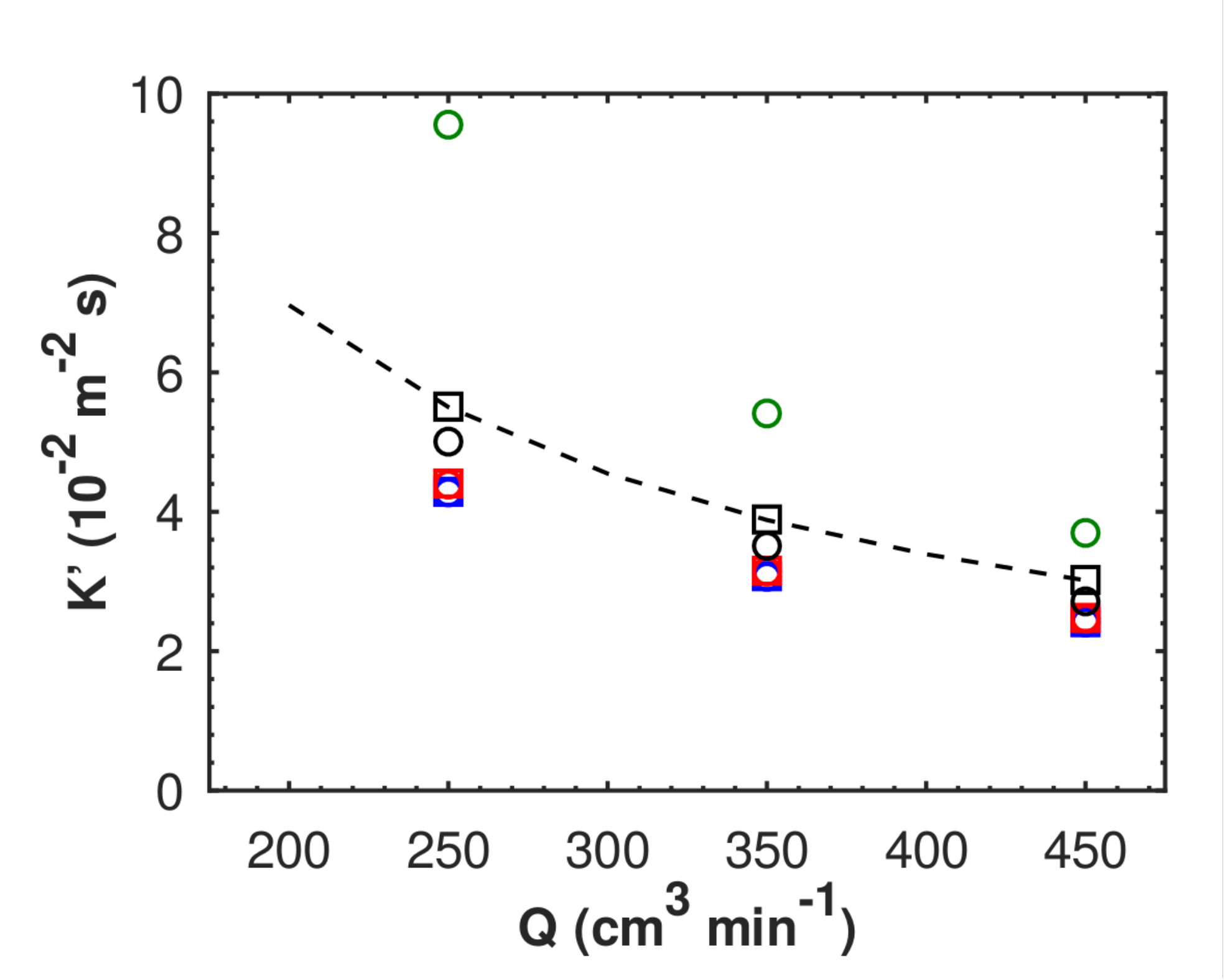}
        \caption{}
        \label{fig:ks_Kapp}
    \end{subfigure}
    ~
    \begin{subfigure}{0.45\textwidth}
        \includegraphics[width=\textwidth]{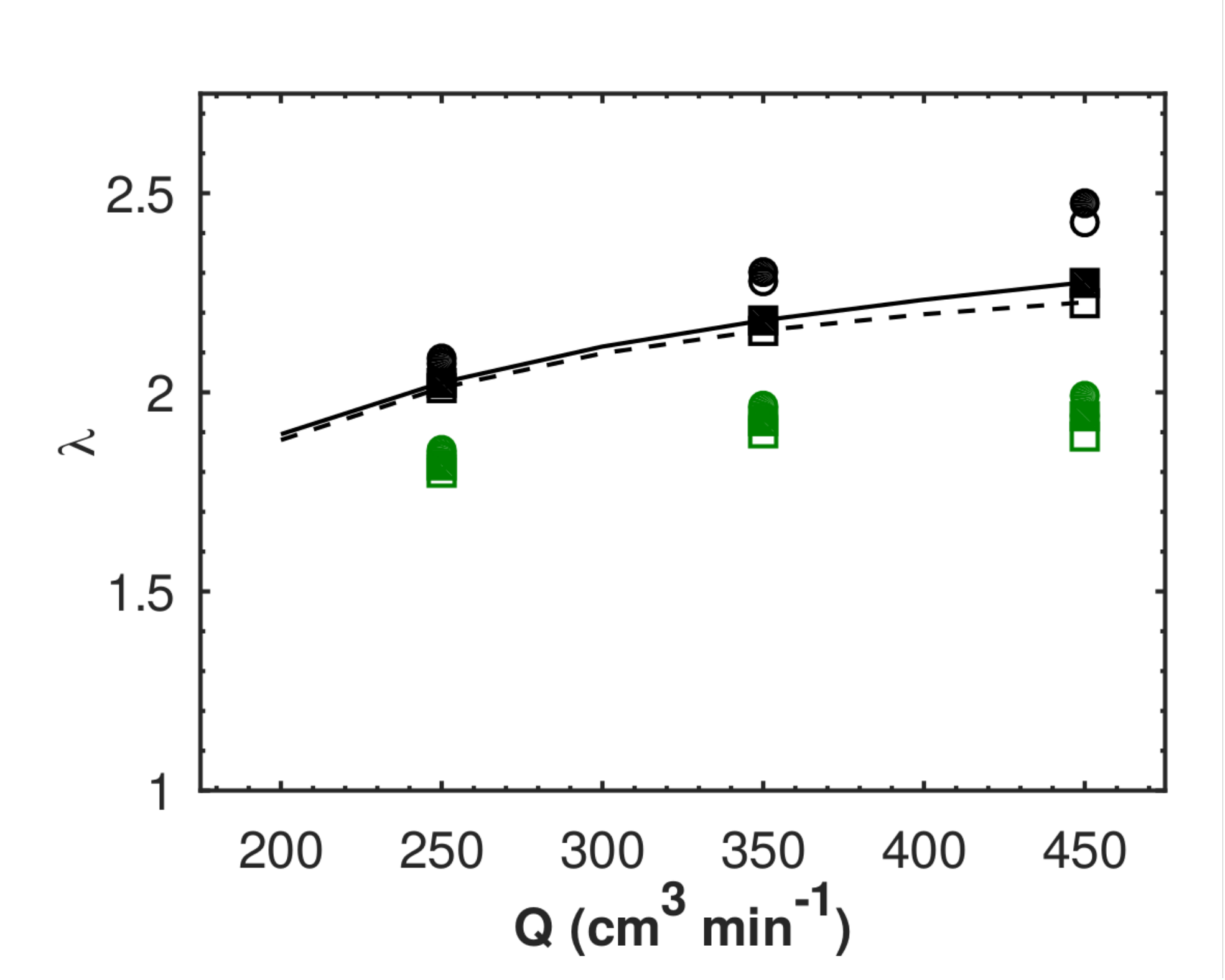}
        \caption{}
        \label{fig:ks_lambdas}
    \end{subfigure}
    \caption{
        Scalar variables as function of the inlet flow rate for the Beta
        model, using different combinations of $k_1$ and $k_2$:
        \textbf{(a)} $\Delta\chi_{\mathrm{O_3}}$, \textbf{(b)} $R'
        _{\mathrm{O_3}}$, \textbf{(c)} $K'=\Delta\chi_{\mathrm{O_3}}/R'
        _{\mathrm{O_3}}$, and \textbf{(d)} $\lambda$ (empty) and $\lambda'$
        (full). Different colors map the values of $k_1$: \textcolor{blue}{$1$},
        \textcolor{red}{$10$}, $10^2$, and \textcolor[rgb]{0 0.5 0}{$10^3$}$\
        \mathrm{s^{-1}}$; while the values of $k_2$ are mapped by different
        symbols: $10^{-1}$ ($\circ$) and $10^3\ \mathrm{s^{-1}}$ ($\Box$). The
        Beta model, using the same reduced geometry, is shown as reference
        (dashed and full lines), with $k_1$ and $k_2$ given in Tab.
        \ref{tab:par2}.
    }
    \label{fig:ks_scalars}
\end{figure}

    Moving to the results proper, Figure \ref{fig:ks_scalars} presents the
scalar response variables: the $K'= \Delta\chi_{\mathrm{O_3}}/R'
_{\mathrm{O_3}}$ ratio, as well as the individual variables, and the
stoichiometries $\lambda$ and $\lambda'$. Analyzing first the individual
variables, $\Delta\chi_{\mathrm{O_3}}$ (Fig. \ref{fig:ks_dchi}) behaves as
expected, with increased values as both $k_1$ and $k_2$ increases. However it
becomes clear that $k_1$ is much more important in defining the absolute value
of $\Delta\chi_{\mathrm{O_3}}$, with $k_2$ mainly contributing at higher $Q$,
providing fine adjustment to the behavior of the curve (see Fig. S21 for the
relative change in $\Delta\chi_{\mathrm{O_3}}$ as $k_2$ increases for fixed
$k_1$). In addition, it is seen that the increase in $\Delta\chi
_{\mathrm{O_3}}$, when moving from $k_1 = 10^2$ to $10^3\ \mathrm{s^{-1}}$, is
significantly smaller than observed for other intervals. This is also expected,
as there is a limit to the mass transport from the Ch to CL domains given
intrinsically by $Q$ and the idiosyncrasies of the device (characterized by
the Damköhler numbers).

    The $R'_{\mathrm{O_3}}$ (Fig. \ref{fig:ks_Rapp}), on the other hand, shows
more nuanced results. Up until $k_1 = 10^2\ \mathrm{s^{-1}}$ the data conforms
with expectations and Fig. \ref{fig:ks_dchi}. However, for $k_1 = 10^3\
\mathrm{s^{-1}}$, there is a sudden decrease in $R'_{\mathrm{O_3}}$ values.
This is interpreted as a lack of reactants reaching the upper CL boundary,
where $R'_{\mathrm{O_3}}$ is calculated, given the high $k_1$ value. Indeed,
Fig. S22 of the SI, which shows the reaction rate integrated over the CL
domain, corroborates this reasoning by showing a trend compatible with Fig.
\ref{fig:ks_dchi}, i.e. with values for $k_1 = 10^3\ \mathrm{s^{-1}}$ being
higher than all others. This naturally explains the results shown in Fig.
\ref{fig:ks_Kapp}, for the $K'$ ratio, which shows significantly higher values
for $k_1=10^3\ \mathrm{s^{-1}}$ in comparison to $k_1 = 10^2\ \mathrm{s^{-1}}$,
whereas the remaining results for $k_1\leq 10^2\ \mathrm{s^{-1}}$ shows similar
values and trends in $Q$ (see Fig. S22 for the full range of $k_1$ and $k_2$).
Interestingly, the values of such a ratio obtained when using the reaction rate
integrated over the CL domain, shown in Fig. S22, are very close regardless of
the values of $k_1$ and $k_2$.

    The stoichiometries $\lambda$ and $\lambda'$ (Fig. \ref{fig:ks_lambdas})
are shown for $k_1 \geq 10^2\ \mathrm{s^{-1}}$, since the range for $k_1 \leq
10\ \mathrm{s^{-1}}$ was significantly larger and out of touch with
experimental results (see Fig. S23 for the full range). Both $\lambda$ and
$\lambda'$ shows the same trends with increasing $k_1$ and $k_2$: reduced
values and slope as $k_1$ increases, and for a given mean value, reduced slope
as $k_2$ increases. It should be pointed out that both $\lambda$ and $\lambda'$
show good sensitivity towards $k_2$, although slightly smaller than $K'$, when
considering the full range of $k_2$ investigated. What is truly noteworthy,
however, is that, relative to the error estimates from Part 2 (shown in Fig.
S13), $\lambda$ and $\lambda'$ shows much larger sensitivity than any other
scalar response variable, particularly at the high end of the $Q$ range where
it is one order of magnitude larger. The stoichiometries thus provide an
excellent opportunity to constrain the values of $k_2$, with other variables
providing constraints for $k_1$. Thus, with further improvements in both
computational and experimental precision, information on the reaction kinetics
may be available without the need to delve in the complexities of modeling
transient effects.

\begin{figure}
    \centering
    \begin{subfigure}{0.45\textwidth}
        \includegraphics[width=\textwidth]{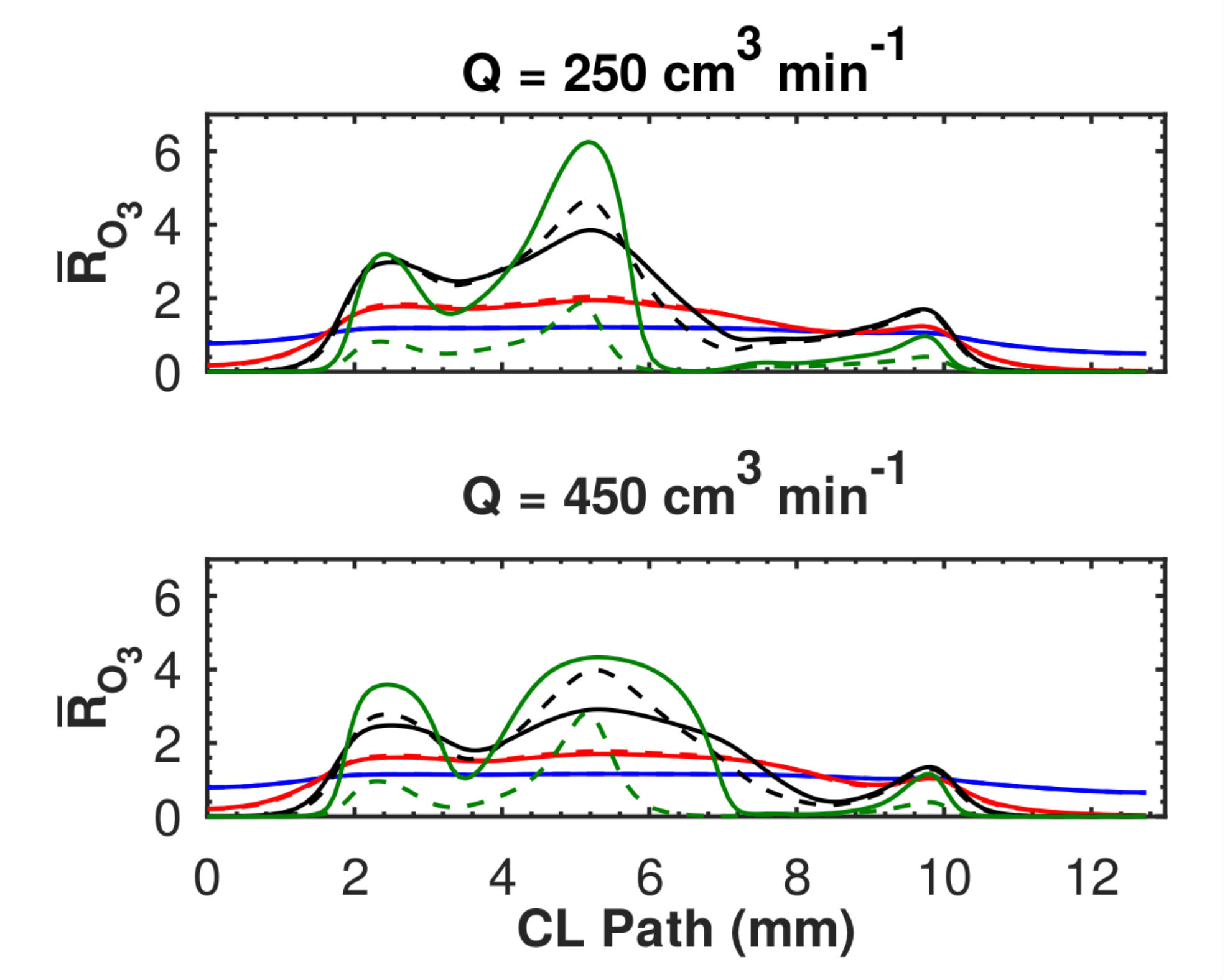}
        \caption{}
        \label{fig:ks1_Ro3x}
    \end{subfigure}
    ~
    \begin{subfigure}{0.45\textwidth}
        \includegraphics[width=\textwidth]{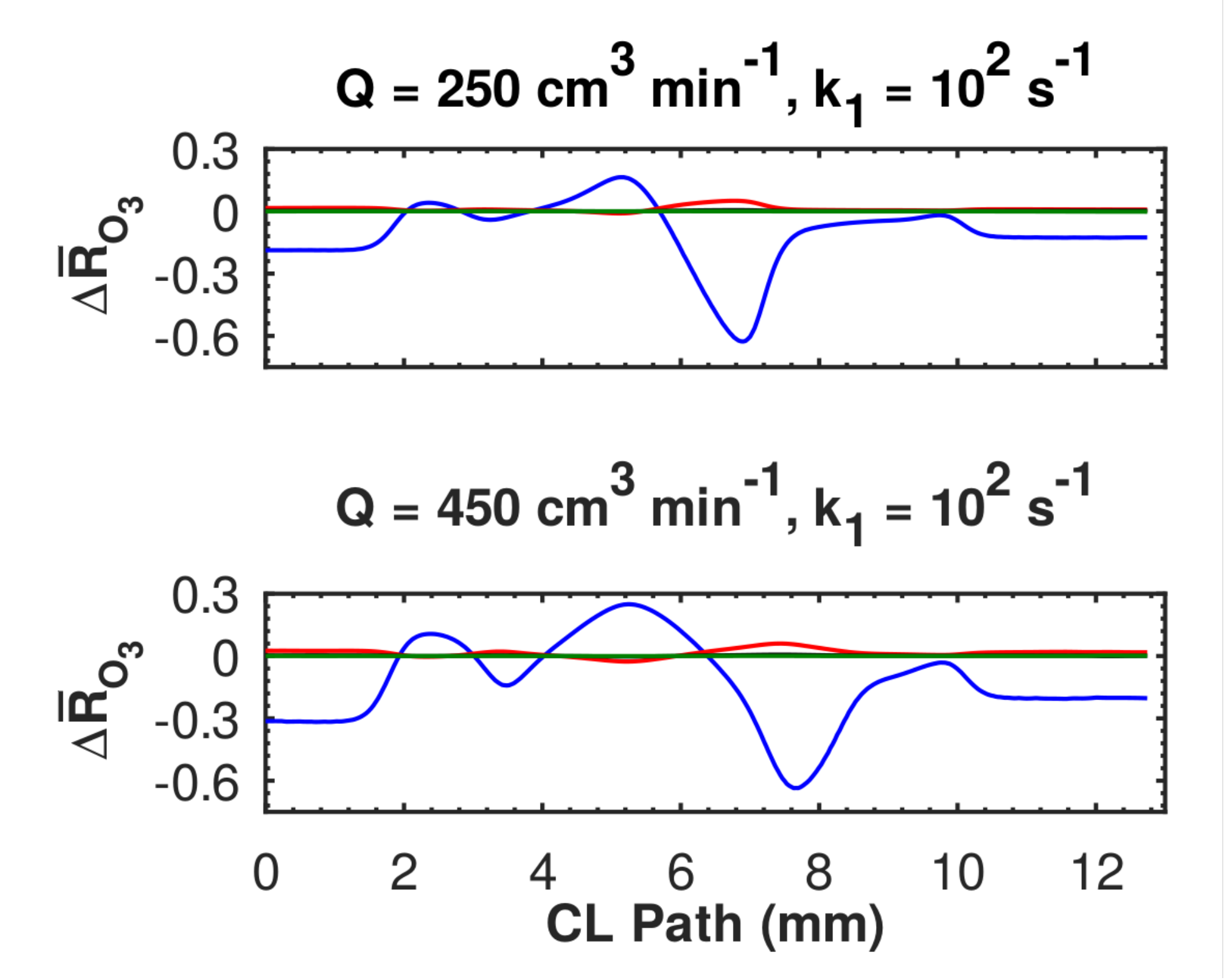}
        \caption{}
        \label{fig:ks2_Ro3x}
    \end{subfigure}
    \caption{
        Profiles associated with the normalized reaction rate, at $Q=250$ (top)
        and $450\ \mathrm{cm^3\  min^{-1}}$ (bottom), for selected values of
        $k_1$ and $k_2$. \textbf{(a)} Normalized reaction rate profiles, with
        $k_1$ = \textcolor{blue}{$1$}, \textcolor{red}{$10$}, $10^2$, and
        \textcolor[rgb]{0 0.5 0}{$10^3$}$\ \mathrm{s^{-1}}$; and $k_2$:
        $10^{-1}$ (full lines) and $10^3\ \mathrm{s^{-1}}$ (dashed lines).
        \textbf{(b)} Relative difference between profiles for increasing values
        of $k_2$, using $k_1 = 10^2\ \mathrm{s^{-1}}$:
        \textcolor{blue}{$k_2=1-k_2'=10^{-1}$}, \textcolor{red}{$k_2=10-k_2'=1$},
        $k_2=10^2-k_2'=10$, and \textcolor[rgb]{0 0.5 0}{$k_2=10^3-k_2'=10^2$}.
    }
    \label{fig:ks_Ro3x}
\end{figure}

    Moving forward, we turn now to the profiles, where once again focus will
be given to the $\bar{R}_{\mathrm{O_3}}$ profiles. Figure \ref{fig:ks_Ro3x}
shows the $\bar{R}_{\mathrm{O_3}}$ profiles for selected values of $k_1$, $k_2$,
and $Q$. A similar trend to Fig. \ref{fig:ks_Rapp} is seen for $\bar{R}
_{\mathrm{O_3}}$ magnitude in Fig. \ref{fig:ks1_Ro3x}, when comparing different
$k_1$ and $k_2$ values, including the inversion seen for $k_1 = k_2 = 10^3\
\mathrm{s^{-1}}$. Contrary to Fig. \ref{fig:ks_scalars}, however, Fig.
\ref{fig:ks1_Ro3x} shows that, for $k_1 \gtrapprox 10^2\ \mathrm{s^{-1}}$, the
value of $k_2$ shows greater influence on $\bar{R}_{\mathrm{O_3}}$ than for the
scalar variables, thus providing the possibility of additional validation for
$k_1$ and $k_2$. Fig. \ref{fig:ks2_Ro3x} provides a full comparison between
values of $k_2$ for $k_1 = 10^2\ \mathrm{s^{-1}}$, showing the relative
difference between increasing values of $k_2$, where it can be seen that
the largest changes are seen when moving from $k_2 = 10^{-1}$ to $1\
\mathrm{s^{-1}}$, with significantly less changes for $k_2 = 1$ to $10\
\mathrm{s^{-1}}$, and virtually no changes for $k_2 = 10$ to $10^2\
\mathrm{s^{-1}}$ ($\Delta\bar{R}_{\mathrm{O_3}}\sim 0.1\%$). Therefore, while
there is an opportunity for the validation of $k_2$, the range of values is
somewhat limited when using $\bar{R}_{\mathrm{O_3}}$ profiles, regardless of
the precision available from either experimental or computational data.

    The partial pressure profiles shows similar results to Fig.
\ref{fig:ks_Ro3x} and are shown in Fig. S24a. While the observed trends are
similar, comparing $P_{\mathrm{O_3}}$ values for model runs with different
values of $k_1$ and $k_2$ demands some caution, as the situation is similar to
the comparison between Alpha and Beta models (Fig. \ref{fig:AxB_prfls}).
Nevertheless, the relative difference in $P_{\mathrm{O_3}}$ profiles for
increasing values of $k_2$, shown in Fig. S24b for $k_1=10^2\ \mathrm{s^{-1}}$,
shows very similar sensitivity to the absolute value of $k_2$ as the
$\bar{R}_{\mathrm{O_3}}$ profiles of Fig. \ref{fig:ks_Ro3x}. Thus, while a
comparison between model runs with different $k_1$ and $k_2$ values might prove
redundant given $\bar{R}_{\mathrm{O_3}}$ profiles, the $P_{\mathrm{O_3}}$
profiles are currently available from experimental data, providing an excellent
opportunity for validation of the $k_1$ and $k_2$ values in addition to the
scalar response variables of Fig. \ref{fig:ks_scalars}.

\begin{figure}
    \centering
    \begin{subfigure}{0.3\textwidth}
        \includegraphics[width=\textwidth]{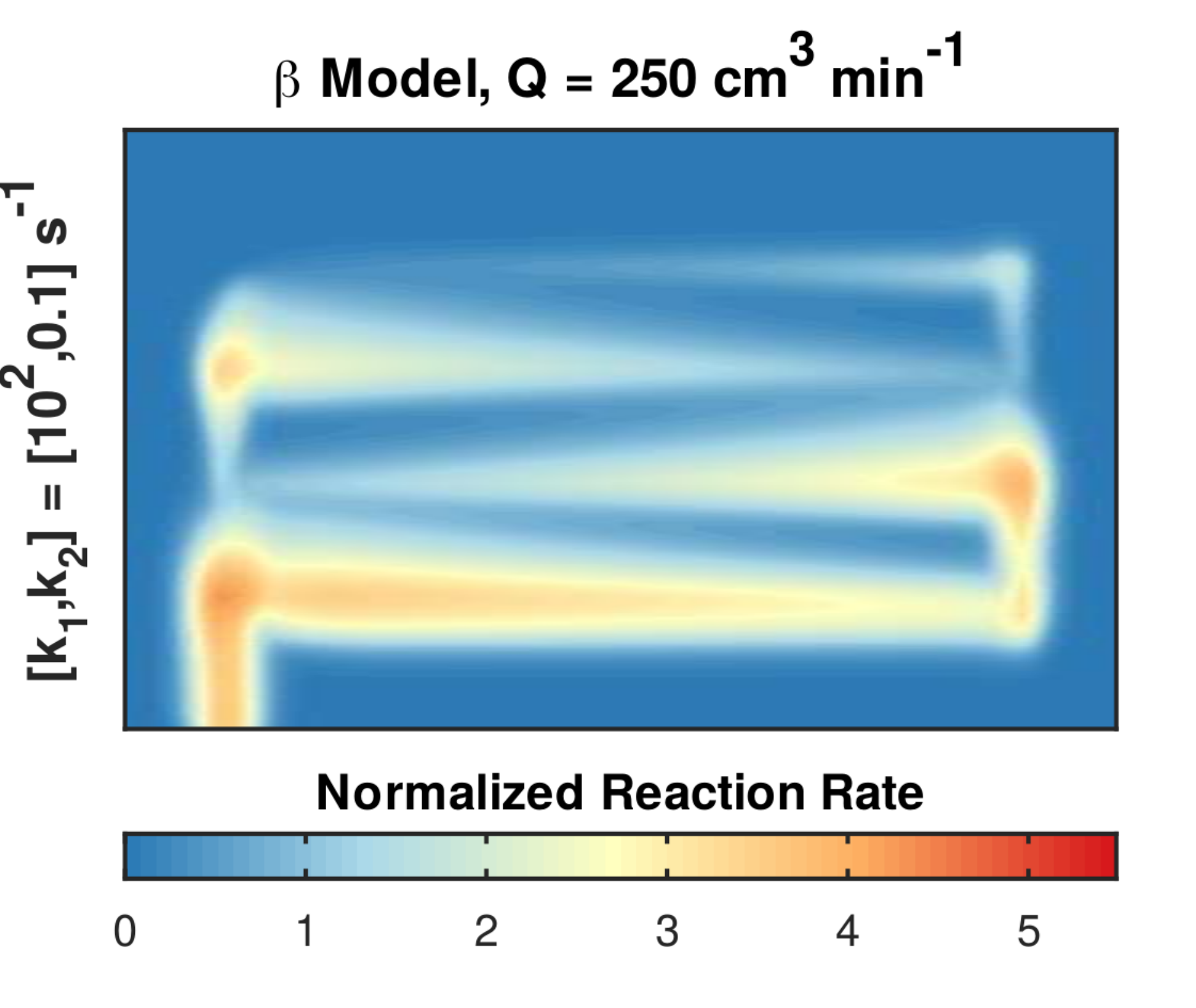}
        \caption{}
    \end{subfigure}
    ~
    \begin{subfigure}{0.3\textwidth}
        \includegraphics[width=\textwidth]{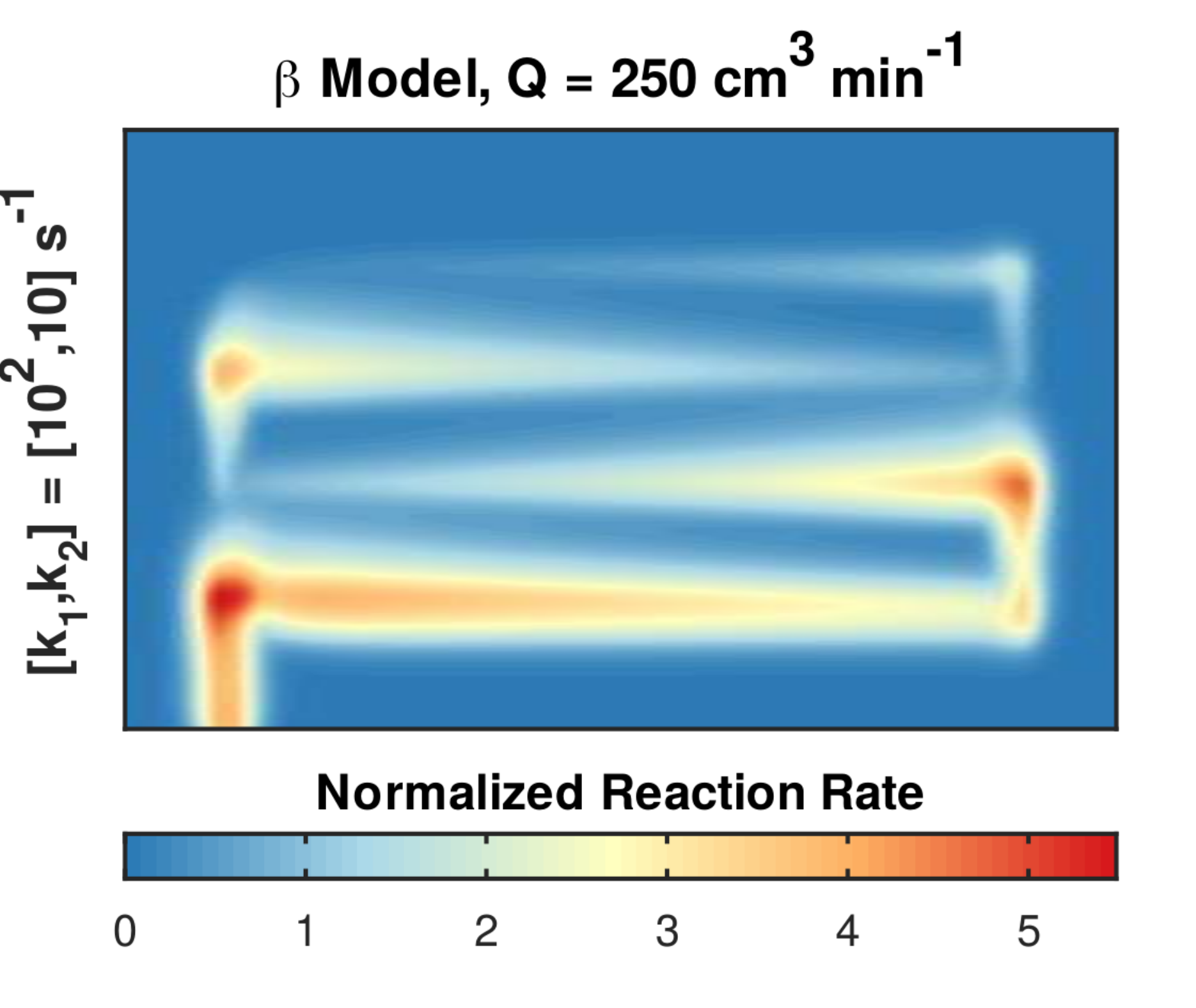}
        \caption{}
    \end{subfigure}

    \begin{subfigure}{0.3\textwidth}
        \includegraphics[width=\textwidth]{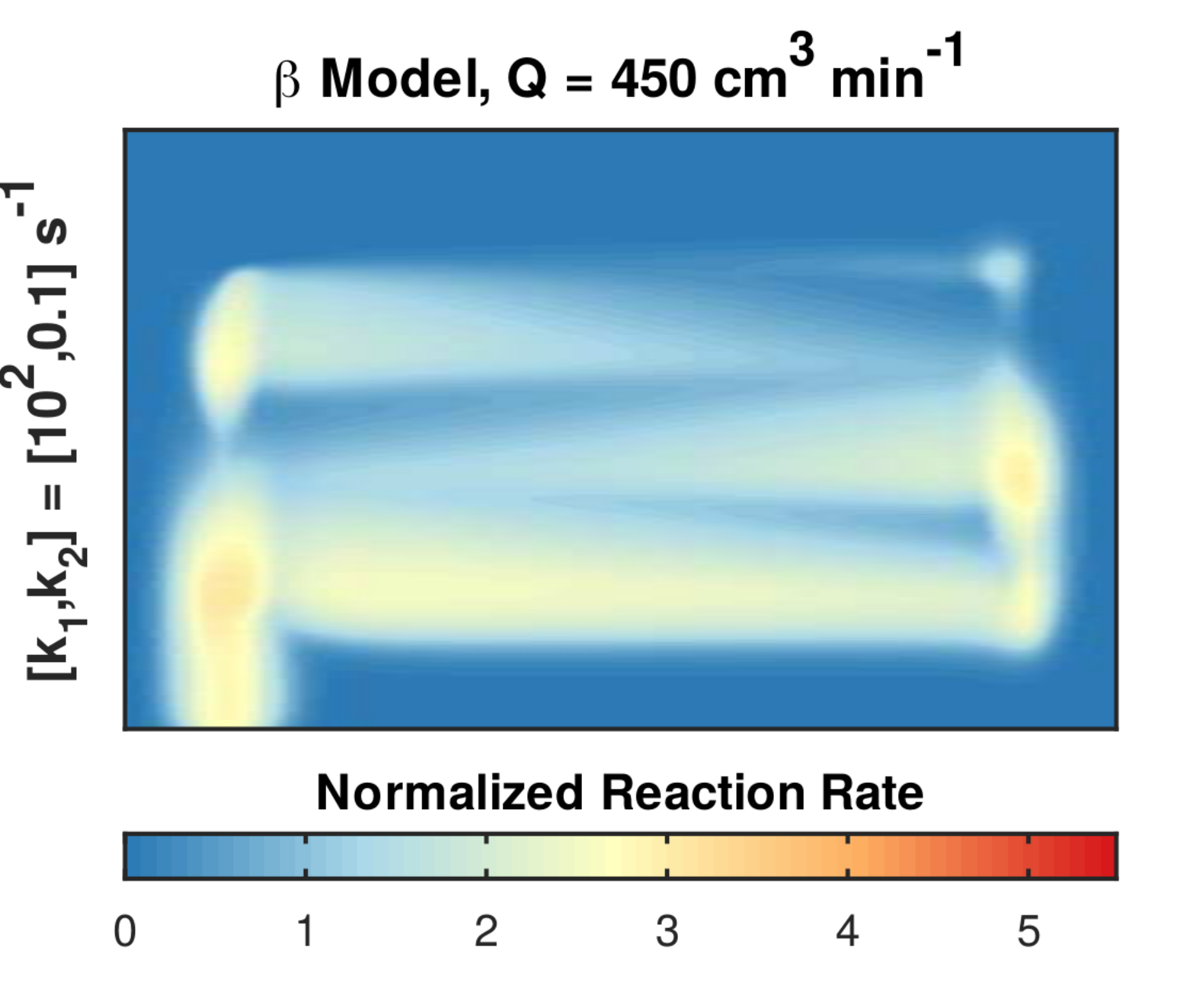}
        \caption{}
    \end{subfigure}
    ~
    \begin{subfigure}{0.3\textwidth}
        \includegraphics[width=\textwidth]{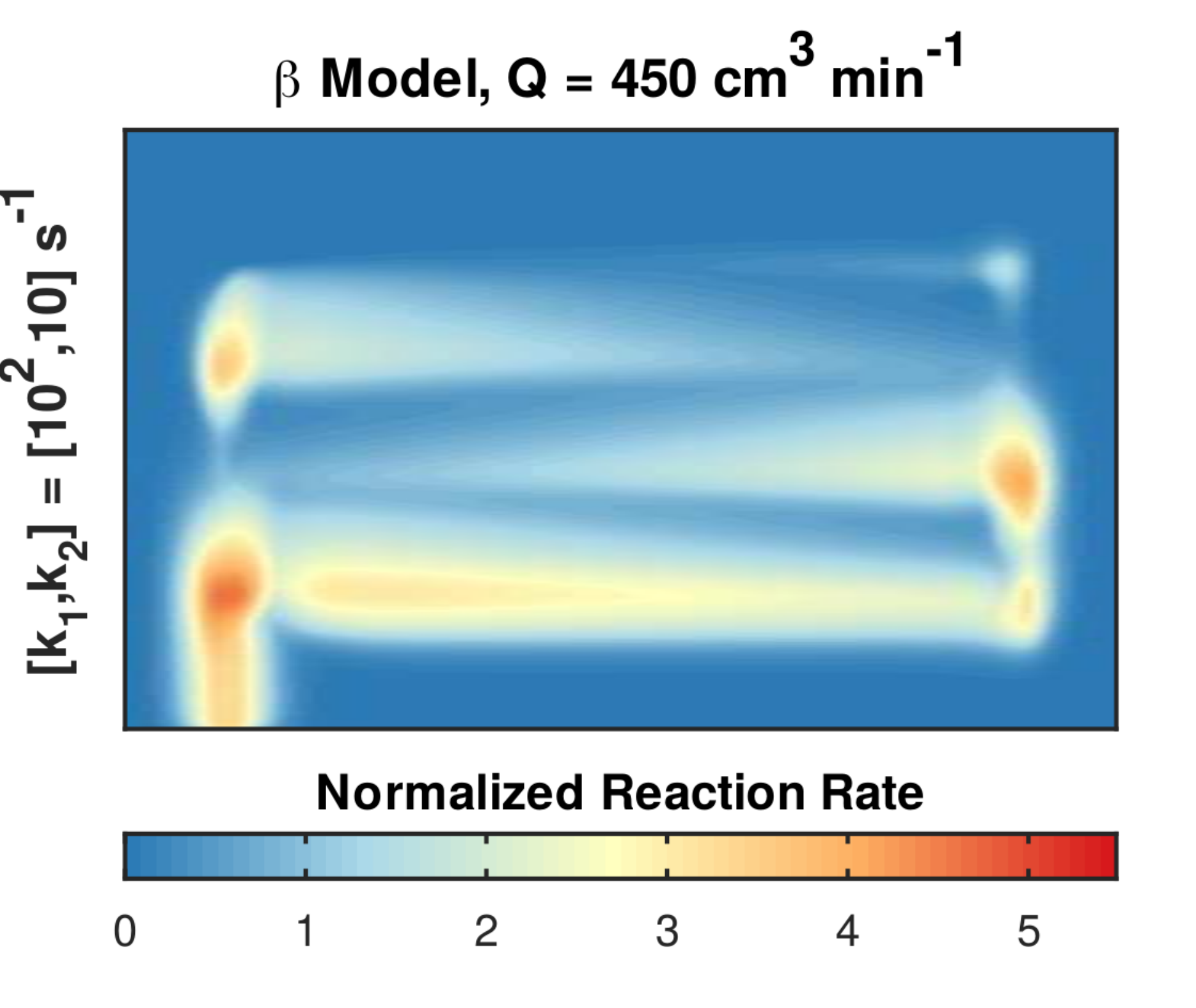}
        \caption{}
    \end{subfigure}
    \caption{
        Normalized reaction rate surfaces for $k_1 = 10^2\ \mathrm{s^{-1}}$,
        with $Q = 250$ (top row) and $Q = 450\ \mathrm{cm^3\ min^{-1}}$ (bottom
        row), and $k_2 = 10^{-1}$ (left column) and $10\ \mathrm{s^{-1}}$
        (right column).
    }
    \label{fig:ks_Ro3xy}
\end{figure}

    Finally, we move on to the spatially-resolved data, where focus will
naturally be given to $\bar{\mathbf{R}}_{\mathrm{O_3}}$ surfaces, in particular
to $k_1 = 10^2\ \mathrm{s^{-1}}$, since Fig. \ref{fig:ks_Ro3x} suggests $k_1
\gtrapprox 10^2\ \mathrm{s^{-1}}$ for the experimental device. Figure
\ref{fig:ks_Ro3xy} shows $\bar{\mathbf{R}}_{\mathrm{O_3}}$ surfaces, for $Q=
250$ and $450\ \mathrm{cm^3\ min^{-1}}$, for $k_1 = 10^2\ \mathrm{s^{-1}}$ and
$k_2 = \left[10^{-1},10,10^3\right]\ \mathrm{s^{-1}}$. Additional surfaces may
be found in the SI (Figs. S25 to S28). The trend observed as $k_2$ increases
follows what has been discussed above, with $\bar{\mathbf{R}}_{\mathrm{O_3}}$
showing virtually no changes when further increasing $k_2$ from $10$ to $10^3\
\mathrm{s^{-1}}$. There is novel information, however, in Fig.
\ref{fig:ks_Ro3xy} (and Figs. S25 to S28): $k_2$ apparently has an important
role in ozone dispersion within reactant plumes. This is clearer for $Q = 450\
\mathrm{cm^3\ min^{-1}}$, where it can be seen that $\bar{\mathbf{R}}
_{\mathrm{O_3}}$ is more homogeneously distributed around the Ch domain region
for $k_2 = 10^{-1}$ than for $10\ \mathrm{s^{-1}}$. Fig. S29 shows these same
surfaces normalized by their respective maximum values, reinforcing the point.
This suggests that, for a given value of $k_1$, $k_2$ modulates the
gradient $\boldsymbol{\nabla}\bar{R}_{\mathrm{O_3}}$ within the reactant plume,
whereas $k_1$ strongly impacts on the spread of the plume itself. It must not
be misunderstood that each reaction, and consequently each reaction rate
constant, is solely responsible for either aspect of $\bar{\mathbf{R}}
_{\mathrm{O_3}}$, as it is very likely that both $k_1$ and $k_2$ play a role in
the final, steady-state shape of $\bar{\mathbf{R}}_{\mathrm{O_3}}$. But rather
that, given the mechanism used to describe the ozone-coumarin interaction,
experimental data on the overall spread of the reaction plume may be used to
validate $k_1$, while data on the gradient may be used to validate $k_2$.

\begin{figure}
    \centering
    \begin{subfigure}{0.3\textwidth}
        \includegraphics[width=\textwidth]{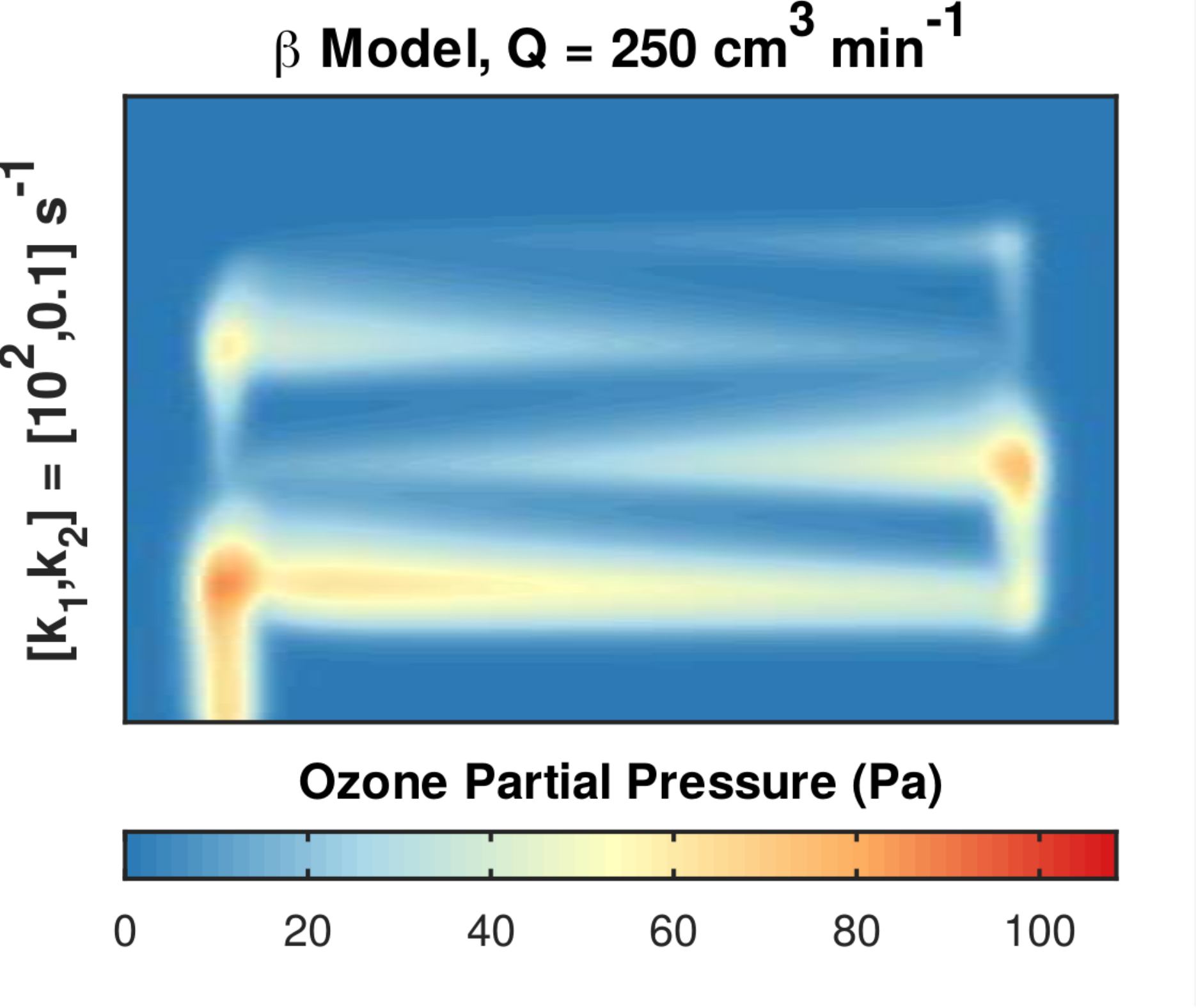}
        \caption{}
    \end{subfigure}
    ~
    \begin{subfigure}{0.3\textwidth}
        \includegraphics[width=\textwidth]{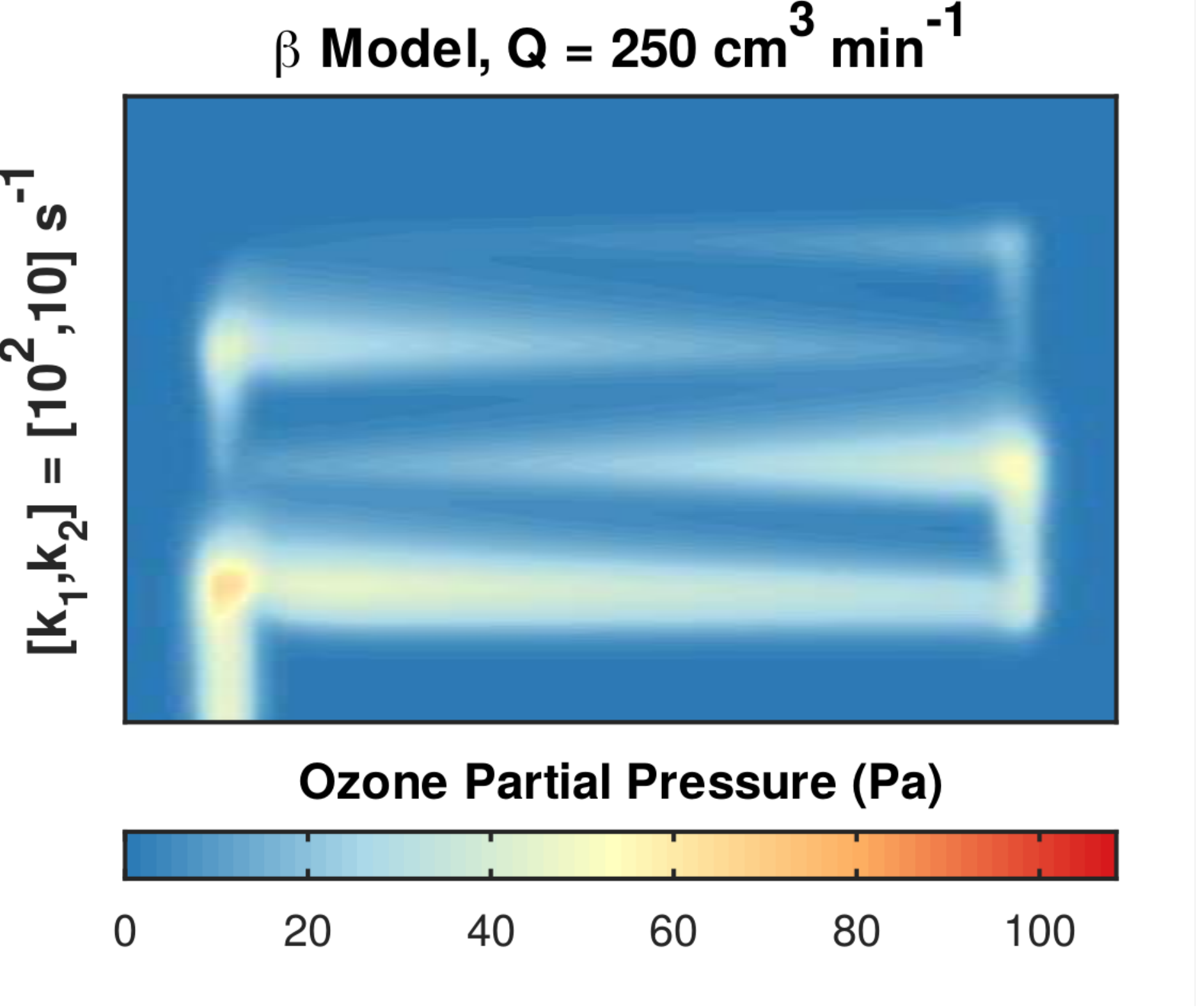}
        \caption{}
    \end{subfigure}

    \begin{subfigure}{0.3\textwidth}
        \includegraphics[width=\textwidth]{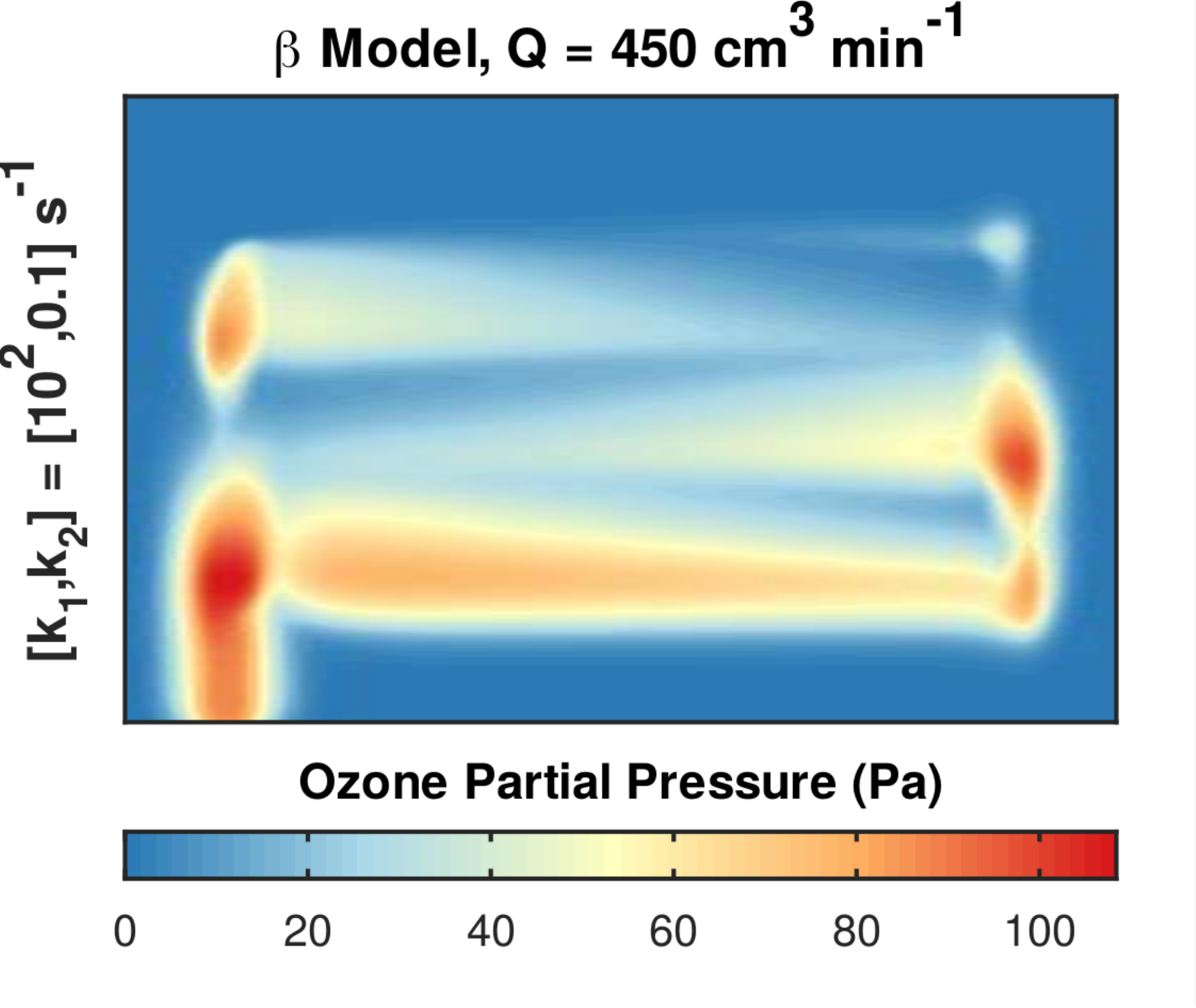}
        \caption{}
    \end{subfigure}
    ~
    \begin{subfigure}{0.3\textwidth}
        \includegraphics[width=\textwidth]{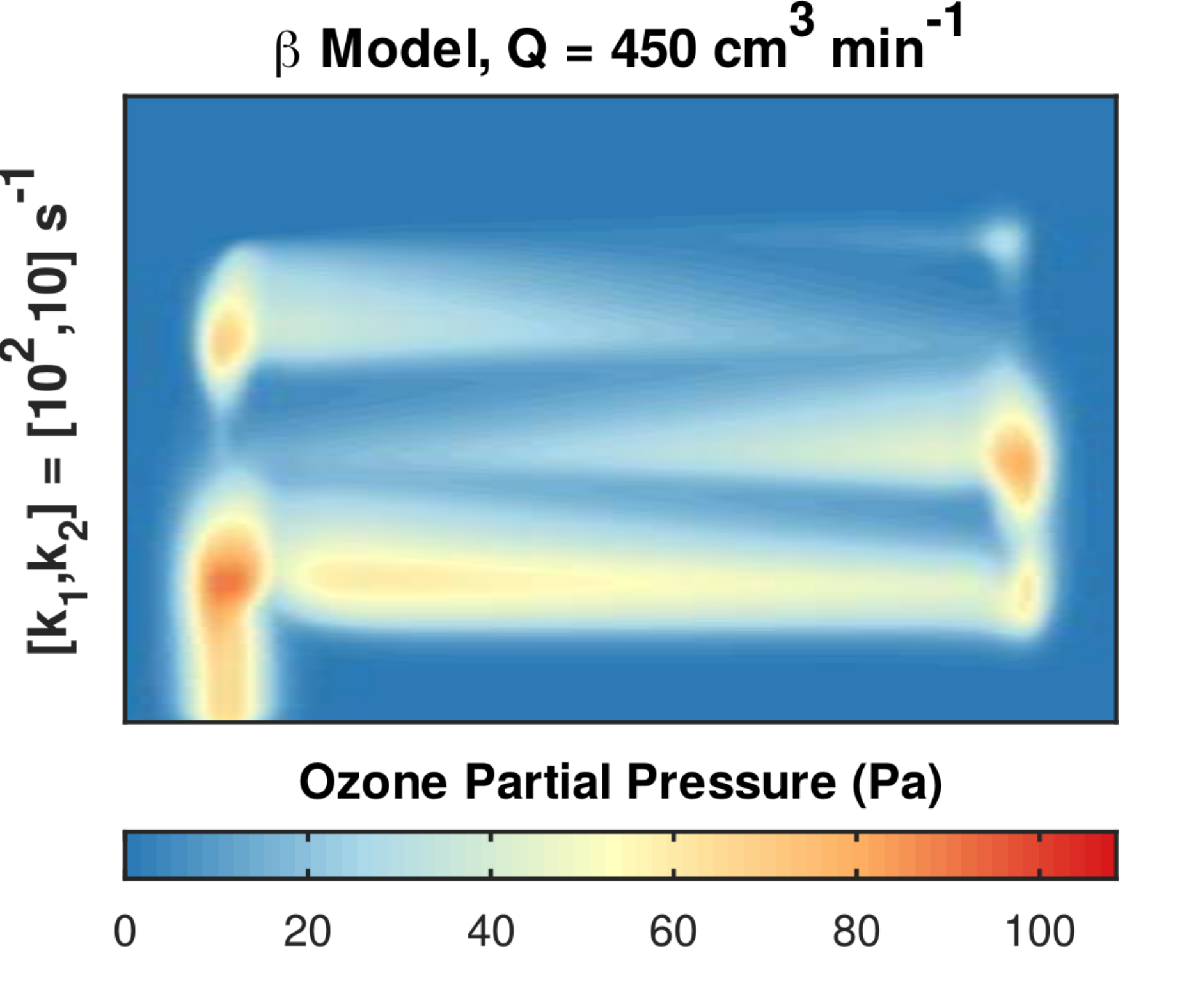}
        \caption{}
    \end{subfigure}
    \caption{
        Ozone partial pressure surfaces for $k_1 = 10^2\ \mathrm{s^{-1}}$,
        with $Q = 250$ (top row) and $Q = 450\ \mathrm{cm^3\ min^{-1}}$ (bottom
        row), and $k_2 = 10^{-1}$ (left column) and $10\ \mathrm{s^{-1}}$
        (right column).
    }
    \label{fig:ks_Po3xy}
\end{figure}

    Considering now the $\mathbf{P}_{\mathrm{O_3}}$ surfaces, Figure
\ref{fig:ks_Po3xy} shows the surfaces corresponding to Fig. \ref{fig:ks_Ro3xy},
i.e. for $k_1 = 10^2\ \mathrm{s^{-1}}$, $k_2=10^{-1}$ and $10\ \mathrm{s^{-1}}$,
and $Q = 250$ and $450\ \mathrm{cm^3\ min^{-1}}$. Additional surfaces may
be found in the SI (Figs. S30 to S33). Once again the trend of $\mathbf{P}
_{\mathrm{O_3}}$ follows that of $\bar{\mathbf{R}}_{\mathrm{O_3}}$, keeping in
mind the caution one must take in comparing results with different values of
$k_1$ and $k_2$. In this case, Fig., \ref{fig:ks_Po3xy} shows absolute values
of $P_{\mathrm{O_3}}$, albeit for shared color scales, in order to reach back
towards the experimental results (such as shown in \cite{lopes15,beruski17,
lopes19}). Thus, while the trend follows that of Fig. \ref{fig:ks_Ro3xy}, it
should be noticed that it is reversed: higher values of $k_2$ leads to smaller
gradients in $P_{\mathrm{O_3}}$, i.e. more homogeneous plumes. It should also
be noticed that the effect is somewhat smaller than seen in Fig.
\ref{fig:ks_Ro3xy}, as further suggested by Fig. S29. Thus, while it may be
more intuitive to look at $\mathbf{P}_{\mathrm{O_3}}$ surfaces for qualitative
information and, perhaps, diagnostics; $\bar{\mathbf{R}}_{\mathrm{O_3}}$
surfaces might provide a more stringent test for model and parameter validation.

   Wrapping up, the range of $k_1$ and $k_2$ explored in the parametric study
was shown to cover a wide range of device responses, from negligible $\mathrm{O
_3}$ usage, to a significant reactant drop. At the lower end of the $k_1$ range,
the reduced geometry uses $4.4\%$ to $2.6\%$ of the inlet $\mathrm{O_3}$ molar
fraction, averaged over the range of $k_2$, while at the upper end it averages
to $55\%$ to $53\%$. Through the $K'$ ratio we can estimate the equivalent
reactant drop for the full geometry (see Part 2 for justification on this),
which at $k_1 = 10^3\ \mathrm{s^{-1}}$ would be $> 1$ for all values of $k_2$
and $Q$ except for one ($k_2 = 10^{-1}\ \mathrm{s^{-1}}$ and $Q = 450\
\mathrm{cm^3\ min^{-1}}$). Considering the proximity between the Beta model,
using the full geometry, and the available experimental results, we consider
safe to state that $k_1 < 10^3\ \mathrm{s^{-1}}$ for the present mathematical
formulation. The scalar response variables also suggest that $k_1 > 10\
\mathrm{s^{-1}}$, given the range of $k_2$ explored and the relatively low
sensitivity towards changes in its value. One noteworthy exception is the
stoichiometries $\lambda$ and $lambda'$, which show good sensitivity towards
changes in $k_2$ at the high end of the $Q$ range. From the comparison with the
available data (Fig. \ref{fig:AxB_lambda}), and the behavior of $\lambda$ with
$k_2$ (Fig. \ref{fig:ks_lambdas}), it seems that $k_1 \lessapprox 10^2\
\mathrm{s^{-1}}$ and $k_2 \gtrapprox 10^3\ \mathrm{s^{-1}}$. The profiles also
appear to show good sensitivity towards $k_2$ for $k_1 = 10^2\ \mathrm{s^{-1}}$,
however the current lack of high-quality experimental data does not allow a
finer constraint on $k_2$ than already obtained with the scalar variables.
Finally, the parametric study allowed us to identify how each reaction rate
constant appears to influence the response surfaces, viz. the $\bar{R}
_{\mathrm{O_3}}$ and $P_{\mathrm{O_3}}$ surfaces. In the range of $k_1$ and
$k_2$ explored, it seems that $k_1$, associated with the adsorption reaction,
controls the spread of the $\mathrm{O_3}$ plume, while $k_2$, associated with
the decomposition reaction, controls the homogeneity of the plume. Thus we
suggest that spatially-resolved data on the reactant partial pressure may be
used to improve the constraints on $k_1$, while data on the local gradient of
$P_{\mathrm{O_3}}$ might be used to constrain $k_2$. While the currently
available experimental results do not allow tighter constraints on $k_1$ or an
upper bound of $k_2$, this study has helped to identify opportunities for
future rounds of model validation following improvements in both the
experimental and computational setups.

\section{\label{sec:sum}Summary}

    Work is presented concerning progress on the numerical modeling of a
prototype polymer electrolyte fuel cell (PEFC)\cite{lopes15}, building on
previously published work\cite{beruski17}. Known limitations of the original
model, Alpha, were addressed in order to increase fidelity with the current
understanding of fuel cell devices, resulting in the Beta model. In this part,
both mathematical frameworks are compared, and approximations and shortcomings
are discussed. Having established the differences between models, a parametric
study is carried out for Beta in order to assess the effects of the order of
magnitude of each reaction rate constant on the global results. In part II,
a grid convergence study was carried out in order to provide an estimate to the
discretization error intrinsic to the Beta model.

    Response variables, some of experimental interest, are compared between the
Alpha and Beta models, viz., the reactant drop $\Delta\chi_{\mathrm{O_3}}$; a
proxy for the total reaction rate, $R'_{\mathrm{O_3}}$; the ratio between the
two, $K' = \Delta\chi_{\mathrm{O_3}}/R'_{\mathrm{O_3}}$; the real and apparent
stoichiometries, $\lambda$ and $\lambda'$ respectively; and spatially-resolved
variables, in the form of profiles and surfaces along the catalyst layer (CL),
given by reactant partial pressure $P_{\mathrm{O_3}}$ and normalized reaction
rate $\bar{R}_{\mathrm{O_3}}$. It should be noticed that each model has degrees
of freedom associated to the reaction kinetics, despite efforts to choose
response variables that minimize such effects, thus care should be taken when
drawing strong conclusions from comparisons. From the scalar response variables,
relatively small differences are seen between models, although $\Delta\chi
_{\mathrm{O_3}}$ and $\lambda'$ show noteworthy results. In both cases, the
Beta model shows a better fit to the available experimental data than the Alpha
model, in particular for $Q\geq 350\ \mathrm{cm^3\ min^{-1}}$. For the
spatially-resolved variables, $\bar{R}_{\mathrm{O_3}}$ profiles and surfaces
show differences between models that might be distinguishable through
experimental validation.

    Shortcomings of the model comparison and the mathematical framework of the
Beta model are discussed. Concerning model comparison, it should be noted that,
given the estimated error of the numerical data in Part 2, the results shown
lack validation power, as both models and experimental data are well within
error bounds of each other. Adding uncertainties in parameterization, these
results call for additional experimental data and denser meshes (i.e. more
computational resources) in order to be able to accurately distinguish between
models, providing further insight into the physical phenomena underlying PEFC
devices. In addition, an apparent divergence from earlier results was also
addressed, where the Alpha model, at the current mesh and solver schemes,
differs from previously published results\cite{beruski17}. New results using
the Stokes-Darcy (SD) and Darcy-Brinkman (DB) formulations, with both Alpha and
Beta models, show similarities between formulations for a given model, with
both models affected similarly. Some response variables do show differences
that might be significant when comparing to experimental results, in particular
$\Delta\chi_{\mathrm{O_3}}$, and consequently $\lambda'$, and $P_{\mathrm{O
_3}}$ profiles. Considering the differences seen and the current errors
estimates for both computational and experimental results, it is reasonable to
expect that small gains in precision should be enough to allow proper
distinction between fluid flow formulations with a range of $Q$ similar to the
explored here. Finally, improving on previous results\cite{beruski17}, flow
speed profiles along the thickness of the Pm domains show a smoother profile
for the DB formulation in both models. These differences are largely
understandable when considering known differences in free and porous media flow
coupling and wall boundary conditions. On the other hand, SD flow speed
profiles show a discontinuity between MPS and CL domains, which is suggested to
be an artifact from the software's approach when coupling media with different
porosity and permeability. Flow speed profiles are taken as increased evidence
that, for the prototype PEFC device under consideration and the way it has been
modeled, the Beta model using the DB formulation provides a better description
of the experimental data available.

    Concerning the mathematical framework of the Beta model, known
approximations and uncertainties in the model were addressed and analysed,
namely the use of Fick diffusivity for $\mathrm{O_3}$ despite the 
Maxwell-Stefan formulation, the use of an approximated $\mathrm{O_3/air}$
mixture, and the coupling between free, porous media, and Knudsen diffusion.
While for the Fick diffusion coefficient there is little to be done about at
the moment, the mixture and coupling between diffusion modes were investigated.
It is shown that there is a negligible difference, considering the error
estimates and order of magnitude of the experimental data, between the
approximate $\mathrm{O_3/air}$ and the proper $\mathrm{O_3/O_2/N_2}$ mixtures.
The different coupling between diffusion modes also shows negligible
differences, however it is shown that the main factor affecting results is the
presence of porous media correction, while the Knudsen regime appears to be of
little significance for the current device.

    Finally, a parametric study of the reaction rate constants $k_1$ and $k_2$,
associated with $\mathrm{O_3}$ adsorption and decomposition reactions, was
carried out. Given the degrees of freedom due to the lack of information on
these reactions, it is important to establish the range of responses the model
can provide given a variation of $k_1$ and $k_2$. The study comprised the
ranges $1\leq k_1\leq 10^3\ \mathrm{s^{-1}}$ and $10^{-1}\leq k_2\leq 10^3\
\mathrm{s^{-1}}$, where the Beta model used $k_1 = 10^2$ and $k_2 = 10\
\mathrm{s^{-1}}$, and with $k_1 = 10^4\ \mathrm{s^{-1}}$ failing to converge.
The chosen range was shown to cover negligible reactant usage up to unreal
projections of the fraction of reactant used $> 1$. In addition, the results
show that $k_1$ largely controls the order of magnitude of the chosen response
variables, with $k_2$ typically providing finer control, being particularly
important at high $Q$. From comparisons with the Beta model, using $k_1 = 10^2$
and $k_2=10\ \mathrm{s^{-1}}$, and its known good correlation with the
available experimental data, the parametric study was able to constrain the
reaction rate constants to $10<k_1<10^3\ \mathrm{s^{-1}}$, with $k_1 \approx
10^2\ \mathrm{s^{-1}}$ showing good fit, and $k_2 \gtrapprox 10^3\ \mathrm{s
^{-1}}$. While the range of values is somewhat broad, the study also identified
opportunities for further validation of the Beta model's reaction kinetics,
viz., the stoichiometries $\lambda$ and $\lambda'$, and the $P_{\mathrm{O_3}}$
profiles at the CL.

    All the shortcomings and opportunities identified in this work would guide
a future Gamma model regarding parameterization and whether to include
additional features, such as solid mechanics, improved porous media description
and fluid flow formulations; in order to settle existing questions and, thus,
further reduce the gap between experimental and computational results. Given
the relatively simplicity of the model, being able to uncouple species and
momentum transport from charge and heat transport and still maintain similarity
to actual PEFC devices, we believe there is much to be gained from further
refining this coupled experimental-numerical approach to fuel cell research.

\begin{acknowledgments}
O.B. acknowledges the Fundação de Apoio à Universidade de São Paulo, FUSP,
grant \#2968. I.K and T.L. acknowledge the Fundação de Ampara à Pesquisa do
Estado de São Paulo, FAPESP, grants \#2016/12397-0, and \#2014/22130-6 and
\#2017/15304-6, respectively. Authors also acknowledge the Research Centre for
Gas Innovation, RCGI, sponsored by FAPESP grant \#2014/50279-4 and Shell
Brasil.\footnotemark[4]
\end{acknowledgments}

\footnotetext[4]{Author contributions: O.B. designed and conducted the
research, analyzed the data and wrote the paper. I.K. contributed with
discussions and by revising the paper. T.L. and F.C.F. contributed by revising
the paper and with funding sources.}

\bibliographystyle{unsrt}
\bibliography{refs}

\end{document}


\title{
    Progress Report on Numerical Modeling of a Prototype Fuel Cell: 2.
    Grid Convergence Study
}

\author{Otávio Beruski}
\email[]{oberuski@alumni.usp.br}
\affiliation{Instituto de Pesquisas Energéticas e Nucleares,
IPEN/CNEN-SP, 05508-000, São Paulo, São Paulo, Brazil}

\author{Ivan Korkischko}
\affiliation{Instituto de Pesquisas Energéticas e Nucleares,
IPEN/CNEN-SP, 05508-000, São Paulo, São Paulo, Brazil}

\author{Thiago Lopes}
\affiliation{Instituto de Pesquisas Energéticas e Nucleares,
IPEN/CNEN-SP, 05508-000, São Paulo, São Paulo, Brazil}

\author{Fabio Coral Fonseca}
\affiliation{Instituto de Pesquisas Energéticas e Nucleares,
IPEN/CNEN-SP, 05508-000, São Paulo, São Paulo, Brazil}

\begin{abstract}
    Progress on the numerical modeling of a prototype fuel cell is reported.
Some known limitations of the previously published Alpha model are addressed,
and the numerical uncertainty due to discretization of the improved model,
Beta, was estimated. In this part, a mesh convergence study is carried out to
estimate the discretization error of Beta. A reduced, proxy geometry and
two extrapolation schemes are used to estimate the exact solution, which is
then used to estimate the model's uncertainty through the Grid Convergence
Index framework. Qualitatively, non-monotonic behavior is seen for the response
variables as function of effective grid spacing. As such, the mixed-order
extrapolation is favored over the generalized Richardson extrapolation in order
to estimate the exact solution. Quantitatively, error estimates are on average
$\sim 10\%$ for the flow rate range simulated, larger than experimental ones
available. Results suggest a difficulty in achieving mesh convergence in fuel
cell-like models, even in simpler cases. Caution is thus suggested during
validation or when devising predictions from numerical models.

\end{abstract}

\keywords{fluid dynamics; modeling; fuel cell; model validation}

\maketitle

\section{\label{sec:int}Introduction}

    Mathematical models are extremely important in basic and applied research,
helping understand physical processes and allowing optimization of processes of
interest. Models are based on observations of natural phenomena or controlled
experiments, and as such must be validated, that is, must be compared to
empirical data in order to assess its accuracy in reproducing the system of
interest. Given this fundamental dependence on observations, it is crucial that
one understands, and preferably quantifies, the uncertainties involved in the
validation process. While it is common to think of uncertainties in
experimental data, it is also imperative to identify, and quantify whenever
possible, the uncertainties in mathematical models. Roache\cite{roache97} gives
an authoritative review on the subject, with more modern aspects discussed by,
for example, Roy and Oberkampf\cite{roy11}.

    Focusing on numerical procedures, particularly computational fluid dynamics
(CFD), a key source of uncertainty is the discretization of the model's
variables, e.g. spatial and time variables\cite{roy10}. This discretization
error arises from the approximation of continuous variables to discrete ones,
for instance by stipulating that time $t$ runs in small, finite steps $\delta
t$. This leads to a dependency of the model's variables of interest, say a
scalar $\phi =\phi(t)$, on the size of $\delta t$, with the exact result being
obtained in the limit of $\delta t\rightarrow 0$, in this example. Thus, it
becomes important to establish that numerical results are independent of the
discretization of independent variables, in order that the model represents the
actual physical processes being studied, and not the approximations used to
solve the underlying equations. Demonstrating this so-called grid convergence
is, in practice, challenging, being untenable for most real-world
scenarios\cite{roache97}, particularly for systems with complex geometry and
physical processes\cite{roy10} (fuel cells come to mind). The accepted practice
is then to treat this as an uncertainty in the model's results.

    There are several methods to estimate discretization error in CFD.
Following Roy\cite{roy10}, two broad classes may be defined: \emph{a priori}
and \emph{a posteriori} methods, with \emph{a posteriori} methods being more
widely used. These are further divided in two branches: i) higher-order
estimates, which relies on estimates of the exact solution, and may be applied
in a post-processing manner after the solution has been computed; and ii)
residual-based methods, which generally require the solution of an adjoint
problem, specific to the system under study. While the latter has the potential
of providing more detailed information on the error and its sources, it is in
general more difficult to implement. The former methods are in general simpler
to perform, however it may be necessary to obtain several solutions using
systematically-refined, or coarsened, meshes. Among the higher-order methods,
the Richardson extrapolation along with the Grid Convergence Index (GCI)
framework\cite{roache94,roache97} are widely used, given its ease of
application and applicability to any discretization scheme.

    In the field of fuel cells, mathematical models play a significant role,
with increasing complexity and spatial and time resolutions (see for instance
\cite{bavarian10,hajimolana11,weber14} and references therein). Given that the
complexity of fuel cell models and the computational resources employed have
increased, while the precision and diversity of experimental data has
also increased\cite{weber14}, it becomes particularly important to assess the
uncertainties that invariably follow in order to properly validate a given
model and use it for prediction. Uncertainty assessments in fuel cells are
relatively uncommon, understandably so given the complexity of these devices
and the costs associated with both empirical and numerical experiments.
Concerning discretization error, when assessed, it is typically done through
mesh refinement methods using a single response quantity, however usually
without the higher-order extrapolate and the error estimate (see for example
\cite{iranzo10,carnes13,duivesteyn13,pourmahmoud17}). One exception, while not
of a full device, is the work of Tranter and colleagues\cite{tranter15}, who
used the GCI framework\cite{roache94} to estimate the discretization error in a
2D model of a fibrous medium, representing the gas diffusion layer of a fuel
cell. On the other hand, Kjeang and colleagues\cite{kjeang06} provide an upper
bound for the error when using a given mesh, possibly using an \emph{a priori}
method, and then proceed to compare the result of a single response quantity to
the analytical solution of an approximate formulation of the problem. It is
also worth noticing that the work of Beruski and colleagues\cite{beruski17},
on which this work is based, also lacks the description of a discretization
error estimate, like so many in the fuel cell literature.

    We build on the previously published model for the aforementioned
prototype fuel cell cathode\cite{beruski17}, with the main goals of improving
the mathematical framework, and therefore its accuracy, and estimating the
discretization error, thus assessing its precision. In this
part\footnotemark[1], a mesh convergence study has been carried out in order
to assess the model's uncertainty due to discretization, and thus provide an
estimate of its precision. A reduced, proxy geometry was used to fully explore
the feasible range of mesh density, and the convergence behavior of the chosen
response variables was evaluated with respect to an effective element spacing.
Two extrapolation schemes were then used and compared in order to assess the
most appropriate given the convergence behavior observed. The error estimates
for the proxy geometry were then calculated within the GCI
framework\cite{roache94,roache97}, while the error for the geometry of interest
was further estimated from correlations with the proxy geometry. In this way,
the remainder of this work is organized as follows. Section \ref{sec:met}
provides an overview of the experimental significance and mathematical
formulation of the model, as well as a description of the mesh convergence
study. Section \ref{sec:res} first presents a qualitative assessment of the
convergence behavior of relevant variables with mesh resolution, proceeding to
a quantitative estimation of the discretization error. Finally, Section
\ref{sec:sum} concludes with a summary of the results and brief remarks on
future studies.

\footnotetext[1]{A joint version of parts 1 and 2 can be found with DOI:
arXiv:2002.04519.}

\section{\label{sec:met}Methods}

    We first present a brief description of the experimental device that is
simulated by our model, first described by Lopes \emph{et al.}\cite{lopes15},
followed by the computational methods used for the development of this work.

    The device is a prototype polymer electrolyte fuel cell (PEFC), more
specifically a cathodic half-cell, which allows \emph{in situ} observation of
reactive fluid flow in a porous medium. Ozone-enriched air is used as tracer in
a carrier fluid, and a coumarin-based dye is used as the sensor to map the
local concentration of $\mathrm{O_3}$. The $\mathrm{O_3}$ interacts with the dye
anchored on silica particles, akin to electrocatalysts dispersed in a porous
layer, resulting in the emission of photons and the degradation of both
reactant and dye. Measuring the light emission from this “catalyst layer”
allows determination of the local $\mathrm{O_3}$ concentration and, thus, the
species and fluid dynamics. Global variables, such as pressure drop and
total reactant drop, provide additional information on the inner workings of
the device.

\begin{figure}
    \centering
    \begin{subfigure}{0.45\textwidth}
        \includegraphics[width=\textwidth]{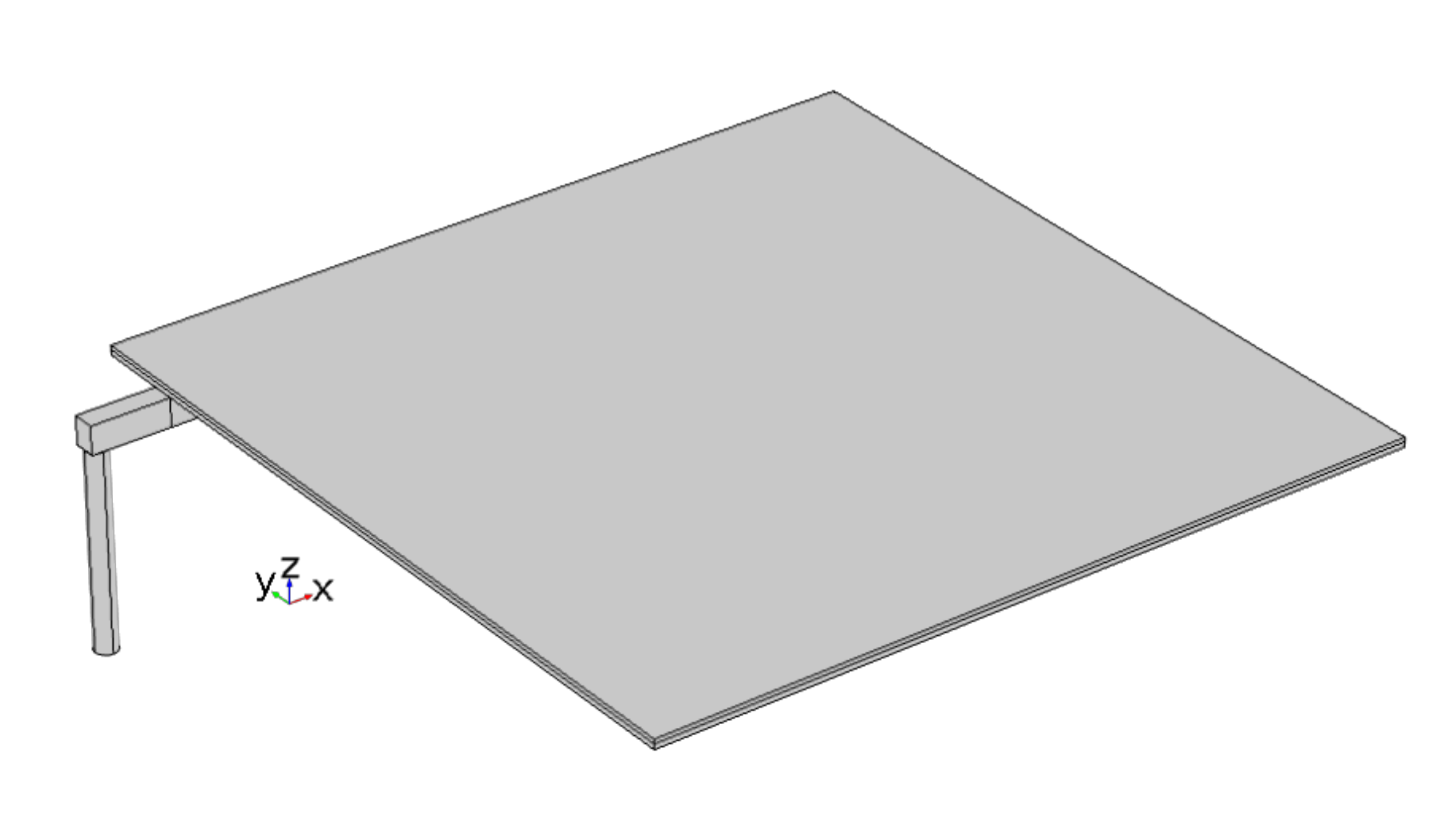}
        \caption{}
    \end{subfigure}
    ~
    \begin{subfigure}{0.45\textwidth}
        \includegraphics[width=0.8\textwidth]{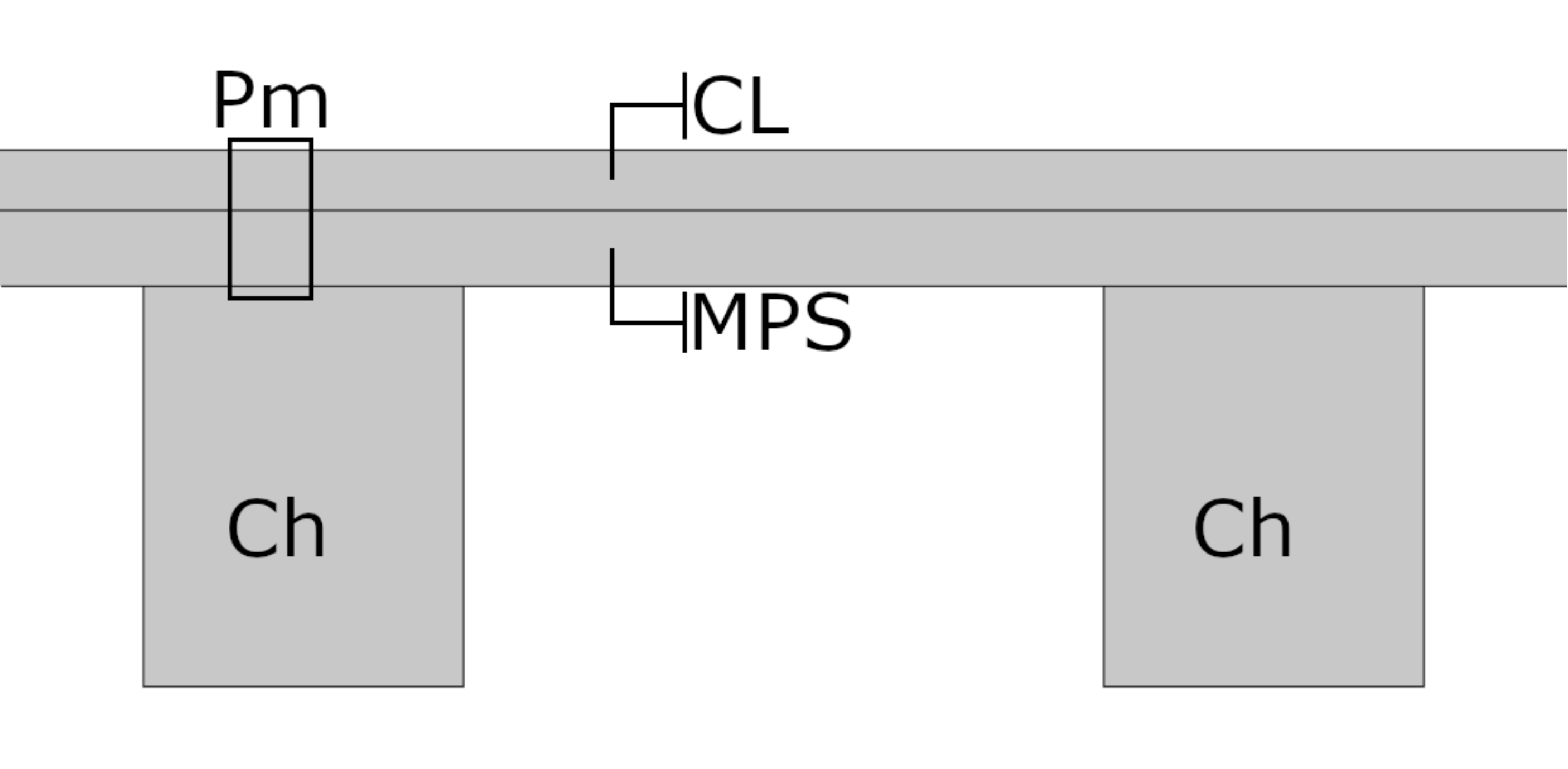}
        \caption{}
    \end{subfigure}

    \begin{subfigure}{0.45\textwidth}
        \includegraphics[width=\textwidth]{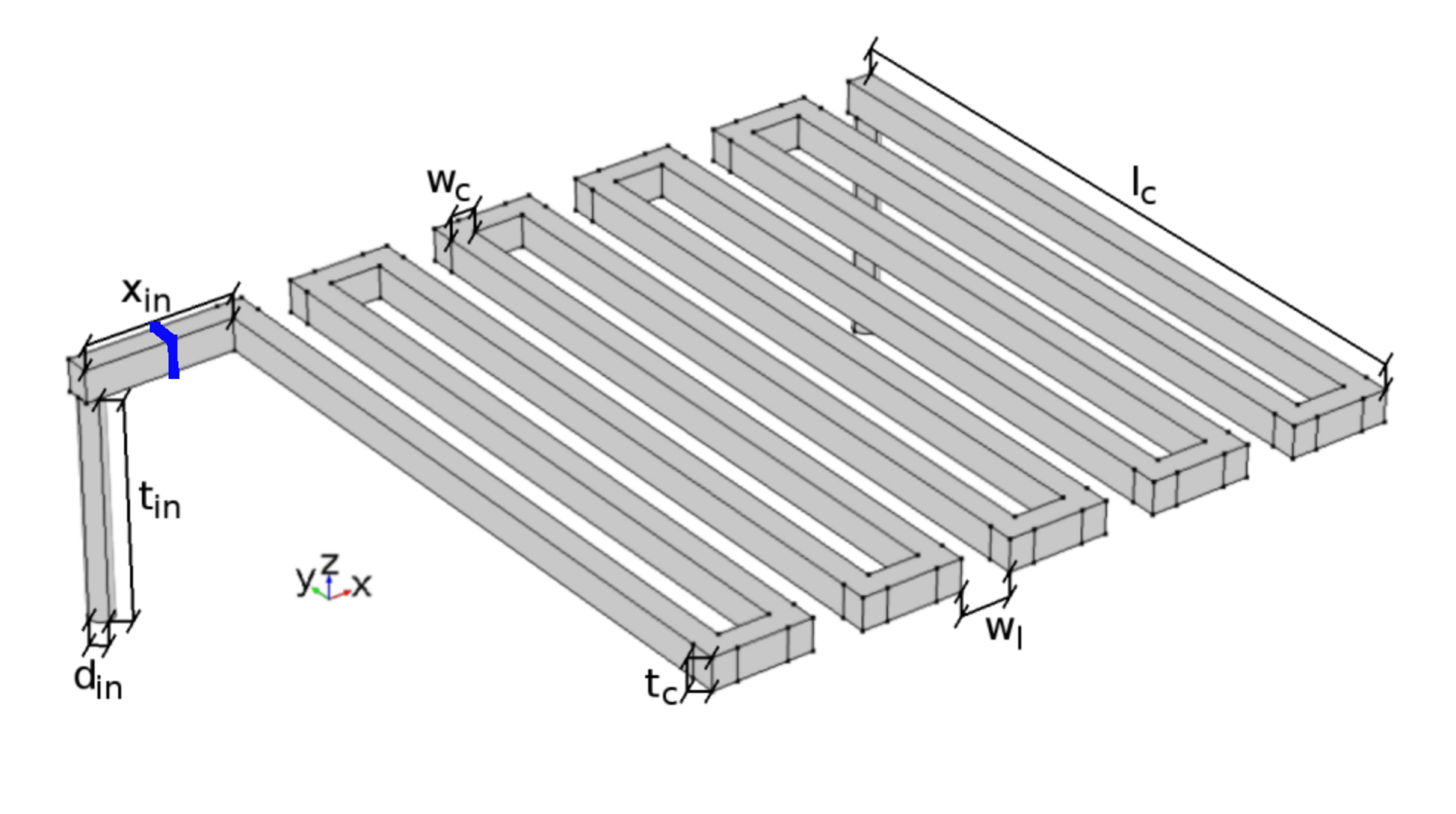}
        \caption{}
        \label{fig:geo_ch}
    \end{subfigure}
    ~
    \begin{subfigure}{0.45\textwidth}
        \includegraphics[width=0.8\textwidth]{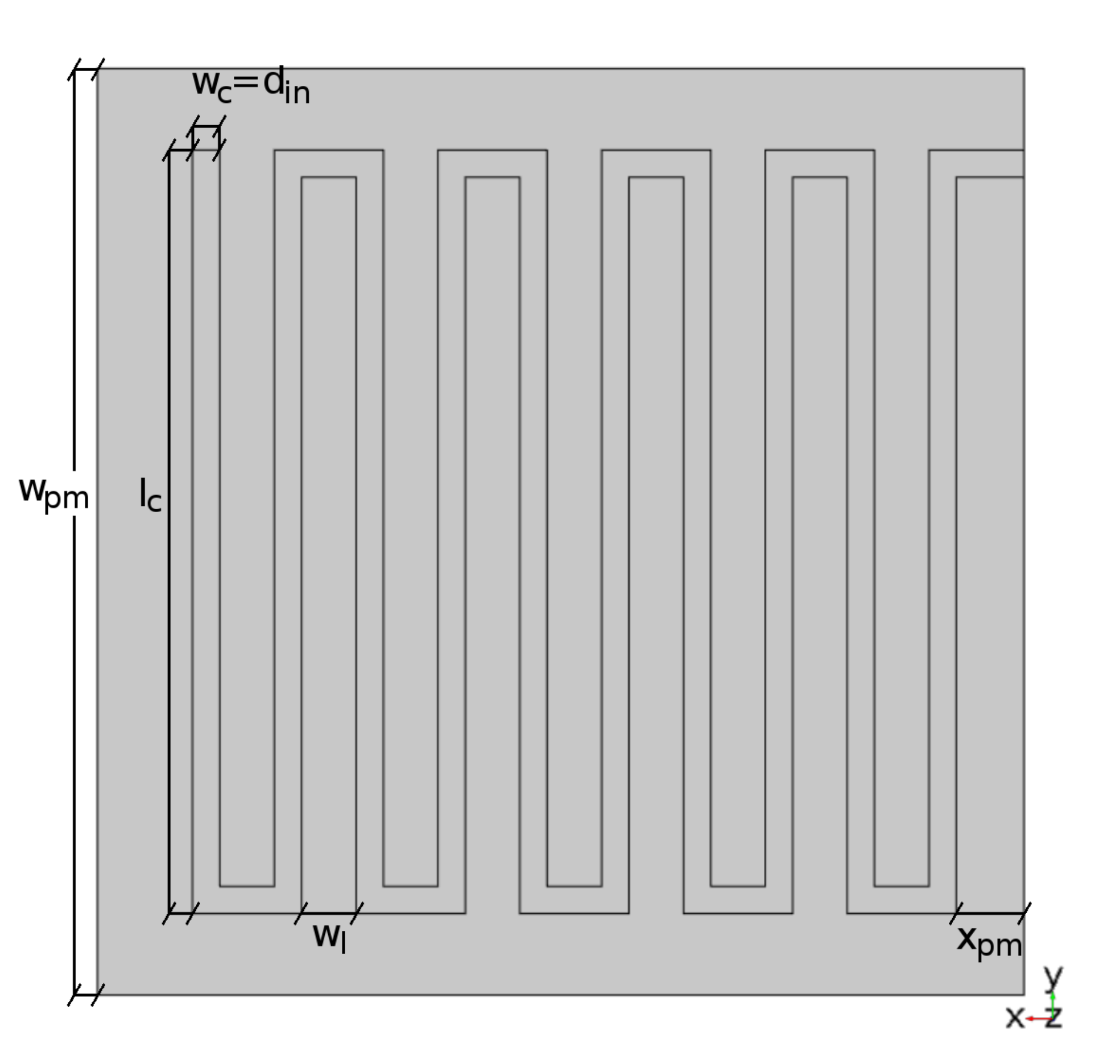}
        \caption{}
        \label{fig:geo_pm}
    \end{subfigure}
    \caption{
        Geometry used in the proto-cell simulations: \textbf{(a)} full geometry,
        showing the flow channel (Ch) and porous media (Pm) domains;
        \textbf{(b)} cross-section in the $xz$ plane, showing the structure
	along the $z$ axis; \textbf{(c)} Ch domain, with the segmentation edges
        shown in blue (see text), and \textbf{(d)} lower boundary of the MPS
        domain, showing the projection of the Ch domain.
    }
    \label{fig:geo}
\end{figure}

    Thus, considering the experimental device, the proposed computational model
covers momentum and species transport, while assuming thermal equilibrium and
steady state. The computational domains included are: i) the flow channel (Ch),
in this case a single serpentine geometry, ii) the porous transport layer, here
comprising only a macroporous substrate (MPS) made of carbon paper
(representing Toray TGH-060 with a 10\% PTFE hydrophobic treatment), and iii)
the catalyst layer (CL), where the catalyst and substrate particles are
considered homogeneously distributed (representing Sigma Aldrich Nano Silica
Gel on TLC plates, see \cite{lopes15} for details). Domains ii) and iii)
comprise the porous media domains (Pm). Figure \ref{fig:geo} shows the domains'
disposition and Table \ref{tab:geo} presents the geometrical parameters.

    The setup used for all simulations was a workstation with two
Intel$^\circledR$ Xeon$^\circledR$ E3 processors and 128GB of RAM, operated
with a 64 bits Debian9 distribution, Linux kernel v. 4.9.0-4. The software used
was the commercial package COMSOL Multiphysics$^\circledR$, v.5.1.0.234, along
with the Batteries and Fuel Cell, CFD, and Chemical Engineering modules. All
data handling and processing was done using GNU Octave v.4.2.1\cite{octave},
while image processing was done using GIMP v.2.10.14\footnotemark[2].

\footnotetext[2]{Available at https://www.gimp.org/.}

\begin{table}[h]
    \centering
    \caption{Geometrical parameters of the computational domains.}
    \label{tab:geo}
    \begin{tabular}{c c c}
        \hline
        \textbf{Parameter} & \textbf{Symbol} & \textbf{Value} \\
        \hline
        Channel width & $w_{\mathrm{c}}$ & $0.8\ \mathrm{mm}$\\
        Land width & $w_{\mathrm{l}}$ & $1.6\ \mathrm{mm}$\\
        Channel depth & $t_{\mathrm{c}}$ & $1.0\ \mathrm{mm}$\\
        Channel section length & $l_{\mathrm{c}}$ & $22.4\ \mathrm{mm}$\\
        Inlet/outlet diameter & $d_{\mathrm{in}}$ & $0.8\ \mathrm{mm}$\\
        Inlet/outlet length & $t_{\mathrm{in}}$ & $7.0\ \mathrm{mm}$\\
        Inlet offset & $x_{\mathrm{in}}$ & $0.188\ \mathrm{in}$\\
        Porous media edge & $w_{\mathrm{pm}}$ & $(l_{\mathrm{c}}+x_{\mathrm{in}})$\\
	    Porous media offset & $x_{\mathrm{pm}}$ & $(x_{\mathrm{in}}-w_{\mathrm{w}})/2$\\
        MPS thickness & $t_{\mathrm{MPS}}$ & $190\ \mathrm{\mu m}$\\
        CL thickness & $t_{\mathrm{CL}}$ & $150\ \mathrm{\mu m}$\\
        \hline
    \end{tabular}
\end{table}

\subsection{\label{ssec:math}Mathematical Formulation}

    The vintage mathematical formulation, Alpha, can be found in
\cite{beruski17}, while the improved formulation, Beta, is detailed in Part 1.
In this part, only the Beta model was used, and a brief description will be
given below.

    The momentum transport is described using the (compressible) Darcy-Brinkman
formulation (DB), which covers both free and porous media flow\cite{lebars06}:
\small
\begin{eqnarray}
    \frac{\rho}{\epsilon}\left(\mathbf{u}\cdot\nabla\right)\left(\frac{\mathbf{u}}{\epsilon}\right) & = & \nabla\left[-P\mathbf{I} + \frac{\mu}{\epsilon}\left(\nabla\mathbf{u}+\left(\nabla\mathbf{u}\right)^{\mathrm{T}}\right) - \frac{2\mu}{3\epsilon}\left(\nabla\cdot\mathbf{u}\right)\mathbf{I}\right] - \frac{\mu}{\kappa}\mathbf{u},
\end{eqnarray}
\normalsize
\noindent
where $\rho$ and $\mu$ are the fluid's density and kinematic viscosity,
respectively, $\mathbf{u}$ is the velocity field vector, $P$ is the relative
pressure, and $\epsilon$ and $\kappa$ are the domain's porosity and
permeability, respectively ($\epsilon = 1$ and $\kappa \rightarrow \infty$ for
free flow). The parameters are shown in Table \ref{tab:par}, chosen in order to
describe the materials and conditions used (see \cite{lopes15} and
\cite{beruski17} for additional information). The inlet boundary condition is a
(standard) mass flow rate condition:
\begin{eqnarray}
    -\int_{\partial\Omega}\frac{\rho}{\rho_{\mathrm{std}}}\left(\mathbf{u}\cdot\mathbf{n}\right)\mathrm{d}S & = & Q
\end{eqnarray}
\noindent
with dry air at $T_{\mathrm{std}}=273.15\ \mathrm{K}$ and $P_{\mathrm{std}}=1\
\mathrm{atm}$ defining the standard density $\rho_{\mathrm{std}} =
P_{\mathrm{std}}M_{\mathrm{air}}/RT_\mathrm{std}$. A constant pressure $P =
P_{\mathrm{out}}$ was used as the outlet boundary condition. The fluid's
density is a function of the composition, calculated using the ideal gas law,
while the viscosity is given by a constitutive relation for air, from the
software:
\begin{eqnarray}
    \mu_{\mathrm{air}}\left(T\right) & = & -8.38278\times 10^{-7} + 8.35717342\times 10^{-8}T  - 7.69429583\times 10^{-11}T^2 \nonumber\\
    & & + 4.6437266\times 10^{-14}T^3 - 1.06585607\times 10^{-17}T^4,
\end{eqnarray}
\noindent
where $\mu_{\mathrm{air}}$ has units of $\mathrm{Pa\ s}$ and $T$ of
$\mathrm{K}$.

\begin{table}[h]
    \centering
    \caption{Parameterization of the model.}
    \label{tab:par}
    \begin{tabular}{c c c c}
        \hline
        \textbf{Parameter} & \textbf{Symbol} & \textbf{Value} & \textbf{Reference}\\
        \hline
        Reference pressure & $P_{\mathrm{ref}}$ & $1.027\ \mathrm{bar}$ & \cite{beruski17} \\
        Cell back pressure & $P_{\mathrm{out}}$ & $1.0994\ \mathrm{bar} - P_{\mathrm{ref}}$ & \cite{beruski17} \\
        Carbon paper porosity & $\epsilon_{\mathrm{MPS}}$ & $0.801$ & \cite{fishman10} \\
        Carbon paper permeability & $\kappa_{\mathrm{MPS}}$ & $9.18\times 10^{-12}\ \mathrm{m}^2$ & \cite{fishman11} \\
        Carbon paper tortuosity & $\tau_{\mathrm{MPS}}$ & $1.199$ & \cite{fishman11} \\
        Catalyst layer porosity & $\epsilon_{\mathrm{CL}}$ & $0.497$ & \cite{beruski17} \\
        Catalyst layer permeability & $\kappa_{\mathrm{CL}}$ & $8.82\times 10^{-11}\ \mathrm{m}^2$ & \cite{beruski17} \\
        $\mathrm{O_3}$ diffusion coefficient in N$_2$ & $D_{O_3}$ & $0.16\ \mathrm{cm}^2\mathrm{s}^{-1}$ & \cite{ono04,massman98} \\
        Average molar mass of dry air & $M_{\mathrm{air}}$ & $28.96546\ \mathrm{g\ mol^{-1}}$ & \cite{nist08}\\
        $\mathrm{O_3}$ adsorption reaction rate constant & $k_1$ & $100\ \mathrm{s^{-1}}$ & see part I \\
        $\mathrm{O_3(ads)}$ decomposition reaction rate constant & $k_2$ & $10\ \mathrm{s^{-1}}$ & see part I \\
        Average silica particle radius & $r_p$ & $6.5\ \mathrm{\mu m}$ & see part I \\
        Quantity of dye deposited on the CL & $\Gamma_{\mathrm{dye}}$ & $3\ \mathrm{\mu mol\ cm^{-2}}$ & \cite{lopes15} \\
        \hline
    \end{tabular}
\end{table}

    The mass fraction formulation of the advection-reaction equation was used
for the species transport, in a concentrated-mixture fashion, of an
$\mathrm{O_3/air}$ gaseous mixture:
\begin{eqnarray}
    \nabla\cdot\mathbf{J}_i + \rho\left(\mathbf{u}\cdot\nabla\right)\omega_i & = & R_i \label{eq:mf_ar}
\end{eqnarray}
\noindent
where $\omega_i$ is the mass fraction of the $i$-th species. In this model, the
relative composition of $\mathrm{O_2}$ and $\mathrm{N_2}$ in the carrier gas,
air, is taken to be constant. As such an effective species, ``air'', with molar
mass $\mathrm{M_{air}}$, was used instead of the explicit components (see Part
1 for additional information on this subject). The mass diffusive flux was
given by the Maxwell-Stefan (MS) model:
\begin{eqnarray}
    \mathbf{J}_i & = & -\left(\rho\omega_i\sum_k D_{ik}\mathbf{d}_k\right) \label{eq:ms} \\
    \mathbf{d}_k & = & \nabla\chi_k + \frac{\left[\left(\chi_k - \omega_k\right)\nabla P_A\right]}{P_A}
\end{eqnarray}
\noindent
where $D_{ik}$ are multi-component diffusivities, $\mathbf{d}_k$ is a so-called
diffusional driving force, $\chi_k$ is the molar fraction and $P_A$ is the
absolute pressure. The Knudsen regime was included for porous media transport:
\begin{eqnarray}
    D_{ij} = \left[\frac{1}{D_{ij}^{\mathrm{Pm}}} + \frac{1}{D_{ij}^K}\right]^{-1} \label{eq:dcs}
\end{eqnarray}
\noindent
Here two effects are added in a parallel-resistances fashion: i) the porous
media diffusion coefficient, $D_{ij}^{\mathrm{Pm}} = f_{\mathrm{Pm}}
D_{ij}^{\mathrm{free}}$, which applies a correction $f_{\mathrm{Pm}}$, function
of the porosity and tortuosity of the porous medium, to the ``free'' diffusion
coefficient $D_{ij}^{\mathrm{free}}$; and ii) the Knudsen regime diffusivity,
obtained through kinetic theory using the average pore diameter, $d_p$, instead
of the mean free path in porous media. In this case, $d_p$ was obtained using
the Millington \& Quirk model for cemented particle beds\cite{quirk61} for
simplicity.

    The reaction kinetics involving ozone degradation, and hence light
emission, was modeled using an adsorption-decomposition scheme:
\begin{eqnarray}
    \mathrm{O_3(g)} + \mathrm{dye} & \rightarrow & \mathrm{O_3(ads)} \\
    \mathrm{O_3(ads)} & \rightarrow & \mathrm{products} + h\nu
\end{eqnarray}
\noindent
where the dye molecule is 7-diethylamino-4-methylcoumarin\cite{lopes15} and
the desorption step was ignored. A differential equation involving the surface
cover of $\mathrm{O_3}$, $\theta_{\mathrm{O_3}}$, was solved along the fluid
and species dynamics:
\begin{eqnarray}
    \frac{\mathrm{d}\theta_{\mathrm{O_3}}}{\mathrm{d}t} & = & k_1 \chi_{\mathrm{O_3}}\left(1-\theta_{\mathrm{O_3}}\right) - k_2\theta_{\mathrm{O_3}} \label{eq:hk}
\end{eqnarray}
\noindent
where $k_1$ and $k_2$ are reaction rate constants for the adsorption and
decomposition reactions, respectively. To couple with Eq. \ref{eq:mf_ar}, one
may write:
\begin{eqnarray}
    R_{\mathrm{O_3}} & = & -M_{\mathrm{O_3}}A_v\Gamma^*_s k_1 \chi_{\mathrm{O_3}}\left(1-\theta_{\mathrm{O_3}}\right) \label{eq:sink}
\end{eqnarray}
where $A_v$ is the total surface area by unit volume and $\Gamma^*_s$ is the
total surface concentration of active sites, i.e. dye molecules. The total
surface concentration of adsorption sites can be obtained with:
\begin{eqnarray}
    \Gamma_s^* & = & \frac{\Gamma_{\mathrm{dye}}}{t_{\mathrm{CL}}A_v}
\end{eqnarray}
\noindent
where $\Gamma_{\mathrm{dye}}$ is the reported\cite{lopes15} quantity of dye
deposited over a given geometric area of CL with thickness $t_{\mathrm{CL}}$.
The $A_v$ is calculated using a simple model:
\begin{eqnarray}
    A_v & = & \frac{3\epsilon_{\mathrm{CL}}}{r_p}
\end{eqnarray}
\noindent
where $r_p$ is the average particle radius, in this case the silica particles
where the dye were anchored on (Nano Silica Gel, Sigma Aldrich).

    The solver configuration used depends on the actual mesh used, however
it followed a general configuration according to the number of mesh elements.
A detailed description, as much as possible, can be found in Section SII A of
the Supplemental Material\footnotemark[3] (SM). The general outline is the
following: a segregated 3-steps solver was employed, with step i) solving for
$\theta_{\mathrm{O_3}}$, step ii) solving for $\mathbf{u}$ and $P$, and step
iii) solving for $\omega_{\mathrm{O_3}}$. Step i) always employed a direct
solver, based on MUMPS\cite{MUMPS:1,MUMPS:2}, while step ii) always employed an
iterative solver based on GMRES\cite{saad86}, a multigrid scheme with a number
of levels and mesh coarsening scheme depending on the number of mesh elements.
Step iii) either employed a direct solver, also MUMPS-based, or a GMRES-based
iterative solver using the domain decomposition method\cite{toselli05}. The
direct solver was employed for the cases where $\lambda \leq 4$, otherwise the
iterative solver was used, with the number of sub-domains depending on the
number of mesh elements. All steps used a constant dampening factor in all
cases, being $1.0$, $1.0$ and $0.7$ for steps i), ii) and iii), respectively.
In addition, a parametric solver was used to vary $Q$ from $200$ to $450\
\mathrm{cm^3\ min^{-1}}$ with a step of $50\ \mathrm{cm^3\ min^{-1}}$, with the
previous solution used as initial value for the next one.

\footnotetext[3]{Supplemental Material is available at 
https://arxiv.org/abs/2002.04519.}

    Finally, the use of an assembly for the computational geometry allowed
greater control of the meshing procedure. As a consequence, explicit coupling
between the geometric entities was needed to ensure the proper continuity of
the model's variables, i.e. the  flow field, $\mathbf{u}$, and ozone mass
fraction, $\omega_{\mathrm{O_3}}$. This was accomplished via an identity pair.
Since the relevant boundaries were sectioned to properly match one another
(at the Ch and MPS domains, see Fig. \ref{fig:geo}), no fallback features were
necessary.

\subsection{\label{ssec:mesh}Mesh Convergence Study}

\subsubsection{\label{sssec:ma}Algorithm and refinement procedure}

    As mentioned above, the model's geometry was an assembly of two geometric
entities, the flow channel (Ch) and the porous media (Pm). This separation
was used in order to achieve a larger degree of control with respect to the
meshing procedure, allowing, for example, meshing the Ch and Pm domains
separately in order to investigate the effects of each mesh individually.
Following much experimentation, a protocol was established in order to
assess mesh convergence in all domains, with respect to a single index,
$\lambda$, which is somewhat connected to the more usual element spacing
$h$ in structured meshes. The meshing algorithm, briefly, had the following
outline:
\begin{enumerate}
    \item Meshing of the Ch domain with tetrahedral elements, with scaling
    factor $\lambda$.
    \item Splicing the boundary-adjacent elements to add 3 hexahedral elements
    (``boundary layer elements'', BLEs).
    \item Copying the upper boundary of the Ch mesh to the projection onto the
    lower MPS boundary.
    \item Meshing the remainder of the lower MPS boundary with triangular
    elements, with scaling factor $1$.
    \item Extruding the lower boundary elements throughout the Pm domains,
    with $\lambda$ elements in each domain.
\end{enumerate}
\noindent
The element size parameters used in steps 1 and 4 were based on software
recommendations, which in turn are believed to be based on the size of the
full geometry. The so-called BLEs of step 2 are a software recommendation when
dealing with relatively coarse meshes. A variant of this meshing algorithm
without step 2 was also used. Details of the algorithm and of each mesh are
given in Section SII B of the SM\footnotemark[3], along with Fig. S1
illustrating the meshes built.

\begin{figure}
    \centering
    \begin{subfigure}{0.4\textwidth}
        \includegraphics[width=\textwidth]{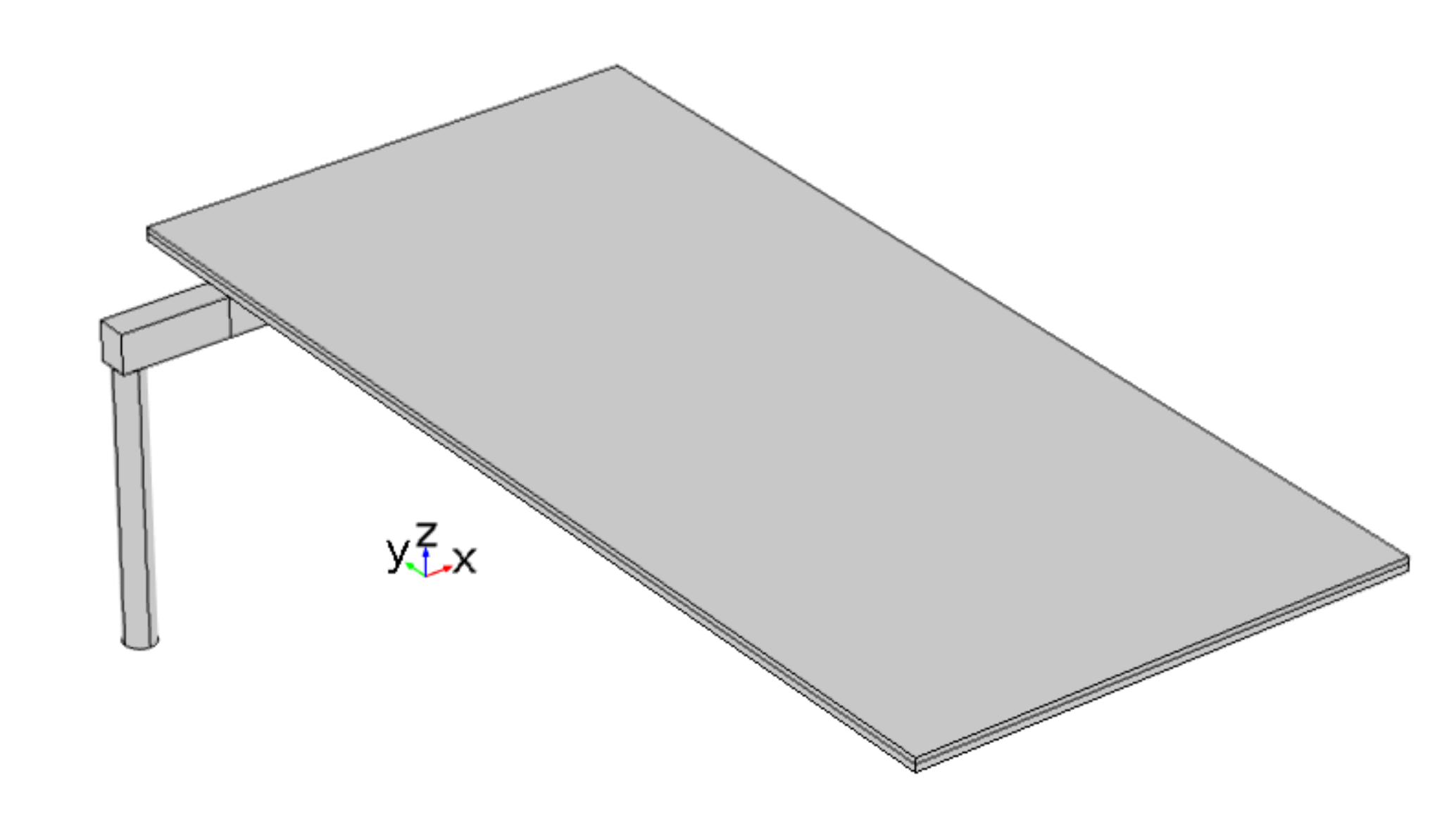}
        \caption{}
    \end{subfigure}
    ~
    \begin{subfigure}{0.4\textwidth}
        \includegraphics[width=\textwidth]{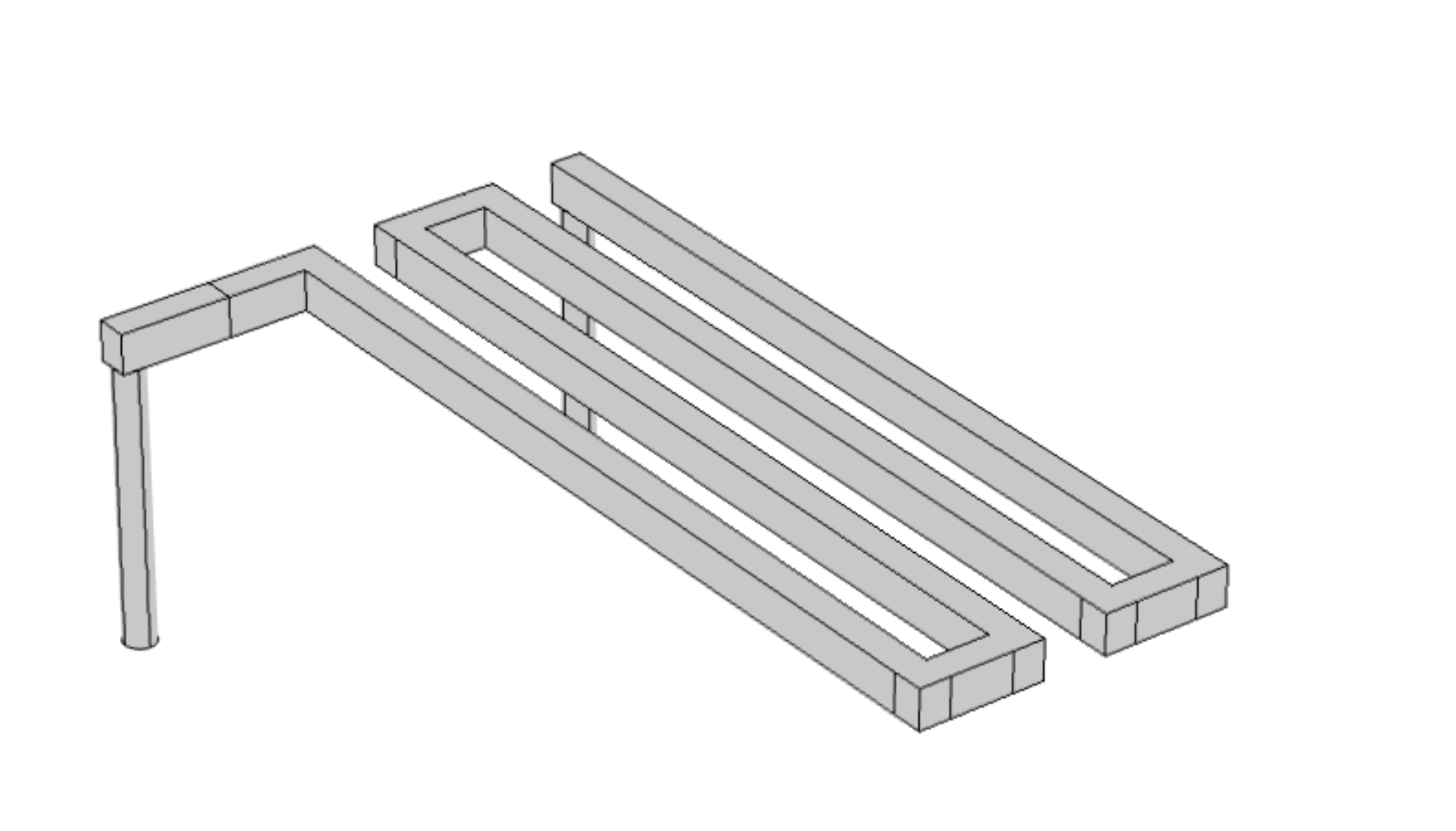}
        \caption{}
    \end{subfigure}
    \caption{
        Reduced geometry used in the proto-cell simulations using dense meshes:
        \textbf{(a)} full geometry, showing both Ch and Pm domains, and
        \textbf{(b)} Ch domain.
    }
    \label{fig:lil_geo}
\end{figure}

    Given the current computational resources available, a reduced version of
the geometry was employed to solve the model when using dense meshes. The Ch
domain was shortened to 4 horizontal sections of the serpentine geometry, down
from 10, as shown in Figure \ref{fig:lil_geo}. The Pm domains were reduced
accordingly, with the remaining geometric parameters kept constant. As such,
the scaling factor $\lambda$ for the reduced geometry takes the values
$\lambda \in [2,4,8,10,12,16]$, while for the full geometry it reduces to
$\lambda \in [2,4,8,12]$.

\subsubsection{\label{sssec:mc}Convergence analysis}

\begin{figure}
    \centering
    \begin{subfigure}{0.4\textwidth}
        \includegraphics[width=\textwidth]{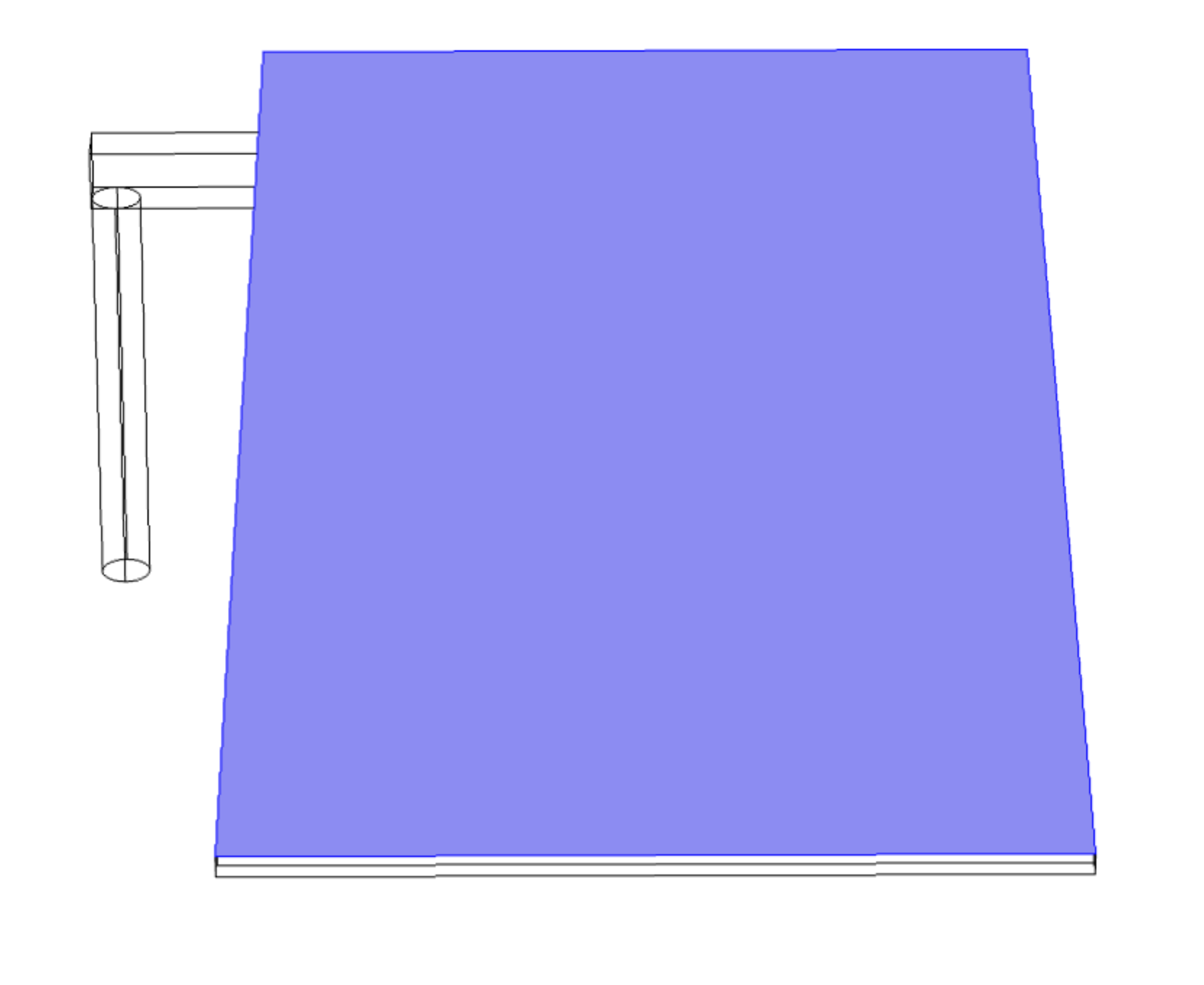}
        \caption{}
        \label{fig:surf}
    \end{subfigure}
    ~
    \begin{subfigure}{0.4\textwidth}
        \includegraphics[width=\textwidth]{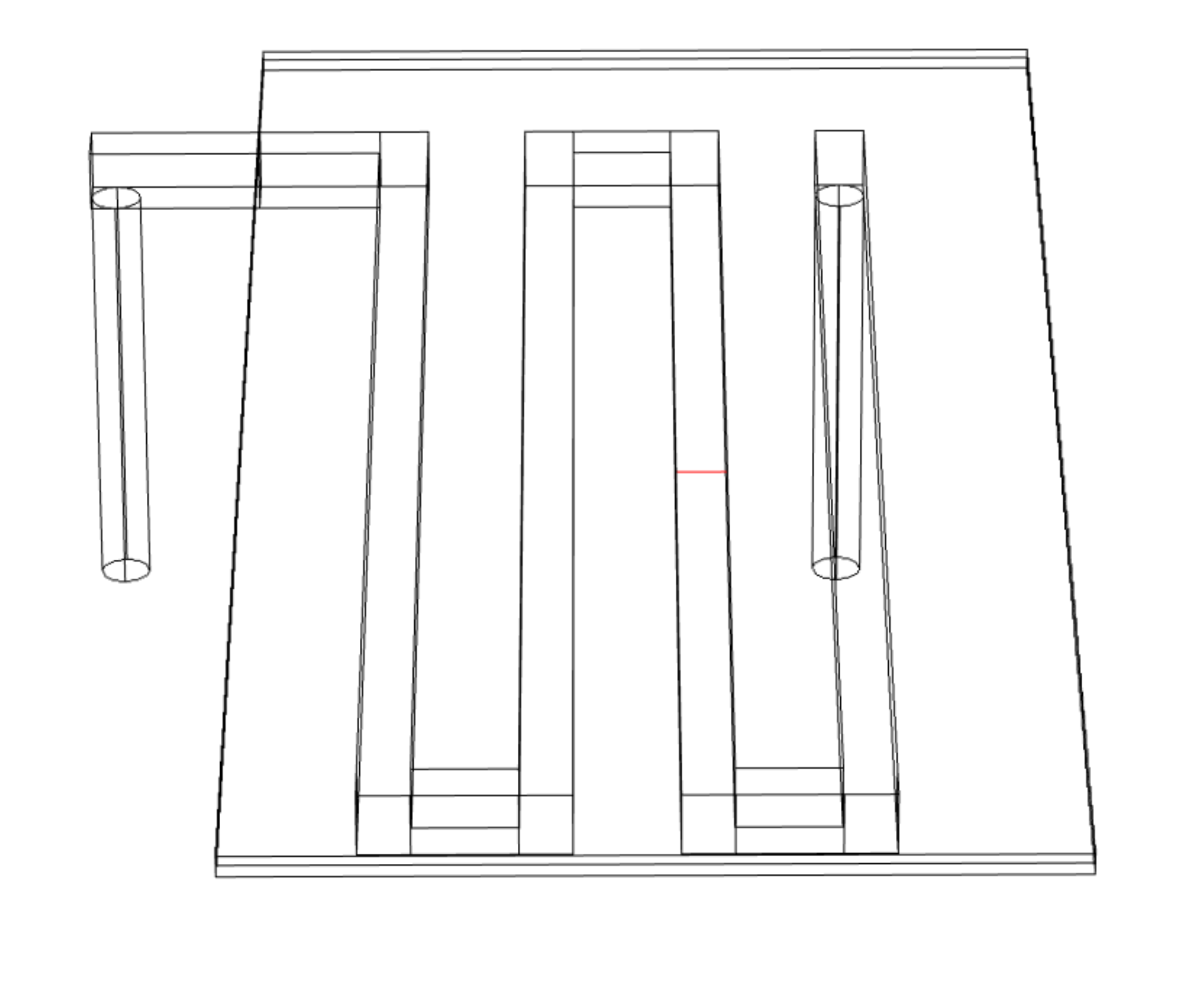}
        \caption{}
        \label{fig:xline}
    \end{subfigure}

    \begin{subfigure}{0.4\textwidth}
        \includegraphics[width=\textwidth]{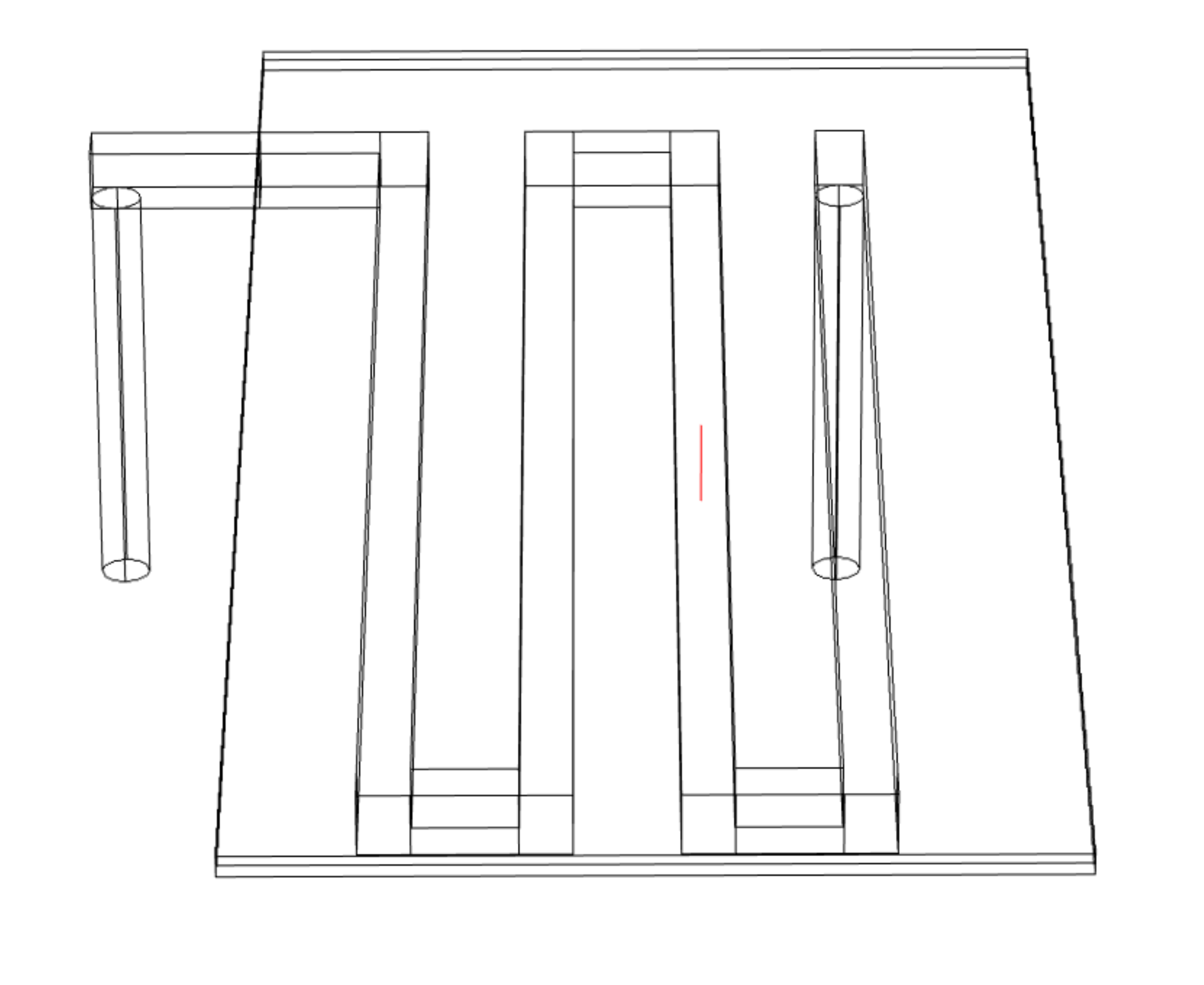}
        \caption{}
        \label{fig:zline}
    \end{subfigure}
    ~
    \begin{subfigure}{0.4\textwidth}
        \includegraphics[width=\textwidth]{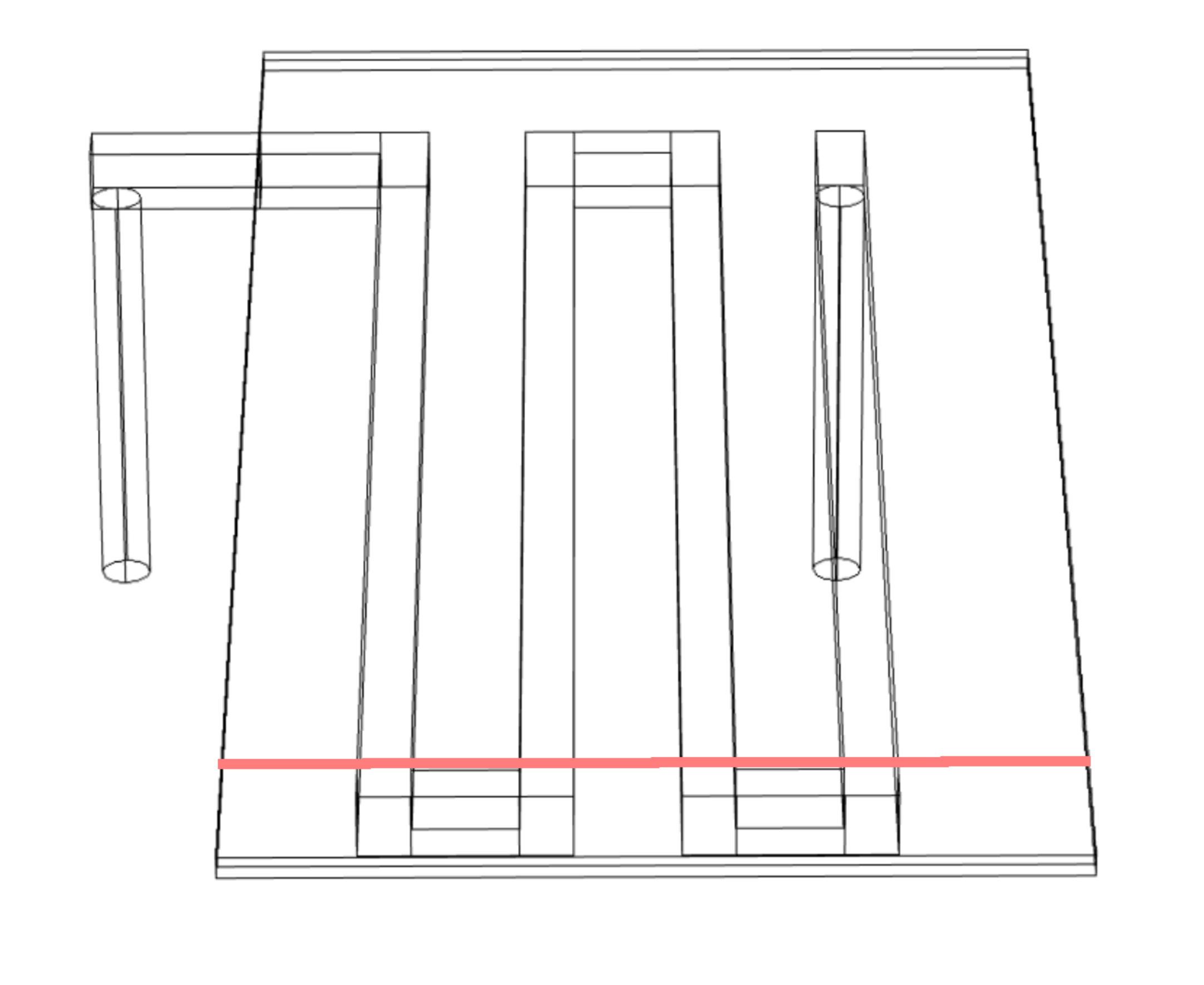}
        \caption{}
        \label{fig:cl_line}
    \end{subfigure}
    \caption{
        Geometrical entities used to obtain the variables relevant for mesh
        convergence, shown for the reduced geometry: \textbf{(a)}
        \textcolor{blue}{surface} used to obtain the apparent reaction rate,
        corresponding to the upper boundary of the CL domain; \textbf{(b,c)}
        \textcolor{red}{lines} used to track the flow speed $x$ and $z$
        profiles, respectively; and \textbf{(d)} \textcolor{red}{line} used to
        track the ozone partial pressure profile atop the CL domain.
    }
    \label{fig:var_track}
\end{figure}

    Four quantities were chosen as response variables in order to assess mesh
convergence. These are of experimental interest, and, with the exception of the
speed profiles, are readily available given the experimental
setup\cite{lopes15} and improvements being sought. These were selected to
allow comparison between the reduced and full geometries, as well as between
previous and present model formulations (see Part 1). It is understood that
some variables are functionals of the base variables solved in the model, and
thus might not be as sensitive to the mesh. The response variables are the
following:
\begin{itemize}
    \item the ratio $K'$ between reactant consumption, $\Delta\chi_{\mathrm{O
    _3}}$, and the apparent reaction rate $R'_{\mathrm{O_3}}$;
    \item the pressure drop, $\Delta P$, normalized by the inlet pressure,
    $P_{\mathrm{in}}$;
    \item the flow speed profile, $U$, along the $x$ and $z$ axes; and
    \item the ozone partial pressure, $P_{\mathrm{O_3}}$, along the $x$ axis.
\end{itemize}
\noindent
Concerning the acquisition of the data from the model, Figure
\ref{fig:var_track} shows, for the reduced model, the relevant geometric
entities used. The same are used for the full model, when appropriately scaled.
A brief description of the adopted procedure follows. $\Delta\chi
_{\mathrm{O_3}}$ and $\Delta P$ are simply the differences in value between
inlet and outlet averages, while $P_{\mathrm{in}}$ is the inlet average
pressure. $R'_{\mathrm{O_3}}$ is given by integration over the upper CL
boundary (Fig. \ref{fig:surf}) of the term $A_v\Gamma^*_s k_2\theta
_{\mathrm{O_3}}$, which is the volume-averaged decomposition term for
$\theta_{\mathrm{O_3}}$. The $U$ profiles were obtained at mid-length of the
third horizontal section (Figs. \ref{fig:xline} and \ref{fig:zline}), where the
$x$ axis profile was obtained at mid-height, while the $z$ axis profile was
obtained at mid-width. The $P_{\mathrm{O_3}}$ was obtained at the upper CL
boundary (Fig. \ref{fig:cl_line}), along a line parallel to the $x$ axis,
passing over the turn sections of the flow channel.

    Additional information and results associated with the response variables
may be found in Part 1. It should be noted that, in this part, the apparent
reaction rate $R'_{\mathrm{O_3}}$ is not normalized by the inlet concentration
as in Part 1, and consequently neither is the $K'$ ratio. This was done in
order to maximize sensitivity to the mesh resolution in the response variables.
In any way, as a matter of completeness, the SM\footnotemark[3] provides
relevant data for the quantification of the discretization error for the
$\mathrm{O_3}$ inlet concentration $C_{\mathrm{O_3,in}}$, the normalized
apparent reaction rate $R'_{\mathrm{O_3}}/C_{\mathrm{O_3,in}}$ and the
corresponding ratio $K'$, as well as the real and apparent stoichiometries
($\lambda$ and $\lambda'$, respectively, in Part 1). Thus, if needed, it is
possible to convert between response variables with the appropriate error
estimate.

    In addition to the response variables mentioned above, some
spatially-resolved quantities were also acquired, albeit mainly as
illustrations of the effects of the grid resolution. These are the ozone
partial pressure $\mathbf{P}_{\mathrm{O_3}}$, obtained at the upper CL boundary,
and $U$ surfaces, obtained with a $xy$ plane mid-height of the Ch domain and a
$yz$ plane cutting the middle of the second turn, before the section where the
$U$ profiles were taken. Finally, some quantities were obtained to explore
secondary issues, not related to mesh convergence. These were i) additional
$P_{\mathrm{O_3}}$ profiles, parallel to the one shown in Fig.
\ref{fig:cl_line}, obtained at different $y$ values: $y_1=\left(l_c-w_c\right)
/2$ (shown in Fig. \ref{fig:cl_line}), $y_2 = y_1 - \left(w_c+w_l\right)$ and
$y_3 = 0$. And ii) components of the ozone total molar flux vector, i.e.
diffusive and convective, given by the 1st and 2nd terms of Eq. \ref{eq:mf_ar}
respectively. These were acquired in the following situations: the
$z$-component of a given molar flux vector, integrated over the boundary
between Ch and MPS domains, or MPS and CL domains; and the magnitude of a given
molar flux vector, averaged over the MPS or CL domains.

    In order to quantitatively assess mesh convergence, the scalar response
variables, i.e. $K'$ and $\Delta P/P_{\mathrm{in}}$, were analyzed within the
framework of the grid convergence index (GCI), as reviewed by
Roache\cite{roache97}. Thus, the generalized Richardson extrapolation was
employed to provide an estimate on the error given by the discretization of the
computational domains. First, the effective refinement factors were calculated
using\cite{roache97}:
\begin{eqnarray}
    r_{\mathrm{eff}} & = & \left(\frac{N_{\mathrm{finer}}}{N_{\mathrm{coarser}}}\right)^{1/D} \label{eq:r_eff}
\end{eqnarray}
\noindent
where $N$ is the number of elements in a given mesh and $D = 3$ is the
dimensionality. Eq. \ref{eq:r_eff} was also used to calculate an effective,
normalized element spacing, $h$, used to plot and visually analyze the mesh
convergence of the selected variables. For such effect, a mesh is chosen
such that $r_{\mathrm{eff}}=h=1$, thus coarser meshes are given by $h>1$. For
the reduced geometry, $\lambda = 16$ was used, while for the full geometry
$\lambda = 12$ was used. The orders of accuracy were all calculated iteratively
following the suggestion given in \cite{roache97} (Eq. 15 of \cite{roache97}),
as the refinement ratios were non-integers and not constant. The generalized
Richardson extrapolation was then used to estimate the exact value for a given
variable $f$, by comparing the finest grid solution with a coarser one:
\begin{eqnarray}
    \bar{f}_{\mathrm{R}} & = & f_1 + \frac{\epsilon_{21}}{1-r_{12}^p}
\end{eqnarray}
\noindent
where we use the commonly employed convention of referencing the finer mesh
with index $1$, and progressively coarser meshes as $2$, $3$ and so on;
$\epsilon_{21} = f_2 - f_1$ is the error, while $r_{12}$ is the (effective)
refinement factor between meshes $1$ and $2$. The GCI is then given by:
\begin{eqnarray}
    \mathrm{GCI} & = & F_s\left|\bar{f}_{\mathrm{R}}-f_1\right| \label{eq:gci}
\end{eqnarray}
\noindent
where $F_s$ is a safety factor, taken as $F_s = 1.25$ in this
case\cite{roache97}. As mentioned above, the GCI is then used as an estimate
to the error of the model due to discretization of the computational domains.

    In addition to the generalized Richardson extrapolation, a mixed-order
extrapolation, described by Roy\cite{roy03}, was also employed. This
extrapolation scheme has the advantages of not requiring the calculation of the
order of accuracy or assuming the variables are in the asymptotic range, thus
potentially allowing non-monotonic behavior and singularities in the grid
convergence analysis. In this case, the discretized solution is given by a
power series in $h$\cite{roy03}:
\begin{eqnarray}
    f_k & = & f_{\mathrm{exact}} + g_1h_k + g_2h_k^2 + \mathcal{O}\left(h_k^3\right) \label{eq:mix_ext}
\end{eqnarray}
\noindent
where $k$ stands for a given mesh index following the convention pointed out
above. In \cite{roy03}, a 1st- and 2nd-order scheme is presented, with
$\bar{f}_{\mathrm{M}} \approx f_{\mathrm{exact}}$ being obtained by solving a
set of three linear equations up to powers in $h_k^2$. It is also
mentioned\cite{roy03} that it is possible to include higher-order terms
by coupling additional equations, and thus meshes, in order to increase the
order of the extrapolation. The set of linear equations was solved using the
\emph{linsolve} function of Octave\cite{octave}, without any additional
options. The error estimate is then calculated following a simple error
calculation:
\begin{eqnarray}
    \epsilon_{\mathrm{M}} & = & F_s\left|\bar{f}_{\mathrm{M}}-f_1\right| \label{eq:meps}
\end{eqnarray}
\noindent
where it can be seen that it is essentially the same as Eq. \ref{eq:gci}, when
using the Richardson extrapolation to estimate the exact value.

\section{\label{sec:res}Results}

    First a qualitative analysis of the convergence behavior of the response
variables will be given in Sections \ref{ssec:mc_res_lil} and
\ref{ssec:mc_res_full}, focusing at the reduced and full geometries
respectively; followed by Section \ref{ssec:mc_res_error} with a quantitative
estimation of the discretization error and, therefore, the convergence of the
variables with respect to the mesh. The different mesh schemes and relevant
response variables analyzed are described in Section \ref{ssec:mesh}, and
additional details can be found in the SM\footnotemark[3].

\subsection{\label{ssec:mc_res_lil}Reduced Geometry}

\begin{figure}
    \centering
    \begin{subfigure}{0.45\textwidth}
        \includegraphics[width=\textwidth]{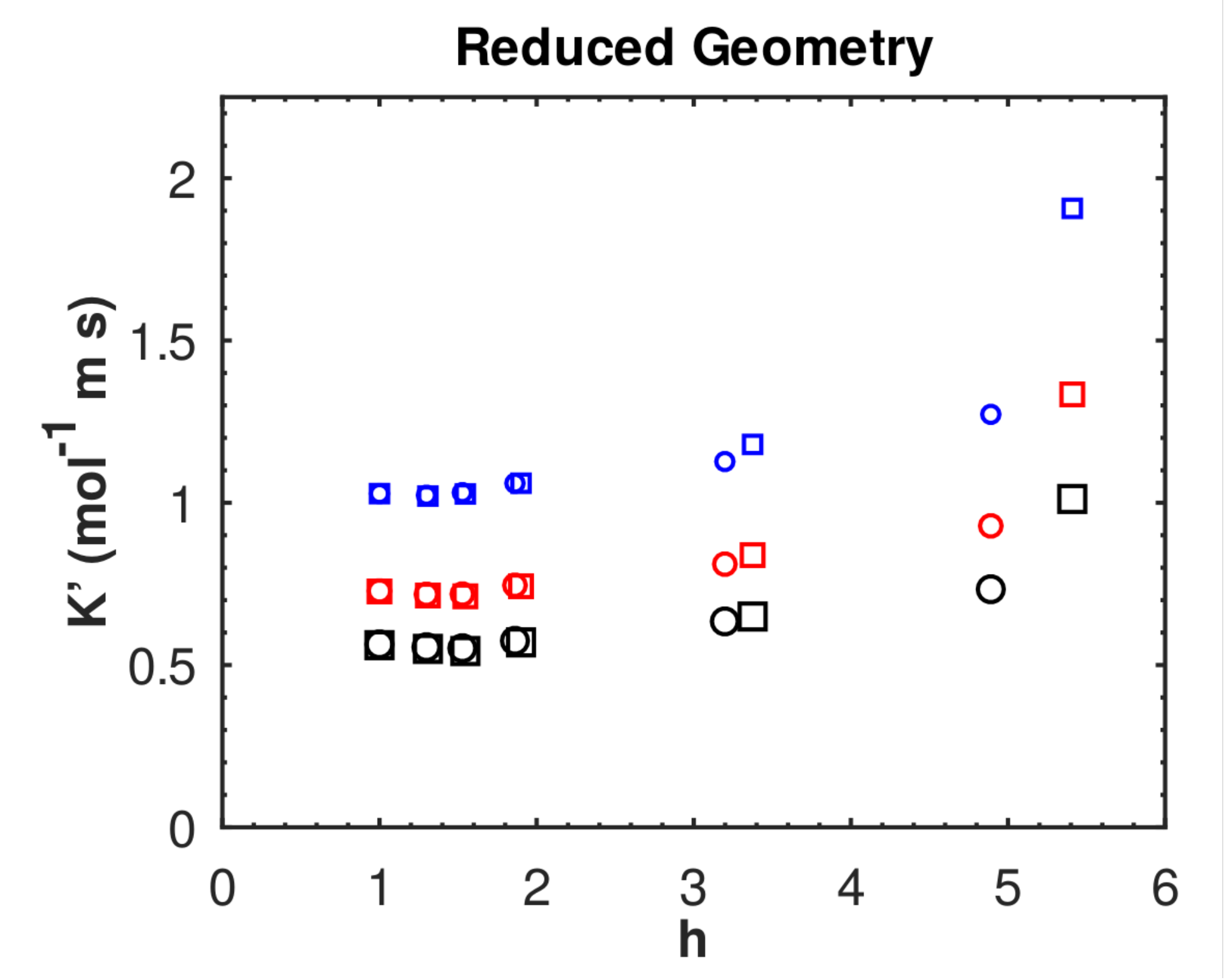}
        \caption{}
        \label{fig:lil_Ks}
    \end{subfigure}
    ~
    \begin{subfigure}{0.45\textwidth}
        \includegraphics[width=\textwidth]{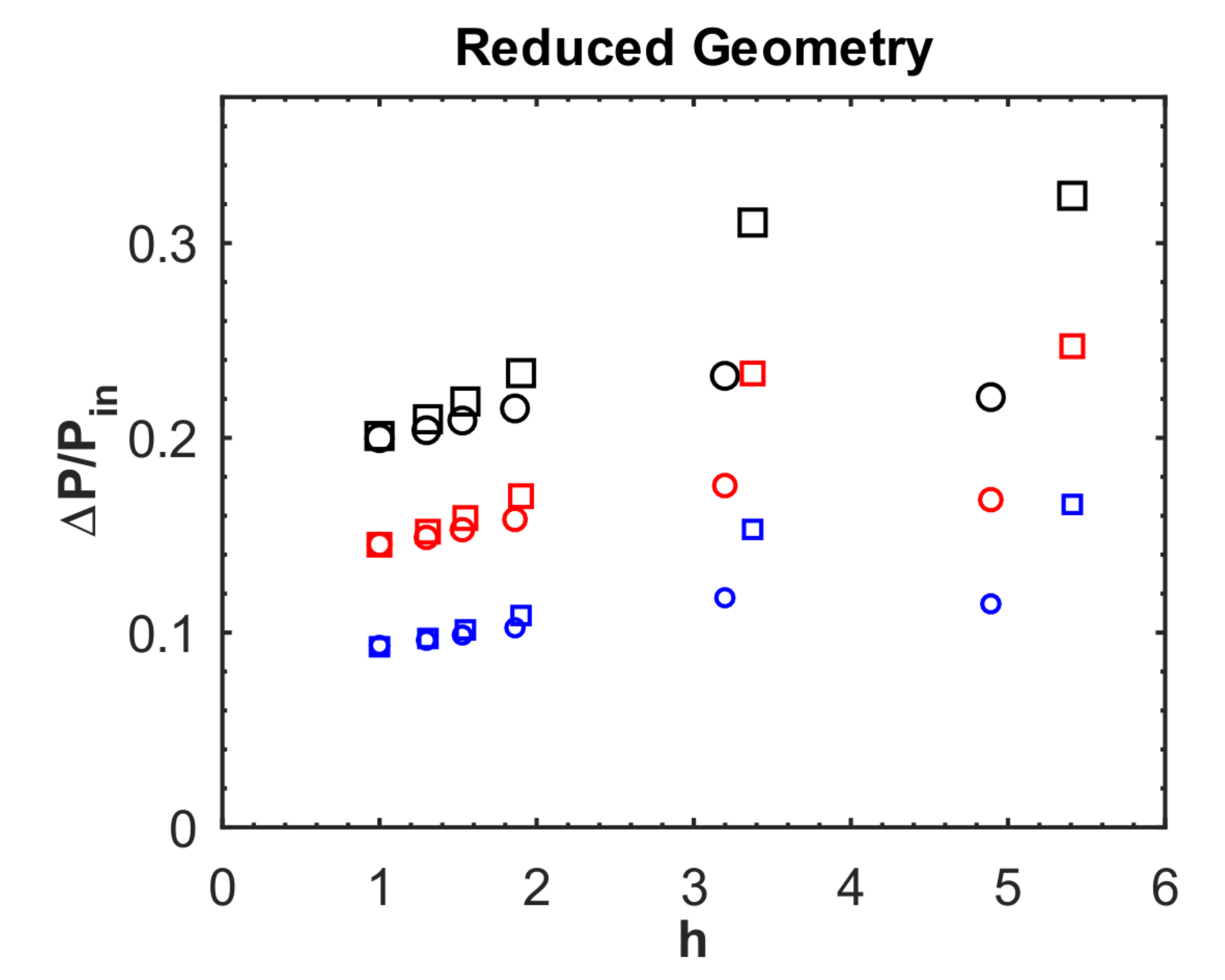}
        \caption{}
        \label{fig:lil_Npds}
    \end{subfigure}
    \caption{
        Scalar response variables as functions of the normalized effective
        element spacing, for the reduced geometry: \textbf{(a)} $K'=\Delta\chi
        _{\mathrm{O_3}}/R'_{\mathrm{O_3}}$ and \textbf{(b)} $\Delta P/P
        _{\mathrm{in}}$. The following values of inlet flow rate $Q$ are
        shown, with marker size increasing with $Q$: \textcolor{blue}{$250$},
        \textcolor{red}{$350$} and $450\ \mathrm{cm^3\ min^{-1}}$. Both mesh
        variants are shown: with ($\circ$) and without ($\Box$) BLEs.
    }
    \label{fig:lil_scalars}
\end{figure}

    We start with the analysis of the reduced model in order to cover the
full range of $\lambda$ studied. Following Section \ref{sssec:mc}, first the
scalar response variables will be analyzed, followed by the speed and partial
pressure profiles and finally the partial pressure surfaces. Thus, Figures
\ref{fig:lil_Ks} and \ref{fig:lil_Npds} present the ratios $K'=\Delta\chi
_{\mathrm{O_3}}/R'_{\mathrm{O_3}}$ and $\Delta P/P_{\mathrm{in}}$,
respectively, as a function of $h$. In each case both BLEs and no-BLEs study
variants are shown (See Section \ref{sssec:ma}). Individual variables are left
to Section SIII of the SM\footnotemark[3] (Fig. S2).

    Readily apparent features of Fig. \ref{fig:lil_scalars} are the large
differences between the BLE and no-BLE variants at coarser meshes, i.e.
high $h$, and the convergence of values at finer meshes, particularly for the
$K'$ ratio. In addition, it is seen that, for the $\Delta P/P_{\mathrm{in}}$
ratio, the existence of BLEs leads to a non-monotonic behavior at high $h$.
Indeed, this is also seen for $\Delta\chi_{\mathrm{O_3}}$ (Fig. S2a) and
$\Delta P$ (Fig. S2c), while, notably, $R'_{\mathrm{O_3}}$ (Fig. S2b) shows
such behavior for both mesh variants. The differences caused by the existence
of BLEs, or likewise the absence of BLEs, are linked to a much better
description of the velocity field $\mathbf{u}$ in coarser meshes, as will be
clear when discussing speed profiles. Likewise, the non-monotonic behavior seen
with $R'_{\mathrm{O_3}}$ is linked to the poor description of $\mathbf{u}$
 along the thickness of the Pm domains, and, consequently, to that of the ozone
mass fraction $\omega_{\mathrm{O_3}}$.

    Additional features of Fig. \ref{fig:lil_scalars} are the non-monotonic
behavior of $K'$ at low $h$, and a relatively slow convergence of $\Delta P/P
_{\mathrm{in}}$. It is not clear why $K'$, and likewise $\Delta\chi_{\mathrm{O
3}}$, shows such behavior, or why is it dependent on $Q$. Suspicion once again
resides with $\mathbf{u}$, as does with the reason behind the slow convergence
of $\Delta P/P_{\mathrm{in}}$ (and also of $\Delta P$), and will be further
explored below. Nevertheless, Roache\cite{roache97} does point out that, far
from the asymptotic regime, a variable can show non-monotonic behavior as
function of element spacing. Indeed, considering the apparent lack of studies
exploring mesh convergence in models of fuel cells and similar devices, an
important result of this work is exactly the demonstration that experimentally
relevant response variables, even functionals of base variables, show
non-monotonic behavior in a relatively wide range of $h$ for 3D unstructured
meshes. Thus, rather unfortunately, demonstrating mesh convergence for
experimentally relevant variables in fuel cells may be a harder case than
usually assumed.

\begin{figure}
    \centering
    \begin{subfigure}{0.45\textwidth}
        \includegraphics[width=\textwidth]{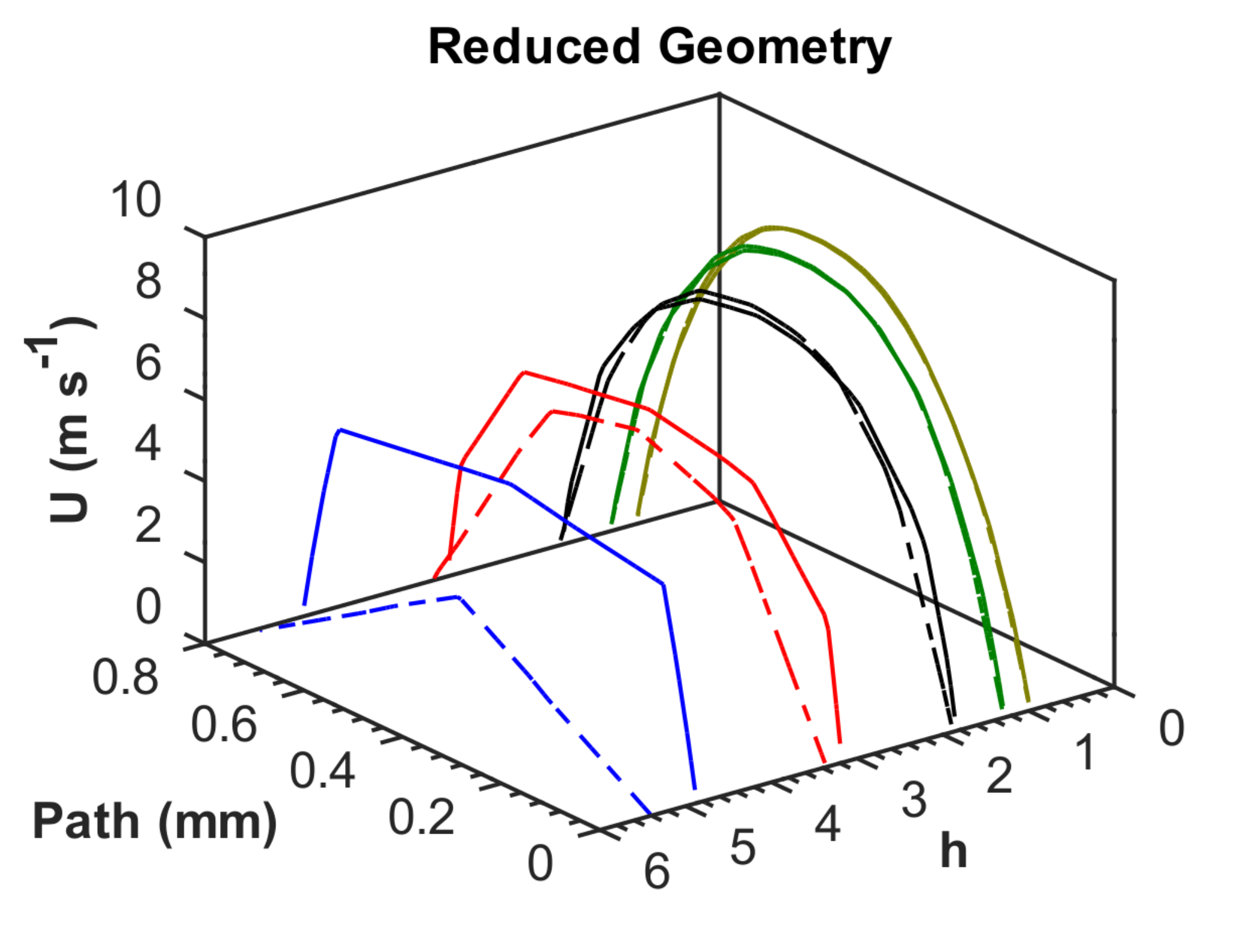}
        \caption{}
        \label{fig:lil_Uxs}
    \end{subfigure}
    ~
    \begin{subfigure}{0.45\textwidth}
        \includegraphics[width=\textwidth]{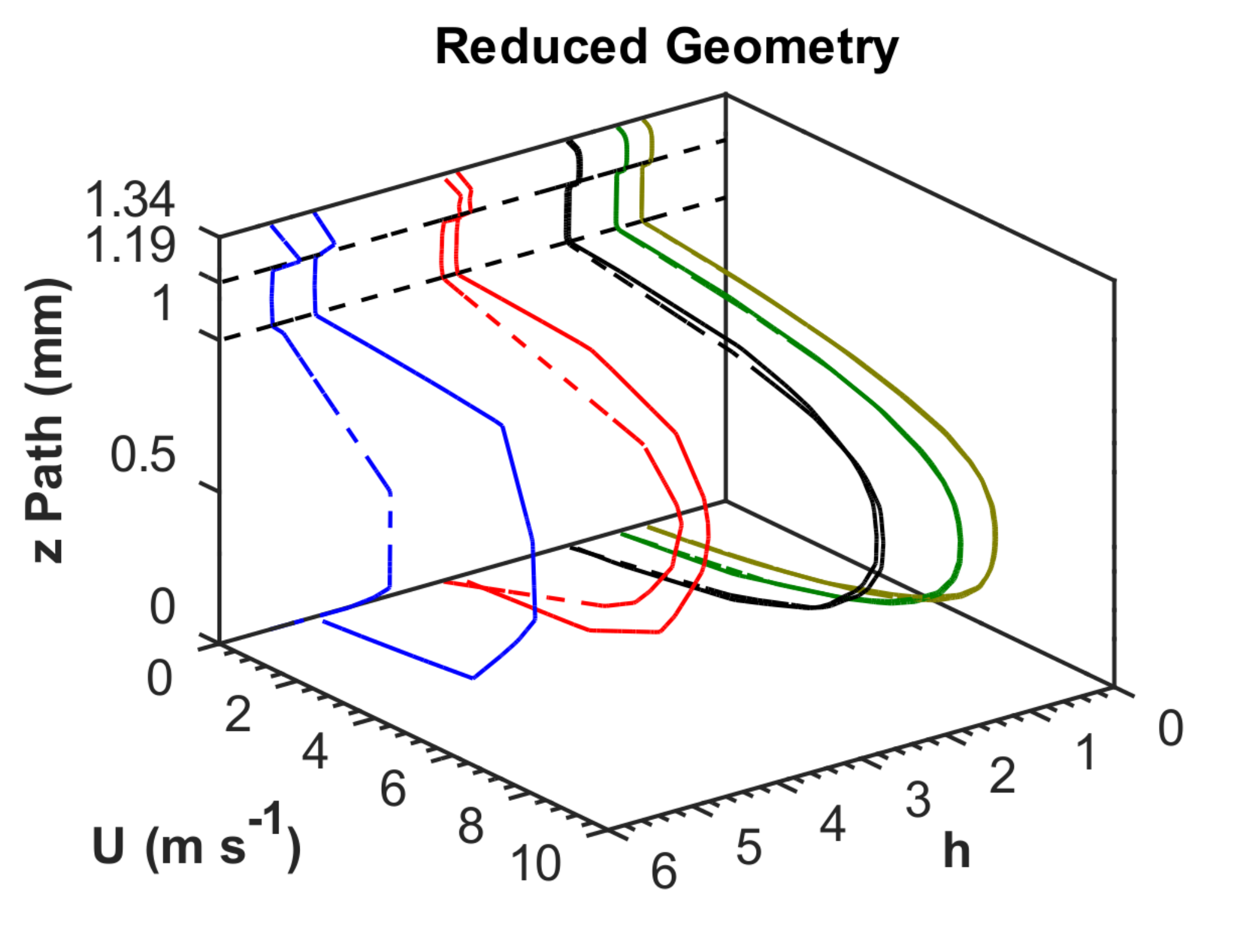}
        \caption{}
        \label{fig:lil_Uzs}
    \end{subfigure}

    \begin{subfigure}{0.45\textwidth}
        \includegraphics[width=\textwidth]{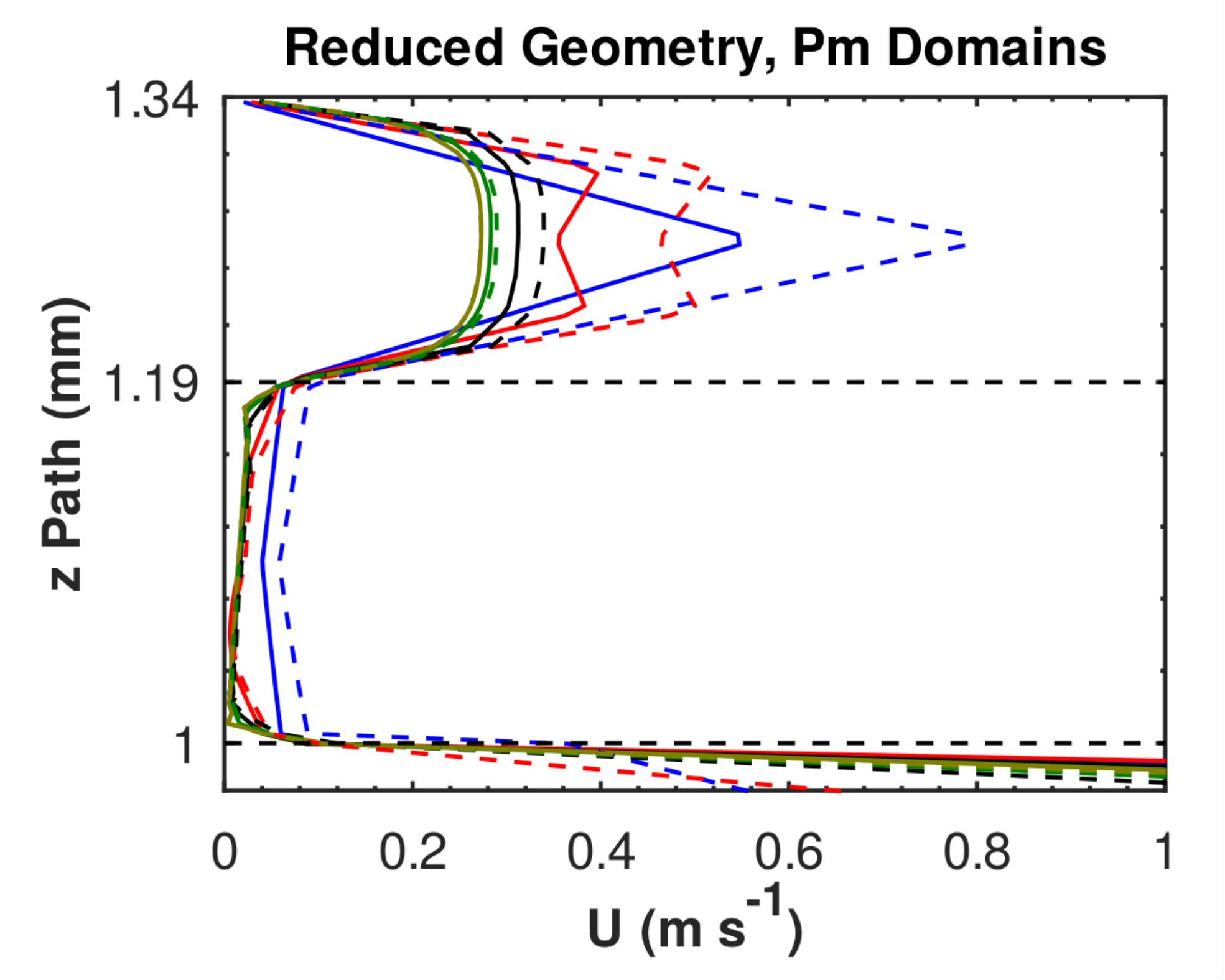}
        \caption{}
        \label{fig:Uzs_zoom}
    \end{subfigure}
    \caption{
        Flow speed profiles as functions of the normalized effective element
        spacing, for $Q = 350\ \mathrm{cm^3\ min^{-1})}$: \textbf{(a)} profile
        along the $x$ axis, \textbf{(b)} along the $z$ axis, with \textbf{(c)}
        zoom in on the Pm domains. Both BLE (full lines) and no-BLEs
        (dashed lines) mesh variants are shown. The following values of
        $\lambda$ are shown: \textcolor{blue}{$2$}, \textcolor{red}{$4$},
        $8$, \textcolor[rgb]{0 0.5 0}{$12$}, and \textcolor[rgb]{0.5 0.5 0}
        {$16$}. In \textbf{(b-c)} the dashed black lines in the $y=0$ plane
        correspond to the boundaries between Ch and MPS, and MPS and CL
        domains, while in \textbf{(c)} the MPS-CL boundary is depicted.
    }
    \label{fig:lil_Us}
\end{figure}

    We now proceed to analyze the profiles, starting with the speed profiles,
$U$. Figure \ref{fig:lil_Us} shows the $U$ profiles parallel to $x$ and $z$
axes, at $Q = 350\ \mathrm{cm^3\ min^{-1}}$. Once again a marked difference is
seen between meshes with and without BLEs at high $h$ (low $\lambda$). As
pointed out above, such marked differences at high $h$ are thought to be
responsible for the distinct behaviors seen for each mesh variant in Figs.
\ref{fig:lil_scalars} (likewise in Fig. S2). The absence of BLEs leads
to a large underestimation of $U$ in the Ch domain, which is known to
contribute significantly to reactant transport to the Pm
domains\cite{beruski17}, thus affecting total reaction rate and reactant
consumption. It is clear, then, from Figs. \ref{fig:lil_Uxs} and
\ref{fig:lil_Uzs} that the presence of BLEs provides a much needed resolution
close to the Ch walls, however that necessity fades as the mesh improves.
Indeed, at $\lambda = 12$ the mesh variants display close similarity, with an
average difference of $\approx 3.5\%$ for both $x$ and $z$ profiles.

    Concerning the dependence of $U$ with $h$, it can be seen that, as $h
\rightarrow 0$, the flow in the Ch domain assumes a Poiseuille-like shape, i.e.
parabolic as function of wall distance. This is of course expected at the
center of the channel, for a well-developed flow field, and suggests that
meshes with $\lambda \geq 12$ provide a good description of the flow field. A
similar trend is seen in Fig. \ref{fig:Uzs_zoom} for the CL, however with an
initial overestimation of $U$ and a sharp gradient along $z$. This poor
description of $U$, when comparing $\lambda = 2$ to $16$, might be related to
the non-monotonic behavior observed for $R'_{\mathrm{O_3}}$ (Fig. S2b) at high
$h$: while $U$ is overestimated at the peak, the sharp gradient leading to $U
=0$ at the CL upper boundary might contribute to lower rates of $\mathrm{O_3}$
transport to the surface, thus reducing the observed value of $R'
_{\mathrm{O_3}}$. For the MPS, on the other hand, $U$ takes a more complex
shape that is quite inappropriately modeled for $\lambda \leq 4$. Such
profile is thought to result from the interaction with the Ch and CL domains,
where both have larger permeabilities than the MPS (with $\kappa_{\mathrm{Ch}}
\rightarrow \infty$). While the case of coupled free and porous media flow
is widely studied in general, when approaching the problem analytically it is
common to assume a semi-infinite porous medium domain (see for instance
\cite{ochoa95a,lebars06}), which leads to a transition zone at the boundary and
a plug-flow-like front in the porous medium that is typical of Darcy's law.
Interactions between porous media, or walls, are less explored and possibly
quite idiosyncratic, leading to unconventional effects (see for instance
\cite{ye15}). Thus, it is difficult to judge if the $U$ profile in the Pm is
well-described, as some changes are still noticeable between $\lambda = 12$ and
$16$.

\begin{figure}
    \centering
    \begin{subfigure}{0.45\textwidth}
        \includegraphics[width=\textwidth]{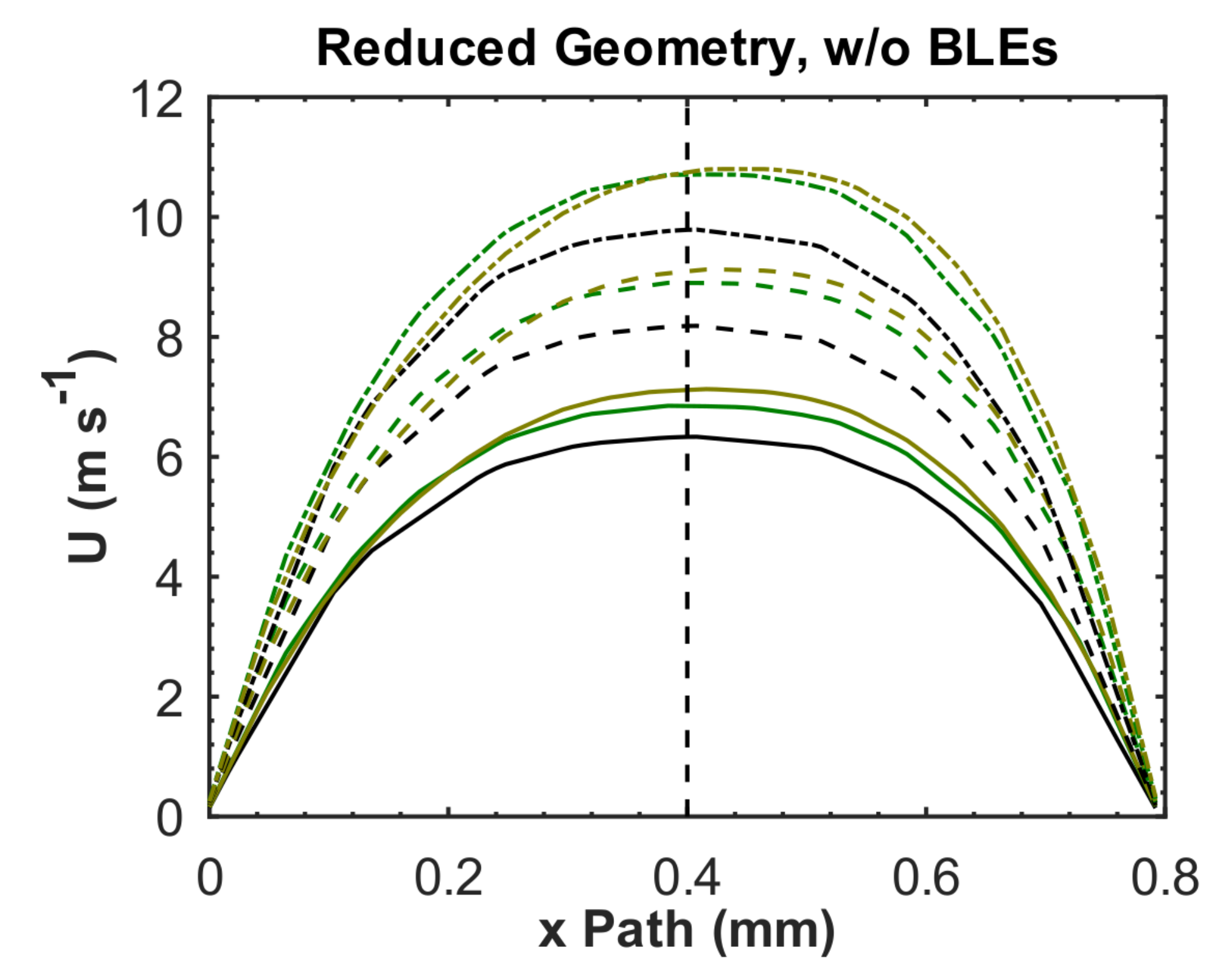}
        \caption{}
        \label{fig:lil_Uxs2}
    \end{subfigure}
    ~
    \begin{subfigure}{0.45\textwidth}
        \includegraphics[width=\textwidth]{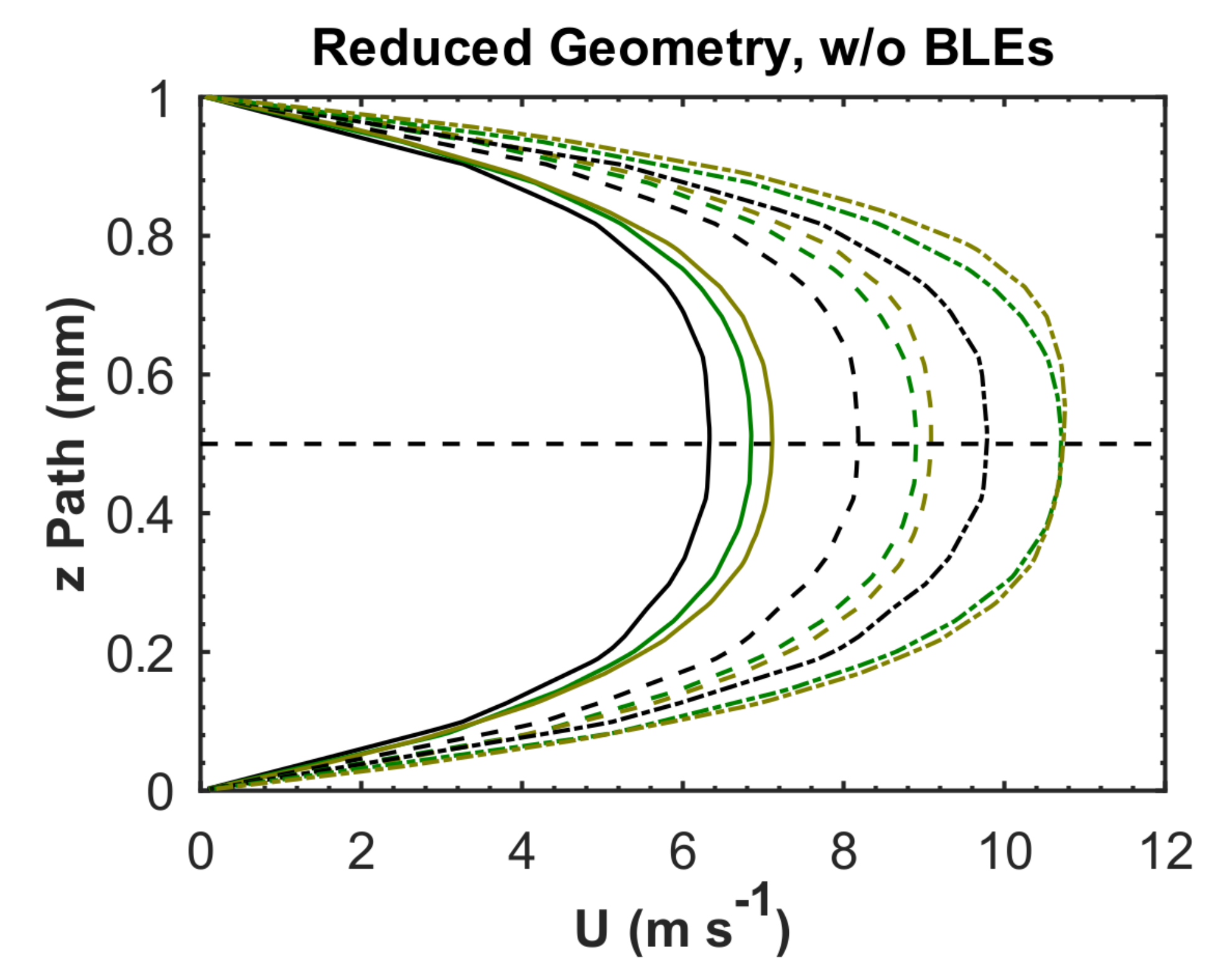}
        \caption{}
        \label{fig:lil_Uzs2}
    \end{subfigure}
    \caption{
        Flow speed profiles for selected values of $\lambda$ and $Q$, using the
        no-BLEs mesh variant: \textbf{(a)} profile along the $x$ axis and
        \textbf{(b)} along the $z$ axis. The following values of $\lambda$ are
        shown: $8$, \textcolor[rgb]{0 0.5 0}{$12$}, and \textcolor[rgb]
        {0.5 0.5 0}{$16$}. The following values of $Q$ are shown: $250$ (full),
        $350$ (dash), and $450\ \mathrm{cm^3\ min^{-1}}$ (dash-dot).
        The dashed black lines parallel to the $U$ axes are visual guides
        stressing the middle of the path along the Ch domain.
    }
    \label{fig:lil_Us2}
\end{figure}

    Another point worth noticing concerns the symmetry of the $U$ profiles with
wall distance. While it is not clear in Fig. \ref{fig:lil_Us}, and the effect
is somewhat subtle, the improved description of $\mathbf{u}$ as $\lambda$
increases allows modeling finer structures that might provide additional
insight on the interpretation of experimental results. Figure \ref{fig:lil_Us2}
shows superimposed $U$ profiles of the no-BLEs mesh variant, for $\lambda\geq 8$
and selected values of $Q$. For the sake of visibility the BLEs variant is not
shown, however it does show the same trends discussed below, as suggested by
the close agreement between mesh variants shown in Fig. \ref{fig:lil_Us}. For
the $x$ profiles (Fig. \ref{fig:lil_Uxs2}) it can be seen that, for $\lambda =
8$, $U$ is virtually symmetric around the center of the channel, even as $Q$
increases. As $\lambda$ increases, however, it becomes increasingly apparent
that $U$ is not symmetric, in particular for high $Q$. It is noteworthy that
for $\lambda = 16$ the $U$ profile is asymmetric throughout the $Q$ range
simulated. This is likely to be an effect of secondary flows at the turns of
the channel, which are better resolved the finer the mesh. This will be further
explored below, but it already points out that the turns of a serpentine flow
channel have far-reaching effects on the flow field, besides the fact that it
greatly contributes to the convective transport in porous
media\cite{beruski17}. Regarding the $z$ profiles (Fig. \ref{fig:lil_Uzs2}),
the asymmetry in $U$ is much more subtle: there are virtually no deviations for
$\lambda = 8$ and $12$ throughout the $Q$ range, while $\lambda = 16$ shows
some asymmetry at $Q \geq 350\ \mathrm{cm^3\ min^{-1}}$, becoming apparent at
$450\ \mathrm{cm^3\ min^{-1}}$. This effect appears to be caused by the porous
interface, between Ch and MPS domains, as no such asymmetry is seen when a
no-slip boundary condition is set instead (not shown). It is not clear,
however, if it is also related to the turns of the flow channel.

\begin{figure}
    \centering
    \begin{subfigure}{0.3\textwidth}
        \includegraphics[width=\textwidth]{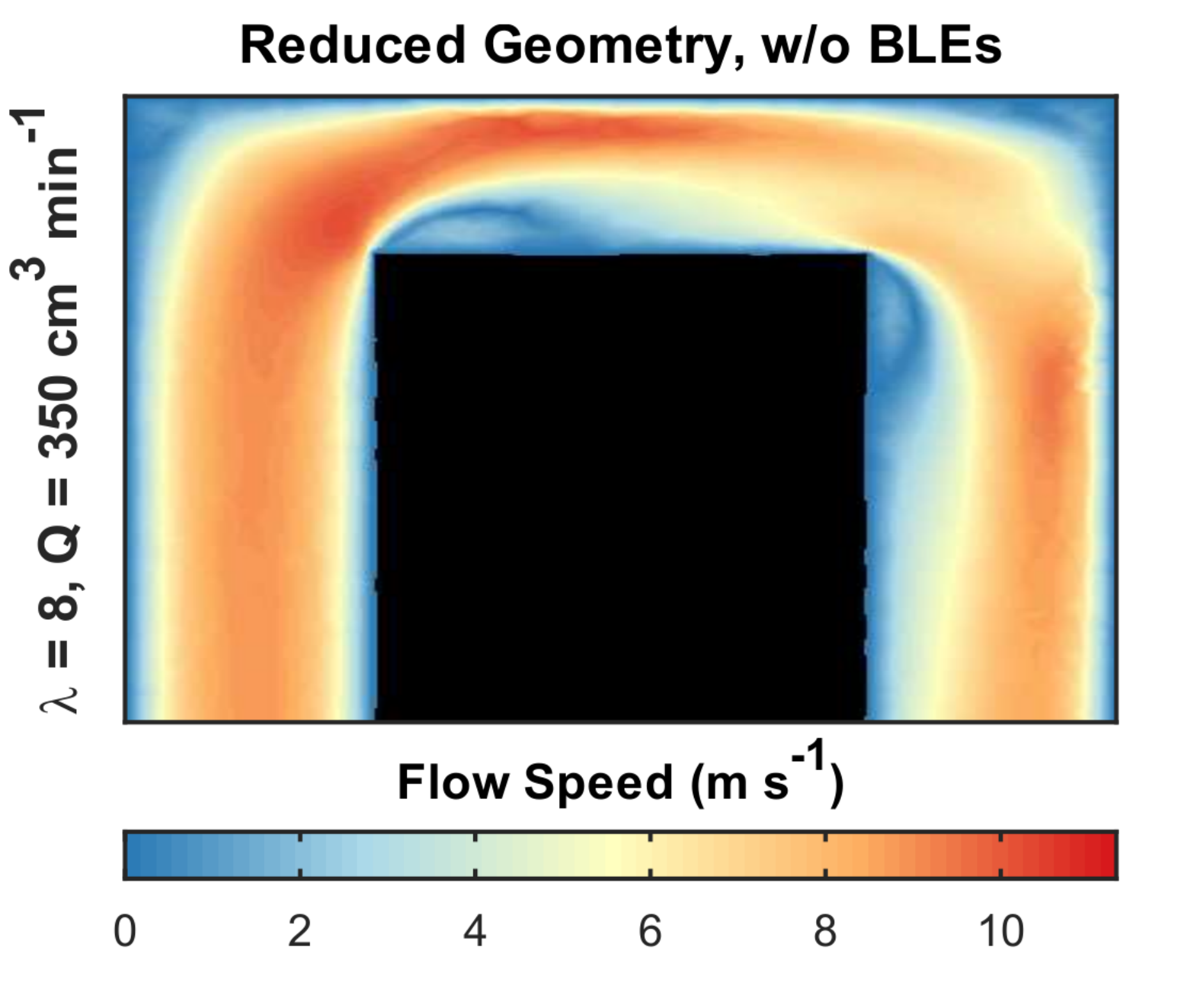}
        \caption{}
        \label{fig:lil_Uxy8}
    \end{subfigure}
    ~
    \begin{subfigure}{0.3\textwidth}
        \includegraphics[width=\textwidth]{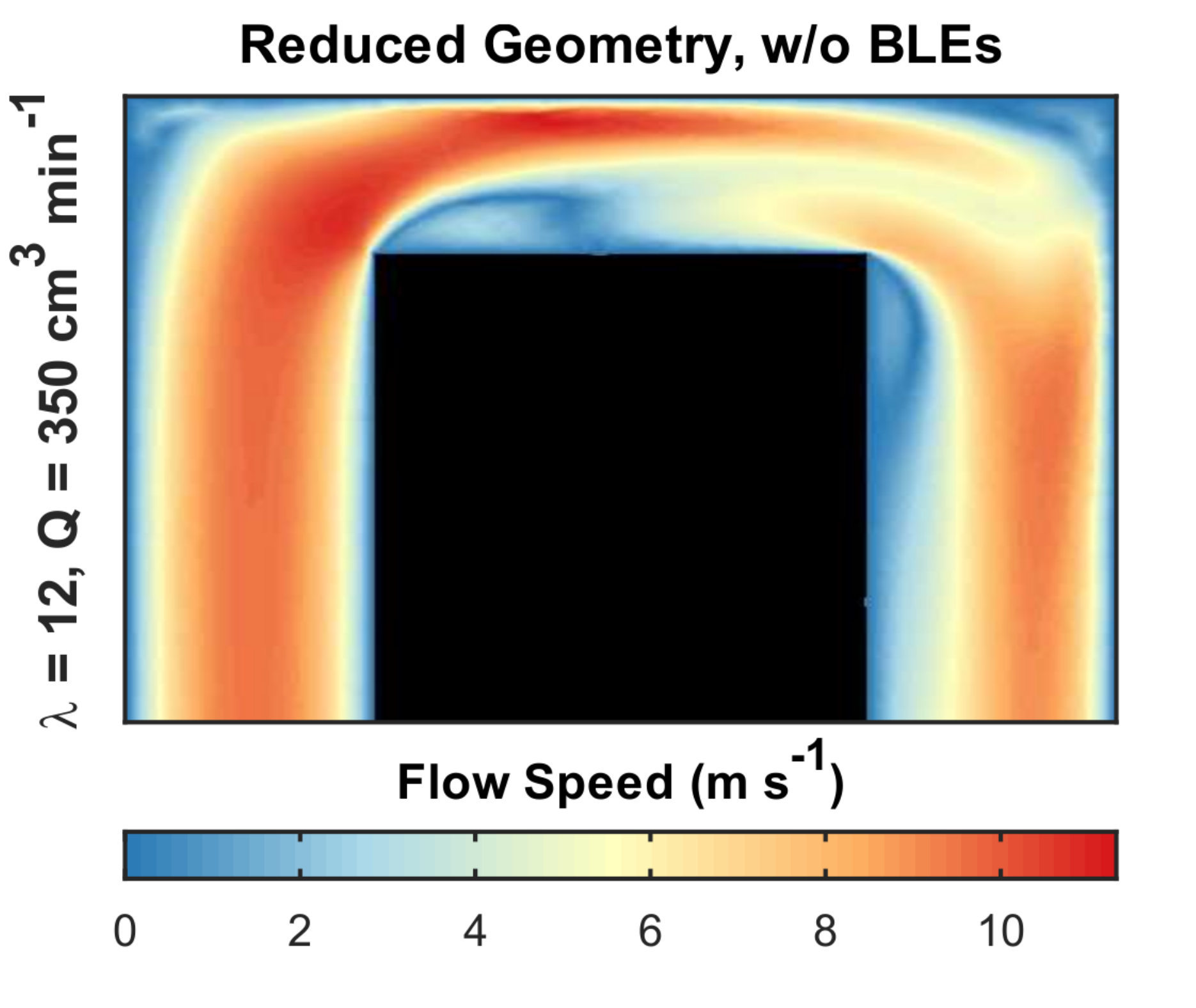}
        \caption{}
        \label{fig:lil_Uxy12}
    \end{subfigure}
~
    \begin{subfigure}{0.3\textwidth}
        \includegraphics[width=\textwidth]{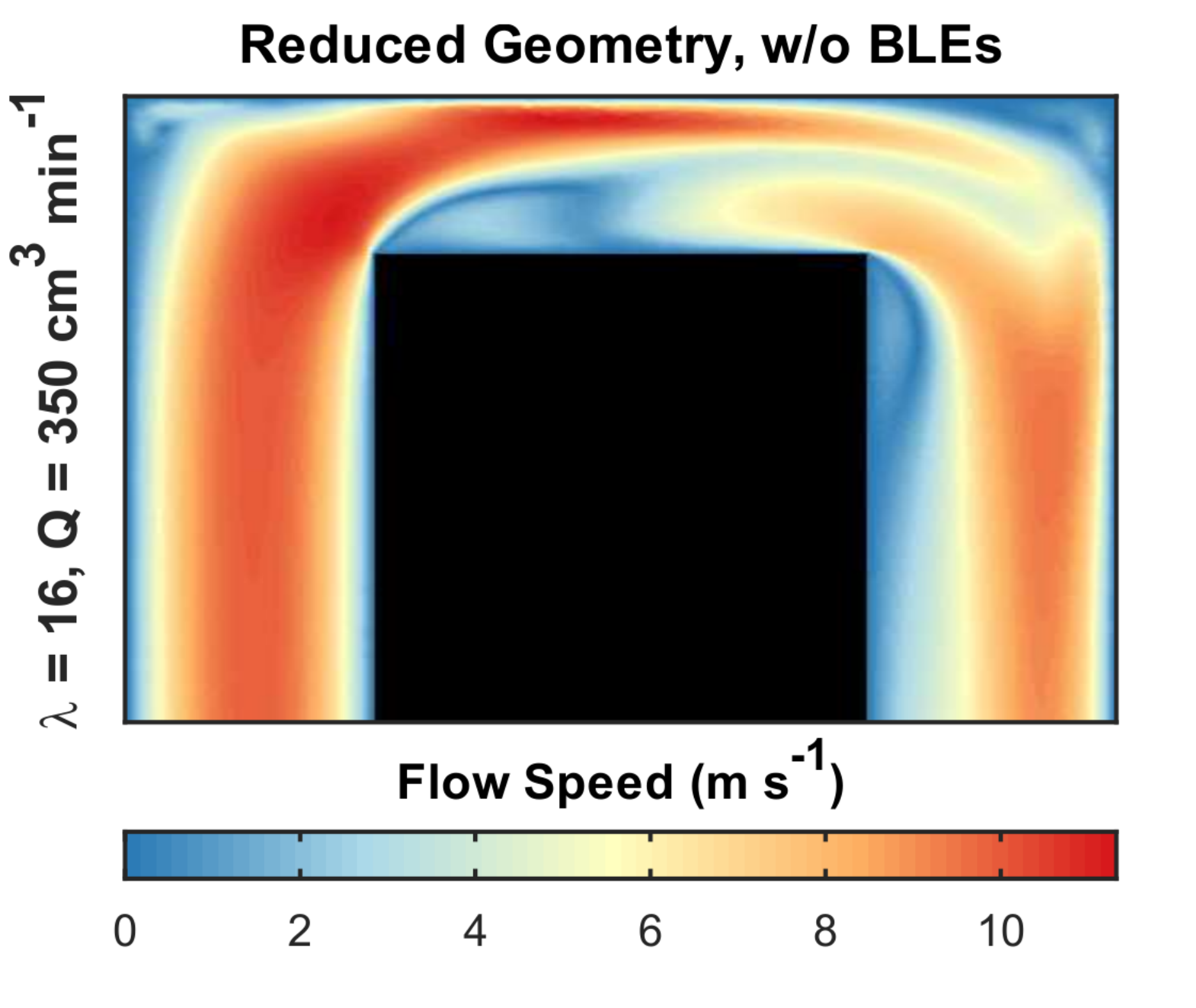}
        \caption{}
        \label{fig:lil_Uxy16}
    \end{subfigure}

    \begin{subfigure}{0.3\textwidth}
        \includegraphics[width=\textwidth]{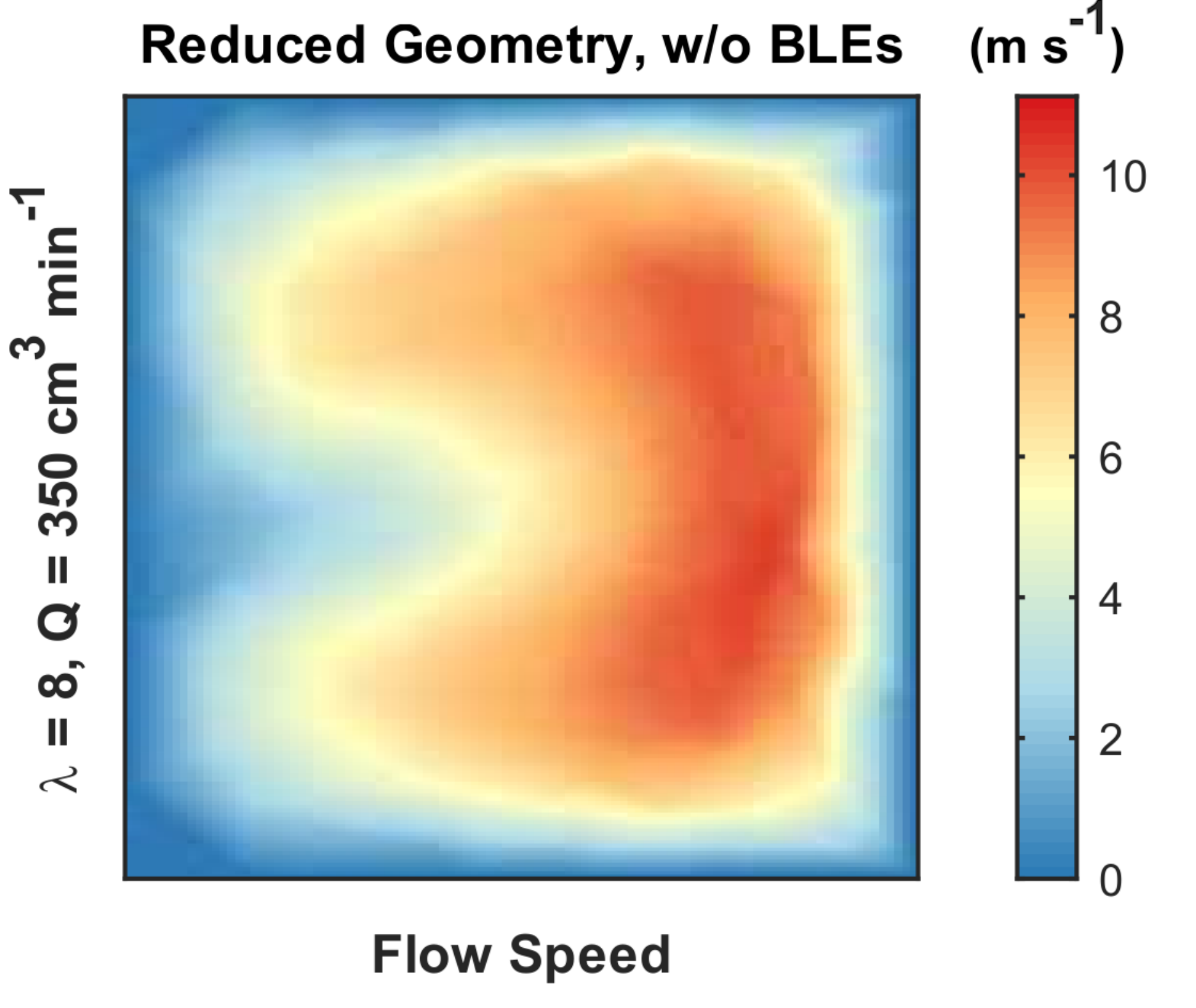}
        \caption{}
        \label{fig:lil_Uyz8}
    \end{subfigure}
    ~
    \begin{subfigure}{0.3\textwidth}
        \includegraphics[width=\textwidth]{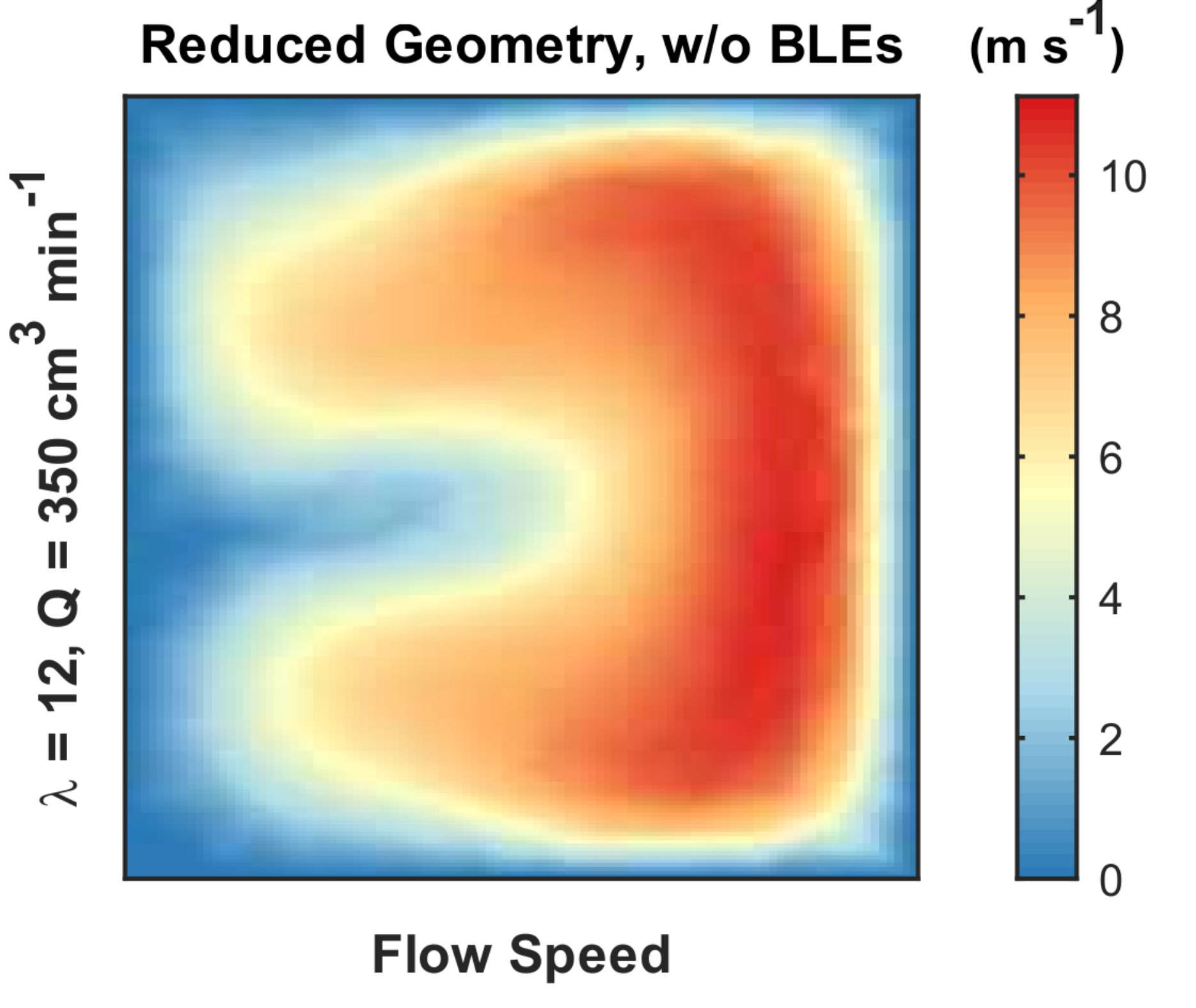}
        \caption{}
        \label{fig:lil_Uyz12}
    \end{subfigure}
~
    \begin{subfigure}{0.3\textwidth}
        \includegraphics[width=\textwidth]{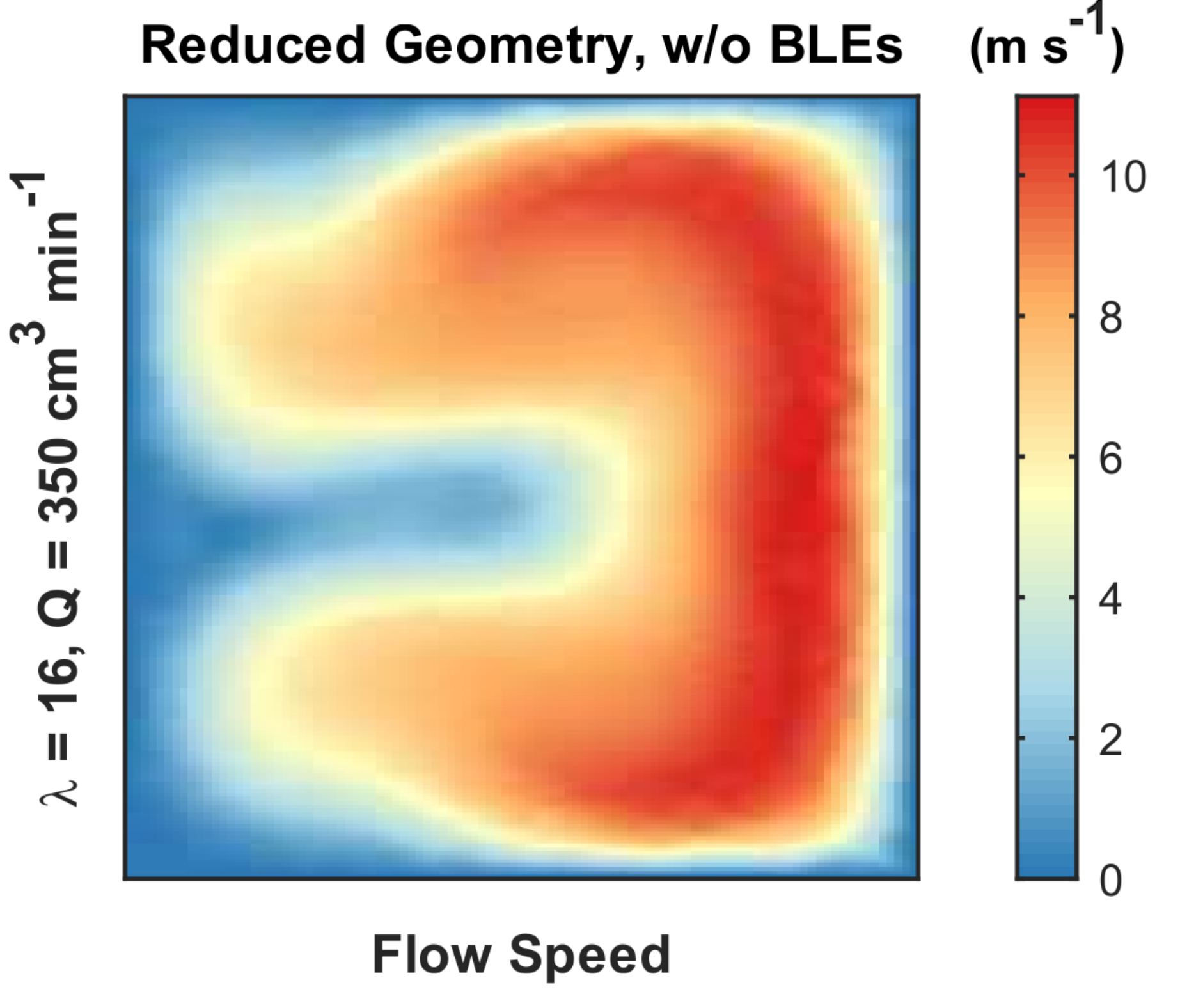}
        \caption{}
        \label{fig:lil_Uyz16}
    \end{subfigure}
    \caption{
        Flow speed surfaces for selected values of $\lambda$ and $Q = 350\
        \mathrm{cm^3\ min^{-1}}$, using the no-BLEs mesh variant: \textbf{(a-c)}
        $xy$ plane mid-height of the 2nd turn of the Ch domain and \textbf{(d-
        f)} $yz$ plane cutting through the middle of the turn. The values of
        $\lambda$ sampled are indicated, and include $\lambda = 8$, $12$, and
        $16$.
    }
    \label{fig:lil_Us3}
\end{figure}

    We sidetrack now for a moment to consider the spatially-resolved $U$
surfaces, in order to get a glimpse of the effects of the mesh on $\mathbf{u}$
as mentioned above. Figure \ref{fig:lil_Us3} shows $U$ surfaces, for $\lambda
\in \left[8,12,16\right]$ and $Q = 350\ \mathrm{cm^3\ min^{-1}}$, at the second
turn of the serpentine Ch domain. Figures \ref{fig:lil_Uxy8} to
\ref{fig:lil_Uxy16} show a mid-height $xy$ plane, where it is readily
noted the existence of known secondary flows, due to the corners and from
boundary layer separation (BLS). While real turns are not $90^\circ$ sharp, it
has been shown, albeit indirectly, that the simulated flow field correlates
quite well with experiments\cite{lopes15,beruski17}. As $\lambda$ increases, a
larger effect of the BLS is noticed, and recirculation at the corners become
better resolved. Better resolution is of course the whole point, however it is
noteworthy how the improved description of the BLS significantly changes the
flow field after the turn: at $\lambda = 8$, the flow field is almost
completely re-developed at the end of the Ch section shown, while at $\lambda =
16$, there is a significant shift of the flow field to the right. This builds
upon the discussion around Fig. \ref{fig:lil_Uxs2}, establishing that BLS is
a major source of the $U$ profile asymmetry observed, and that fine grid
resolution is needed to properly describe such effects. In addition to effects
at the Ch domain, BLS is thus expected to impact on reactant distribution at
the Pm domains. As shown in \cite{beruski17}, secondary flows at the turns of
serpentine Ch are major drivers of reactant distribution, therefore one
would expect that the effect caused by BLS just uncovered might have
implications as well. One such implication is a discrepancy between
experimental and simulated $\mathbf{P}_{\mathrm{O_3}}$ surfaces, noted in
\cite{beruski17}, concerning a broader plume of $\mathrm{O_3}$ along each
horizontal section of the Ch in the experimental device. We shall come back to
this point below, when analyzing $\mathrm{O_3}$ partial pressure.

    It is also worth looking at the cross-section of the turn. It was shown in
\cite{beruski17} that the turns of the serpentine Ch creates vortices, which
are thought to contribute to reactant transport to the Pm domains. Figures
\ref{fig:lil_Uyz8} to \ref{fig:lil_Uyz16} show $U$ at a $yz$ plane at the
middle of the turn, also for $\lambda \in \left[8,12,16\right]$ and $Q=350\
\mathrm{cm^3\ min^{-1}}$. Far from being a numerical artifact, it is indeed
seen that these vortices are robust, and that increased grid resolution shows
better defined structures, with higher $U$ and possibly greater contribution to
reactant transport.

    In all, Figs. \ref{fig:lil_Us2} and \ref{fig:lil_Us3} provide further
evidence that the flow field in fuel cell flow channels, in particular
serpentine ones, are rather complex and require good grid resolution to
properly capture its features. It also sheds some light on the convergence
behavior of the scalar variables analyzed above (Fig. \ref{fig:lil_scalars}):
i) there is a significant change in $\mathbf{u}$ at $\lambda > 8$, which might
underlie the non-monotonic behavior of the $K'$ ratio at low $h$ (or the
individual variables), and ii) the grid resolution necessary to resolve such
relatively small-scale structures might underlie the slow convergence rate of
$\Delta P/P_{\mathrm{in}}$. Thus, it is of utmost importance that grid
convergence studies be performed before validating and actually employing fuel
cell models for predictions.

\begin{figure}
    \centering
    \begin{subfigure}{0.45\textwidth}
        \includegraphics[width=\textwidth]{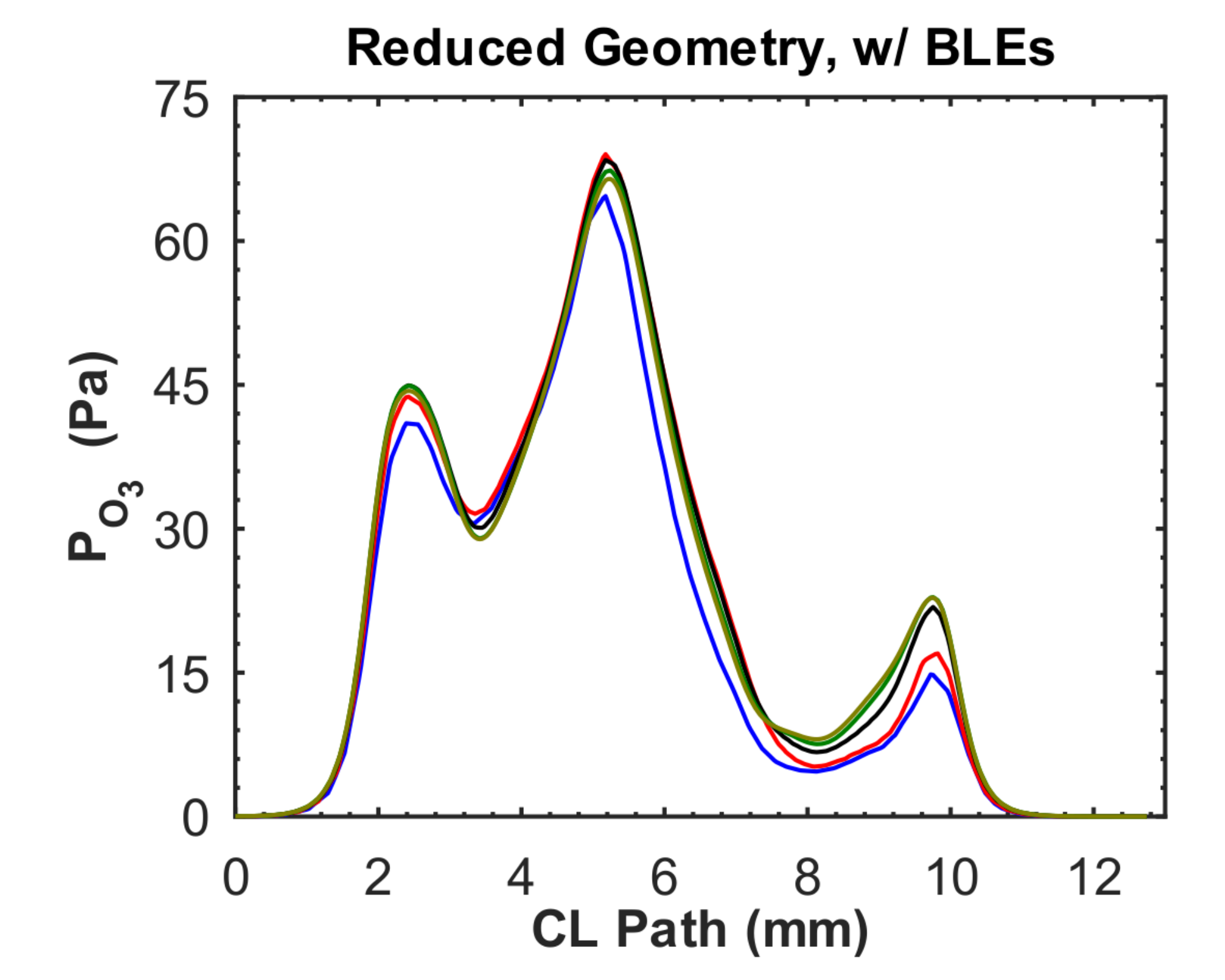}
        \caption{}
        \label{fig:lil_Po3_w}
    \end{subfigure}
    ~
    \begin{subfigure}{0.45\textwidth}
        \includegraphics[width=\textwidth]{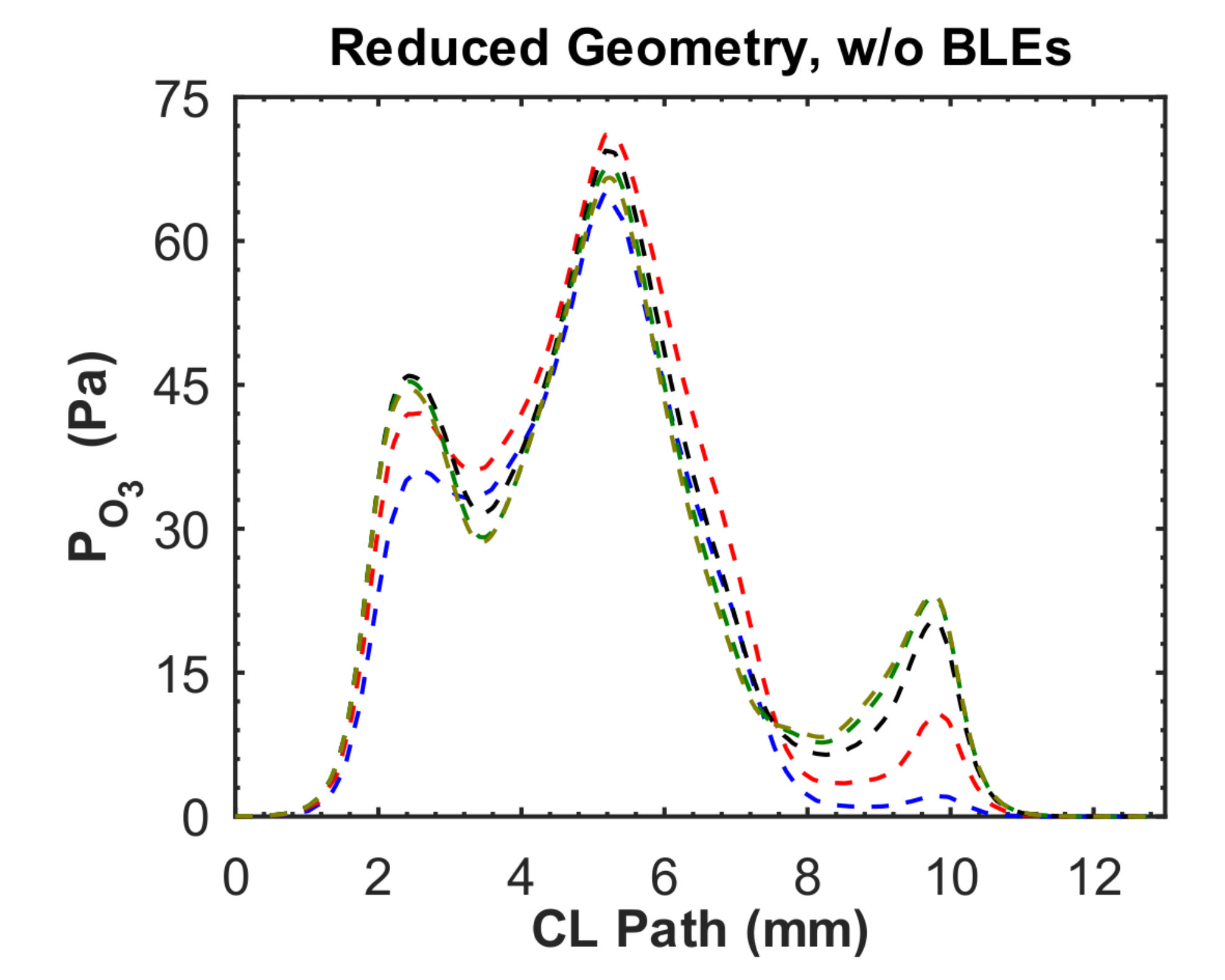}
        \caption{}
        \label{fig:lil_Po3_wo}
    \end{subfigure}
    \caption{
        Ozone partial pressure profiles, at $Q = 350\ \mathrm{cm^3\ min^{-1}}$,
        for both mesh variants: \textbf{(a)} with BLEs (full line) and
        \textbf{(b)} without BLEs (dashed line). The following values of
        $\lambda$ are shown: \textcolor{blue}{$2$}, \textcolor{red}{$4$},
        $8$, \textcolor[rgb]{0 0.5 0}{$12$}, and \textcolor[rgb]{0.5 0.5 0}
        {$16$}.
    }
    \label{fig:lil_Po3s}
\end{figure}

    Resuming our analysis, we move to the ozone partial pressure profiles,
$P_{\mathrm{O_3}}$. Figure \ref{fig:lil_Po3s} shows the profiles obtained at
the upper CL boundary, along the $x$ axis, for $Q = 350\ \mathrm{cm^3\
min^{-1}}$. Analyzing each mesh variant individually at first, it is observed
that the BLEs variant (Fig. \ref{fig:lil_Po3_w}) shows relatively little
qualitative change, and progressively small changes as $\lambda$ increases. On
the other hand, the no-BLEs variant (Fig. \ref{fig:lil_Po3_wo}) show a marked
increase in $P_{\mathrm{O_3}}$ at the end of the flow path, as $\lambda$
increases. Nevertheless, when comparing both mesh variants, it can be noticed
that their profiles are very similar at $\lambda = 16$, although their
dependency on $h$ is somewhat different. This is further illustrated on Fig.
S10, where fewer values of $\lambda$ are shown, and very small differences are
seen between the variants at $\lambda = 16$. It is thus seen that the
$P_{\mathrm{O_3}}$ profiles benefit from BLEs at high $h$, particularly at the
end of the flow path.

\begin{figure}
    \centering
    \begin{subfigure}{0.4\textwidth}
        \includegraphics[width=\textwidth]{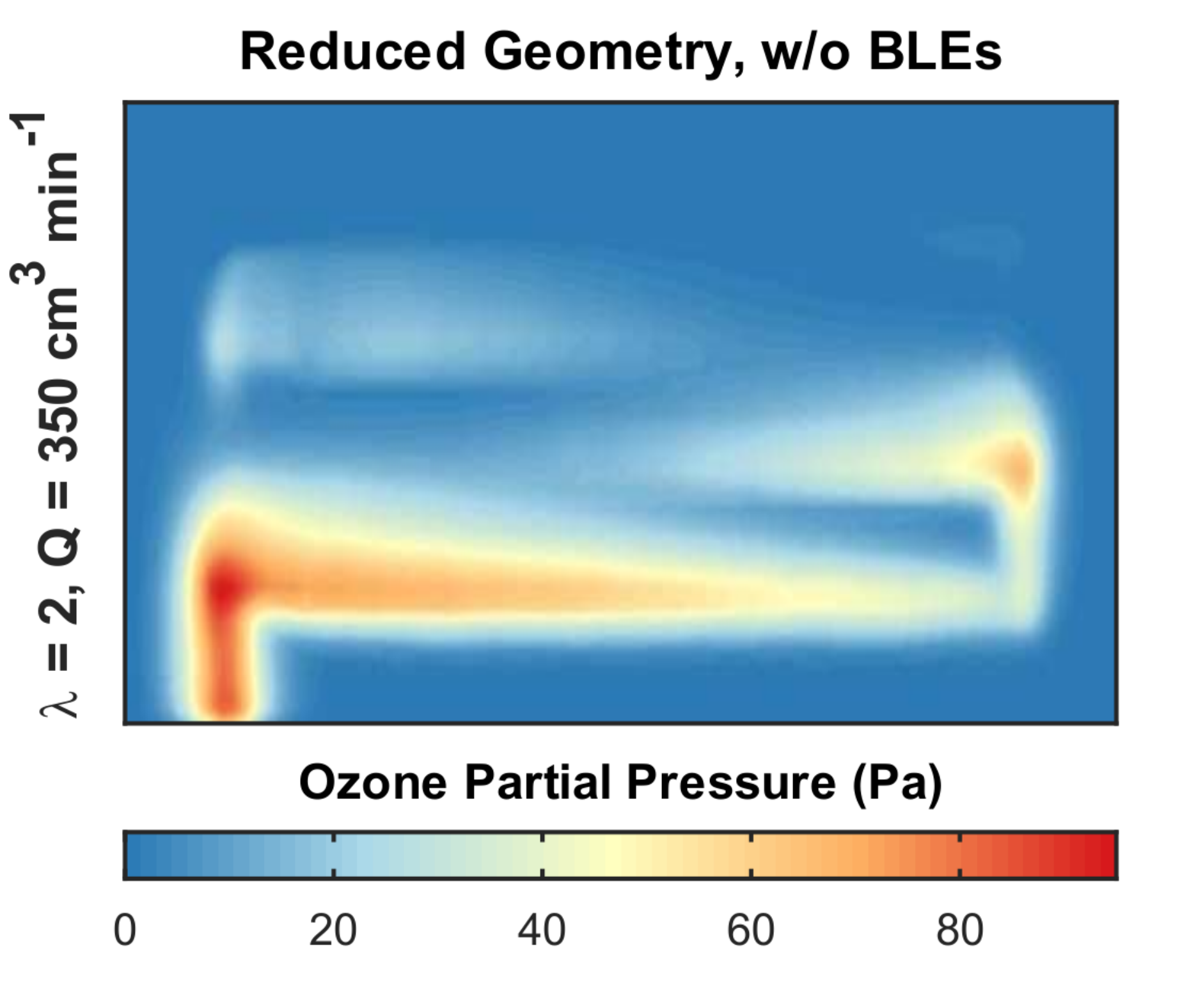}
        \caption{}
        \label{fig:lil_Po32}
    \end{subfigure}
    ~
    \begin{subfigure}{0.4\textwidth}
        \includegraphics[width=\textwidth]{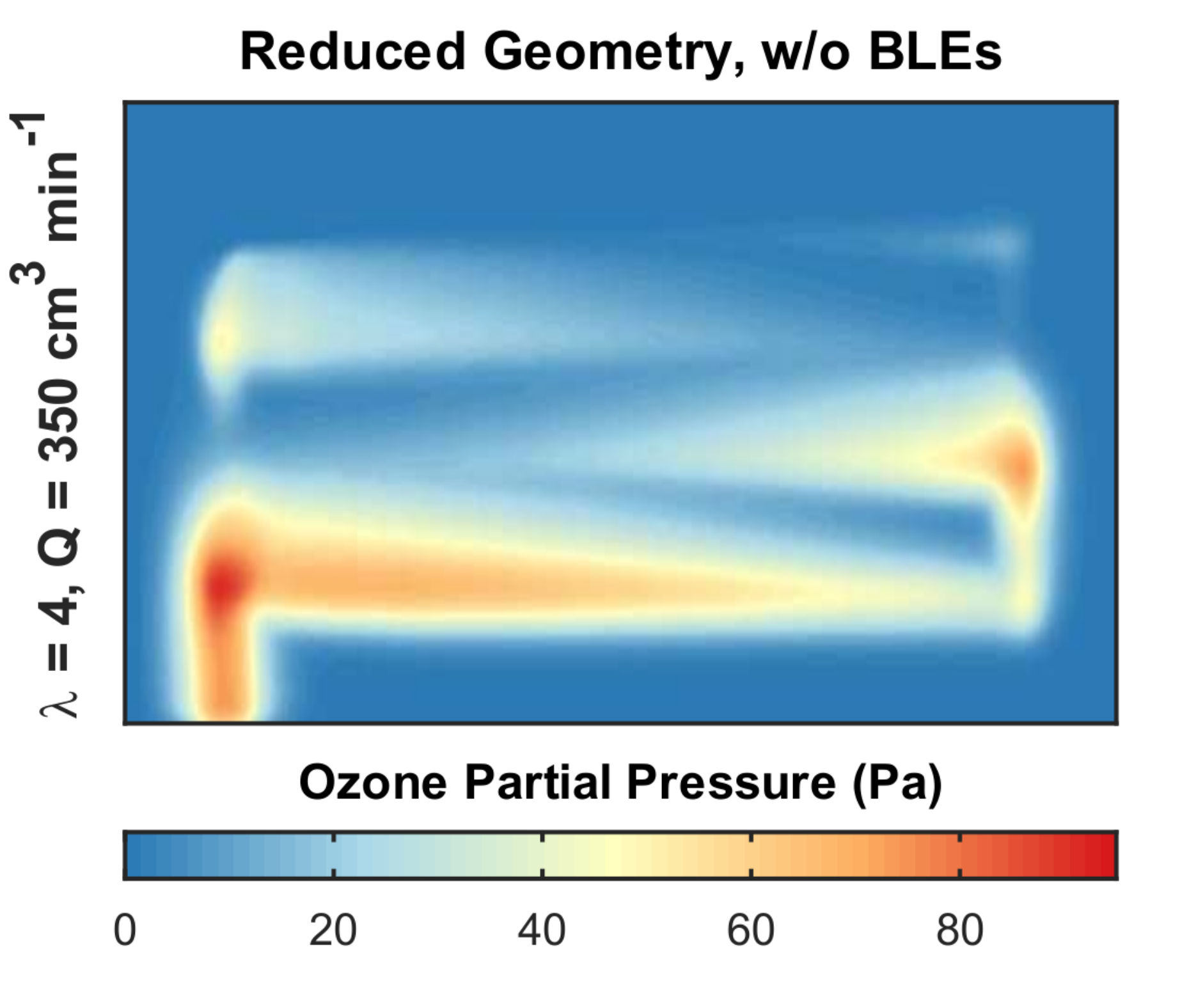}
        \caption{}
        \label{fig:lil_Po34}
    \end{subfigure}

    \begin{subfigure}{0.4\textwidth}
        \includegraphics[width=\textwidth]{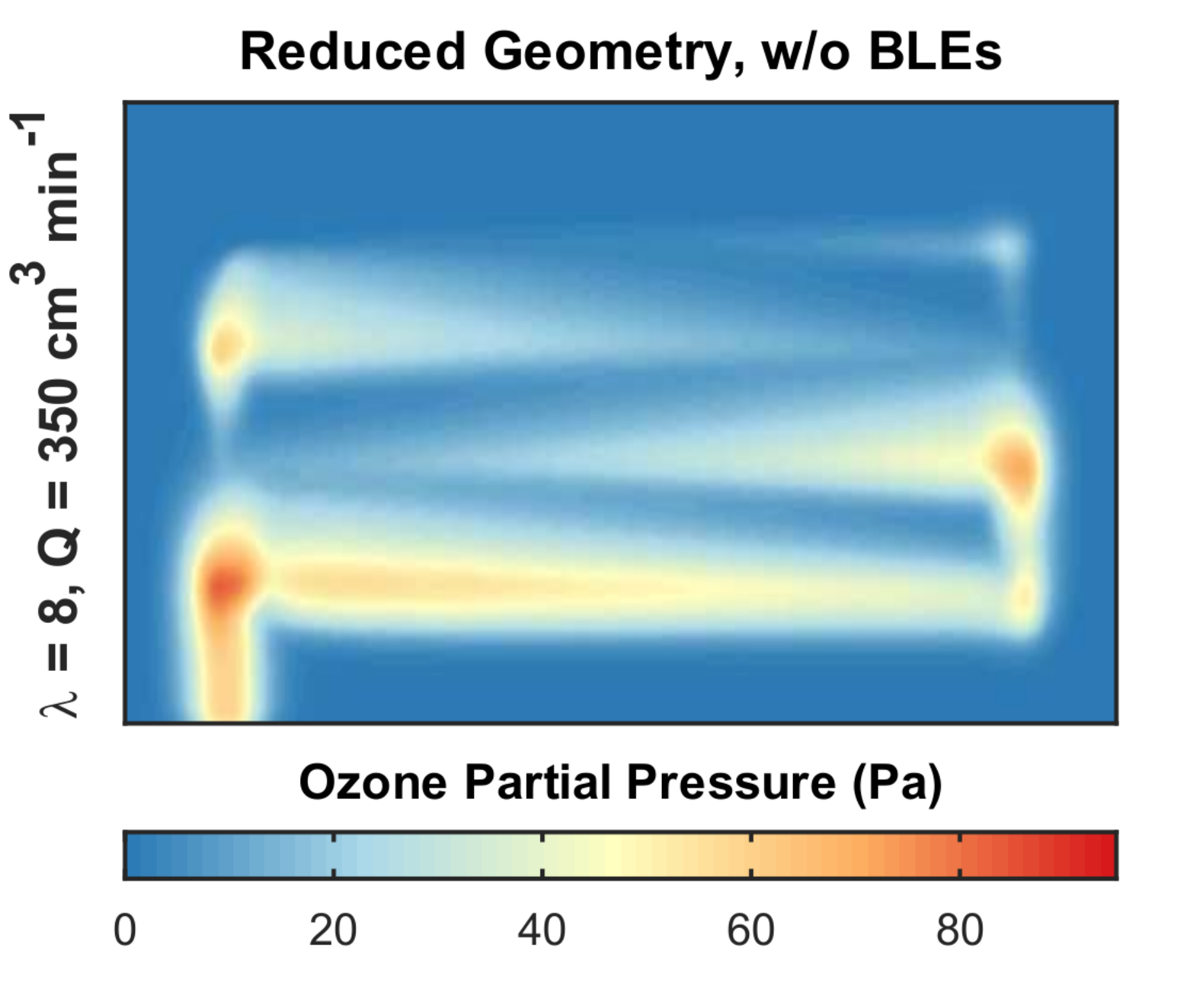}
        \caption{}
        \label{fig:lil_Po38}
    \end{subfigure}
    ~
    \begin{subfigure}{0.4\textwidth}
        \includegraphics[width=\textwidth]{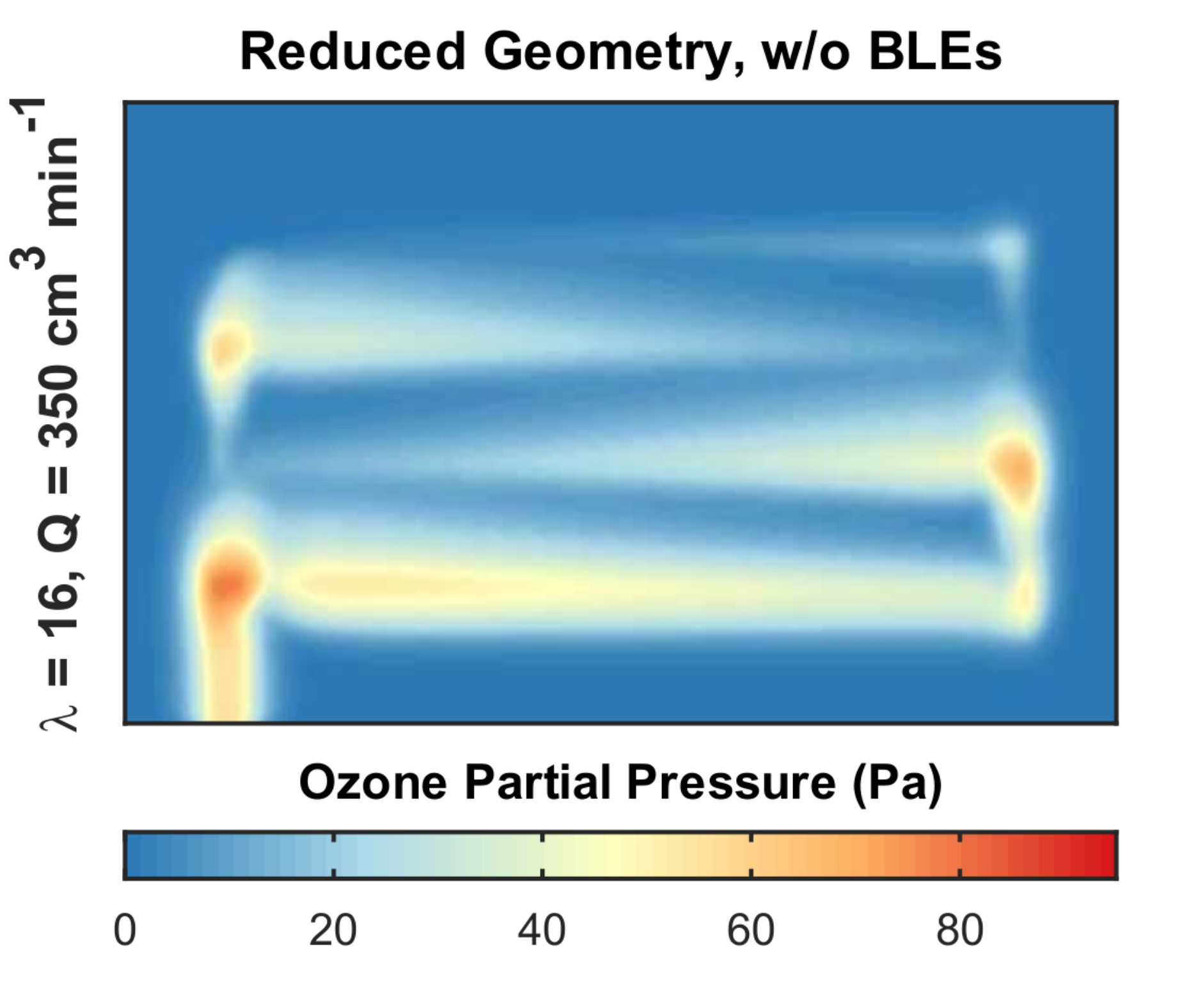}
        \caption{}
        \label{fig:lil_Po316}
    \end{subfigure}
    \caption{
    Ozone partial pressure surfaces for selected values of $\lambda$ and $Q=
    350\ \mathrm{cm^3\ min^{-1}}$, using the no-BLEs mesh variant: \textbf{(a)}
    $\lambda = 2$, \textbf{(b)} $\lambda = 4$, \textbf{(c)} $\lambda = 8$, and
    \textbf{(d)} $\lambda = 16$.
    }
    \label{fig:lil_Po3s2}
\end{figure}

    Finally, in order to provide a broader, albeit qualitative view on the
effects of the mesh on $P_{\mathrm{O_3}}$, we will briefly analyze the ozone
partial pressure surfaces, $\mathbf{P}_{\mathrm{O_3}}$. Figure
\ref{fig:lil_Po3s2} presents such surfaces for selected values of $\lambda$,
with $Q = 350\ \mathrm{cm^3\ min^{-1}}$. The no-BLEs mesh variant is shown in
order to highlight the changes as $\lambda$ increases, however, in general, the
changes follow those seen in Fig. \ref{fig:lil_Po3s}. There are several trends
that can be observed as $\lambda$ increases: i) the increased $P
_{\mathrm{O_3}}$ values at the end of the flow path, after the last turn; ii)
an improved description of the local $P_{\mathrm{O_3}}$ maxima after each Ch
turn; iii) a narrowing of the ozone plume along each horizontal section; and
iv) a decrease of maximum $P_{\mathrm{O_3}}$ values. Trends i) and ii) are
directly related to the improved description of $\mathbf{u}$, as seen in Figs.
\ref{fig:lil_Us} and \ref{fig:lil_Us3}, which impacts not only $P
_{\mathrm{O_3}}$ through better resolution, but also by reducing $\Delta\chi
_{\mathrm{O_3}}$ (Figs. \ref{fig:lil_scalars} and S9) and consequently
allowing increased concentrations of ozone to reach the end section of the Ch
domain. Skipping to trend iv), reduced $P_{\mathrm{O_3}}$ values might also
be connected to $\mathbf{u}$, although it is thought that the increased
resolution at the Pm domains might play a larger role in this case. In
addition, smaller $P_{\mathrm{O_3}}$ values might also be related to the
reduced $\Delta\chi_{\mathrm{O_3}}$ values at high $\lambda$, suggesting that
increased influence of $\mathbf{u}$, possibly in the form of convective
transport, impacts reactant distribution, thus reducing local $P
_{\mathrm{O_3}}$ maxima.

    Trend iii) is also thought to be connected to the improved description of
$\mathbf{u}$, as well as improved resolution across the Pm domains' thickness.
However it provides a counter-evidence to the point raised when analyzing Figs.
\ref{fig:lil_Us2} and \ref{fig:lil_Us3}. It was suggested that the effect the
BLS has on $\mathbf{u}$ after each turn, that of deforming its $x$ profile
towards the wall, could underlie the broader reactant plume observed in the
experimental $P_{\mathrm{O_3}}$ surfaces (see \cite{lopes15} for the original
data and \cite{beruski17} for a comparison between experimental and numerical
data). While $\mathbf{u}$ at the Ch domain has a large influence on the Pm
domains, and refining the grid indeed corroborates this point, it appears that
this discrepancy between experimental and numerical data is not yet settled.

\begin{figure}
    \centering
    \begin{subfigure}{0.45\textwidth}
        \includegraphics[width=\textwidth]{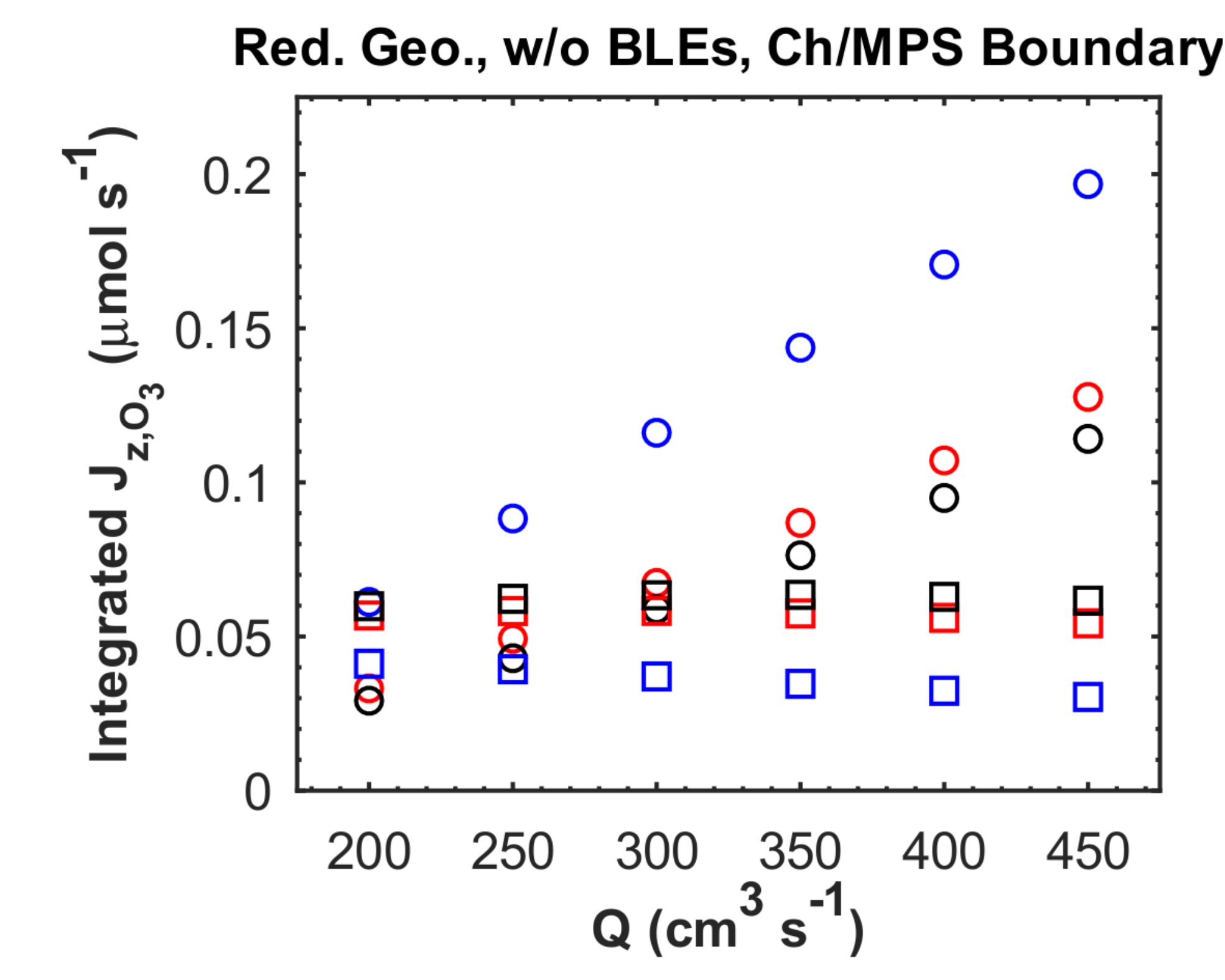}
        \caption{}
        \label{fig:lil_intJ_Ch}
    \end{subfigure}
    ~
    \begin{subfigure}{0.45\textwidth}
        \includegraphics[width=\textwidth]{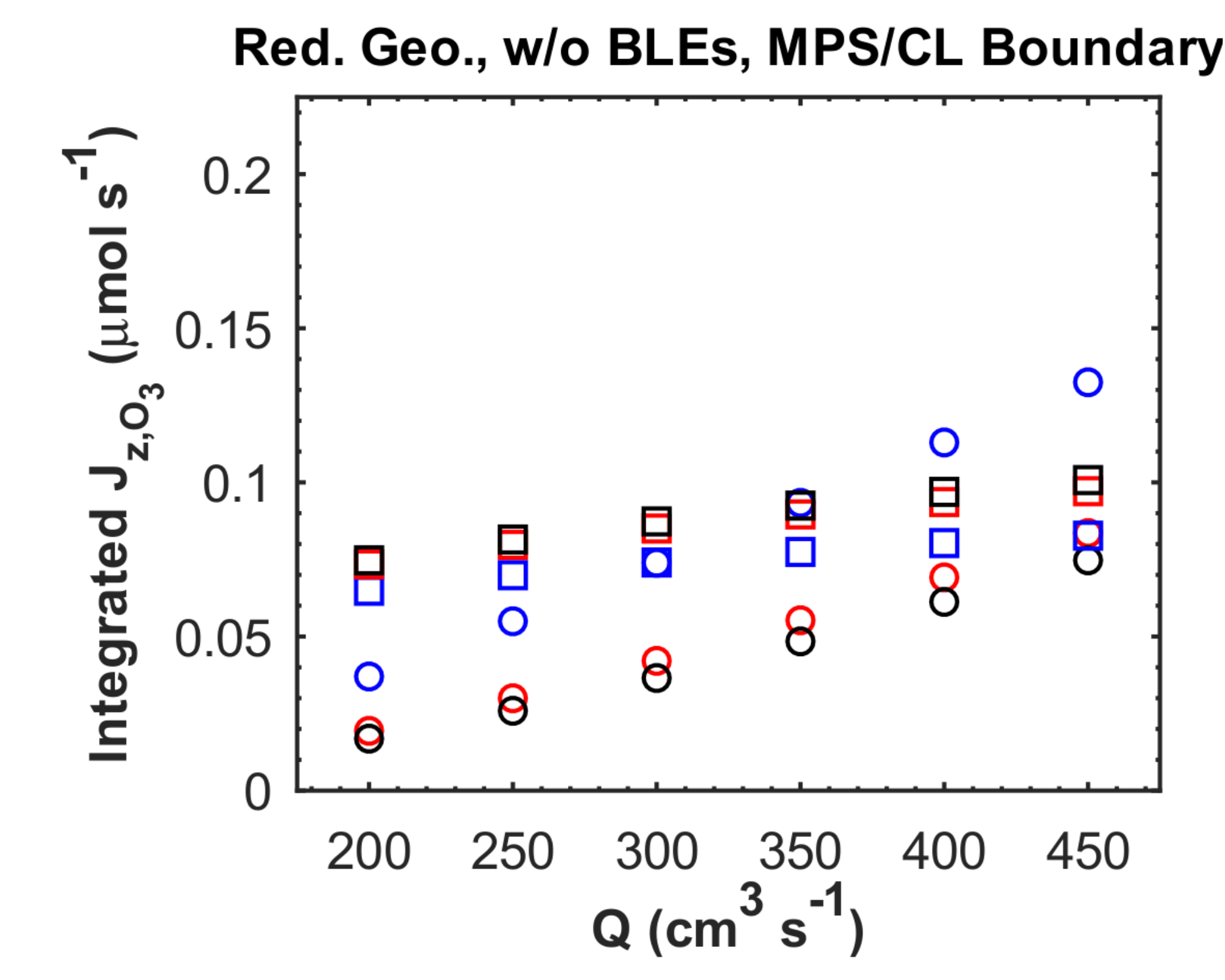}
        \caption{}
        \label{fig:lil_intJ_Pm}
    \end{subfigure}

    \begin{subfigure}{0.45\textwidth}
        \includegraphics[width=\textwidth]{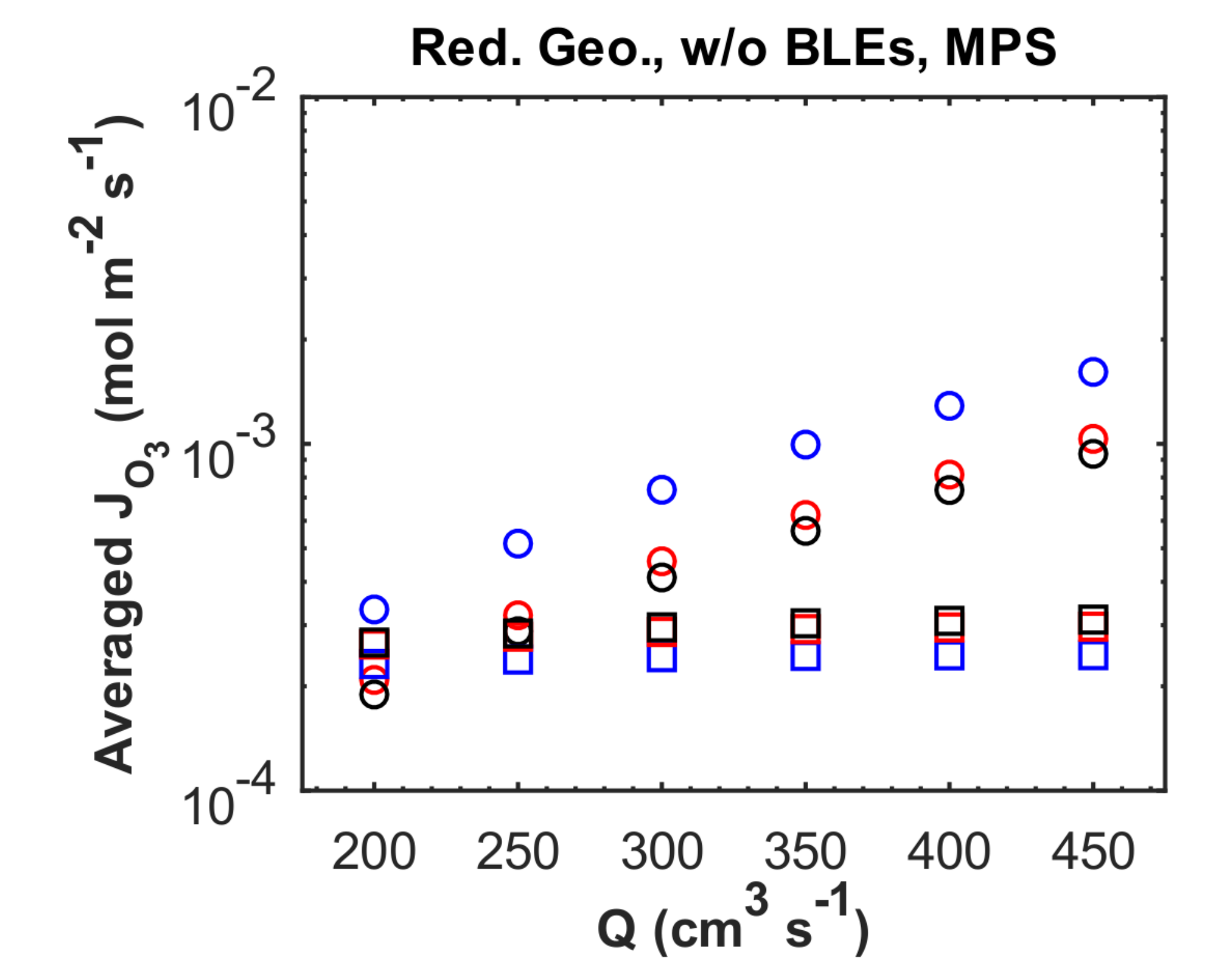}
        \caption{}
        \label{fig:lil_avgJ_MPS}
    \end{subfigure}
    ~
    \begin{subfigure}{0.45\textwidth}
        \includegraphics[width=\textwidth]{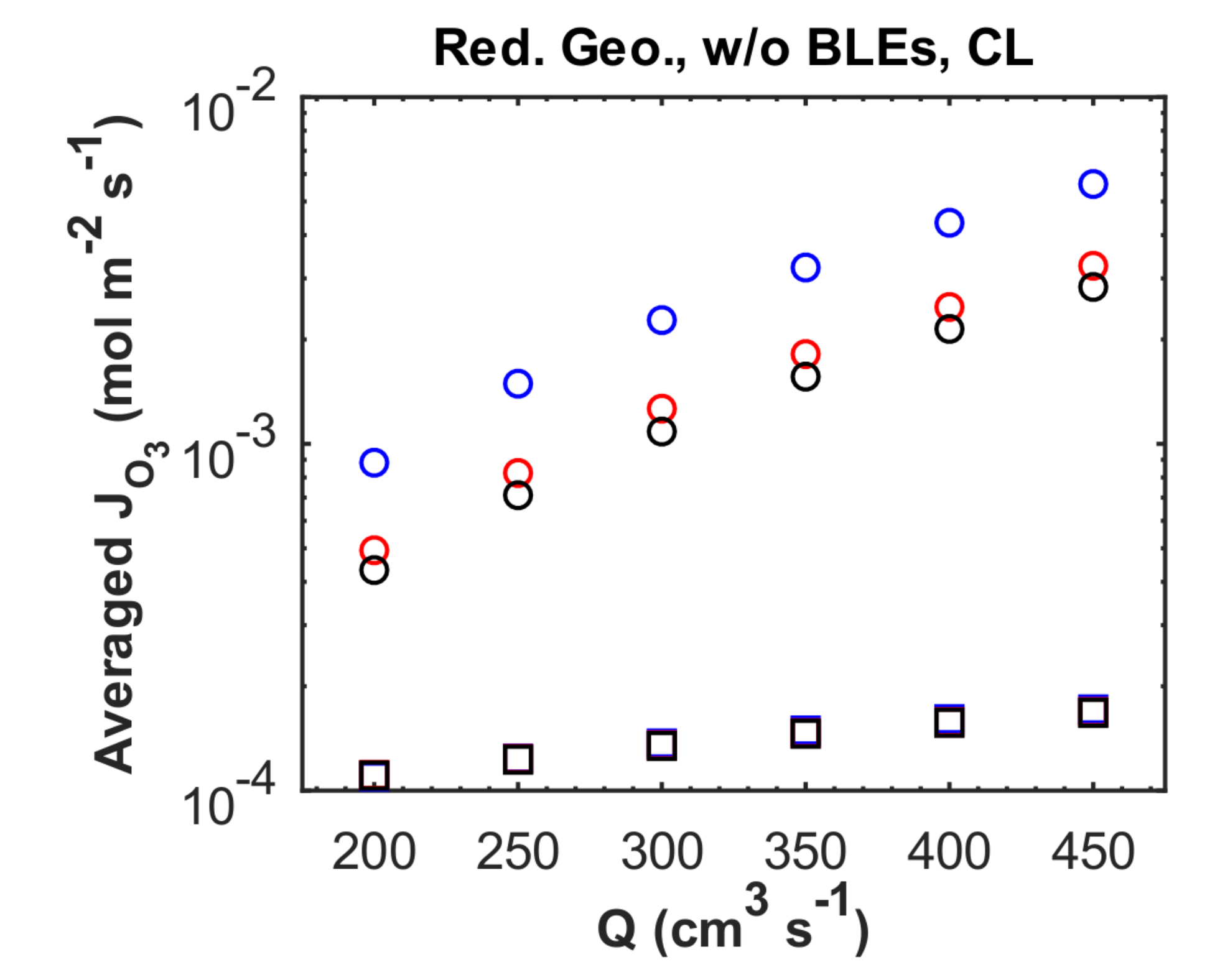}
        \caption{}
        \label{fig:lil_avgJ_CL}
    \end{subfigure}
    \caption{
        Contributions to ozone molar flux as function of the flow rate, using
        the no-BLEs mesh variant, for selected values of $\lambda$:
        \textcolor{blue}{$4$}, \textcolor{red}{$8$}, and $16$. \textbf{(a)}
        integrated $z$ component of the molar flux over the boundary between
        Ch and MPS domains, and \textbf{(b)} between MPS and CL domains;
        \textbf{(c)} average magnitude of the molar flux over the MPS and
        \textbf{(d)} domains. Both convective ($\circ$) and diffusive ($\Box$)
        contributions to the total molar flux are shown.
    }
    \label{fig:lil_rts}
\end{figure}

    Trends iii) and iv) warrant some additional inspection, as it might be
useful when considering numerical results with lower grid resolution. Following
the inverted analysis of trends iii) and iv), we start by further analyzing
trend iv), by looking into the reactant transport. Figure \ref{fig:lil_rts}
presents the diffusive and convective contributions to $\mathrm{O_3}$ transport
for selected values of $\lambda$, for the cases described in Section
\ref{sssec:mc}. From Figs. \ref{fig:lil_intJ_Ch} and \ref{fig:lil_intJ_Pm} it
is seen that convective transport between domains increases with $Q$, as
expected, while diffusive transport shows a maximum around $Q = 300\
\mathrm{cm^3\ min^{-1}}$, between the Ch and MPS domains, and a slight increase
with $Q$ between MPS and CL domains. Concerning molar flux inside each Pm
domain, Figs. \ref{fig:lil_avgJ_MPS} and \ref{fig:lil_avgJ_CL} show that
convective transport is the dominant process, observed the difference
in permeability between domains and thus $\mathbf{u}$ (see Fig.
\ref{fig:lil_Us}), showing large increase with $Q$, while the diffusive
transport shows relatively little increase with $Q$. Overall, this picture
corroborates the conclusions drawn from the previous study\cite{beruski17},
where the convective transport is found to be a large contributor to reactant
transport in all domains of the device. Concerning this work, Fig.
\ref{fig:lil_rts} adds to the picture, showing that, in general, increased
grid resolution ($h\rightarrow 0$) leads to a decrease in convective transport,
while the diffusive contribution slightly increases. This has the overall
effect of reducing total $\mathrm{O_3}$ transport to the CL, and thus reducing
$P_{\mathrm{O_3}}$ at the CL, as well as $\Delta\chi_{\mathrm{O_3}}$ and
$R'_{\mathrm{O_3}}$ (see Fig. S4 for the explicit dependency on $h$).

\begin{figure}
    \centering
    \begin{subfigure}{0.3\textwidth}
        \includegraphics[width=\textwidth]{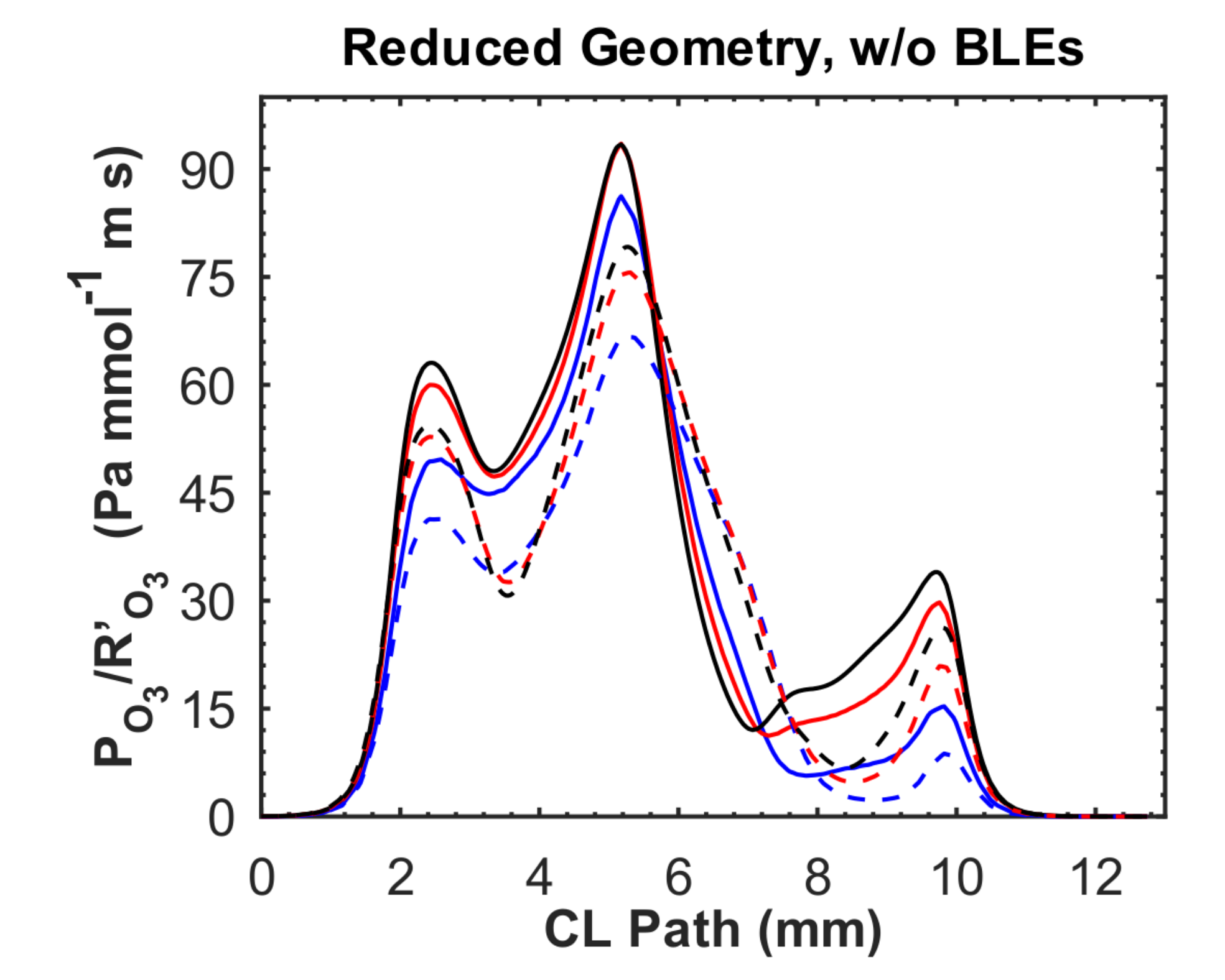}
        \caption{}
        \label{fig:lil_Po3s_1}
    \end{subfigure}
    ~
    \begin{subfigure}{0.3\textwidth}
        \includegraphics[width=\textwidth]{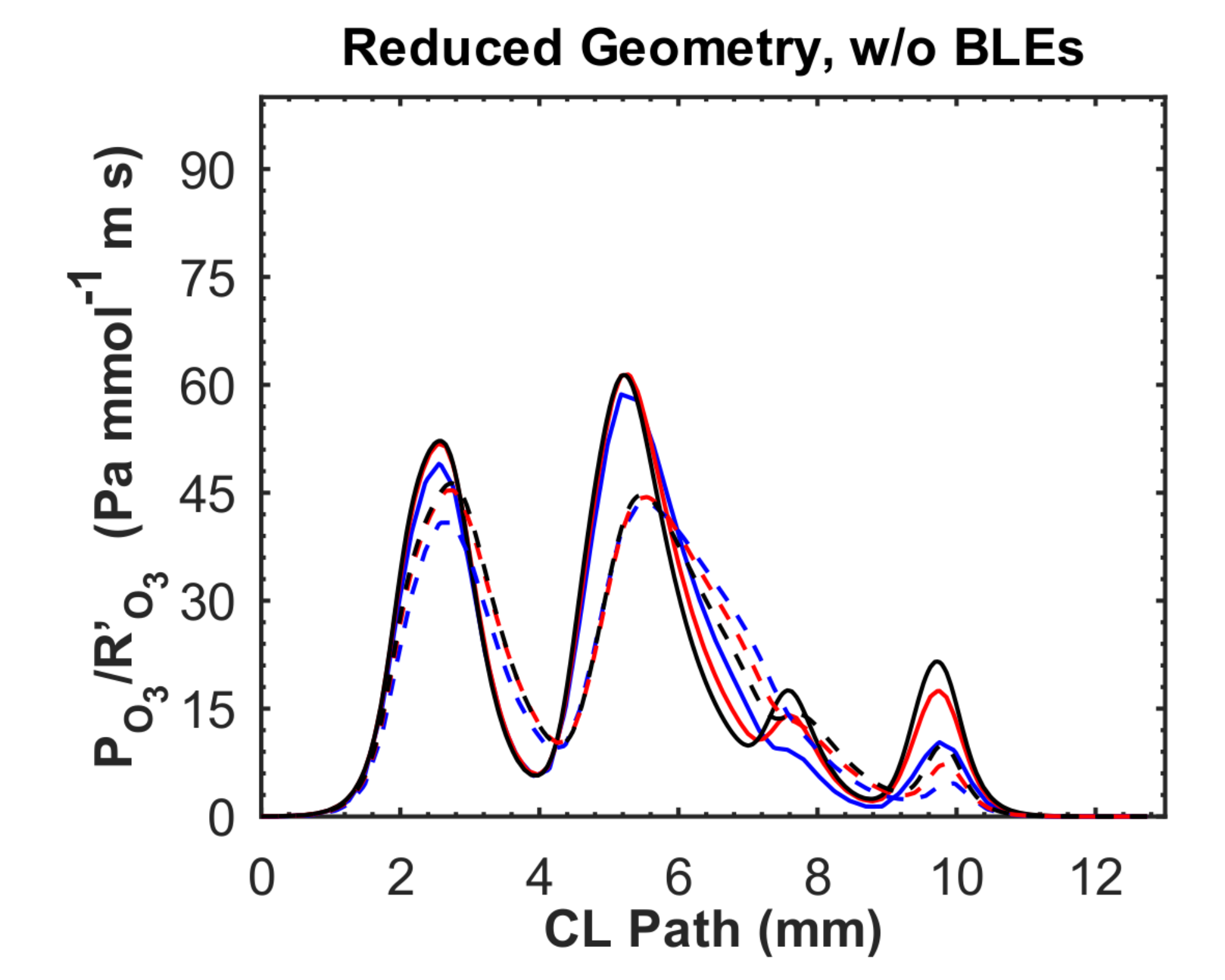}
        \caption{}
        \label{fig:lil_Po3s_2}
    \end{subfigure}
    ~
    \begin{subfigure}{0.3\textwidth}
        \includegraphics[width=\textwidth]{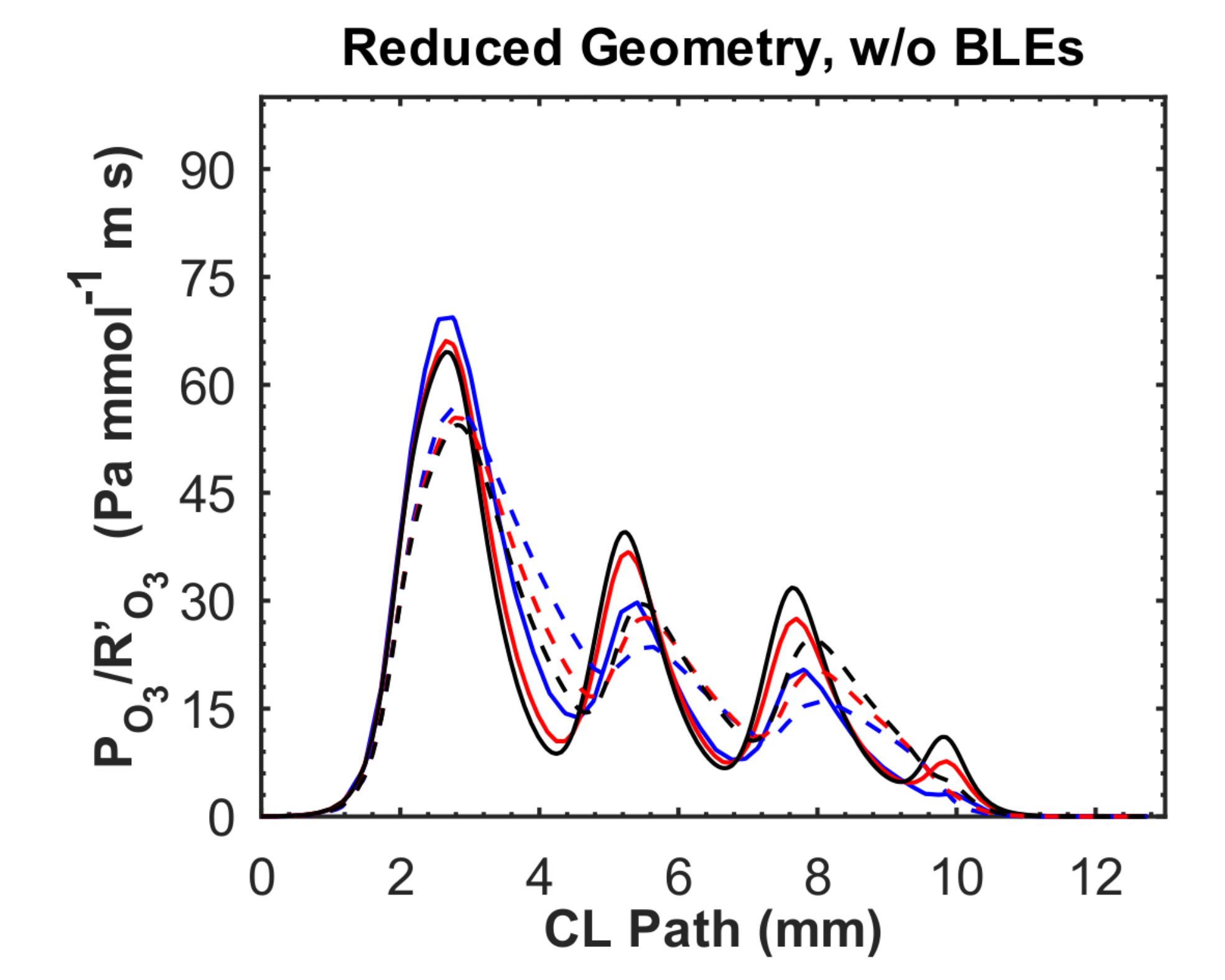}
        \caption{}
        \label{fig:lil_Po3s_0}
    \end{subfigure}
    \caption{
        Ozone partial pressure profiles, using the no-BLEs variant, for
        selected values of $\lambda$: \textcolor{blue}{$4$}, \textcolor{red}
        {$8$}, and $16$; and $Q$: $250$ (full line) and $450\ \mathrm{cm^3\
        min^{-1}}$ (dashed line). The profiles were obtained at the upper
        boundary of the $CL$, parallel to the $x$ axis, according to Section
        \ref{sssec:mc}: \textbf{(a)} $y_1$, shown in Fig. \ref{fig:cl_line};
        \textbf{(b)} $y_2$; and \textbf{(c)} $y_3$.
    }
    \label{fig:lil_Po3s3}
\end{figure}

    Returning once again to trend iii), additional insight may be obtained by
considering $P_{\mathrm{O_3}}$ profiles normalized by $R'_{\mathrm{O_3}}$.
Figure \ref{fig:lil_Po3s3} shows the $P_{\mathrm{O_3}}/R'_{\mathrm{O_3}}$
profiles at three parallel lines along the $x$ axis, described in Section
\ref{sssec:mc} by $y_1$, $y_2$, and $y_3$. Shown are selected values of
$\lambda$ and $Q$ for the no-BLEs mesh variant, in order to illustrate the
changes as either $\lambda$ or $Q$ increases. In all profiles it is seen that,
contrary to Fig. \ref{fig:lil_Po3s}, the local maxima corresponding to the Ch
domain increases: while the absolute value of $P_{\mathrm{O_3}}$ may decrease
as $\lambda$ increases, the value relative to $R'_{\mathrm{O_3}}$ increases.
This corroborates the scenario that, while the magnitude of the convective
transport of reactant may decrease with $\lambda$, the improved description of
$\mathbf{u}$ given by the finer grids better capture the reactant distribution
in the device. This is seen to be particularly important at the end of the
flow path, where $\lambda = 4$ greatly underestimates the local reactant
distribution. Regarding trend iii) and the hypothesis discussed before, Fig.
\ref{fig:lil_Po3s3} settles the issue, showing that in fact the spread of the
reactant plume, between Ch sections, decreases with $\lambda$. Increasing $Q$
also increases the plume spread, as already seen in Fig. \ref{fig:lil_Po3s2},
as well as slightly shifting the local maxima towards higher $x$ values,
corroborating the effects of the shifted $U$ maximum seen in Fig.
\ref{fig:lil_Us2}. Thus, while the improved description of $\mathbf{u}$ in the
Ch domain has effects throughout the Pm domains, it does not seem to be
responsible for the reactant plume spread seen in the experimental data.

\subsection{\label{ssec:mc_res_full}Full Geometry}

    We complete this first analysis with a brief comparison between the
behavior of the reduced geometry, analyzed above, and that of the full
geometry. Focus will be given to the largest values of $\lambda$ used in
the full geometry, $8$ and $12$, as well as the largest value used in the
reduced geometry, $\lambda =16$. In addition, only the no-BLEs mesh variant
will be used, as the analysis above already shows that both mesh variants
provide very similar results for $\lambda \geq 12$. The entire range of
$\lambda$ for the full geometry, as well as the BLEs mesh variant can be found
in Section SIII B of the SM\footnotemark[3] (Figs. S5 to S12).

\begin{figure}
    \centering
    \begin{subfigure}{0.45\textwidth}
        \includegraphics[width=\textwidth]{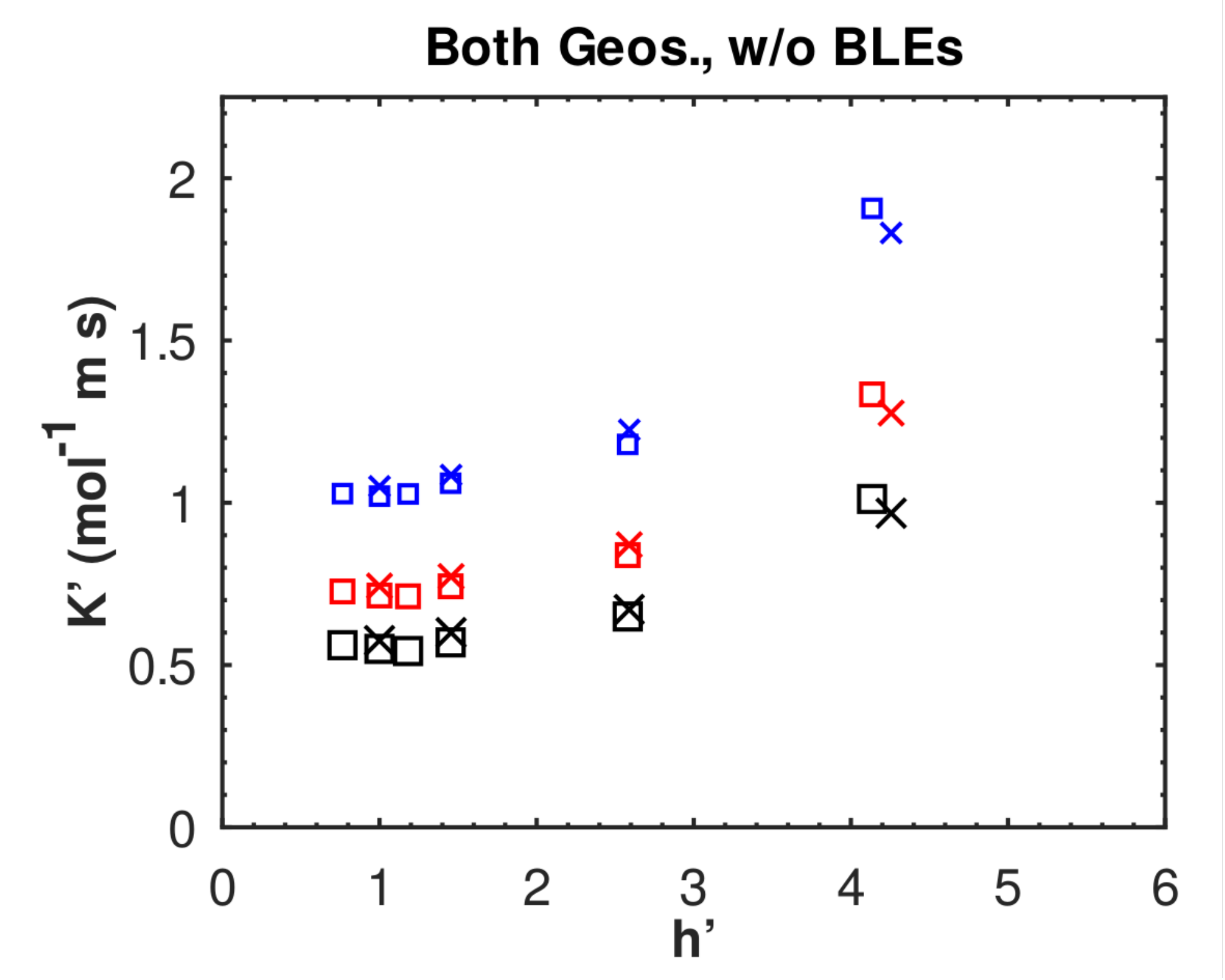}
        \caption{}
        \label{fig:big_Ks}
    \end{subfigure}
    ~
    \begin{subfigure}{0.45\textwidth}
        \includegraphics[width=\textwidth]{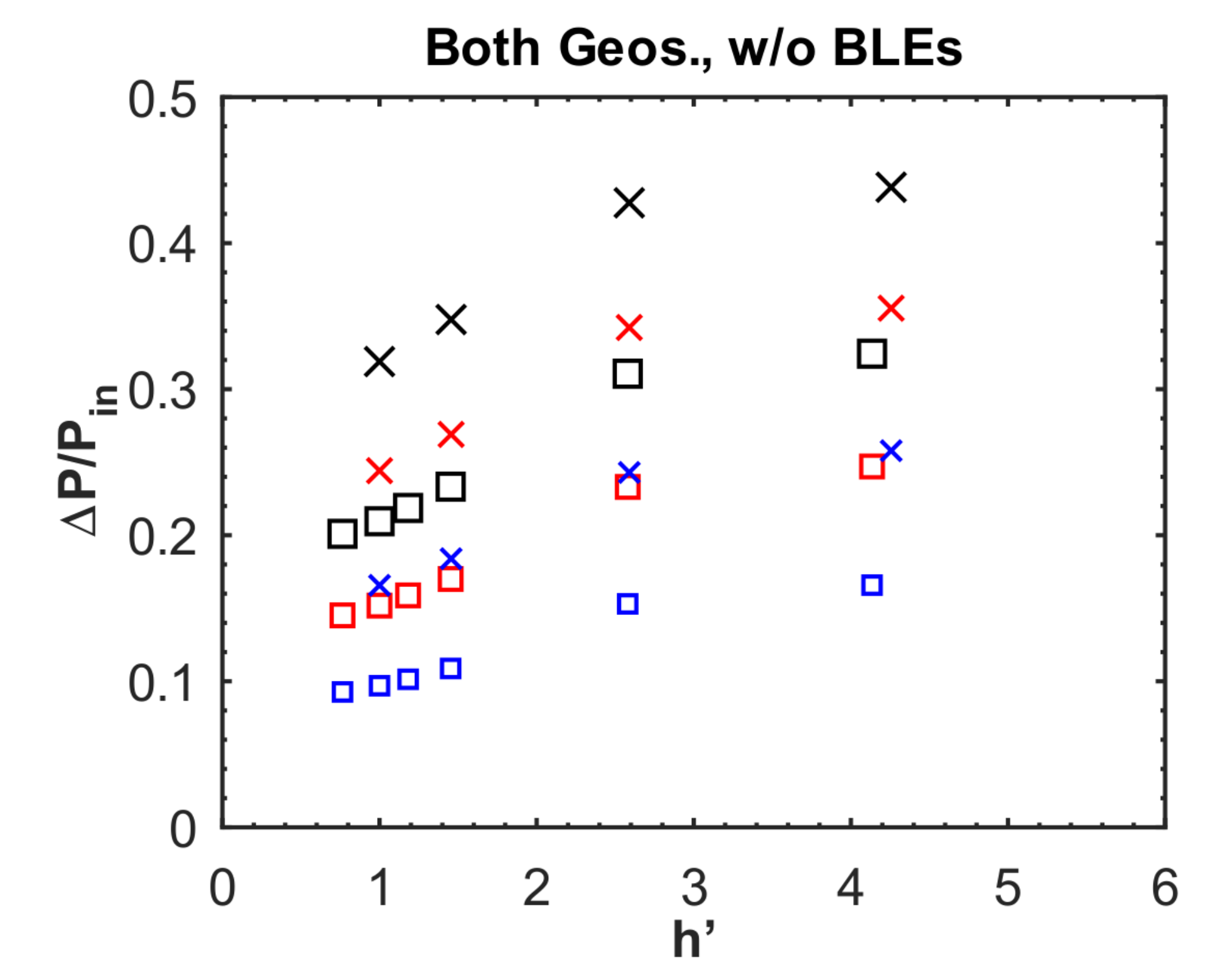}
        \caption{}
        \label{fig:big_Npds}
    \end{subfigure}
    \caption{
        Scalar variables as functions of the normalized effective element
        spacing, obtained for both reduced ($\Box$) and full ($\times$)
        geometries: \textbf{(a)} $K'=\Delta\chi_{\mathrm{O_3}}/R'_{\mathrm{O
        _3}}$ and $\Delta P/P_{\mathrm{in}}$. The following values of inlet
        flow rate $Q$ are shown, with markers increasing in size with $Q$:
        \textcolor{blue}{$250$}, \textcolor{red}{$350$} and $450\ \mathrm{cm^3\
        min^{-1}}$.
    }
    \label{fig:big_scalars}
\end{figure}

    Once again we follow the order previously established, starting with the
scalars described on Section \ref{sssec:mc}. Figure \ref{fig:big_scalars}
presents the $K'=\Delta\chi_{\mathrm{O_3}}/R'_{\mathrm{O_3}}$ and $\Delta P/P
_{\mathrm{in}}$ ratios for both reduced and full geometries, as functions
of a normalized effective element spacing $h'$, calculated with reference to
the meshes with $\lambda = 12$. It can be seen that, for both scalars, the
convergence behavior is very similar between reduced and full geometries. It
should be pointed out, however, that the $K'$ ratio displays a monotonic
behavior for the full geometry. An additional mesh with $\lambda =10$ might
help to draw this behavior out, although it is unnecessary given that $R'
_{\mathrm{O_3}}$ display non-monotonic behavior with $h'$ (Fig. S5b). Care must
be taken, therefore, in order to realize a fuller picture before drawing
conclusions on grid convergence. Small differences in value, mainly seen in
Fig. \ref{fig:big_Npds}, are likely due to the fact that the corners in the Ch
domain contribute non-linearly to the overall results of the device,
particularly for $\Delta P$. This is another evidence to the well-known fact
that the scaling of fuel cells is not linear, in particular for serpentine-type
flow channels.

\begin{figure}
    \centering
    \begin{subfigure}{0.3\textwidth}
        \includegraphics[width=\textwidth]{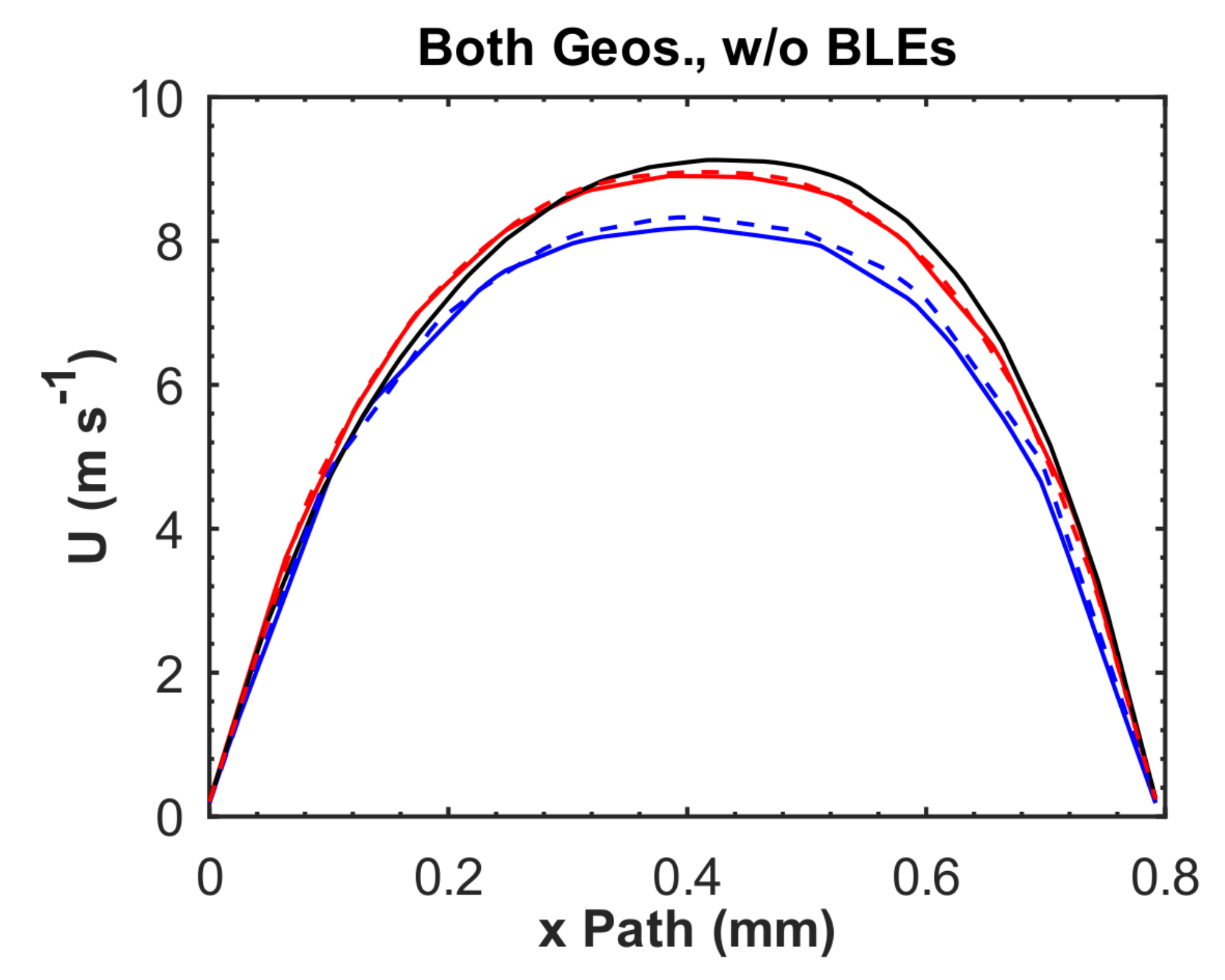}
        \caption{}
    \end{subfigure}
    ~
    \begin{subfigure}{0.3\textwidth}
        \includegraphics[width=\textwidth]{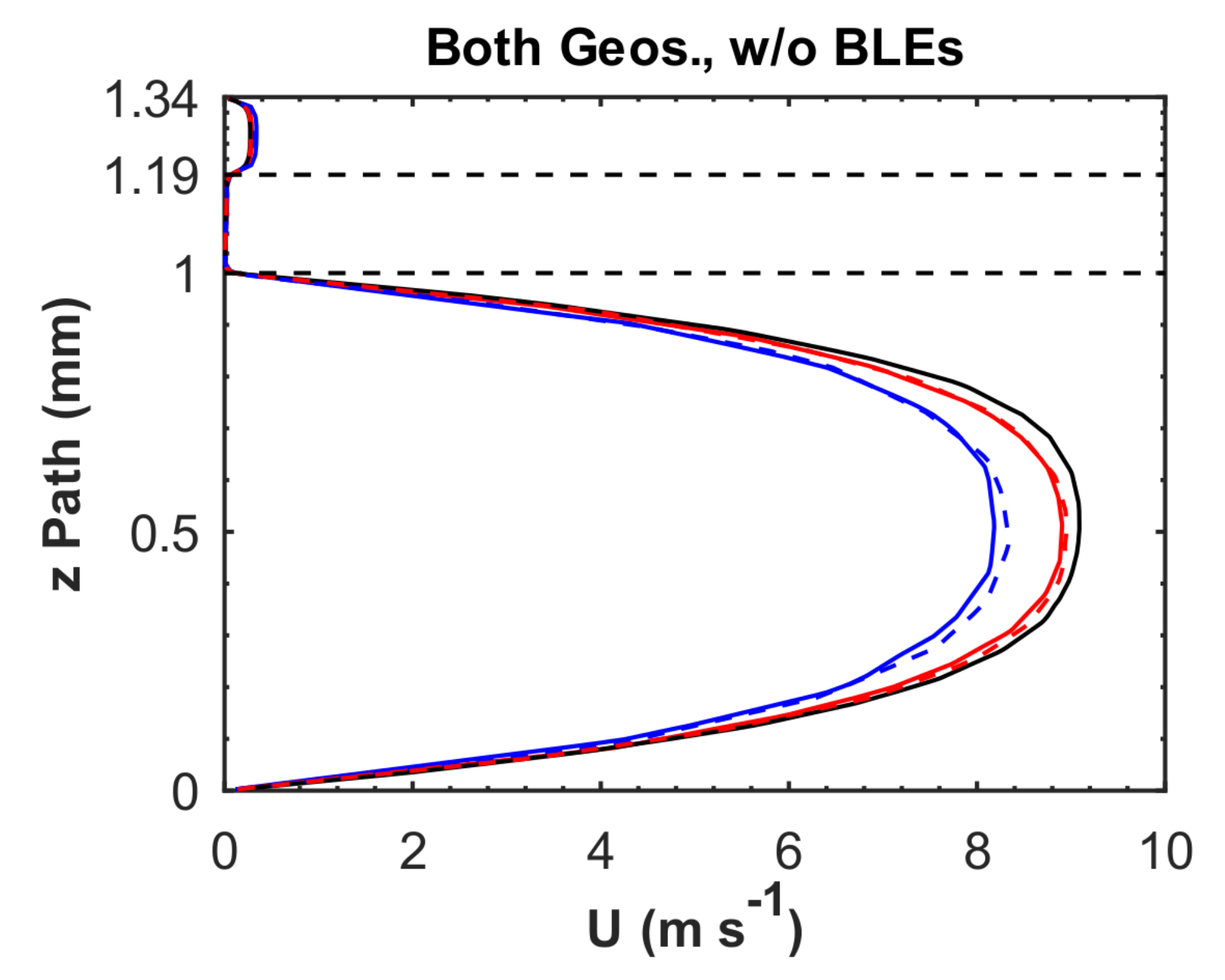}
        \caption{}
    \end{subfigure}
    ~
    \begin{subfigure}{0.3\textwidth}
        \includegraphics[width=\textwidth]{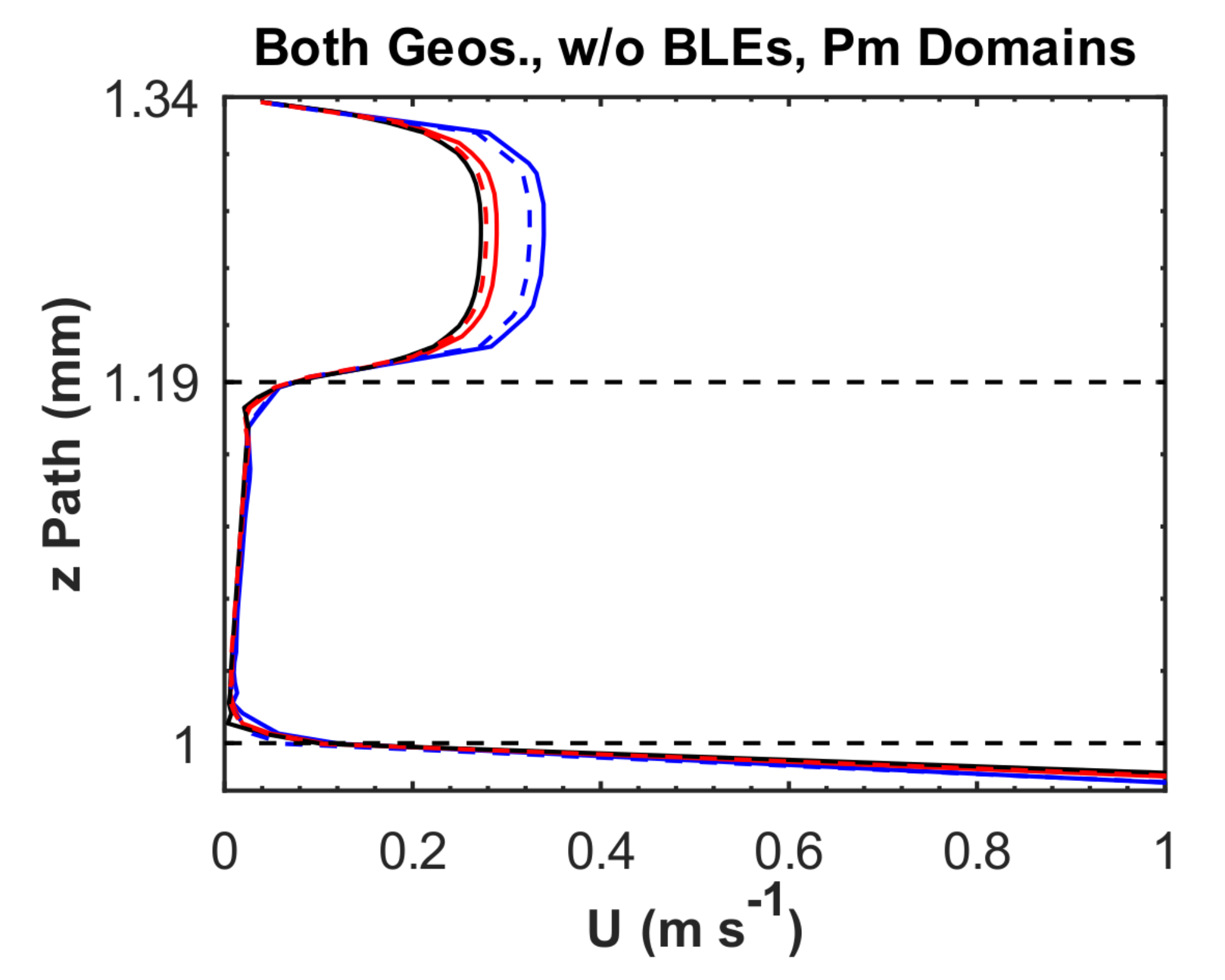}
        \caption{}
    \end{subfigure}
    \caption{
        Flow speed profiles for the reduced (full lines) and full (dashed lines)
        geometries, using the no-BLEs mesh variant, for $Q = 350\ \mathrm{cm^3\
        min^{-1})}$: \textbf{(a)} profile along the $x$ axis, \textbf{(b)}
        along the $z$ axis, with \textbf{(c)} zoom in on the Pm domains. The
        following values of $\lambda$ are shown: \textcolor{blue}{$8$},
        \textcolor{red}{$12$}, and $16$.
    }
    \label{fig:big_Us}
\end{figure}

    Moving on to the $U$ profiles, Figure \ref{fig:big_Us} shows the $x$ and
$z$ profiles for both geometries, superimposing values of $\lambda$ and with
$Q = 350\ \mathrm{cm^3\ min^{-1}}$. It can be observed that both geometries
have very similar $U$ profiles, both in $x$ and $z$, with the largest
differences at the CL domain, while the MPS domain shows the smallest. This
result is expected, given that the characteristic dimensions of the flow and
$Q$ are the same for both geometries. It is important nevertheless to be
assured of this, given the already pointed out non-linear scaling of fuel
cells, but also due to possible numerical artifacts.

\begin{figure}
    \centering
    \begin{subfigure}{0.45\textwidth}
        \includegraphics[width=\textwidth]{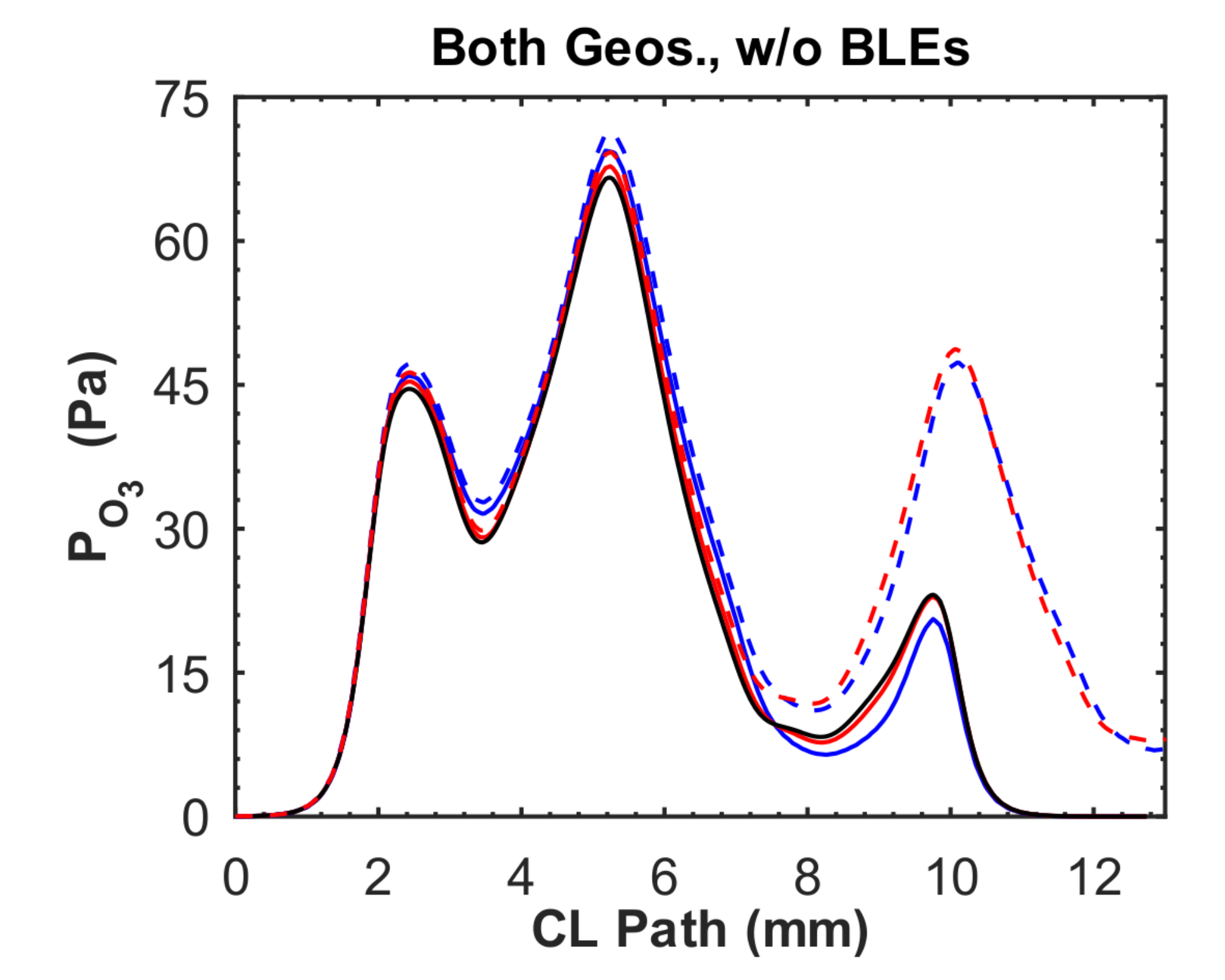}
        \caption{}
        \label{fig:big_Po3_lil}
    \end{subfigure}
    ~
    \begin{subfigure}{0.45\textwidth}
        \includegraphics[width=\textwidth]{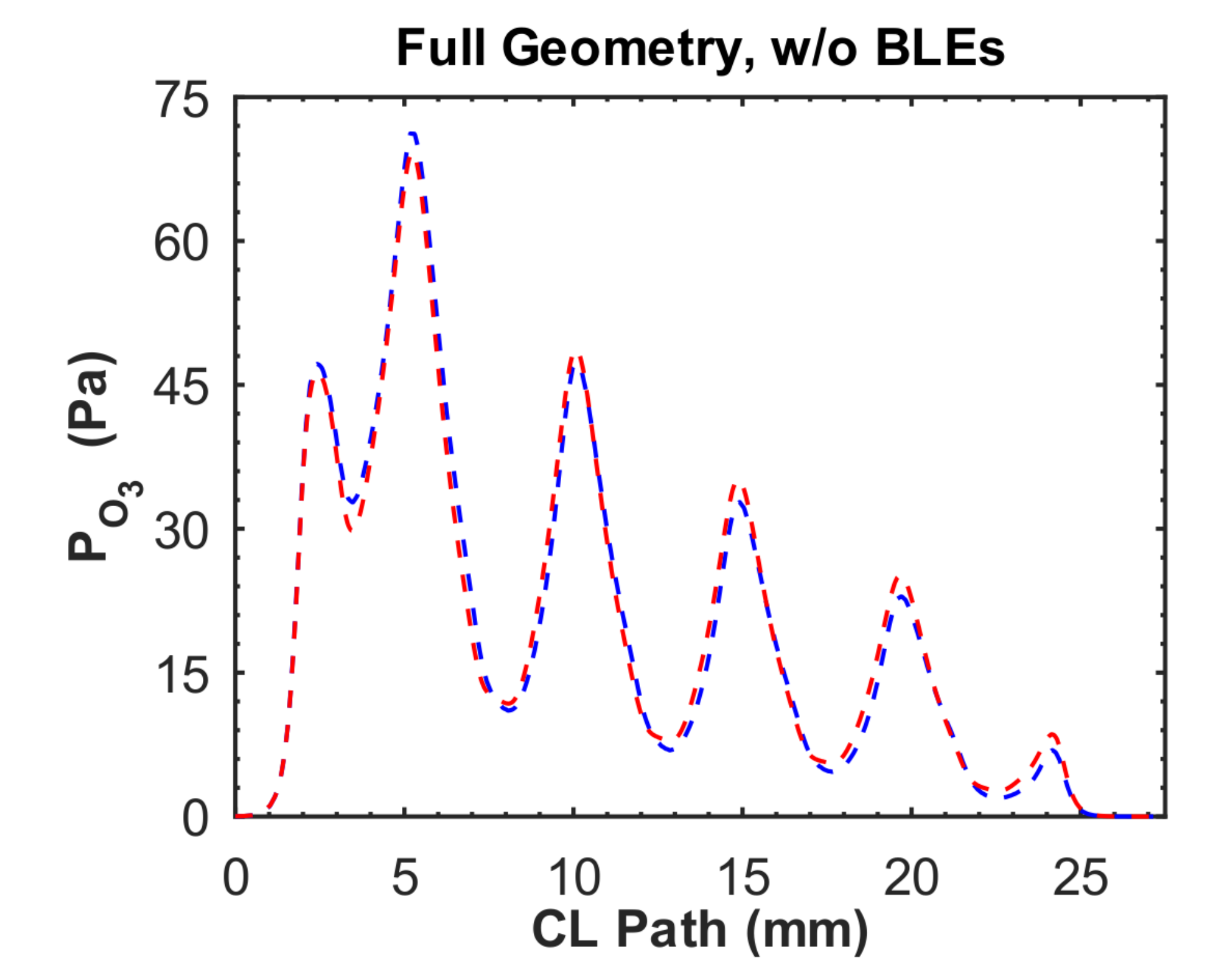}
        \caption{}
        \label{fig:big_Po3_woBLE}
    \end{subfigure}
    \caption{
        Ozone partial pressure profiles, at $Q = 350\ \mathrm{cm^3\ min^{-1}}$,
        using the no-BLEs mesh variant for selected values of $\lambda$:
        \textcolor{blue}{$8$}, \textcolor{red}{$12$}, and $16$. \textbf{(a)}
        Comparison between reduced (full lines) and full (dashed lines)
        geometries; \textbf{(b)} full view of the profile for the full
        geometry.
    }
    \label{fig:big_Po3s}
\end{figure}

\begin{figure}
    \centering
    \begin{subfigure}{0.3\textwidth}
        \includegraphics[width=\textwidth]{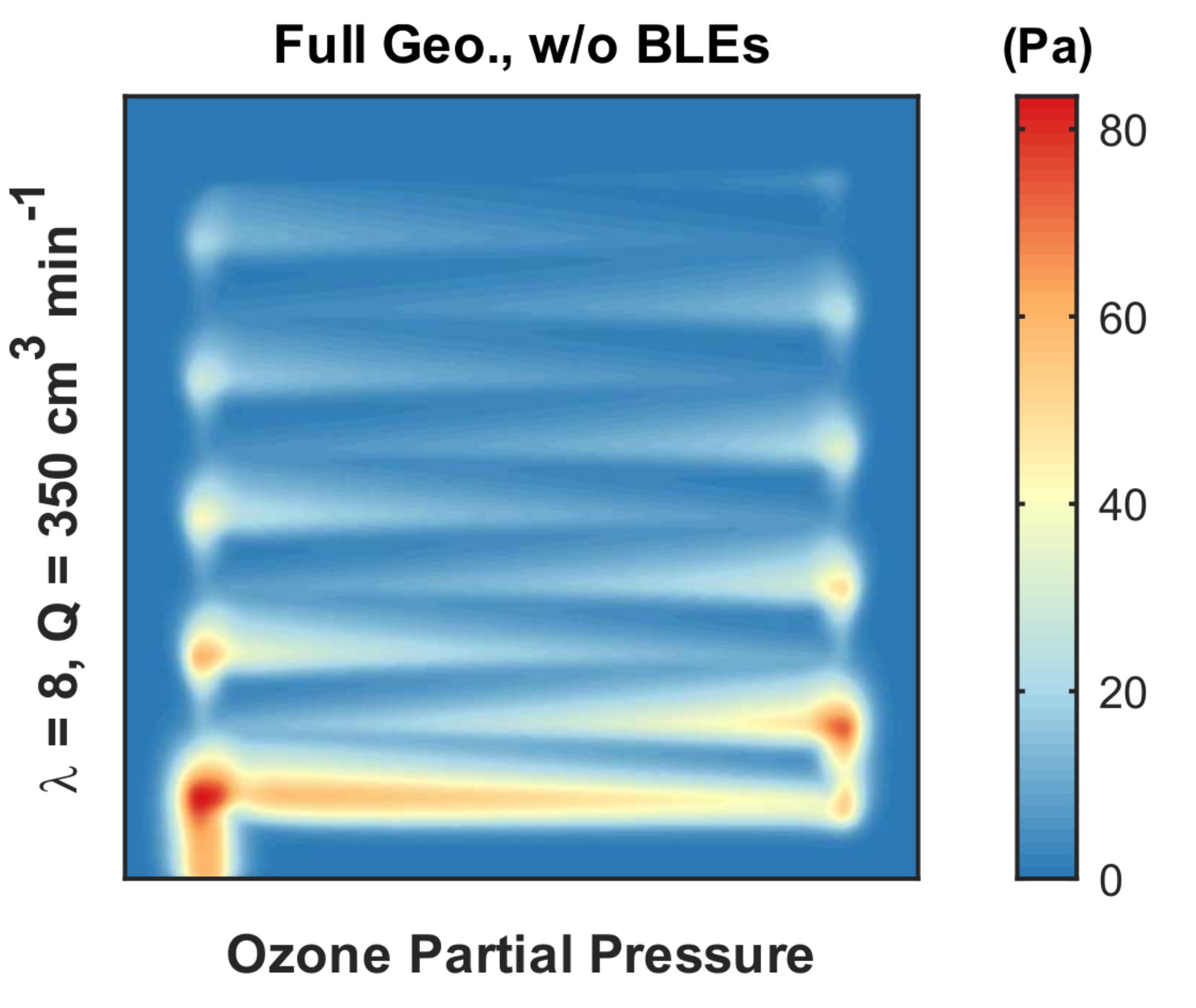}
        \caption{}
    \end{subfigure}
    ~
    \begin{subfigure}{0.3\textwidth}
        \includegraphics[width=\textwidth]{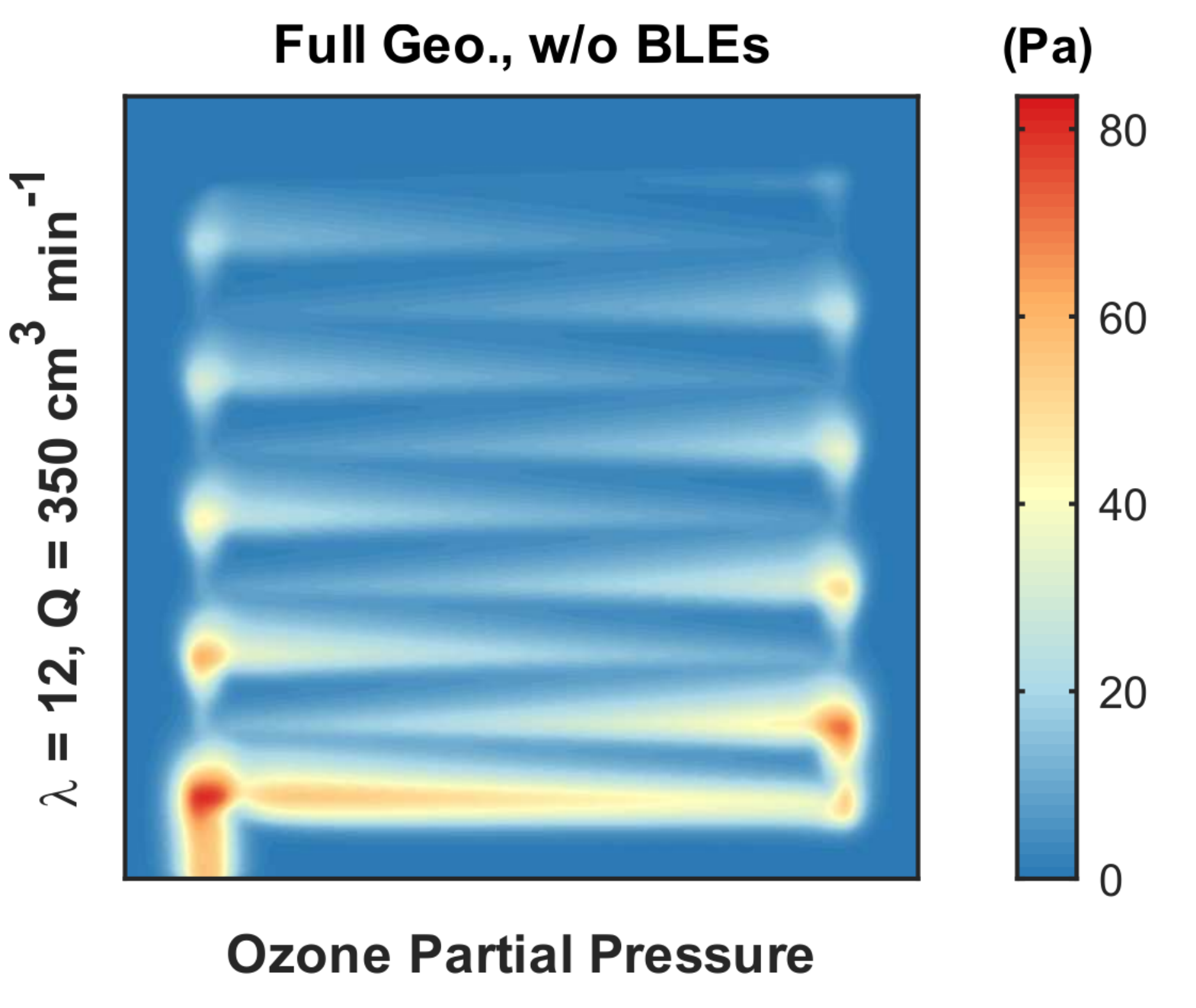}
        \caption{}
    \end{subfigure}
    ~
    \begin{subfigure}{0.3\textwidth}
        \includegraphics[width=\textwidth]{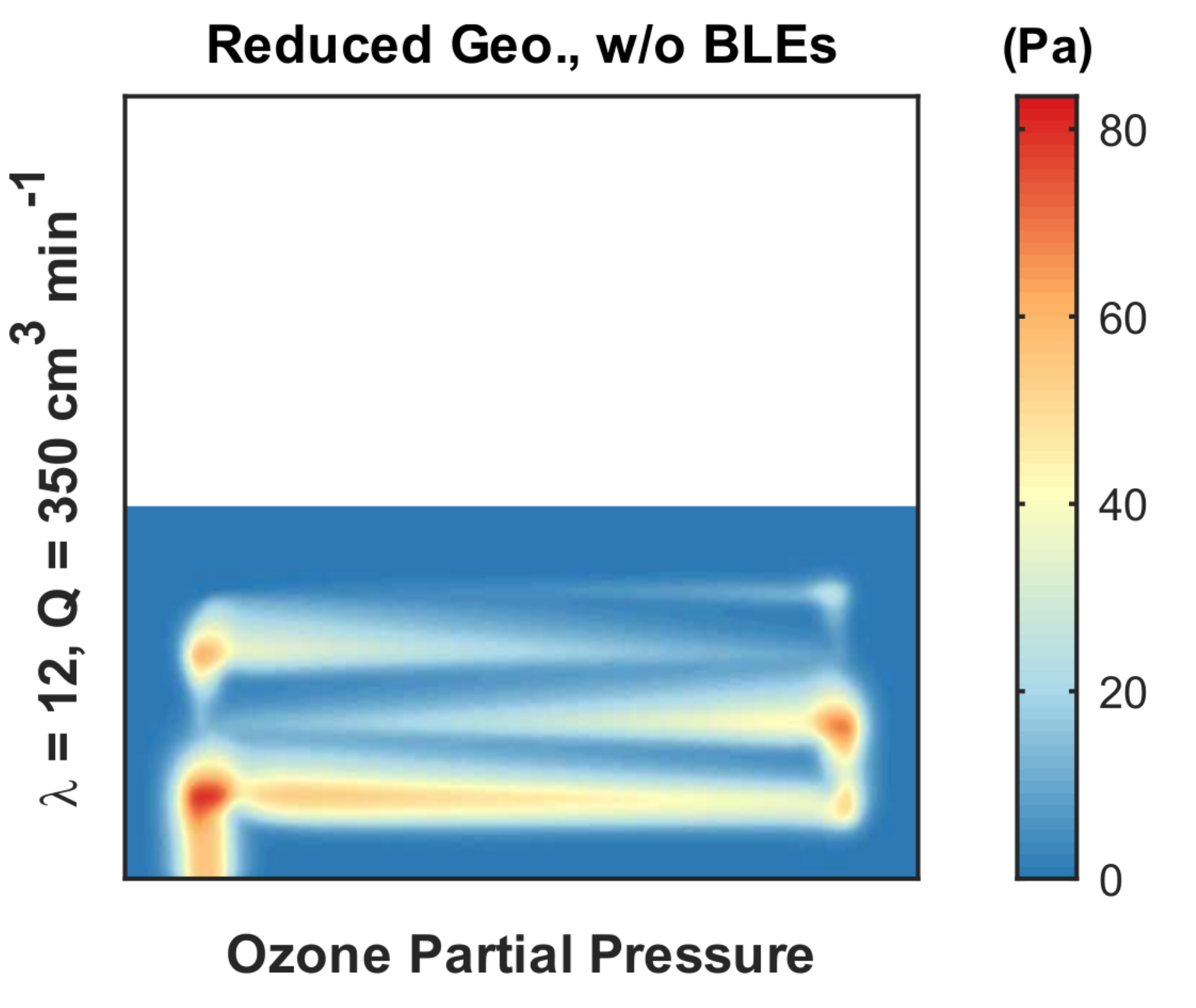}
        \caption{}
        \label{fig:lil_Po3_12}
    \end{subfigure}
    \caption{
        Ozone partial pressure surfaces for selected values of $\lambda$ and
        $Q=350\ \mathrm{cm^3\ min^{-1}}$, using the no-BLEs mesh variant:
        \textbf{(a)} full geometry, $\lambda =8$ and \textbf{(b)} $\lambda =
        12$, and \textbf{(c)} reduced geometry, $\lambda = 12$.
    }
    \label{fig:big_Po3s2}
\end{figure}

    Finally, the $P_{\mathrm{O_3}}$ profiles are compared in Figure
\ref{fig:big_Po3s}, while the $\mathbf{P}_{\mathrm{O_3}}$ surfaces for the full
geometry are shown in Figure \ref{fig:big_Po3s2}, comparing them with the
reduced geometry for $\lambda = 12$ (see also Fig. \ref{fig:lil_Po3s2}).
Regarding the $P_{\mathrm{O_3}}$ profiles, Fig, \ref{fig:big_Po3_lil} shows
close similarities between reduced and full geometries at the beginning of the
flow path, with a marked distinction at the last local peak. The similarities
are once again to be expected, given the geometrical and operational parameters
of each model. The significant difference at the last peak in Fig.
\ref{fig:big_Po3_lil}, while at first may stand out, is readily understood when
noticing that in Fig. \ref{fig:big_Po3_woBLE} the same happens for the full
geometry: the last channel of the Ch domain has no gradients driving the
reactant forward in the CL domain, thus acting only as the final sink. Fig.
\ref{fig:big_Po3s} thus shows once again that both reduced and full geometries
have similar grid convergence behavior, while at the same time further
illustrating the non-linear scaling of fuel cell devices.

    Regarding the $\mathbf{P}_{\mathrm{O_3}}$ surfaces, shown in Fig.
\ref{fig:big_Po3s2}, the same behavior observed for the reduced geometry is
seen when comparing $\lambda =8$ and $12$ for the full geometry, however in
smaller magnitude, given the smaller grid refinement between them. The clearer
aspect to be seen is the decrease in maximum $P_{\mathrm{O_3}}$, which follows
the behavior seen in Fig. \ref{fig:lil_Po3s2}. When comparing both geometries,
with the reduced one shown scaled in Fig. \ref{fig:lil_Po3_12} for better
comparison, it can be seen that the $\mathbf{P}_{\mathrm{O_3}}$ surfaces are
very similar. The major difference is at the spread of the plume after the
third peak in the reduced geometry, however that is clearly due to the
interaction with the last section of the Ch, as pointed out above when
analyzing Fig. \ref{fig:big_Po3s}. Indeed, the same behavior, albeit with much
smaller $P_{\mathrm{O_3}}$ values, is seen at the ninth peak.

    The data show that the reduced geometry is a good proxy to study mesh
convergence in the fuel cell prototype's model, given that we are currently
unable to simulate meshes with $\lambda > 12$ for the full geometry. Having
shown this correlation between geometries, it opens up the possibility of
further studying the numerical aspects of the model as function of the mesh,
extending the current analysis to structured meshes and improved unstructured
mesh schemes that may allow denser meshes with comparable computational
resources.

\subsection{\label{ssec:mc_res_error}Estimation of Discretization Error}

    Given the qualitative analysis of Sections \ref{ssec:mc_res_lil} and
\ref{ssec:mc_res_full}, it might be argued that the variables chosen to track
mesh convergence do not, in fact, show convergence. Not only the convergence
appears to be relatively slow, some variables also show non-monotonic behavior
depending on the mesh scheme. While a robust sign of convergence would be
ideal, given the computational resources and accuracy of experimental data
available, it is important to point out, as did Roache\cite{roache97}, that
most of the time actual mesh convergence is impractical. Thus, instead of
going further down this rabbit hole, we chose to accept the fact that the model
has some error due to discretization, and tried to estimate that. We will now
present the results from the generalized Richardson extrapolation (GRE),
following Roache\cite{roache97}, and the mixed-order extrapolation (MOE),
following Roy\cite{roy03}, used to estimate the discretization error of the
scalar variables, according to Section \ref{sssec:mc}.

    First and foremost, it should be pointed out that, given the convergence
behavior observed in Section \ref{ssec:mc_res_lil} and the refinement ratios
used in this study, it was not clear which set of meshes should be used in the
extrapolations in order to accurately reproduce the behavior seen. It is
pointed out by Roache\cite{roache97} that the refinement ratio between grids,
$r$, should be $r > 1.3$, in order to be able to minimize other possible
numerical errors when performing the convergence study. Thus, some tests were
done with the reduced geometry following this suggestion, and the following was
concluded: i) the extrapolation schemes show different results depending on the
mesh scheme, i.e. with or without BLEs, although the differences are relatively
small and mainly concerns the region at high $h$; ii) when the dataset used to
calculate the extrapolate display non-monotonicity, the GRE returns complex
values for the accuracy order and, consequently, to the extrapolate, although
the magnitude of the imaginary term is relatively small; iii) the GRE and the
MOE show different behavior according to which scalar is being analyzed, with
both schemes showing similar results for the $\Delta P/P_{\mathrm{in}}$ only;
iv) the GRE appears to converge as finer meshes are used to calculate it, in
particular for the $K'$ ratio, while the MOE appears to exaggerate the
non-monotonic behavior in these cases; and v) including a third-order term in
the MOE improves its results over the whole $h$ range. The trends observed when
evaluating the extrapolation schemes are likely to be related to the
non-monotonicity of the data, as suggested in \cite{roy03}, particularly the
fact that the MOE deals, in principle, with such data sets, while the GRE
assumes the data are in the asymptotic range. In these cases where the GRE
returned complex values, only the real part was used to calculate the GCI.
Additional information can be found in Section SIV A of the SM\footnotemark[3],
while Figs. S13 and S14 show selected results of the tests done over the
extrapolation schemes.

    From the conclusions drawn, it was established that for the GRE the meshes
to be used, for both mesh variants, were $\lambda\in\left[8,12,16\right]$,
while for the MOE $\lambda\in\left[2,4,10,16\right]$ were used, with $\left[4,
10,16\right]$ used for the MOE up to $h_k^2$. The same schemes were also used
to estimate the discretization error on the individual variables, $\Delta\chi
_{\mathrm{O_3}}$, $R'_{\mathrm{O_3}}$, and $\Delta P$. We proceed now to
evaluate the results of using the GRE and MOE schemes with the aforementioned
meshes, in order to establish the best estimate of the discretization error of
the reduced and full geometries of the model under study.

\begin{figure}
    \centering
    \begin{subfigure}{0.45\textwidth}
        \includegraphics[width=\textwidth]{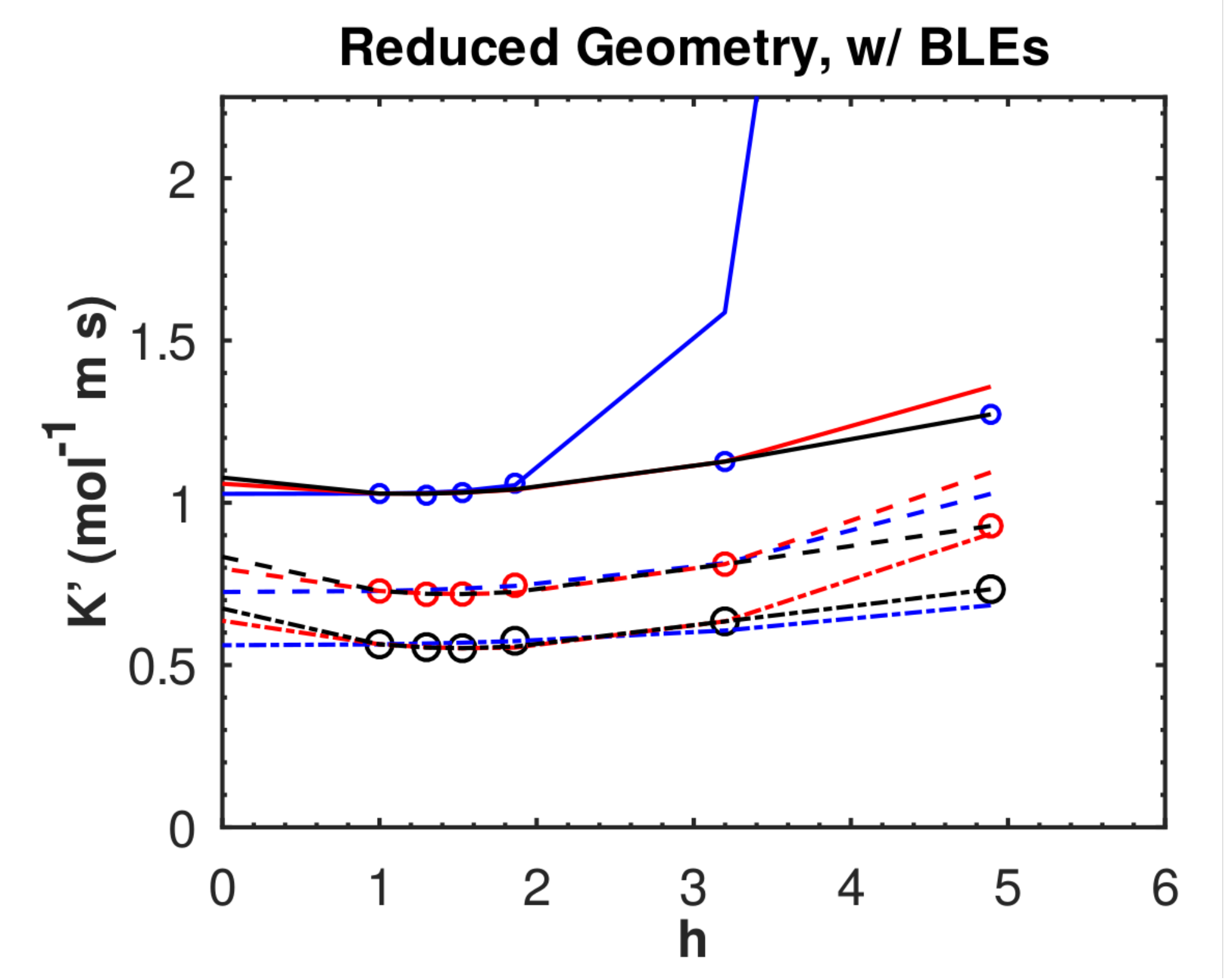}
        \caption{}
        \label{fig:lil_K_ext1}
    \end{subfigure}
    ~
    \begin{subfigure}{0.45\textwidth}
        \includegraphics[width=\textwidth]{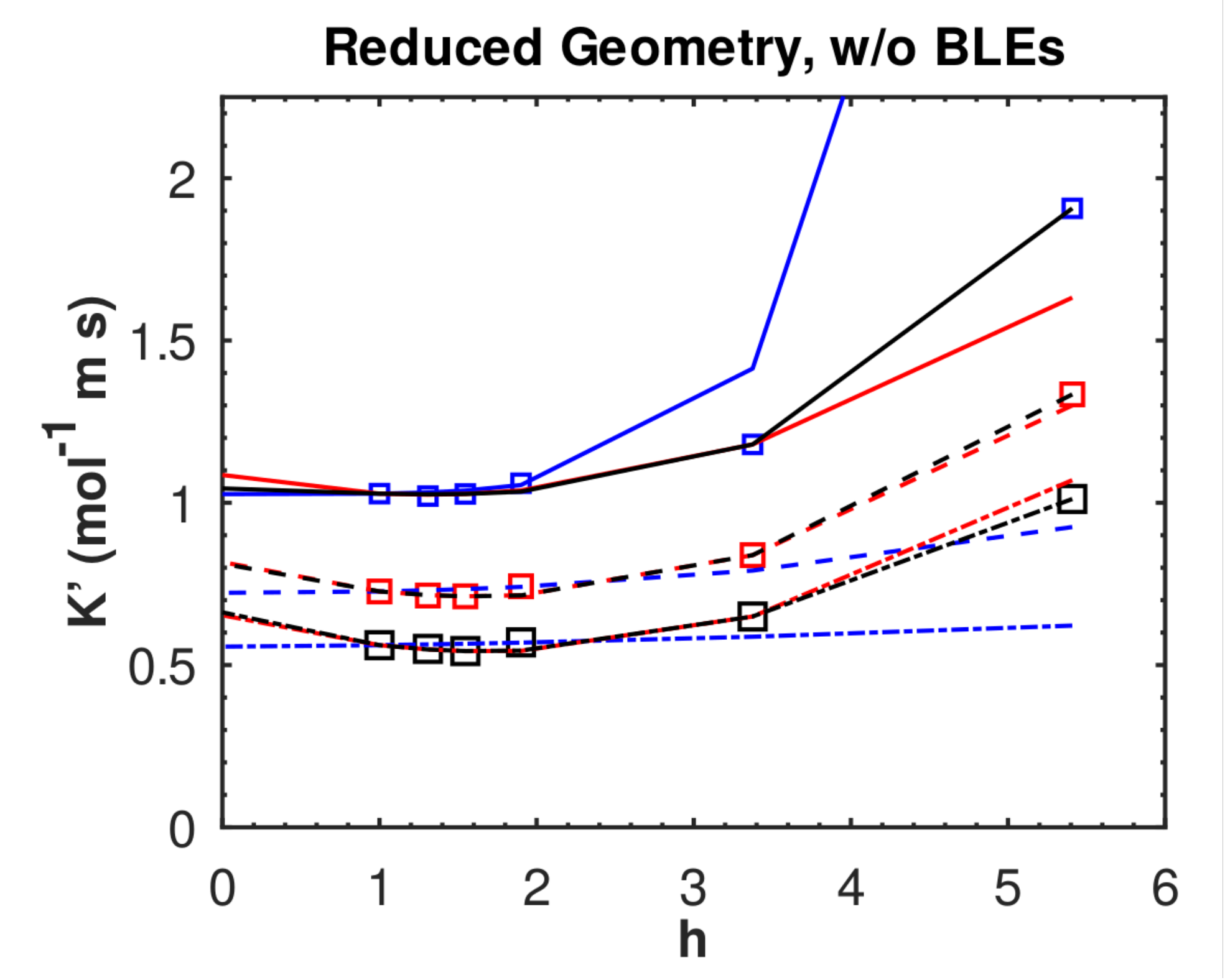}
        \caption{}
        \label{fig:lil_K_ext2}
    \end{subfigure}

    \begin{subfigure}{0.45\textwidth}
        \includegraphics[width=\textwidth]{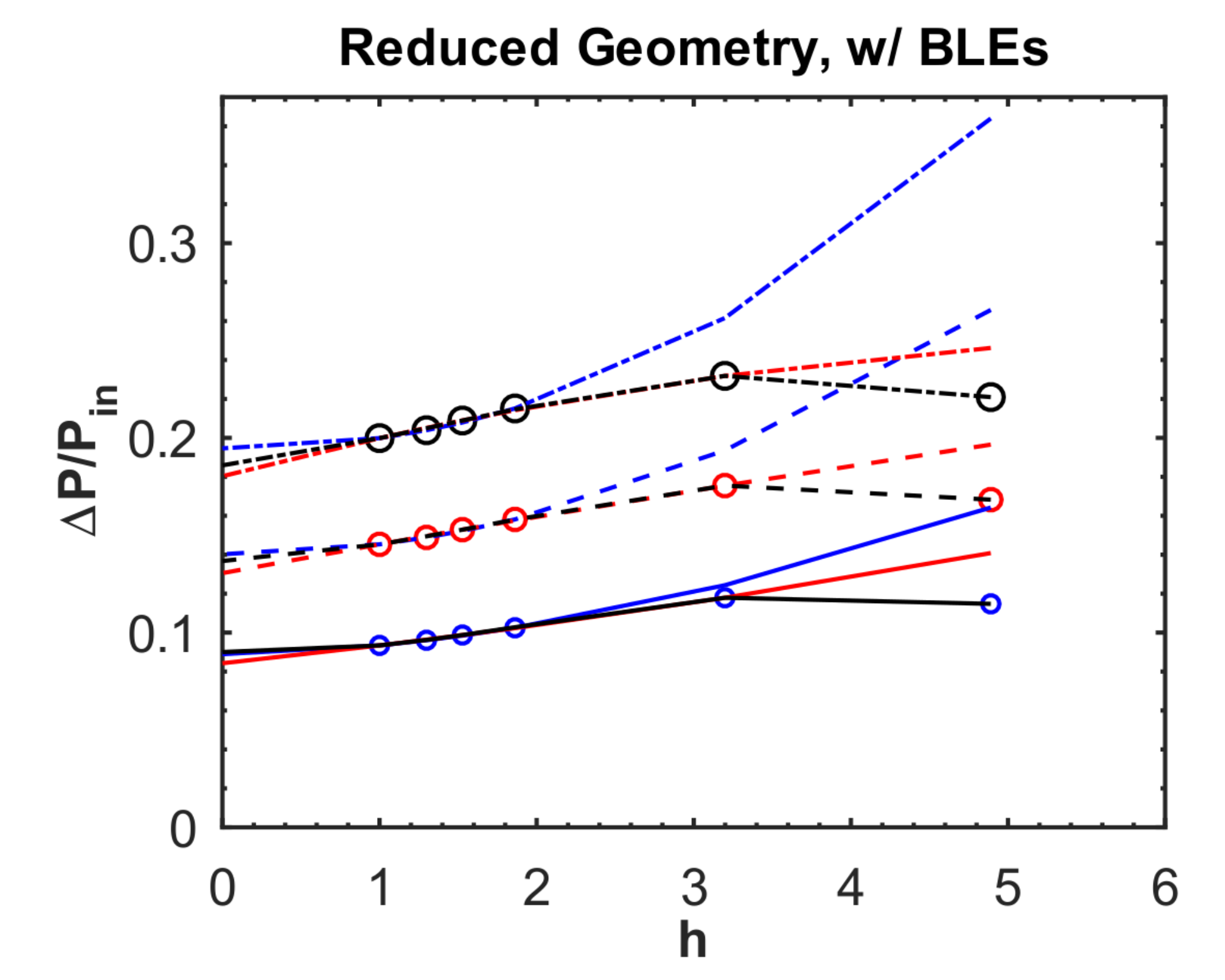}
        \caption{}
        \label{fig:lil_Npd_ext1}
    \end{subfigure}
    ~
    \begin{subfigure}{0.45\textwidth}
        \includegraphics[width=\textwidth]{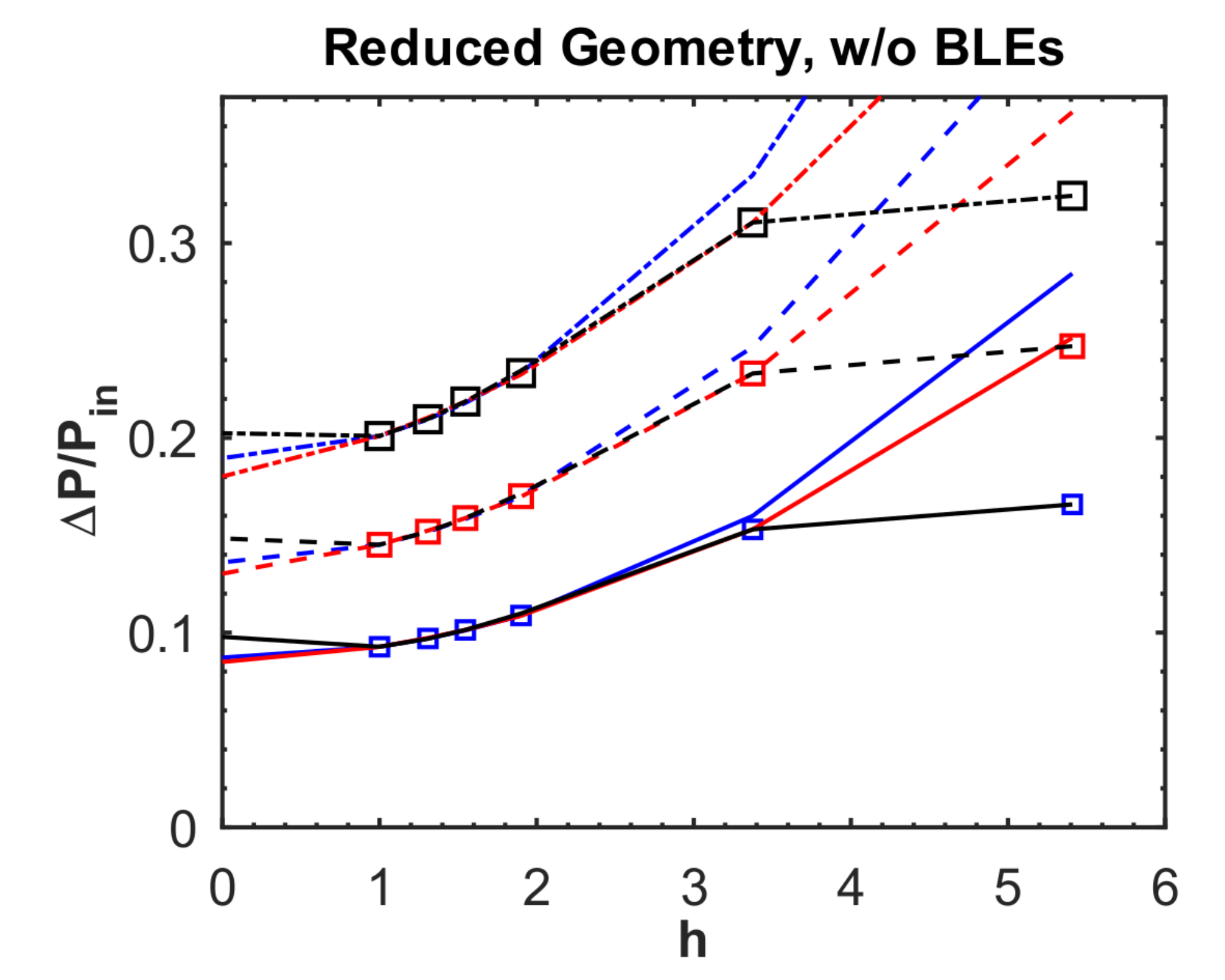}
        \caption{}
        \label{fig:lil_Npd_ext2}
    \end{subfigure}
    \caption{
        Scalar response variables and their respective extrapolations as
        functions of the normalized effective element spacing, obtained at the
        reduced geometry: GRE (\textcolor{blue}{full line}), MOE-12
        (\textcolor{red}{dashed line}), and MOE-123 (dash-dot line)
        \textbf{(a)} $K'=\Delta\chi_{\mathrm{O_3}}/R'_{\mathrm{O_3}}$ for the
        BLEs ($\circ$) and \textbf{(b)} no-BLEs ($\Box$) mesh variants; and
        \textbf{(c)} $\Delta P/P_{\mathrm{in}}$ for the BLEs and \textbf{(d)}
        no-BLEs mesh variants. The following values of inlet flow rate $Q$ are
        shown, with marker size increasing with $Q$: \textcolor{blue}{$250$},
        \textcolor{red}{$350$} and $450\ \mathrm{cm^3\ min^{-1}}$.
    }
    \label{fig:lil_scalars_ext}
\end{figure}

   We start with the scalar response variables already discussed in Sections
\ref{ssec:mc_res_lil} and \ref{ssec:mc_res_full}. Figure
\ref{fig:lil_scalars_ext} shows the scalars as function of $h$, with the three
extrapolation schemes: GRE, 1st- and 2nd-order MOE (MOE-12), and 1st-, 2nd- and
3rd-order MOE (MOE-123). Concerning the $K'$ ratio, Figs. \ref{fig:lil_K_ext1}
and \ref{fig:lil_K_ext2} show that the MOEs show better agreement with the
numerical results, while the GRE clearly overestimates the convergence of the
variable, with very small differences between $h=1$ to $0$. In particular, the
MOE-123 shows the best fit to the entire $h$ range, which is expected given the
choice of meshes to calculate the coefficients and extrapolate. MOE-12 shows
good agreement, despite the poor description at high $h$, in particular for the
BLEs variant. However, when comparing MOE-123 to MOE-12 for the BLEs mesh
variant, it appears that the derivative between $h=1$ and $0$ is somewhat
overestimated, however that is based on the assumption that for $\lambda\geq
10$ ($h\approx 1.5$) the variable is in the asymptotic range.

    For the $\Delta P/P_{\mathrm{in}}$ ratio, Figs. \ref{fig:lil_Npd_ext1} and
\ref{fig:lil_Npd_ext2} show better agreements between the GRE and MOE schemes,
however it is still clear that the GRE greatly overestimates the convergence
of the variable. Like with the $K'$ ratio, both MOE schemes show good agreement
with the numerical results, with MOE-12 showing relatively large differences
for $h>3$. Likewise, MOE-123 once again predicts a seemingly odd behavior at
$h<1$, in particular for the no-BLEs variant. It is not inconceivable that for
$h<1$ a faster rate of convergence might be observed for $\Delta P/P
_{\mathrm{in}}$, as predicted by both GRE and MOE-123 when compared to MOE-12.
However given the slow rate observed for the numerical results available, and
the monotonic behavior at $h < 4$, it appears that the MOE-12 estimate is more
plausible, particularly for the no-BLEs variant.

    From Fig. \ref{fig:lil_scalars_ext} it is concluded that the MOE-12 scheme
applied to the no-BLEs mesh variant provides the best agreement between
extrapolation scheme and numerical results available. In addition, it was
already shown that the results of both BLEs and no-BLEs mesh variants converge
at low $h$ (Fig. \ref{fig:lil_scalars}). Thus, given the smaller, albeit
slightly, computational cost of using the no-BLEs mesh variant, we now
establish it as the standard for further studies and to estimate discretization
error for the individual variables and the ratios analyzed so far. Estimation
of discretization errors for the BLEs mesh variant can be found in Section
SIV B of the SM\footnotemark[3] (Fig. S15 and Table S5).

\begin{figure}
    \centering
    \begin{subfigure}{0.45\textwidth}
        \includegraphics[width=\textwidth]{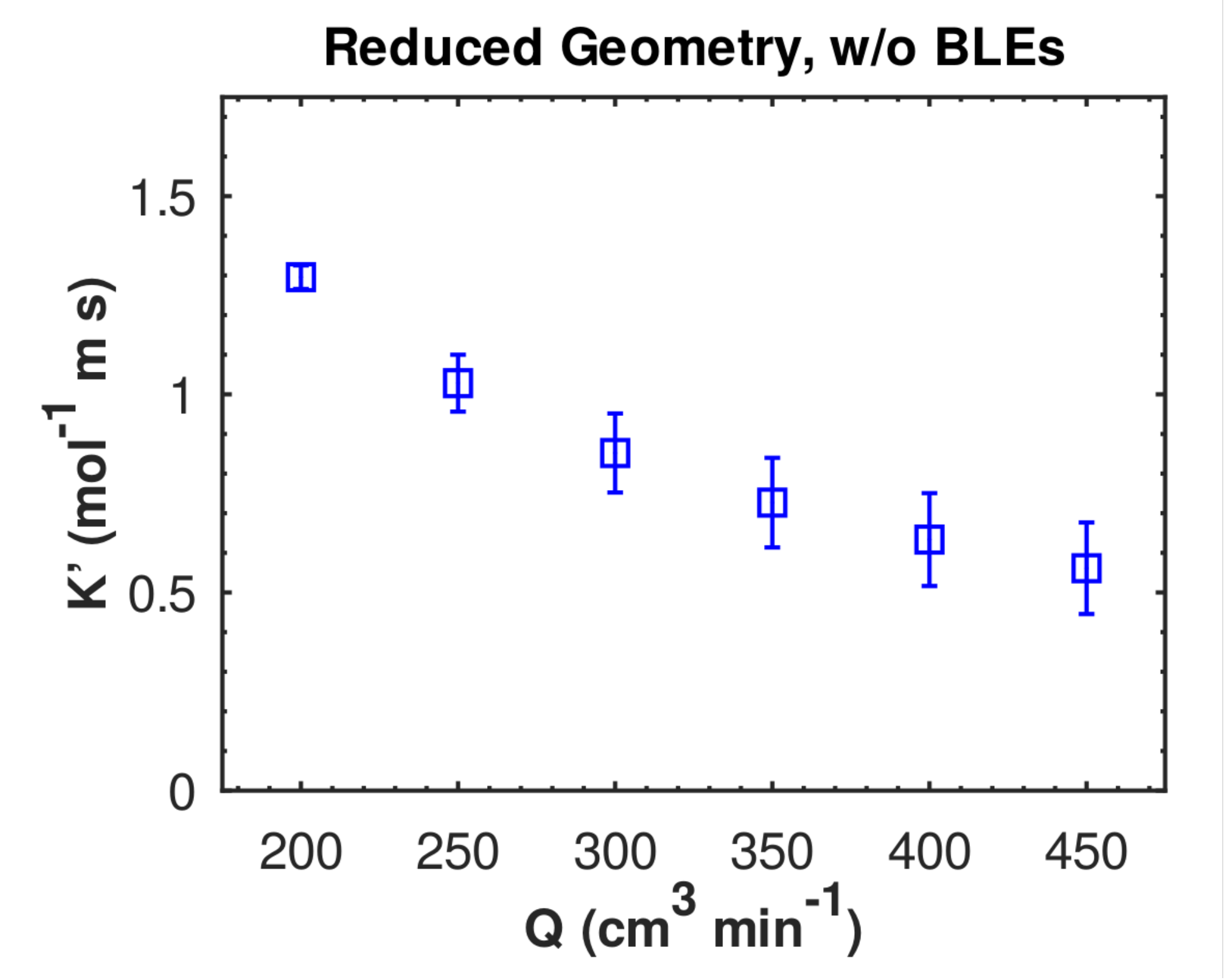}
        \caption{}
        \label{fig:lil_K_err}
    \end{subfigure}
    ~
    \begin{subfigure}{0.45\textwidth}
        \includegraphics[width=\textwidth]{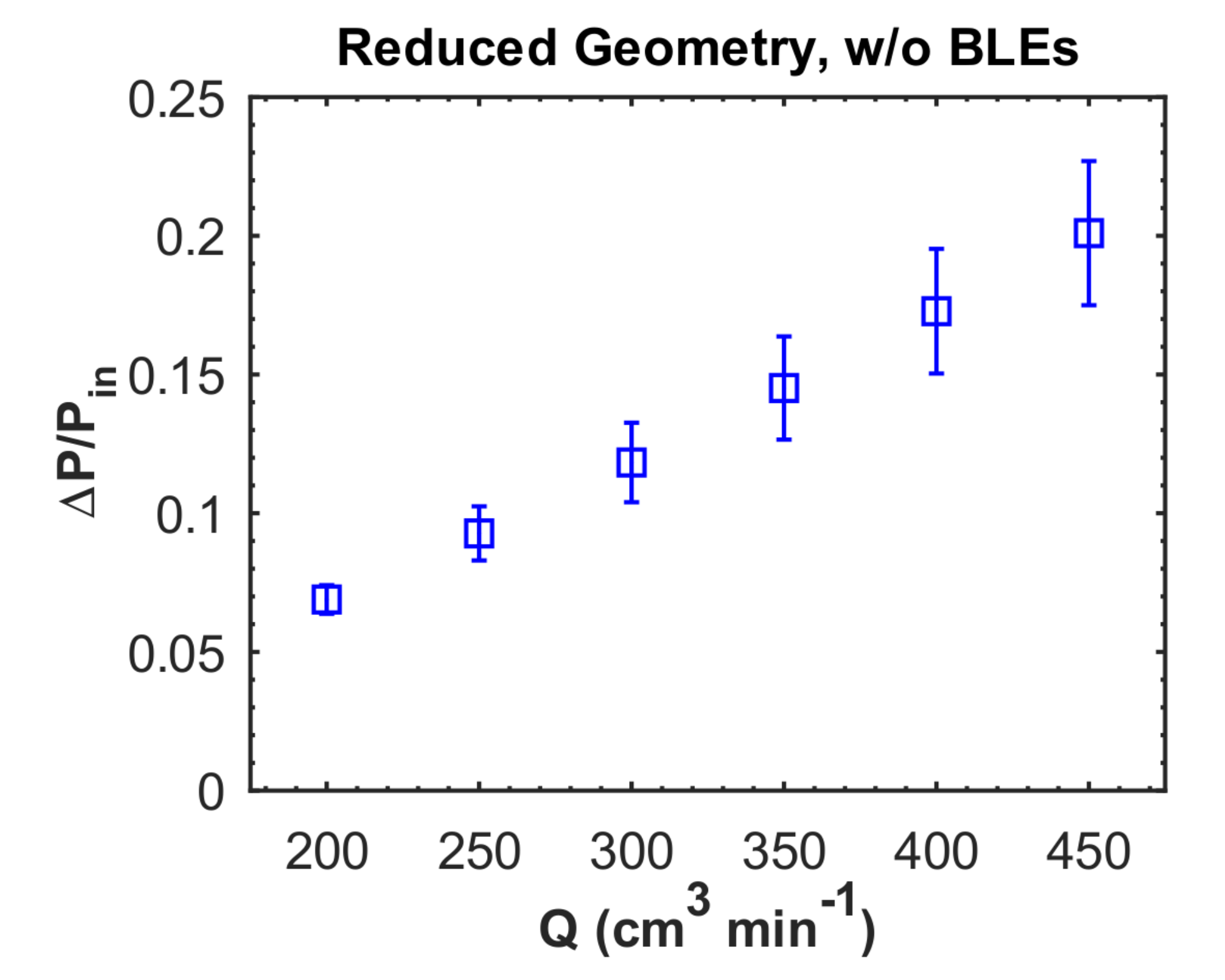}
        \caption{}
        \label{fig:lil_Npd_err}
    \end{subfigure}
    \caption{
        Scalar response variables as functions of the inlet flow rate, using
        the no-BLEs mesh variant with the reduced geometry: \textbf{(a)} $K'=
        \Delta\chi_{\mathrm{O_3}}/R'_{\mathrm{O_3}}$ and \textbf{(b)} $\Delta P
        /P_{\mathrm{in}}$. Error bars show the estimated discretization error
        as given by Eq. \ref{eq:meps}.
    }
    \label{fig:lil_scalars_error}
\end{figure}

    Carrying on, then, with the error estimate, Figure
\ref{fig:lil_scalars_error} shows the $K'$ and $\Delta P/P_{\mathrm{in}}$
ratios as function of $Q$, with the estimated error bars, as given by Eq.
\ref{eq:meps}. As it can be seen, the estimated error interval increases with
$Q$ for both variables, as already observed from Fig. \ref{fig:lil_scalars_ext}.
The estimated relative errors span the following intervals, with the respective
averages:
\begin{eqnarray}
    2.25\%\leq\%\epsilon_{\mathrm{M}}\left[K'\right]\leq 17.7\%, &\ \%\bar{\epsilon}_{\mathrm{M}} = 11.2\% \nonumber\\
    7.98\%\leq\%\epsilon_{\mathrm{M}}\left[\Delta P/P_{\mathrm{in}}\right]\leq 14.5\%, &\ \%\bar{\epsilon}_{\mathrm{M}} = 12.7\% \nonumber
\end{eqnarray}

\begin{figure}
    \centering
    \begin{subfigure}{0.45\textwidth}
        \includegraphics[width=\textwidth]{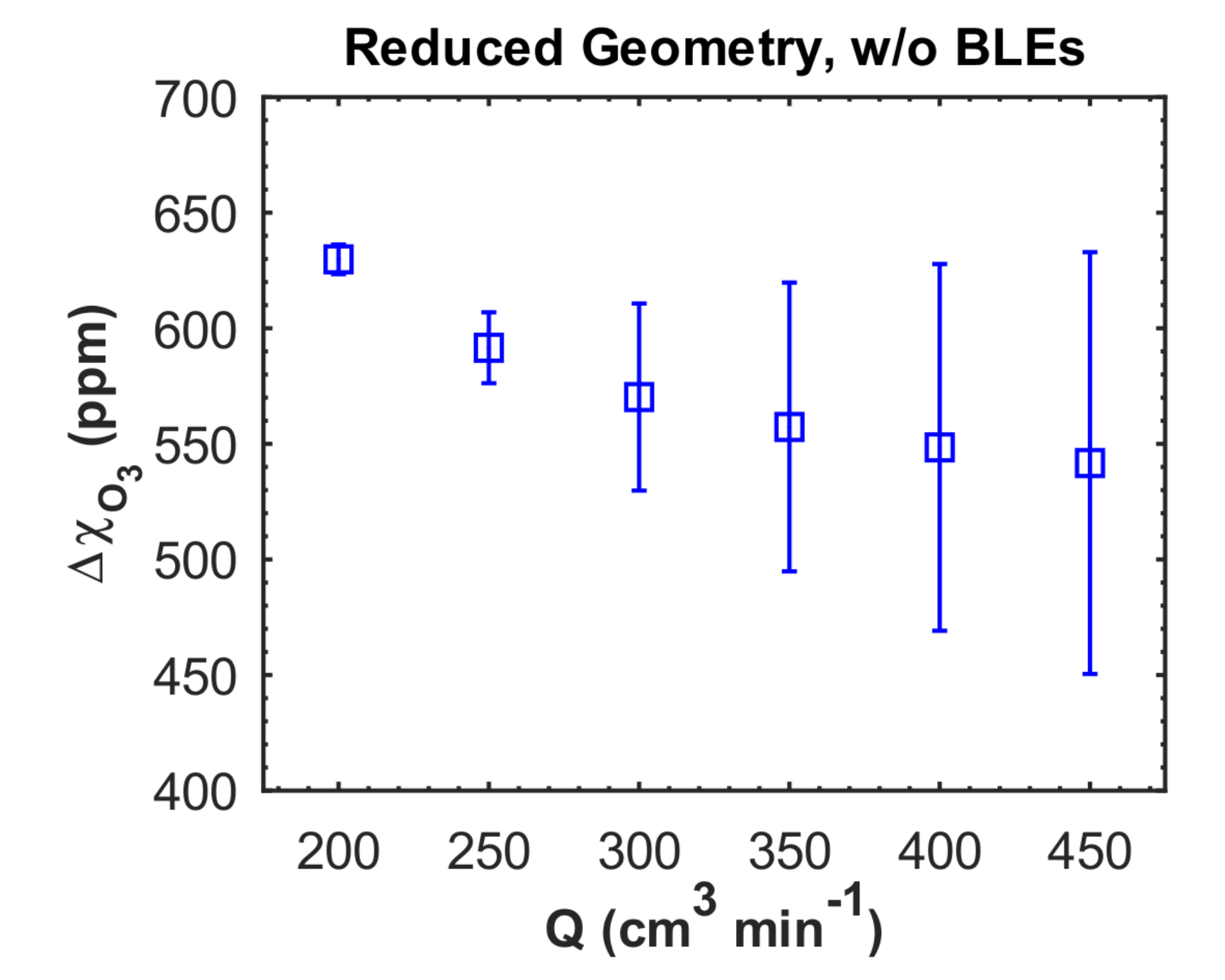}
        \caption{}
        \label{fig:lil_dchi_err}
    \end{subfigure}
    ~
    \begin{subfigure}{0.45\textwidth}
        \includegraphics[width=\textwidth]{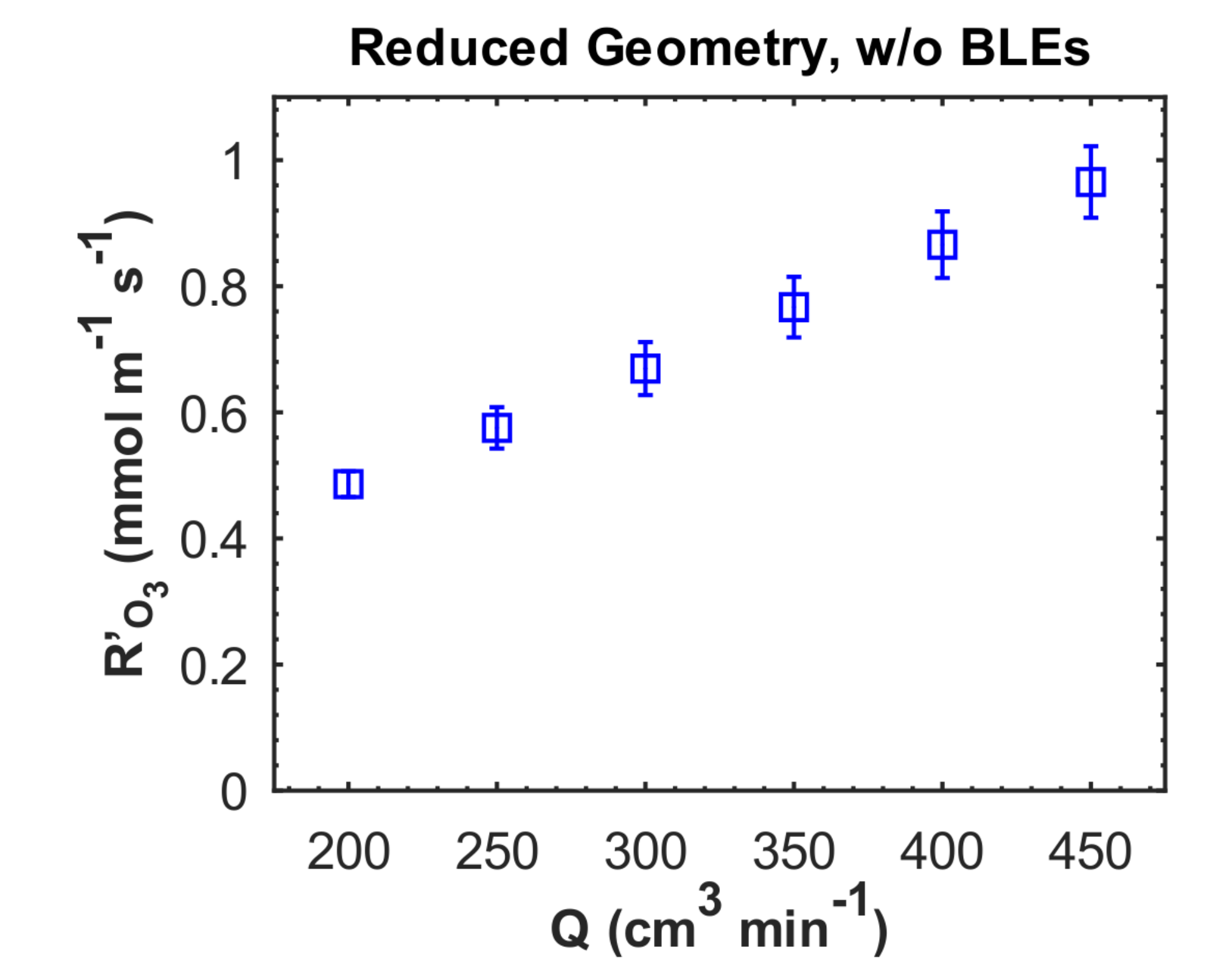}
        \caption{}
        \label{fig:lil_Rapp_err}
    \end{subfigure}

    \begin{subfigure}{0.45\textwidth}
        \includegraphics[width=\textwidth]{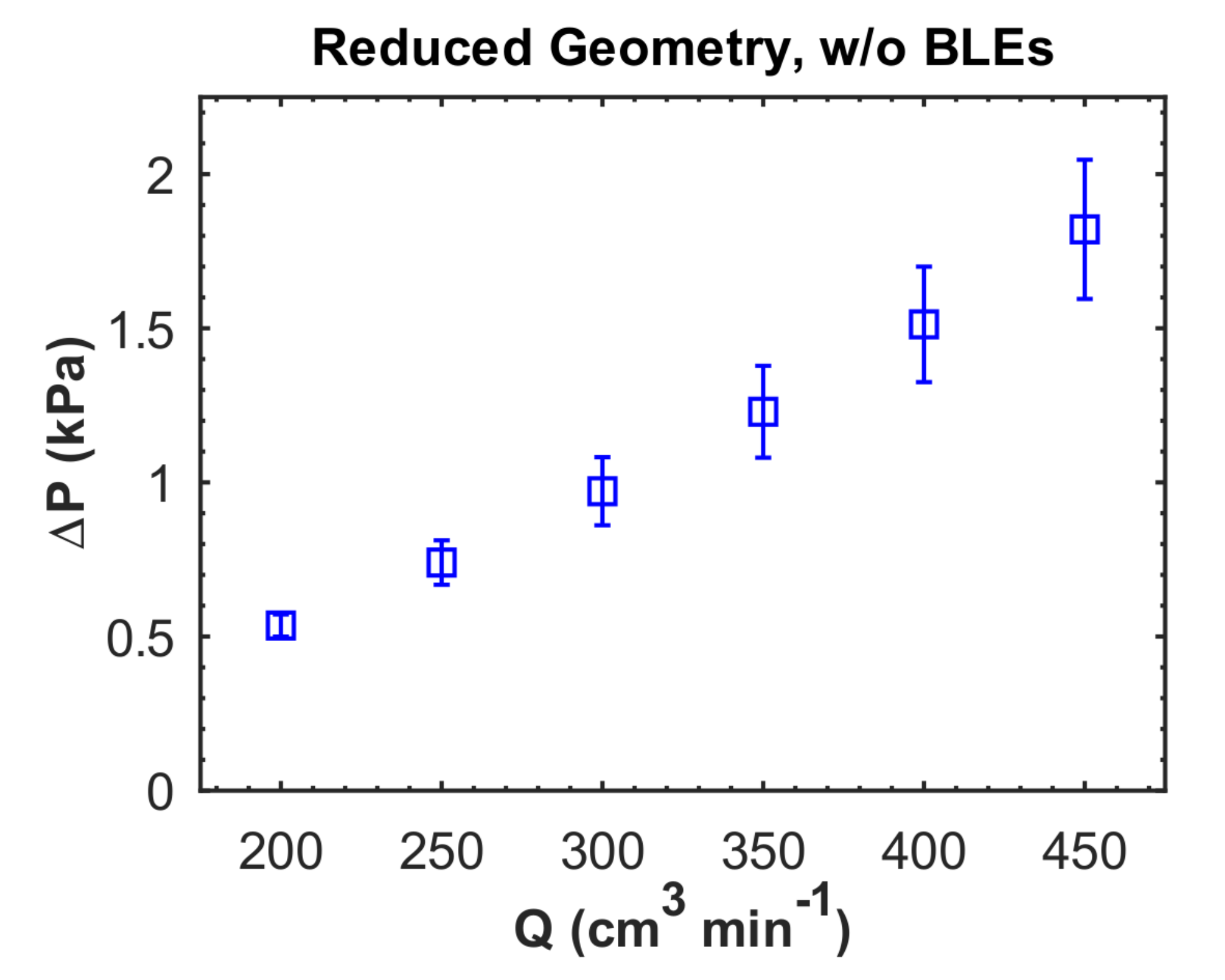}
        \caption{}
        \label{fig:lil_pd_err}
    \end{subfigure}
    \caption{
        Scalar response variables as functions of the inlet flow rate, using
        the no-BLEs mesh variant with the reduced geometry: \textbf{(a)}
        $\Delta\chi_{\mathrm{O_3}}$, \textbf{(b)} $R'_{\mathrm{O_3}}$, and
        \textbf{(c)} $\Delta P$. Error bars show the estimated discretization
        error as given by Eq. \ref{eq:meps}.
    }
    \label{fig:lil_scalars_error2}
\end{figure}

    For the sake of completeness, and also due to experimental interest, we
also briefly report the error estimate for the individual variables. Figure
\ref{fig:lil_scalars_error2} shows the estimates for $\Delta\chi_{\mathrm{O
_3}}$, $R'_{\mathrm{O_3}}$, and $\Delta P$, as function of $Q$. It appears at
once that the error estimates for $\Delta\chi_{\mathrm{O_3}}$ (Fig.
\ref{fig:lil_dchi_err}) are rather large. However that is only because $\Delta
\chi_{\mathrm{O_3}}$ spans a relatively small interval for the range of $Q$
simulated. The calculated relative error spans the interval
\begin{eqnarray}
    1.03\%\leq\%\epsilon_{\mathrm{M}}\left[\Delta\chi_{\mathrm{O_3}}\right]\leq 14.8\%, &\ \%\bar{\epsilon}_{\mathrm{M}} = 8.06\%, \nonumber
\end{eqnarray}
\noindent
smaller than for the $K'$ ratio. The remaining error estimates are the
following:
\begin{eqnarray}
    4.33\%\leq\%\epsilon_{\mathrm{M}}\left[R'_{\mathrm{O_3}}\right]\leq 6.61\%, &\ \%\bar{\epsilon}_{\mathrm{M}} = 6.01\% \nonumber \\
    7.11\%\leq\%\epsilon_{\mathrm{M}}\left[\Delta P\right]\leq 13.8\%, &\ \%\bar{\epsilon}_{\mathrm{M}} = 11.9\% \nonumber
\end{eqnarray}
\noindent
While the magnitude of the error estimates might be one order too high, errors
of $\sim 10\%$ are not absurd, in particular considering the lower values
estimated for the directly comparable scalar variables, $\Delta\chi
_{\mathrm{O_3}}$, $R'_{\mathrm{O_3}}$, and $\Delta P$. Two points should be
noticed, however. First, it is once again pointed out that the goal of this
work is, in part, to estimate the discretization error of the model under
discussion, not to prove a point regarding its accuracy. Mathematical framework
aside, these errors are in principle improvable with additional computational
resources.

    Secondly, these are the estimates for the reduced geometry, which
allows a significantly finer mesh and, given the behavior seen in Fig.
\ref{fig:lil_scalars_ext}, should present smaller error estimates than the full
geometry. Estimating the errors for the full geometry is not straightforward,
however, given what is now known of the behavior of the scalar variables as
functions of $h$ and the effect different mesh schemes have on the
extrapolates. An option is to use the estimates for the reduced geometry: given
the estimate of the exact value using the MOE-12 scheme, one may estimate the
error at a given $h$ for the reduced geometry using Eq. \ref{eq:meps}, and thus
for the full geometry. Scaling between the reduced and full geometries might
be necessary, however, as the scalar ratios used to track convergence are not
entirely independent of the geometry. Fig. S16 shows that, for a good
approximation, the $K'$ and $\Delta P/P_{\mathrm{in}}$ ratios for the full
geometry are linear functions of the ones for the reduced geometry. In
particular, as can be expected from Fig. \ref{fig:big_scalars}, the $K'$ ratio
is, to a very good approximation, related by $f_{\mathrm{full}} = f
_{\mathrm{reduced}} + C$, where in this case $C \approx 0.028$ (red line in
Fig. S16). Thus, the error estimates obtained for the reduced geometry, at
$\lambda = 12$, are good estimates for the full geometry. For $\Delta P/P
_{\mathrm{in}}$, $f_{\mathrm{full}} \approx 0.030 + 1.39f_{\mathrm{reduced}}$,
thus a scaling coefficient of $1.39$ should provide an estimate for the full
geometry based on the reduced one, albeit not a rigorous one. Similar
procedures were taken for the individual variables, were it was found that
$\Delta\chi_{\mathrm{O_3}}$ shows a scaling coefficient of $0.96$, while for
both $R'_{\mathrm{O_3}}$ and $\Delta P$ it is $1.72$ (See Section SV of the
SM\footnotemark[3]).

\begin{figure}
    \centering
    \begin{subfigure}{0.45\textwidth}
        \includegraphics[width=\textwidth]{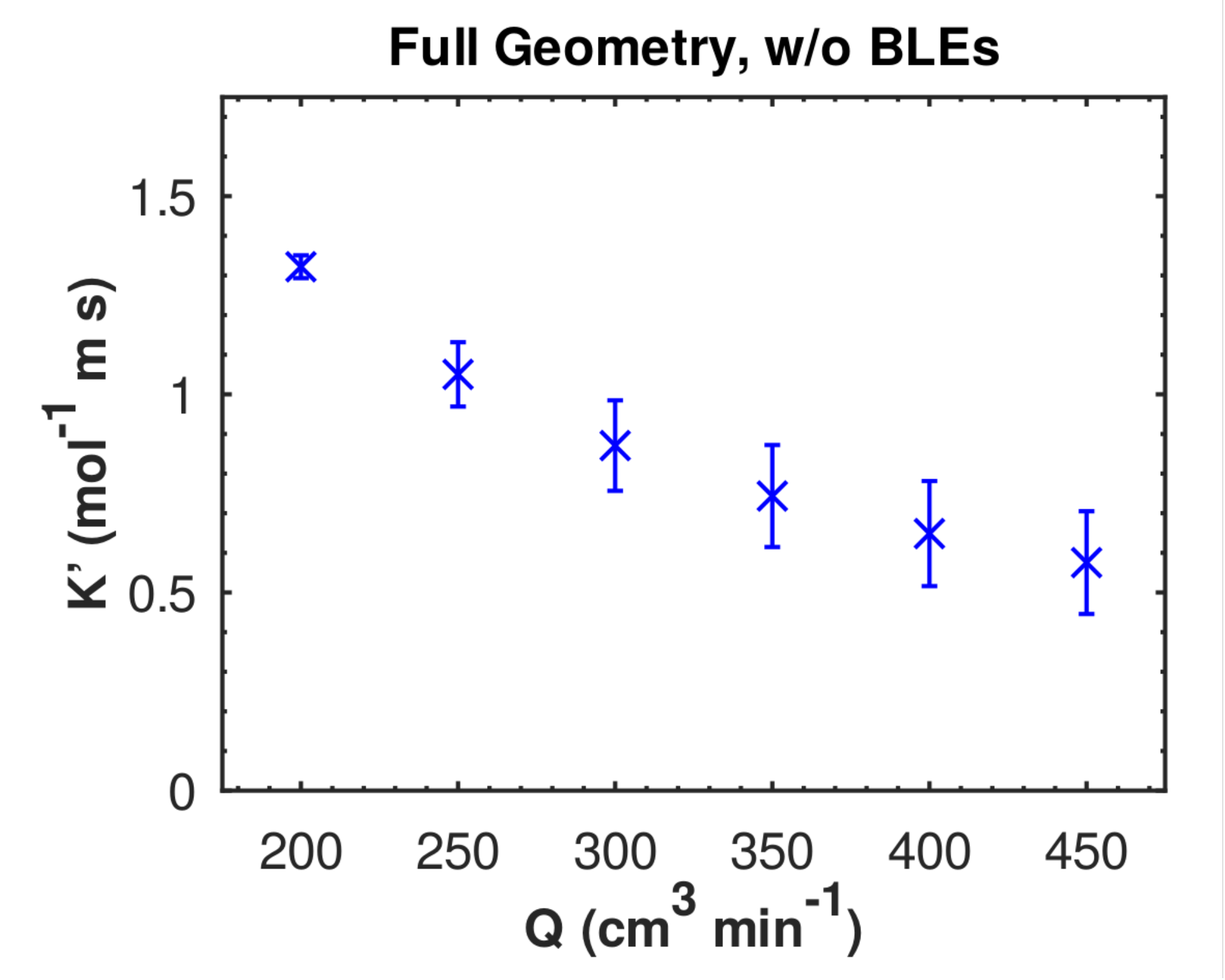}
        \caption{}
        \label{fig:big_K_err}
    \end{subfigure}
    ~
    \begin{subfigure}{0.45\textwidth}
        \includegraphics[width=\textwidth]{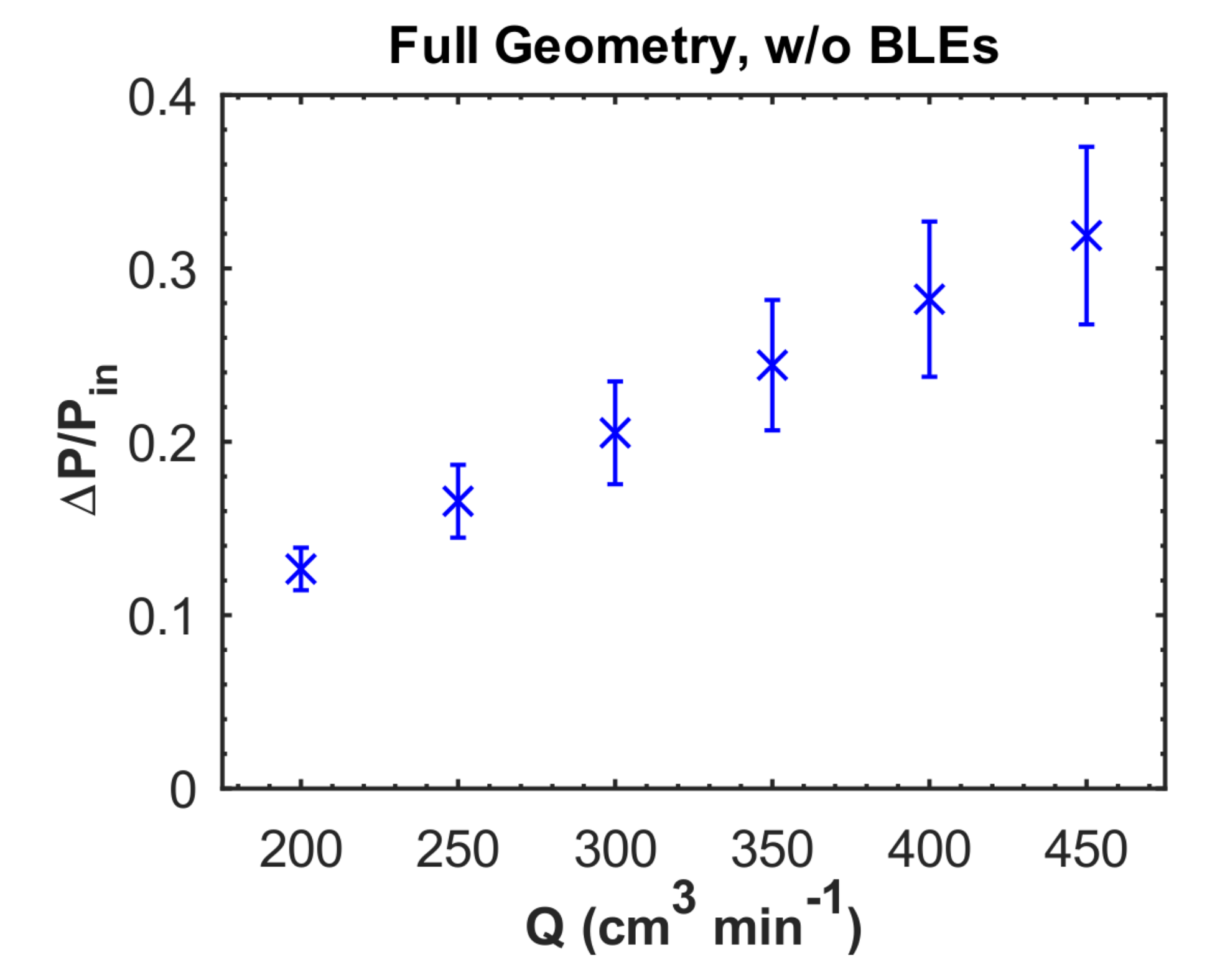}
        \caption{}
        \label{fig:big_Npd_err}
    \end{subfigure}
    \caption{
        Scalar response variables as function of the inlet flow rate, using the
        no-BLEs mesh variant with the full  geometry: \textbf{(a)} $K'=\Delta
        \chi_{\mathrm{O_3}}/R'_{\mathrm{O_3}}$ and \textbf{(b)} $\Delta P/P
        _{\mathrm{in}}$. Error bars show the estimated discretization error as
        given by Eq. \ref{eq:meps}.
    }
    \label{fig:big_scalars_error}
\end{figure}

\begin{figure}
    \centering
    \begin{subfigure}{0.45\textwidth}
        \includegraphics[width=\textwidth]{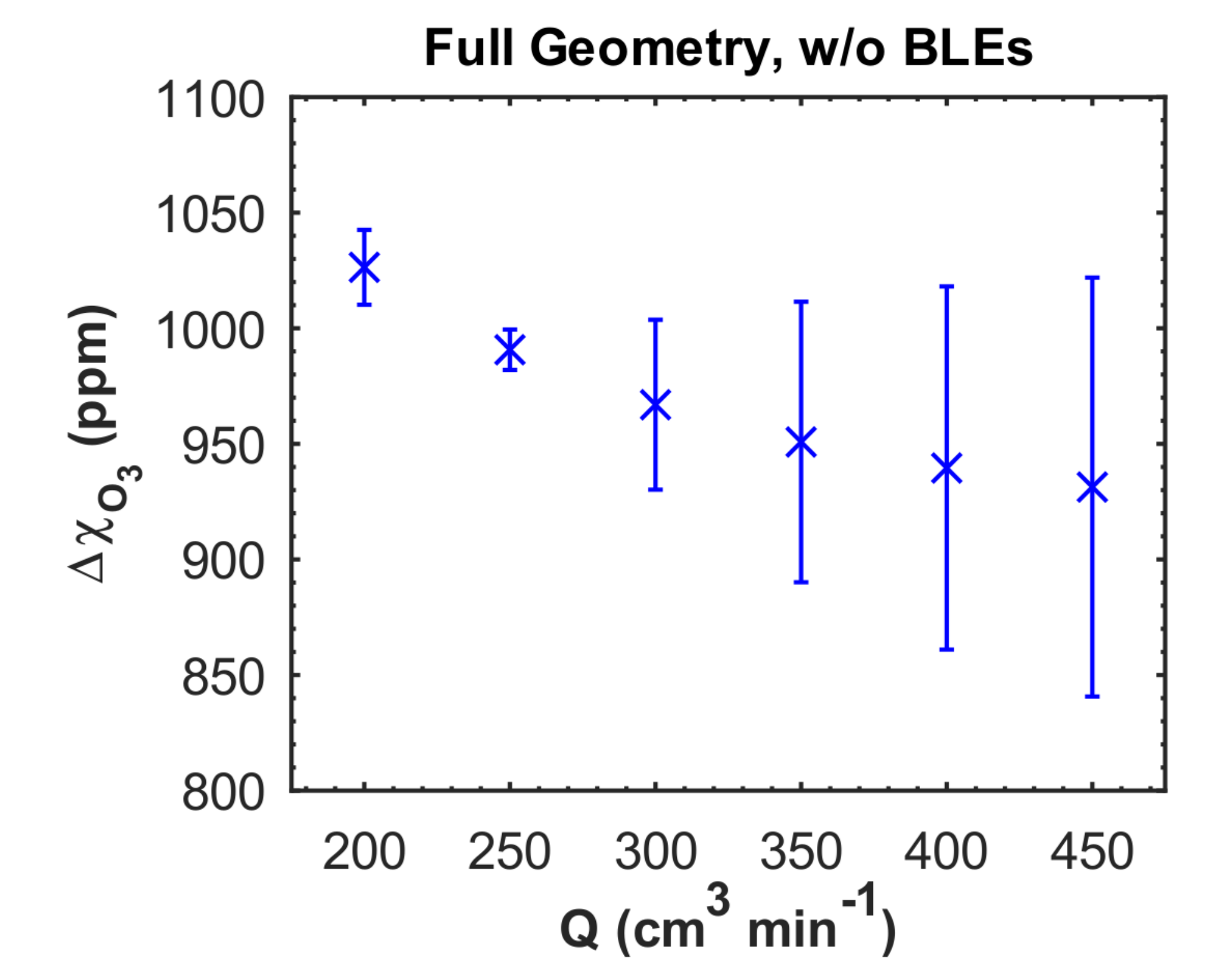}
        \caption{}
        \label{fig:big_dchi_err}
    \end{subfigure}
    ~
    \begin{subfigure}{0.45\textwidth}
        \includegraphics[width=\textwidth]{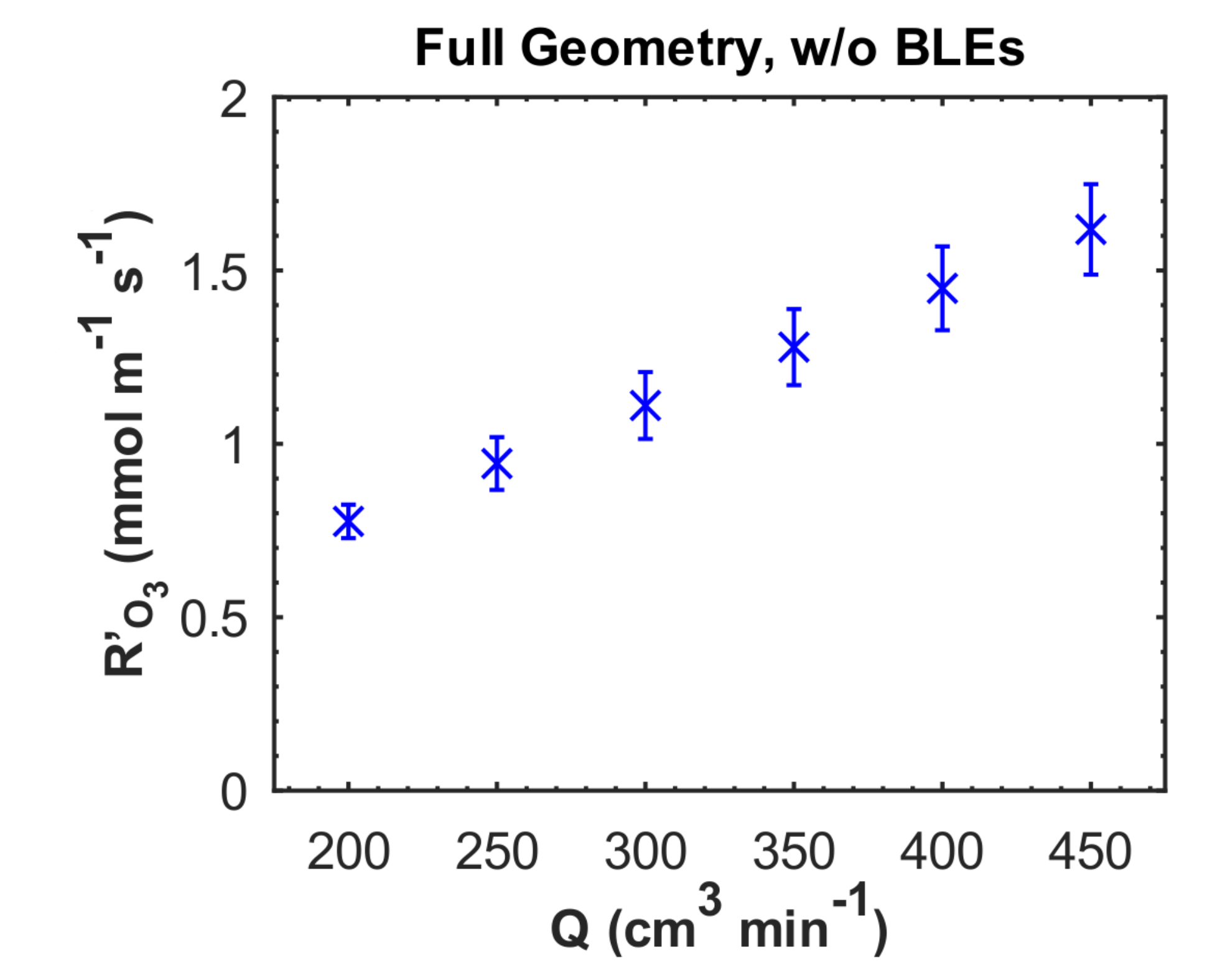}
        \caption{}
        \label{fig:big_Rapp_err}
    \end{subfigure}

    \begin{subfigure}{0.45\textwidth}
        \includegraphics[width=\textwidth]{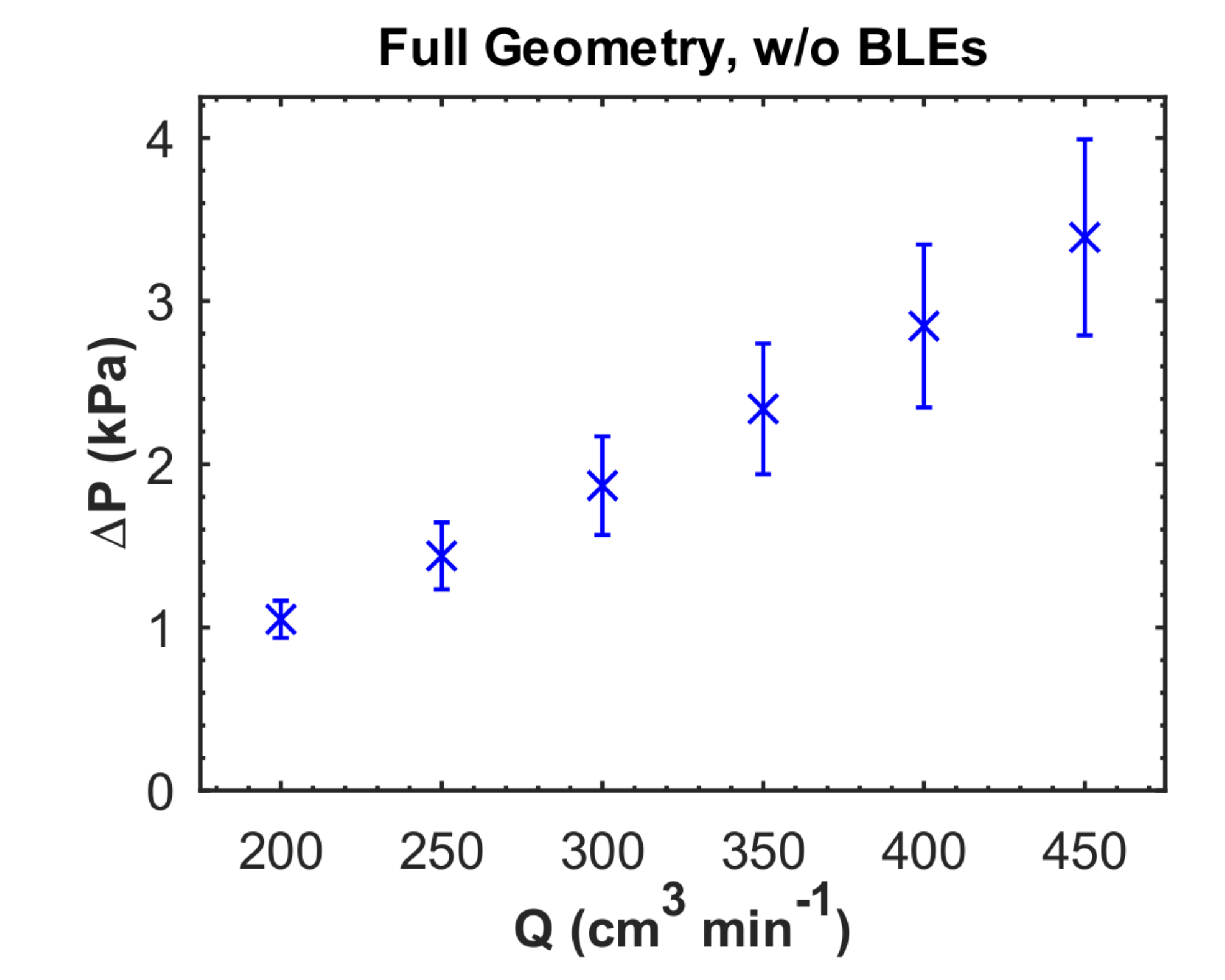}
        \caption{}
        \label{fig:big_pd_err}
    \end{subfigure}
    \caption{
        Scalar response variables as function of the inlet flow rate, using the
        no-BLEs mesh variant with the full geometry: \textbf{(a)} $\Delta\chi
        _{\mathrm{O_3}}$, \textbf{(b)} $R'_{\mathrm{O_3}}$, and \textbf{(c)}
        $\Delta P$. Shown are the numerical results (\textcolor{blue}{$\times$})
        and available experimental data\cite{beruski17} (\textcolor{red}
        {$\Box$}). Error bars show the estimated discretization error as
        given by Eq. \ref{eq:meps}, for the numerical results, and one standard
        deviation for the experimental data.
    }
    \label{fig:big_scalars_error2}
\end{figure}

    With these points in mind, Figures \ref{fig:big_scalars_error} and
\ref{fig:big_scalars_error2} present the scalar variables of interest,
respectively the ratios and individually, for the full geometry using the
no-BLEs mesh variant. In Fig. \ref{fig:big_scalars_error} it can be seen that
the error estimates for the ratios are quite similar for both geometries,
albeit higher. This is expected given how the estimate was taken and the small
differences already seen between meshes with $\lambda = 12$ and $16$ (Fig.
\ref{fig:lil_scalars}). The error estimates span the intervals:
\begin{eqnarray}
    2.14\%\leq\%\epsilon_{\mathrm{M}}\left[K'\right]\leq19.0\%, &\ \%\bar{\epsilon}_{\mathrm{M}} = 12.2\% \nonumber \\
    10.2\%\leq\%\epsilon_{\mathrm{M}}\left[\Delta P/P_{\mathrm{in}}\right]\leq 18.2\%, &\ \%\bar{\epsilon}_{\mathrm{M}} = 15.9\% \nonumber
\end{eqnarray}

    For the individual variables, Fig. \ref{fig:big_scalars_error2} also shows
the available experimental data, previously reported in \cite{beruski17}. Again
the pattern follows closely that of the reduced geometry, with slightly higher
error estimates. The calculated error estimates span the following intervals:
\begin{eqnarray}
    0.878\%\leq\%\epsilon_{\mathrm{M}}\left[\Delta\chi_{\mathrm{O_3}}\right]\leq 9.01\%, &\ \%\bar{\epsilon}_{\mathrm{M}} = 4.85\% \nonumber \\
    6.49\%\leq\%\epsilon_{\mathrm{M}}\left[R'_{\mathrm{O_3}}\right]\leq 9.37\%, &\ \%\bar{\epsilon}_{\mathrm{M}} = 8.54\% \nonumber \\
    11.7\%\leq\%\epsilon_{\mathrm{M}}\left[\Delta P\right]\leq 20.6\%. &\ \%\bar{\epsilon}_{\mathrm{M}} = 17.9\% \nonumber
\end{eqnarray}
\noindent
with increased error estimates for $R'_{\mathrm{O_3}}$ and $\Delta P$, in line
with what was presented above, and a slight reduction in maximum and average
error estimates for $\Delta\chi_{\mathrm{O_3}}$. This might be related to the
higher absolute values of $\Delta\chi_{\mathrm{O_3}}$, for the full geometry,
while using a scaling similar to the reduced geometry when solving the model.
Despite such fact, Fig. \ref{fig:big_dchi_err} still shows that, compared to
the available experimental data, the absolute error is much greater,
approximately $10$ times for $Q \geq 300\ \mathrm{cm^3\ min^{-1}}$. For $\Delta
P$, the situation is somewhat better, with errors showing at least the same
order of magnitude. Nevertheless, it is clear that the meshes being used have
great need for improvement, given the precision already available to the
experimental device, and the expected increase with future developments.

    Additional response variables of interest, concerning the context of Part
1, can be found in Section SVI of the SM\footnotemark[3]. As mentioned in
Section \ref{sssec:mc}, the $\mathrm{O_3}$ inlet concentration, the normalized
$R'_{\mathrm{O_3}}$ and $K'$ ratios, and the real and apparent stoichiometries
had their discretization errors estimated following the approach described in
Section \ref{sssec:mc} and above. Thus, Figs. S17 through S19 show their
dependency on $h'$, the linearity between full and reduced geometries, and the
variables as function of $Q$ with their estimated errors.

    We conclude this grid convergence analysis of the PEFC prototype model by
summarizing the results shown and discussed above. We reiterate the need to
assess the uncertainty in a numerical model, be it low or not. The reduced and
full geometries used to estimate the discretization error of the model show
that the it has great need of mesh improvement, since the estimated relative
errors, even for the best meshes, are $\sim 10\%$ and above. However, the grid
convergence analysis shows that the experimentally relevant variables show
relatively slow convergence, particularly $\Delta P$, and non-monotonic
behavior that depends not only on the element size, but also on the mesh
scheme. Thus, in addition to finer meshes and additional computational
resources, it could be helpful to search for alternatives that might contribute
to faster grid convergence, such as structured grids, which might facilitate
convergence analysis, but also higher-order interpolation schemes, that,
while with increased computational costs, might allow improved error estimates.

\section{\label{sec:sum}Summary}

    Work is presented concerning progress on numerical modeling of a prototype
PEFC\cite{lopes15}, building on previously published work\cite{beruski17}.
Known limitations of the original model, Alpha, were addressed in order to
increase fidelity with the current understanding of fuel cell devices,
resulting in the Beta model. In part I of this work, both mathematical
frameworks are compared, and approximations and potential shortcomings are
discussed (see Part 1). In this part, a grid convergence study has been carried
out in order to provide an estimate to the discretization error intrinsic to
the updated model.

    Grid convergence study was carried out with two mesh schemes, differing
only by the existence of so-called ``boundary layer'' elements (BLEs). Four
response variables were tracked in this study, viz. the ratio $K'$ between the
drop in reactant molar fraction and the apparent reaction rate, $\Delta\chi
_{\mathrm{O_3}}/R'_{\mathrm{O_3}}$; the ratio between pressure drop and inlet
pressure, $\Delta P/P_{\mathrm{in}}$; flow speed profiles, $U$; and reactant
partial pressure profiles, $P_{\mathrm{O_3}}$. Convergence was assessed
qualitatively using plots of the variables as a function of an effective,
normalized element spacing $h$, and quantitatively by estimating the
discretization error through continuum extrapolates using two different
extrapolation schemes\cite{roache97,roy03}.

    A qualitative assessment of the grid convergence study showed that the use
of BLEs leads to non-monotonic behavior in some variables, notably the
$\Delta P/P_{\mathrm{in}}$ ratio, while the $K'$ ratio shows non-monotonic
behavior for both mesh schemes. Despite such differences at large $h$ (i.e.
coarse grids), the mesh schemes converged at low $h$, suggesting the results
were truly approaching mesh independence. In this way, it was established that
the no-BLEs mesh scheme was to be used for further assessments and in future
studies, and that the finest mesh available was to be used. Prior to the
quantitative assessment, analysis suggested that the so-called Mixed 1st and
2nd order extrapolation scheme\cite{roy03} (MOE-12) provided the best estimate
for the continuum extrapolate, possibly due to the non-monotonic behavior shown
by the scalar response variables analyzed. Thus, using MOE-12 on a proxy
geometry (fewer channel turns), the estimates of the discretization error, in
the inlet flow rate range of $200\leq Q\leq 450\ \mathrm{cm^3\ min^{-1}}$,
returned average errors of $13.8\%$ for the $K'$ ratio (range of $2.18\%$ to
$22.4\%$) and of $14.0\%$ for the $\Delta P/P_{\mathrm{in}}$ ratio (range of
$9.67\%$ to $16.1\%$).

    The grid convergence study also allowed the corroboration of previously
established points\cite{beruski17}, such as the importance of the turns of the
flow channel (Ch) domain on the global flow field, as well as the contribution
of convective transport in total molar flux to, and in the porous media (Pm)
domains. It was found that secondary flows in the corners of the Ch domain are
quite mesh-sensitive, particularly the ones caused by boundary-layer separation
(BLS). At low $h$, BLS produces a pronounced effect on the flow field along the
straight sections of the Ch domain, deforming the $U$ profile up to the point
that, at $Q = 450\ \mathrm{cm^3\ min^{-1}}$, the profile is asymmetric with
respect to wall distance into the next corner, i.e. it appears that the flow
does not fully develops along the straight section. Concerning reactant molar
flux, integrated $z$-component of the molar flux at the Ch-MPS (macroporous
substrate) and MPS-CL (catalyst layer) boundaries corroborate a significant
contribution of convection to reactant influx to the Pm domains, while
average molar flux magnitude inside the Pm domains shows dominance of
convective over diffusive transport. Increased mesh resolution ($h\rightarrow
0$) was seen to reduce convective contribution, while simultaneously
increasing diffusive transport, although both remain at similar magnitude.

    Overall it must be pointed out that the model is not considered to have
converged with respect to the discretization. Relative errors of order $\sim
10\%$ in the numerical data are larger than the experimental ones available,
which makes actual model validation impossible at the moment. Adding other
common sources of uncertainty, such as in parameterization, it is clear that
there is a need for denser meshes and schemes with faster convergence, as well
as improved methods for error estimation, in the fuel cell modeling community.
Given the non-monotonic behavior of some variables of experimental relevance,
this work calls for increased caution when comparing numerical to experimental
data and when making predictions. On the other hand, uncertainty is only
problematic if not acknowledged, and we are of the opinion that a major part
of modeling is knowing its limitations and uncertainties. In this way, the use
of proxy models and simpler systems, as exemplified here, may be particularly
important to advance on the fundamental understanding of complex devices such
as fuel cells.

\begin{acknowledgments}
O.B. acknowledges the Fundação de Apoio à Universidade de São Paulo, FUSP,
grant \#2968. I.K and T.L. acknowledge the Fundação de Ampara à Pesquisa do
Estado de São Paulo, FAPESP, grants \#2016/12397-0, and \#2014/22130-6 and
\#2017/15304-6, respectively. Authors also acknowledge the Research Centre for
Gas Innovation, RCGI, sponsored by FAPESP grant \#2014/50279-4 and Shell
Brasil.\footnotemark[4]
\end{acknowledgments}

\footnotetext[4]{Author contributions: O.B. designed and conducted the
research, analyzed the data and wrote the paper. I.K. contributed with
discussions and by revising the paper. T.L. and F.C.F. contributed by revising
the paper and with funding sources.}

\bibliographystyle{unsrt}
\bibliography{refs}